\documentclass[timesnewroman,12pt, twoside]{report}
\usepackage[titletoc]{appendix}
\usepackage[a4paper]{geometry}
\usepackage{authblk, lineno, color, lscape, url, sectsty}
\usepackage[autostyle]{csquotes}
\usepackage{amsmath}
\usepackage{mathrsfs, wasysym} 
\usepackage{slashed}
\usepackage{tikz}
\usepackage{dsfont}
\usepackage{enumitem}
\usepackage{graphicx}
\usepackage[dvipsnames]{xcolor}
\usepackage[labelfont=bf]{caption}
\usepackage{emptypage}
\usepackage{array, float, multirow, graphicx, setspace, cite, url, titlesec, tocloft}
\usepackage{amsmath, amsfonts, mathtools, notoccite}
\usepackage{bm}
\usepackage{chngcntr, rotating, pdflscape, enumitem, algpseudocode, algorithm, subcaption}
\usepackage[square,sort,comma,numbers,sort&compress]{natbib}
\usepackage[colorlinks=True,citecolor=blue,linkcolor=blue,anchorcolor=blue,filecolor=blue,urlcolor=blue]{hyperref}
\allowdisplaybreaks
\usepackage{tensor}
\usepackage{braket}
\usepackage{latexsym}
\usepackage{color}
\usepackage[dvipsnames,svgnames,table]{xcolor}
\usepackage{graphicx}
\usepackage{epsfig}  
\usepackage{epsf}   
\usepackage{dcolumn}
\usepackage{bm}
\usepackage{amssymb}
\usepackage{textcomp}
\usepackage{float}
\usepackage{hypcap}
\usepackage[]{hyperref}
\usepackage{booktabs}
\usepackage{multirow}
\usepackage{diagbox}
\usepackage{alphalph}
\usepackage{textcomp}
\usepackage{epstopdf}
\usepackage{soul} 
\usepackage[normalem]{ulem}
\usepackage{booktabs}
\usepackage{lipsum}
\usepackage[Sonny]{fncychap}
\def\blankpage{%
	\clearpage%
	\thispagestyle{empty}%
	\addtocounter{page}{-1}%
	\null%
	\clearpage}
\begin{document}
\pagenumbering{roman}
\thispagestyle{empty}
\thispagestyle{empty}
\graphicspath{{Figures/PNG/}{Figures/}}
\begin{center}
{\bf {\Large Precision measurements of 2-3 oscillation parameters in the next-generation long-baseline experiments} }
\end{center}
\vspace{0.3 cm}
\begin{center}
    {\it \large By}
\end{center}
\begin{center}
    {\bf {\Large Ritam Kundu } \\ PHYS07201804011}
\end{center}
\begin{center}
\bf {{\large Institute of Physics, Bhubaneswar, India }}
\end{center}
\vskip 2.0 cm
\begin{center}
\large{
{ A thesis submitted to the }  \\
 {Board of Studies in Physical Sciences }\\

In partial fulfillment of requirements \\
For the Degree of } \\
{\bf  DOCTOR OF PHILOSOPHY} \\
\emph{of} \\
{\bf HOMI BHABHA NATIONAL INSTITUTE}
\vskip 1.0 cm
\begin{figure}[H]
	\begin{center}
    \includegraphics[width=4.0cm, height= 4.0cm]{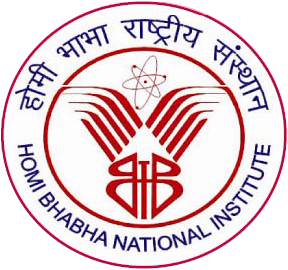}
	\end{center}
\end{figure}
\vskip 0.2 cm
{\bf \today}
\end{center}
\thispagestyle{empty}
\blankpage     
\begin{center}
{\Large\textbf{Homi Bhabha National Institute \\}}
{\large\textbf{\ \\Recommendations of the Viva Voce Committee}}
\end{center}
As members of the Viva Voce Committee, we certify that we have read the
dissertation prepared by \textbf{Ritam Kundu} entitled \textbf{``Precision measurements of 2-3 oscillation parameters in the next-generation long-baseline experiments"}, and recommend that it may be accepted as fulfilling the dissertation requirement for the award of Degree of Doctor of Philosophy.
\begin{center}
\begin{tabular}{p{0.74\linewidth}p{0.19\linewidth}}
\hline
Chairman - \textbf{Prof. Suresh Kumar Patra} & \textbf{Date:07.10.2025} \\[10pt]
 & \\ \hline
Guide / Convener - \textbf{Prof. Sanjib Kumar Agarwalla} & \textbf{Date:07.10.2025}\\[10pt]
 & \\ \hline
Examiner - \textbf{Prof. Manibrata Sen}  & \textbf{Date:07.10.2025}\\ [8pt]
 & \\ \hline
Member 1 - \textbf{Prof. Aruna Kumar Nayak}& \textbf{Date:07.10.2025}\\[10pt]
 & \\ \hline
Member 2 - \textbf{Prof. Kirtiman Ghosh}  & \textbf{Date:07.10.2025}\\[10pt]
 & \\ \hline
 Member 2 - \textbf{Prof. Sanjay Kumar Swain} & \textbf{Date:07.10.2025}\\[12pt]
\end{tabular}
\end{center}
\rule{\linewidth}{0.5pt}

\noindent Final approval and acceptance of this thesis is contingent upon the candidate's submission of the final copies of the thesis to HBNI.\\
\noindent I/We hereby certify that I/We have read this thesis prepared under my direction and recommend that it may be accepted as fulfilling the thesis requirement.\\
\vspace{-2mm} 
\begin{flushleft}
	\begin{tabular}{p{0.185\linewidth}p{0.26\linewidth}>{\centering\arraybackslash}p{0.46\linewidth}}
		\makebox[0.185\linewidth][l]{\textbf{Date:} 07.10.2025} & & \\
		\makebox[0.185\linewidth][l]{\textbf{Place:} Bhubaneswar} & & \\
	\end{tabular}
\end{flushleft}

\vspace*{-1.0cm} 
\hfill
\begin{minipage}[t]{0.4\linewidth}
	\begin{tabular}{@{}l@{}}
		\textbf{Prof. Sanjib Kumar Agarwalla} \\
		\textbf{Guide} 
	\end{tabular}
\end{minipage}

\newpage

\begin{center}
\large \bf
STATEMENT BY AUTHOR
\end{center}
This dissertation has been submitted in partial fulfillment of the requirements
for an advanced degree at Homi Bhabha National Institute (HBNI) and deposited
in the Library to be made available to borrowers under the rules of the HBNI.
\paragraph{}
Brief quotations from this dissertation are allowed without special
permission, provided that accurate acknowledgment of the source is made.
Requests for permission for extended  quotation from or reproduction
of this manuscript in whole or in part may be granted by the competent
Authority of HBNI when in his or her judgment the proposed use of the
material is in the interests of scholarship. In all other instances,
however, permission must be obtained from the author.
\vspace{1.2cm}
\begin{flushright}
	\textbf{(Ritam Kundu)}
\end{flushright}
\newpage
\begin{center}
\large \bf
DECLARATION
\end{center}
I, {\bf Ritam Kundu}, hereby declare that the investigations
presented in the thesis have been carried out by me. The matter
embodied in the thesis is original and has not been submitted earlier
as a whole or in part for a degree/diploma at this or any other
Institution/University.
\vspace{1.2cm}
\begin{flushright}
	\textbf{(Ritam Kundu)}
\end{flushright}
\newpage
     
\begin{center}
\textbf{\large \underline{List of Publications for the thesis}}
\end{center}
\textbf{\large \underline{{Journal}}}\\
\textbf{\underline{Published}}
\begin{enumerate}
\item \textit{A close look on 2-3 mixing angle with DUNE in light of current neutrino oscillation data} \\
Sanjib Kumar Agarwalla, \textbf{Ritam Kundu}, Suprabh Prakash, Masoom Singh,\\
\href{https://link.springer.com/article/10.1007/JHEP03(2022)206}{\textit{Journal of High Energy Physics 03 (2022) 206,}} \href{https://arxiv.org/pdf/2111.11748}{arXiv: 2111.11748 [hep-ph].}

\item  \textit{Improved precision on 2-3 oscillation parameters using the synergy between DUNE and T2HK}\\
Sanjib Kumar Agarwalla, \textbf{Ritam Kundu}, Masoom Singh\\
\href{https://link.springer.com/article/10.1007/JHEP10(2024)243}{\textit{Journal of High Energy Physics 10 (2024) 243,}} \href{https://arxiv.org/pdf/2408.12735}{arXiv: 2408.12735 [hep-ph].}
\end{enumerate}
\textbf{\underline{Work in progress}}
\begin{enumerate}
\item \textit{Measuring Neutrino oscillation parameters in long-baseline experiments in the presence of flavor-dependent long range neutrino interactions}\\ 
Sanjib Kumar Agarwalla, \textbf{Ritam Kundu}, Masoom Singh \\
\textcolor{blue}{Preprint: IOP/BBSR/2024-08.}\\ 
\end{enumerate}  
\textbf{\large\underline{Conference Proceedings}}
\begin{enumerate}
	
\item \textit{Neutrino Oscillation Parameters: Present and Future}\\Sanjib Kumar Agarwalla, \textbf{Ritam Kundu}, Masoom Singh\\\href{https://pos.sissa.it/462/022}{Proceedings of Science (HQL 2023) 022, Conference Proceeding at 16th International Conference on Heavy Quarks and Leptons, TIFR, Mumbai, November 28 - December 2, 2023.}
	
\item \textit{Establishing non-maximal 2-3 mixing with DUNE in light of current neutrino oscillation data}\\Sanjib Kumar Agarwalla, \textbf{Ritam Kundu}, Suprabh Prakash, Masoom Singh\\\href{https://zenodo.org/records/6784883}{Zenodo, Conference Proceeding at 30th International Symposium on Lepton Photon Interactions at High Energies, Manchester, USA, January 10-14, 2022}.
 
\item \textit\textit{Utmost Precision on 2-3 Oscillation Parameters Using DUNE}\\
Sanjib Kumar Agarwalla, \textbf{Ritam Kundu}, Suprabh Prakash, Masoom Singh.\\
\href{https://link.springer.com/chapter/10.1007/978-981-97-0289-3_327}{\textcolor{blue}{XXV DAE-BRNS HEP Symposium; IISER Mohali; December 12-16, 2022.}}
\end{enumerate}
\textbf{\large\underline{Talk/Poster}}
\begin{enumerate}

\item {\bf Talk:} \textcolor{blue}{\textit{Bridging Experiments, Narrowing Uncertainties: When DUNE Meets Hyper-K to unveil insights into 2–3 Oscillation Sector}}\\
EPS HEP 2025 - The 42nd European Physical Society Conference on High Energy Physics; July 7-11, 2025; Marseille, France (In Person).
\vspace{0.5cm}	
\item {\bf Poster:} \textcolor{blue}{\it Precision Measurement of Neutrino Oscillation Parameters Exploiting the Complementarity between DUNE and T2HK}\\
1st DAE Conclave; NISER, Bhubaneswar; October 22-26, 2024.
\vspace{0.5cm}
\item {\bf Talk:} \textcolor{blue}{\textit{High precision measurements of oscillation parameters exploiting the complementarity between DUNE and T2HK}}\\
NuFact 2024 - The 25th International Workshop on Neutrinos from Accelerators; September 16-21 2024; Argonne National Laboratory (Online).
\vspace{0.5cm}	
\item {\bf Talk:} \textcolor{blue}{\textit{Ultimate Precision on 2-3 Oscillation Parameters using the Synergy between DUNE and T2HK}}\\
Neutrino Workshop at IFIRSE; Quy Nhon, Vietnam; July 19, 2023 (Online).
\vspace{0.5cm}	
\item {\bf Poster:} \textcolor{blue}{\textit{A close look on 2-3 mixing angle with DUNE in light of current neutrino oscillation data}}\\
FRONTIERS IN PARTICLE PHYSICS 2023; IISC Bangalore; March 10-12, 2023.

\vspace{0.5cm}
\item {\bf Poster:} \textcolor{blue}{\textit{Utmost Precision on 2-3 Oscillation Parameters Using DUNE}}\\
XXV DAE-BRNS HEP Symposium; IISER Mohali; December 12-16, 2022.
\vspace{0.5cm}
\item {\bf Poster:} \textcolor{blue}{\textit{Precision in Atmospheric Oscillation Parameters and Octant Resolution of $\theta_{23}$ through DUNE's eye}}\\XXX International conference on Neutrino Physics and Astrophysics Neutrino 2022, (Online); Seoul, South Korea; May 30 - June 4, 2022.
\vspace{0.5cm}
\item {\bf Talk:} \textcolor{blue}{\textit{Establishing non-maximal 2-3 mixing with DUNE in light of current neutrino oscillation data}}\\
30th International Symposium on Lepton Photon Interactions at High Energies; Manchester, USA; January 10-14, 2022 (Online).

\end{enumerate}
\vspace{0.5cm}
\begin{flushright}
	\textbf{(Ritam Kundu)}
\end{flushright}
\blankpage
\begin{center}
	{\large\textbf{DEDICATIONS}}
\end{center}
\vspace*{3.0in}
\hspace*{2.0in}
\begin{center}
        {\Large \emph{Dedicated To my teacher Mr. Surojit Dhole and my beloved friend Mr. Abhishek Nandy}\\}
\hrule
\end{center}
    
\blankpage  
\begin{center}
\large \bf
ACKNOWLEDGMENTS
\end{center}
\vspace{1.5cm}
I am feeling really happy to get an opportunity to acknowledge people in my journey through the whole academic life till now. At first, I want to express my respect and gratitude to all Indian soldiers and all taxpayers of India, for whom I could conduct research safely. I am offering my pranam to the Almighty for always staying beside me. I want my cordial respect to my Ph.D. supervisor, Prof. Sanjib Kumar Agarwalla, for holding my hand in the path of harvesting knowledge and giving me the opportunity to conduct research while I faced a very harsh situation in the early days of my Ph.D. \\
I want to bid my gratitude to my doctoral committee members: Prof. Suresh Kumar Patra, Prof. Aruna Kumar Nayak, Prof. Kirtiman Ghosh, and Prof. Sanjay Kumar Swain. The active participation of Prof. Nayak in my annual review seminar and the outstanding teaching of the Particle Physics course of Prof. Ghosh has enriched my knowledge by removing several misconceptions. I want to acknowledge Prof. Debottam Das and Prof. Arijit Saha for providing me the continuous support (both morally and academically) during my Ph.D. tenure. I want to thank Prof. Manimala Mitra for her teaching in QFT course in my pre-doctoral session.\\
I want to thank each of my group members and collaborators, but some members' contributions played exceptional roles in my Ph.D. journey. I want to offer my respect and gratitude to Dr. Masoom Singh. Without her help, I could not start my Ph.D. journey. At every step of my research work, the inspirations offered by Dr. Singh and Dr. Anil Kumar cannot be limited to words. I want to thank my collaborator, Dr. Suprabh Prakash, for giving remarkable inputs during our work. I want to thank Dr. Sudipta Das (Sudipta da) not only for helping me academically but also for showering selfless affection on me as an elder brother since the first day of my Ph.D. life and for supporting me at this difficult time of my life. I want to thank Mr. Pragyanprasu Swain for helping me both academically and non-academically. I want to thank Dr. Ashish Narang, Gopal, Sadashiv, Anuj, Krishnamoorthy, Sharmishtha, and Dr. Mehedi Masud for spending a joyful time with me. I am acknowledging Dr. Akhila Kumar Pradhan for giving suggestions on my thesis. I want to acknowledge all of my batchmates: Arpan, Chitrak, Ayesha, Mousam, Harish, Siddharth, Abhishek, Samir, Sandhya, and Pritam da. I am really grateful to the cordial, generous, and immediate help of Mr. Makrand Siddhabhatti (even beyond the office hours and at the weekends and holidays) and the entire team controlling the High-Performance Computing (HPC) facility of IOP, SAMKHYA, and the library facility. I am honored by getting the affiliation of the Department of Atomic Energy (DAE) and Homi Bhabha National Institute (HBNI), Mumbai. I am thankful to all the trees, mango garden, dogs, birds, and the serenity of the IOP campus.\\
I want to thank my psychologist, Dr. Jayita Saha, for helping me handle the challenging situation during the last 6 years. I feel lucky to have some true friends like Abhishek Nandy, Soumyadipta Ray, Soumallya Banerjee, Sayan Jana, and Uma Das who always stood beside me. I want to thank Dr. Songshaptak De, Dr. Ramita Sarkar, Dr. Antara Dey, Dr. Sudipta Moshat, and Mr. Sunish Kumar Deb for supporting me during the hard time of my Ph.D. I extend my heartfelt love to all my seniors and juniors: Mr. Abhishek Bag, Dr. Ganesh Paul, Dr. Bibhabasu De, Dr. Saiyad Ashanujjaman, Dr. Debasish Saha, Dipak, Moonsun, Subhadip, and Alok Kumar. Mr. Debabrata Dey's academic suggestions helped me immensely, and I want to exceptionally thank him. I want to show my respect to Prof. Amitava Raychowdhury, Prof. Anindya Datta, Prof. Anirban Kundu, Prof. Indrajit Mitra, Prof. Jayashree Saha, and Prof. Sobhan Sounda. I will always be grateful to Prof. Poonam Mehta (JNU) for standing beside me at the dark time of my Ph.D. and helping me both academically and morally. Lastly, I bid my pranam to my physics teachers, Mr. Surojit Dhole and Mr. Vedakrishna Dey, and our Headmaster, Swami Kalyaneshananda, whose inspirations are still motivating me in carrying out the research on Physics and hopefully will illuminate the future path of my life.

\vspace{1.0cm}
\begin{flushright}
	\textbf{(Ritam Kundu)}
\end{flushright}
\newpage

 \hypersetup{linkcolor=blue}
 \tableofcontents
 \blankpage 
\addcontentsline{toc}{chapter}{List of Figures}
\listoffigures
\blankpage 
\addcontentsline{toc}{chapter}{List of Tables}
\listoftables
\newpage
\setcounter{page}{1}
\pagenumbering{arabic}
\chapter{Introduction}
\label{C1} 
\graphicspath{{Figures/Chapter-1figs/PDF/}{Figures/Chapter-1figs/}}

The quest to know the mysteries of the Universe is an age-old hunt of mankind till the primitive stage of civilization. The primordial question, ``What do we consist of?" drives people to search from the prehistoric age to the present day. At a certain point of the discoveries by human beings, it wonders them that all the materialistic objects, even though different by their outside appearance, consist of the same kind of things. This notion drives them to bring the concept of symmetries in Nature. As time went on, particle physicists proposed the Standard Model (SM) by bringing all the elementary particles under the umbrella of \textit{SU}$(3)_C\,\otimes\,$\textit{SU}$(2)_L\,\otimes\,$\textit{U}$(1)_Y$ gauge group and incorporating the four main force carriers (gluons, $W^\pm$ boson, Z boson, and photon). The last recognized parameter of the SM was the mass of the Higgs Boson (2012). This well-established model, built since 1960 (with the hands of Glashow, Salam, and Weinberg by the discovery of electroweak theory \cite{Pati:1974yy, Weinberg:1967tq}) to 1964 (Postulation of spontaneous symmetry breaking invoking Higgs boson by Peter Higgs \cite{Higgs:1964ia, Higgs:1966ev}) upto 2012 (experimental detection of Higgs Boson in LHC \cite{ATLAS:2012yve}), could explain several natural phenomena successfully. \\

However, later on, the experimental results show some phenomena that do not obey the prescriptions of the SM. These motivate people to search for Beyond the Standard Model (BSM) physics. Along with the others, there was a special type of particle in the SM named neutrino \cite{Ramond:2019fsr}. According to the SM, its mass is zero. But, the non-zero mass of the neutrinos was the first experimental proof of the physics beyond the Standard Model. Neutrino Oscillation is such a phenomenon that leads us to the finite, non-zero mass of neutrinos. We start this chapter with some introduction to this amazing mysterious particle, neutrinos, from the earlier days of its discovery.
\section{Neutrino saga: from its birth story to till date}
The birth story of introducing the elusive particle, neutrinos, to the community can be started if we go back almost 100 years ago with the hand of Prof. Wolfgang Pauli \cite{Pauli:1930pc, Pauli:2000ak, Jarlskog:2019axp}. He first postulated neutrinos (1930) to explain the energy, momentum, and angular momentum conservation of beta decay spectrum (3-body decay). In the experimental results of beta decay, it was seen that the decay spectrum is continuous, which couldn't be explained by the theory of two-body beta decay. Pauli proposed that there should be a third particle as a decay product, contributing to energy conservation and angular momentum \cite{Pauli:1988dp}. It was a great alarm when early measurements of $\beta$ decay appeared to distort this pillar of scientific understanding, as the resultant electron was observed to have consistently less energy than the expected Q-value of the decay. The incorporation of this new candidate to the theory of beta decay was executed by the physicist Enrico Fermi (1934) to illustrate the continuous nature of the decay spectrum. Prof. Enrico Fermi named the new particle as ``neutrino", which means ``little neutral one" in Italian language. Pauli also mentioned that this new candidate is invisible and hard to detect.\\

In 1956, Clyde Cowan and Frederick Reines \cite{Cowan:1956rrn} first observed this new candidate experimentally in a nuclear reactor, producing neutrinos artificially with large intensity as a bi-product of nuclear fission. They captured this tiny candidate with a 10-ton detector.\\

After their detections, the query regarding the properties and types of neutrinos came to people's minds and motivated them to search for it. The neutrinos revealed by Cowan and Reines were electron-type neutrinos produced by the reactor experiments. In 1962, the existence of a second neutrino flavor, the muon neutrino ($\nu_\mu$), was confirmed by a team of physicists \cite{Danby:1962nd}. Utilizing a high-energy particle accelerator at Brookhaven National Laboratory, they generated a neutrino beam and observed interactions that provided evidence for the distinct identity of the muon neutrino. In 1975, the existence of third-generation neutrino ($\nu_\tau$) was first theoretically anticipated following the discovery of its charged lepton counterpart, the tau lepton, at the SLAC National Laboratory. Finally, tau neutrino ($\nu_\tau$) was experimentally observed in the DONUT (Direct Observation of Nu Tau) experiment \cite{DONUT:2000fbd}. After touching one of the milestones, people thought what else neutrino types remain in nature. In 1989, the LEP experiment \cite{ALEPH:2005ab} confirmed through the precise measurement of the decay width of Z-boson at the $e^+e^-$ collider that, there are only three fundamental types of active neutrinos in Nature ($i.e.$, $\nu_e$, $\nu_\mu$, and $\nu_\tau$) that take part in weak interaction with the matter (standard interaction). But, still then, many more interesting properties of neutrinos were awaiting to be revealed.\\

Neutrinos are omnipresent and interact with matter via weak interaction through the massive $W^\pm$ and $Z^0$ gauge bosons \cite{Watkins:1986va, Belusevic:1987cw, McFarland:2008xd}. It is the second most abundant particle in the Universe after photon. There are different types of neutrinos based on their sources and energies, $i.e.$, Relic Neutrinos ($\sim 10^{-4}$ to $10^{-6}$ eV) \cite{Lesgourgues:1999wu, Betts:2013uya, Eberle:2005uk}, Solar Neutrinos ($\sim$ sub-MeV to a few MeV) \cite{Bahcall:1964gx, Davis:1968cp, Bahcall:1995am}, Atmospheric Neutrinos (few hundred MeV to TeV) \cite{Kamiokande-II:1992hns, Super-Kamiokande:1998kpq, Super-Kamiokande:2004orf, Akhmedov:1993nn, Sinegovsky:2010dq}, Astrophysical Neutrinos (TeV to PeV and EeV range) \cite{Zatsepin:1966jv, Kamiokande-II:1987idp, Learned:2000sw, Athar:2002im, Bahcall:2000ue}, Geoneutrinos ($\sim$ a few MeV) \cite{Araki:2005qa, Krauss:1983zn, Borexino:2010dli, Bellini:2013wsa}, Accelerator Neutrinos (a few hundreds of MeV to tens of GeV) \cite{Danby:1962nd, Parke:1993dp, Dore:2018ldz}, Reactor Neutrinos ($\sim$ MeV) \cite{Cowan:1956rrn, Baba:2004piw, Qian:2018wid}. In this thesis, we will focus our interest on the accelerator neutrinos produced in the long baseline setup, which will be discussed in the subsequent chapters with more details.\\ 

1998 was a revolutionary year in the roadmap of the journey of neutrino discovery when a novel Japanese experiment named Super-Kamiokande first 
detected a significantly lower flux of upward-going atmospheric neutrinos after traversing the Earth than predicted \cite{Super-Kamiokande:1998kpq}. Around that same time domain, several solar neutrino experiments (e.g., SAGE \cite{SAGE:2000icz}, GALLEX \cite{GALLEX:1992gcp}, etc.) showed similar types of anomalies in the detected solar neutrino flux than the predicted one by the theoretical standard solar model \cite{Bahcall:1987jc}. These two anomalies \cite{Barbieri:1998mq} regarding the neutrinos ($i.e.$, atmospheric anomaly \cite{Super-Kamiokande:1998kpq, LoSecco:2019lxv} and solar anomaly \cite{Davis:1968cp}) gave birth to a new phenomenon in particle physics named ``Neutrino Oscillation". Using the concept of neutrino oscillation, the two above-mentioned discrepancies could be explained theoretically. This phenomenon is still spreading its glory to uncover several unanswered questions in particle physics. Hence, being motivated in this field, this thesis explains some aspects of neutrinos in the light of neutrino oscillations, the theory of which is described in the subsequent chapters. 

\section{Introduction of Neutrinos through the Standard Model framework}
The Standard Model of particle physics started to be constructed by the hands of Glashow \cite{GLASHOW1961579}, Salam \cite{Salam:1968rm}, and Weinberg \cite{Weinberg:1967tq} in 1970 through quantum electrodynamics (QED) and the electroweak theory \cite{Goldstone:1962es}. It is a very successful theory in particle physics. The purpose of this model was to unify all the elementary particles and the gauge bosons under a single gauge group ($i.e.$, \textit{SU}$(3)_C\,\otimes\,$\textit{SU}$(2)_L\,\otimes\,$\textit{U}$(1)_Y$). The elementary particles were allotted their places in this model according to their basic properties ($i.e.$, mass, electric charge, spin, parity, isospin, hypercharge, etc.) \cite{ParticleDataGroup:2024cfk}. Adorned with 19 parameters, the particles belonging to the SM enlightened the path of research in particle physics for many years. \\

In the SM, the three active neutrinos ($i.e.$, $\nu_e$, $\nu_\mu$, and $\nu_\tau$) occupy their positions in the lepton segment, along with their charged cousins ($i.e.$, $e$, $\mu$, and $\tau$, respectively). According to the SM, neutrinos are considered as completely massless and electrically neutral \cite{Weinberg:1967tq}. The assumption of neutrinos being massless implies that they should possess a well-defined parity. The first experimental confirmation of this could be seen in Wu’s experiment \cite{Wu:1957my}, which observed an asymmetric distribution of the electrons in the beta decay spectrum of $^{60}\mathrm{Co}$, providing the direct evidence of parity violation in weak interactions. This pioneering experimental confirmation (Goldhaber experiment in 1957) \cite{Goldhaber:1958nb} established the fact that neutrinos are left-handed in Nature, and it violates the parity maximally \cite{Lee:1956qn}. \\

This experimental confirmation was instrumental to the V-A theory of weak interactions in the SM \cite{Feynman:1958ty, Das:2009zzd}. According to the SM, the three generations of leptons ($i.e.$, $e$, $\mu$, and $\tau$) interact via weak interactions through $W^\pm$ and $Z^0$ gauge bosons and form $SU(2)$ doublets with their neutral cousins ($i.e.$, three corresponding neutrinos). The SM Lagrangian density of these interactions can be written as \\
\begin{equation}
\mathcal{L}  = - i \bar{L}_\beta (ig\slashed{W}^b\tau^b - ig'Y_L\slashed{B})L_\beta,                   
\end{equation}
where, $L_\beta$ is the $SU(2)$ doublets (or lepton fields), $W^b_\mu$ and $B_\mu$ are the fields of the $SU(2)_L$ and $U(1)_Y$ gauge bosons, g and $g^\prime$ are corresponding coupling strengths, $Y_L$ is the hypercharge corresponding to $U(1)_Y$ gauge group. Here, $\tau^b=\frac{1}{2}\sigma^b$ (b = 1, 2, 3), where $\sigma$ are the Pauli matrices, generators of $SU(2)$, b is the index of Pauli Matrices, running from 1 to 3; $\mu$ is the Lorentz index, running from 0 to 3; and $\beta$ denotes the generation index of the leptons ($L_\beta = L_1,\, L_2,$ and $L_3$ denotes $e, \, \mu, \,$ and $\tau$, respectively). The corresponding charged current and neutral current interacting Lagrangian density of neutrinos can be written as,\\
\begin{equation}
    \mathcal{L}_{CC} = \frac{g}{2\sqrt{2}} [\bar{l}\gamma^\rho(\mathds{1}-\gamma_5)\nu]W^-_\rho + h.c.,
\end{equation}
\begin{equation}
    \mathcal{L}_{NC} = \frac{g}{2\cos\theta_W} [\bar{\nu}\gamma_\rho(\mathds{1}-\gamma_5)\nu]Z^\rho.
\end{equation}
Here, ($\frac{\mathds{1}-\gamma_5}{2}$) represents the left projection operator ensuring the neutrinos as left-handed candidates from the experimental confirmation,  $l$ is the charged lepton field, $\theta_W$ represents the weak-mixing angle \cite{Weinberg:1967tq,  Fairlie:1979at, Moroi:1993zj} (Weinberg angle, where tan $\theta_W = \frac{g'}{g}$),  $W^-$ and $Z$ are the gauge bosons playing the role of the mediators of the weak interactions. \\

The particles of the SM mainly acquire their masses by breaking the $\textit{SU}(2)_L\,\otimes\,\textit{U}(1)_Y$  gauge symmetry spontaneously \cite{Higgs:1964pj, Masson:2010vx, Bachu:2023fjn}. But, the lepton (and quark) family can acquire mass via Yukawa interaction after the spontaneous symmetry breaking \cite{Yukawa:1935xg, Tonasse:2001txk, Neznamov:2006yu}. The mass term of the charged fermions of the corresponding Lagrangian density can be written as,\\
\begin{equation}
\mathcal{L} = -m_e\bar{e}_Le_R - m_d\bar{d}_Ld_R - m_u\bar{u}_Lu_R + h.c.
\end{equation}
Here, $m_e$ depicts the mass of the electrons, where $m_e= \dfrac{y\,v}{\sqrt{2}}$, $y$ represents the Yukawa coupling constant (for electron, its value is 2.9$\times10^{-6}$), and $v$ is the vacuum expectation value (VEV) of the Higgs boson of the SM, after breaking symmetry. Here, $m_u$ and $m_d$ stand for the masses of up and down quarks and can be written as the first diagonal entry of $\dfrac{v}{\sqrt{2}}\,M_u$ and $\dfrac{v}{\sqrt{2}}\,M_d$ matrix, respectively. $M_u$ and $M_d$ are the mass matrices of quarks in the diagonal basis. From the above equation, it is clear that to write the mass terms of the lepton, both the right-handed and left-handed candidates of the corresponding lepton are essential. But, neutrinos don't have any right-handed counterpart, and hence it cannot gain any mass by the SM's prediction. Now, apart from the Dirac mass term as shown in the above equation, neutrinos could gain mass, if it is considered as a Majorana particle \cite{Majorana:1937vz, Minkowski:1977sc, Dolinski:2019nrj, McElrath:2009fn}, given in the following equation.\\
\begin{equation}
    \mathcal{L}_M = -\frac{1}{2} m_M \overline{\nu_L^c} \nu_L + h.c.,
\end{equation}

where $m_M$ is the Majorana mass, $\nu_L$ is the left-handed neutrino field and $\overline{\nu_L^c}$ stands for the chrge-conjugated left-handed neutrino field, where $\overline{\nu_L^c} = -\nu^T_LC^{-1}$, and $\nu^T_L$ is the transpose of the left-handed neutrino spinor field. But, it violates the lepton number conservation by two  (2) units and is hence ruled out by the SM's constructions. Hence, in the SM, neutrinos are completely massless. \\

Subsequent observations in particle physics provided compelling evidence suggesting a non-zero neutrino mass, reshaping our understanding of neutrinos and unveiling new possibilities in the field. This discovery encouraged the exploration of BSM physics. However, in BSM physics, there are several types of neutrino mass models (Seesaw models \cite{Minkowski:1977sc, Akhmedov:1999tm, Mohapatra:2004zh}, Scotogenic model \cite{Ma:2006km, Escribano:2020iqq, Ahriche:2022bpx, Vicente:2022zey}, Radiative neutrino mass model \cite{Zee:1980ai, Zee:1985id, Saad:2019bqf, Cai:2017jrq, Ma:2006km}, etc.), giving the explanation of the mechanism of mass generation of neutrinos. Neutrino oscillation \cite{Super-Kamiokande:1998kpq} was the first experimental motivation to conduct the BSM research. Like several mass models, physicists proposed several mixing models of neutrinos in the BSM landscape, which will be discussed next.\\

\section{Neutrino mixing models}
The neutrino oscillation phenomenon is a pioneering discovery that unveils neutrinos should have non-zero mass and they should mix with each other. Although the construction of the neutrino oscillation theory will be discussed in the subsequent chapter, it is essential to mention some basic properties of the oscillation phenomenology in the neutrino sector to get an idea of mixing \cite{Bilenky:1978nj}. According to this theory, neutrinos, while produced or detected, remain in their flavor eigenstates (or gauge states/ weak eigen states, $i.e.$, $\nu_e$ or $\nu_\mu$ or $\nu_\tau$). But, while they propagate through vacuum or matter, they are in the mass eigenstates (or physical state, $i.e.$, $\nu_1$, $\nu_2$, and $\nu_3$). In 1962, a group of physicists, Pontekorvo, Maki, Nakagawa, and Sakata (PMNS)\cite{Maki:1962mu, Pontecorvo:1967fh}, first linked these two sets of eigenstates by a unitary mixing matrix (PMNS matrix). Although, there are some attempts in probing the non-unitarity of the mixing matrix \cite{Schechter:1980gr, Agarwalla:2021owd} in recent studies, still at that time, people tried to construct a theory that would preserve the conservation of probability. There are other choices also to construct this linking matrix between the flavor and mass eigenstates maintaining the unitarity, but PMNS's prescription is really handy in this regard and hence is used subsequently in the neutrino oscillation research. If we take the ratio of some specific matrix elements, we can get the value of a distinct oscillation parameter. The PMNS matrix can be illustrated as shown below.

\begin{align}   
\begin{pmatrix}
\nu_e \\
\nu_\mu \\
\nu_\tau
\end{pmatrix}
=
\begin{bmatrix}
U_{e1} & U_{e2} & U_{e3} \\
U_{\mu1} & U_{\mu2} & U_{\mu3} \\
U_{\tau1} & U_{\tau2} & U_{\tau3}
\end{bmatrix}
\begin{pmatrix}
\nu_1 \\
\nu_2 \\
\nu_3
\end{pmatrix},
\end{align}
\begin{align}
   & = 
   	\begin{pmatrix}
   	   {c_{12}c_{13}} & {s_{12}c_{13}} & {s_{13}e^{-i\delta}}\\
   	  {-s_{12}c_{23}-c_{12}s_{13}s_{23}e^{i\delta}} & {c_{12}c_{23}-s_{12}s_{13}s_{23}e^{i\delta}} & {c_{13}s_{23}}\\
   	   {s_{12}s_{23}-c_{12}s_{13}c_{23}e^{i\delta}} & {-c_{12}s_{23}-s_{12}s_{13}c_{23}e^{i\delta}} & {c_{13}c_{23}}
   	   \end{pmatrix}  
       \begin{pmatrix}
 \nu_1\\
 \nu_2\\
 \nu_3\\
\end{pmatrix}.
\label{eq:PMNS matrix}
   \end{align}

Equation \ref{eq:PMNS matrix} shows all the elements of the PMNS matrix. The PMNS matrix can be parameterized by four (4) oscillation parameters, $i.e.$, $\theta_{12}$ (solar mixing angle), $\theta_{13}$ (reactor mixing angle), $\theta_{23}$ (atmospheric mixing angle), and the leptonic CP phase $\delta_{\mathrm{CP}}$. In equation \ref{eq:PMNS matrix}, $c_{ij}\,(s_{ij})$ stands for $\cos\theta_{ij}\,(\sin\theta_{ij})$. The mixing angles are nothing but the coefficient of the linear combination of the mass eigenstates to form the flavor eigenstate. Physically, they tell us the amount of the superpositions of the mass eigenstates over a flavor eigenstate. There are two remaining parameters that govern the neutrino oscillation, $\Delta m^2_{21}\,(m^2_2-m^2_1$, solar mass-splitting) and $\Delta m^2_{31}\,(m^2_3-m^2_1$, atmospheric mass-splitting), which are the constituent of the mass matrix or the kinetic term of the Hamiltonian of neutrino oscillation. \\

Revisiting the trend of the data obtained from several well-precise experiments, it was observed that the value of the two neutrino mixing angles ($\theta_{12}$ and $\theta_{23}$) are larger compared to the third mixing angle ($\theta_{13}$). Before discovering the neutrino oscillation (1998), the oscillation phenomenon was observed in the meson sector of the SM ($K^0-\bar{K}^0$ oscillation in 1960) \cite{Christenson:1964fg, Gell-Mann:1955ipe, Lee:1957qq}. Similar kind of oscillation parameters exist there, constructing the CKM matrix (Cabibbo-Kobayashi-Maskawa Matrix) \cite{Kobayashi:1973fv, Cabibbo:1963yz} that connects between quark's flavor eigenstates ($i.e.$, $d^\prime$, $s^\prime$, and $b^\prime$) to its mass eigenstate ($d,\, s,\,$ and $b$). Note that, the mass and mixing parameters in the quark sector are well determined \cite{ParticleDataGroup:2024cfk}. But, the mixings in the quark sector and in the neutrino sector are completely different. These particular mixings in the neutrino sector have been a puzzle to the scientific community and it is difficult to understand the origin of these mixings without the help of any symmetry. There are many symmetry groups (e.g., $A_4$ \cite{Ma:2004zv, He:2006dk, Morisi:2009sc, Altarelli:2010gt}, $S_3$ \cite{GonzalezCanales:2012blg}, $S_4$ \cite{Zhang:2006fv}, etc.) that can explain the peculiar mixing structure in the neutrino sector ($i.e.$, two large and one small mixing angle).  The $A_4$ discrete symmetry group (or Flavor symmetry group) \cite{Morisi:2009sc, Altarelli:2010gt} is a subgroup of the $S_4$, which is nothing but an even permutation of four objects. \\

Using this flavor symmetry group, several neutrino mixing models were proposed till the neutrino oscillation research entered into the precision era. Some of the well-known neutrino mixing models are discussed below.

\underline{\textbf{Tri-bimaximal mixing (TBM):   }}
  The tri-bimaximal mixing pattern, elegantly formulated by Harrison, Perkins, and Scott \cite{Harrison:1999cf}, emerged as a remarkably successful framework for describing neutrino mixing patterns in the light of early oscillation data. According to this mixing model \cite{Harrison:1999cf, Abbas:2010jw, King:2014nza, Ma:2009wi}, $\theta_{13} = 0^\circ$, $\theta_{23}= 45^\circ$, $\theta_{12}= 35.3^\circ$. The corresponding structure of the PMNS matrix is,
\begin{align}
    U_{\mathrm{TBM}} = \begin{pmatrix}
   	\dfrac{2}{\sqrt 6} & \dfrac{1}{\sqrt 3} & 0\\
   	-\dfrac{1}{\sqrt 6} & \dfrac{1}{\sqrt 3} & \dfrac{1}{\sqrt 2}\\
   	-\dfrac{1}{\sqrt 6} & \dfrac{1}{\sqrt 3} & -\dfrac{1}{\sqrt 2}
   	\end{pmatrix}.
\end{align}
 This model fits the solar neutrino data well, but could not explain the non-zero value of the $\theta_{13}$ as confirmed by the modern experiments like Daya Bay, RENO, and T2K. Before the discovery of non-zero value of $\theta_{13}$, this was a leading mixing model and well-acknowledged by the community.\\
 Later on, people tried to correct this model to match with the data and added the correction terms as \cite{King:2014nza, Plentinger:2005kx},
 \begin{align}
    U_{\mathrm{TBM}}^{\mathrm{modified}} = \begin{pmatrix}
   	\sqrt{\dfrac{2}{3}}(1-\dfrac{1}{2}s)  & \dfrac{1}{\sqrt 3}(1+s) & \dfrac{1}{\sqrt {2}}re^{-i\delta}\\
   	-\dfrac{1}{\sqrt 6}(1+s-a+re^{i\delta}) & \dfrac{1}{\sqrt 3}(1-\dfrac{1}{2}s-a-\dfrac{1}{2}re^{i\delta}) & \dfrac{1}{\sqrt 2}(1+a)\\
   	\dfrac{1}{\sqrt 6}(1+s+a-re^{i\delta}) & -\dfrac{1}{\sqrt 3}(1-\dfrac{1}{2}s+a+\dfrac{1}{2}re^{i\delta}) & \dfrac{1}{\sqrt 2}(1-a)
   	\end{pmatrix},
    \label{equn:Modified_TBM}
\end{align}
with the corrected parameters, $\sin\theta_{12}=\dfrac{1}{\sqrt{3}}(1+s)$, $\sin\theta_{23}=\dfrac{1}{\sqrt{2}}(1+a)$, and $\sin\theta_{13}=\dfrac{r}{\sqrt{2}}$, where $s$, $a$, and $r$ are the corrections in the solar, atmospheric, and reactor sector, respectively. In equation \ref{equn:Modified_TBM}, $s=a=r=0$ will provide us with the standard TBM pattern.\\

 \underline{\textbf{Tri-maximal mixing (TM):   }}
The tri-maximal neutrino mixing \cite{King:2014nza, Albright:2008rp, King:2011zj} was a proposed framework in which the solar correction term ($s=0$ in the equation \ref{equn:Modified_TBM}) is exactly vanishing. The structure of the PMNS matrix takes the form,
\begin{align}
    U_{\mathrm{TM}} = \begin{pmatrix}
   	\sqrt{\dfrac{2}{3}}  & \dfrac{1}{\sqrt 3} & \dfrac{1}{\sqrt {2}}re^{-i\delta}\\
   	-\dfrac{1}{\sqrt 6}(1-a+re^{i\delta}) & \dfrac{1}{\sqrt 3}(1-a-\dfrac{1}{2}re^{i\delta}) & \dfrac{1}{\sqrt 2}(1+a)\\
   	\dfrac{1}{\sqrt 6}(1+a-re^{i\delta}) & -\dfrac{1}{\sqrt 3}(1+a+\dfrac{1}{2}re^{i\delta}) & \dfrac{1}{\sqrt 2}(1-a)
   	\end{pmatrix},
\end{align}
where $s$, $a$, and $r$ are the correction parameters. Using the unitarity property, we can get two limiting conditions from the above equation.
 $a= r\cos\delta$, which is the TM1 scheme, where the 1st column of the PMNS matrix is identical to the TBM pattern. The second limiting condition is $a=-\dfrac{1}{2}r\cos\delta$ depicting TM2 scheme, where the 2nd column of the PMNS matrix takes the shape of the second column of the matrix given by TBM model. However, TM mixing model is less predictive to the CP-violating phase $\delta_{\mathrm{CP}}$ in the neutrino sector. Also, the data from the ongoing and upcoming oscillation experiments may probe these correction parameters with better precision.

 \underline{\textbf{Bimaximal mixing (BM):   }}\\
 According to this mixing pattern \cite{Barger:1998ta, King:2014nza, Xing:1999wz, Dutta:2002nq, Frampton:2004ud}, both solar ($\theta_{12}$) and atmospheric mixing angles ($\theta_{23}$) take the maximal mixing value ($i.e.$, $45^\circ$) and $\theta_{13}=0^\circ$ giving the shape of the PMNS mixing matrix as below.
\begin{align}
    U_{\mathrm{BM}} = \begin{pmatrix}
   	\dfrac{1}{\sqrt 2}  & -\dfrac{1}{\sqrt 2} & 0\\
   	\dfrac{1}{2} & \dfrac{1}{2} & -\dfrac{1}{\sqrt 2}\\
   	\dfrac{1}{2} & \dfrac{1}{2} & \dfrac{1}{\sqrt 2}
   	\end{pmatrix}.
\end{align}
 
It also cannot explain the non-zero $\theta_{13}$ value. The values of the solar mixing angle obtained from the high-precise solar experiments also do not match with the maximal mixing. It also cannot predict the value of $\delta_{\mathrm{CP}}$ precisely. \\
\newpage
\underline{\textbf{Tri-bimaximal Cabibbo (TBC) mixing:   }}\\
The tri-bimaximal Cabibbo (TBC) mixing framework \cite{King:2012vj, Hu:2012eb, Ahluwalia:2012gw} is a modified neutrino mixing pattern obtained from the tri-bimaximal (TBM) mixing scheme, with adjustments to account for the influence of the Cabibbo angle ($\theta_C$) from quark mixing. The TBC model modifies TBM by shifting $\theta_{12}$ and $\theta_{13}$ based on $\theta_C$ for better fitting to the experimental data. In TBC, the neutrino mixing angles are modified as,
$\theta_{12}\approx45^\circ-\theta_C$, $\theta_{23}=45^\circ$, and $\theta_{13}\approx\dfrac{\theta_C}{\sqrt{2}}$. The corresponding PMNS matrix is,

\begin{align}
    U_{\mathrm{TBC}} \approx \begin{pmatrix}
   	\cos\left(45^\circ-\theta_C\right)  & \sin\left(45^\circ-\theta_C\right) & \dfrac{\theta_C}{\sqrt{2}}\\
   	-\dfrac{1}{\sqrt{2}}\left(\sin\left(45^\circ-\theta_C\right)+\dfrac{\theta_C}{\sqrt{2}}\right) & \dfrac{1}{\sqrt{2}}\left(\cos\left(45^\circ-\theta_C\right)-\dfrac{\theta_C}{\sqrt{2}}\right) & \dfrac{1}{\sqrt 2}\\
   	\dfrac{1}{\sqrt{2}}\left(\sin\left(45^\circ-\theta_C\right)-\dfrac{\theta_C}{\sqrt{2}}\right) & -\dfrac{1}{\sqrt{2}}\left(\cos\left(45^\circ-\theta_C\right)+\dfrac{\theta_C}{\sqrt{2}}\right) & \dfrac{1}{\sqrt 2}
   	\end{pmatrix}.
\end{align}

Here, $\theta_C$ is the Cabibbo mixing angle ($\sim 13^\circ$) \cite{Cabibbo:1963yz, Fritzsch:1977za, Blucher:2005dc}. This theoretical neutrino mixing model was highly appreciated by the physicists as it incorporated non-zero value of $\theta_{13}$, showed the deviation of $\theta_{12}$ from the maximal mixing, and gave a connection between the quark and the neutrino mixing patterns (by linking $\theta_C$ to the neutrino mixing) which is famously known as ``Quark-Lepton Complementarity" \cite{Minakata:2004xt}. However, this model also cannot predict much regarding the $\delta_{\mathrm{CP}}$.

The above-mentioned neutrino mixing models could explain several natural phenomena emerging from the neutrino oscillation and tried to explain the peculiar pattern of the neutrino mixing matrix. However, the recent data obtained from the neutrino oscillation experiments show the deviation of the values of the oscillation parameters at their best-fit from the same predicted by the above mixing models. The study performed in this thesis primarily focuses on how the present and upcoming neutrino experiments may provide precise measurements of neutrino mixing angles, which in turn, may shed more light on the above mixing models. 
\section{The structure of the thesis}
We plan the thesis in the following fashion. In Chapter \ref{C2}, we introduce the quantum mechanical approach for constructing the formalism of neutrino oscillations. We provide fundamental insights into two-flavor and three-flavor neutrino oscillations, both in vacuum and Earth's matter. Furthermore, we conduct an in-depth exploration of the degeneracies associated with the neutrino oscillation parameters and highlight the prevailing tensions among various neutrino oscillation experiments. Chapter \ref{C3} presents a comprehensive overview of the evolution of long-baseline experiments, tracing their trajectory from inception to their current advancements. In Chapter \ref{C4}, we illustrate how the contemporary and next-generation long-baseline (LBL) neutrino experiments hold immense potential in probing the deviation of the atmospheric mixing angle ($\theta_{23}$) from maximal mixing, resolving the octant ambiguity ($\theta_{23}>45^\circ$ or $<45^\circ$), and significantly enhancing the precision of the oscillation parameters $\theta_{23}$ and $\Delta m^2_{31}$. We particularly emphasize the role of the upcoming Deep Underground Neutrino Experiment (DUNE) in surpassing present measurements within the three-neutrino framework. Chapter \ref{C5} delves into the synergistic interplay between DUNE and T2HK (Tokai-to-Hyper-Kamiokande) in resolving degeneracies among neutrino oscillation parameters. We investigate how the combined sensitivity of these experiments strengthens key aspects of neutrino oscillation phenomenology, including the establishment of non-maximal $\theta_{23}$, exclusion of the incorrect octant, and precision measurements of the 2-3 oscillation parameters of the PMNS matrix. These pursuits are particularly compelling as neutrino oscillation research transitions into the precision era. Additionally, we demonstrate how the complementary features of DUNE and T2HK facilitate an enhanced physics reach, achieving substantial sensitivity at lower exposures compared to their individual baseline configurations. Lastly, in Chapter \ref{C6}, we explore how the next-generation LBL experiments can uncover signatures of the flavor-dependent long-range neutrino interactions within the neutrino sector. Specifically, we analyze how these experiments can probe the presence of the gauged $(L_e - L_\mu)$, $(L_e - L_\tau)$, and $(L_\mu - L_\tau)$ symmetries through the aforementioned oscillation phenomenology, thereby establishing compelling evidence for such interactions with a high degree of confidence. Finally, Chapter \ref{C7} ends this thesis with some insightful concluding remarks, encapsulating the key takeaways, and their importance in the neutrino oscillation frontier.

\chapter{Fundamentals and Theoretical Framework of Neutrino Oscillations}
\label{C2} 
Neutrino oscillation is one of the pioneering discoveries in the history of particle physics. It has opened a new horizon for the community to utilize this second most abundant particle to reveal several mysteries of the Universe \cite{Tamborra:2024fcd}. The cornerstone of the neutrino oscillation was first laid in the Super-Kamiokande detector (Japan) in 1998 \cite{Super-Kamiokande:1998kpq} by providing experimental evidence in their data. Non-zero neutrino mass is the first experimental proof of BSM physics, for which the Nobel Prize was awarded to Takaaki Kajita and Arthur McDonald in 2015 \cite{Kajita:2016cak, McDonald:2016ixn}.\\
Before entering into the formalism of neutrino oscillation, it is essential to know why neutrinos show the flavor transition (named ``neutrino oscillation") \cite{Rajasekaran:2000nn, Agarwalla:2008jin}, whereas their charged counterparts do not show oscillations. Neutrino oscillation is a mass-induced phenomenon. Due to the non-zero value of neutrino mass and the non-degenerate values of the mass eigenstates \cite{Raychaudhuri:2000zn}, the flavor eigenstates and mass eigenstates of neutrinos are quite different. The tiny mass-squared differences between the neutrino mass eigenstates ($\sim 10^{-5}$ to $10^{-3} \, \mathrm{eV}^2$) protect the coherence of the superposition \cite{LIPKIN2004355, Ravari:2022yfd} of mass eigenstates to the flavor states. In contrast, the mass differences between the leptons are so huge ($\sim 100$ MeV to GeV scale) that any superposition of the mass eigenstates of the leptons decoheres almost immediately before their significant propagation.\\
This chapter is planned in the following way. Section \ref{sec:2.1} outlines the fundamental theoretical framework of neutrino oscillations in the vacuum, examining the formalism in the two-flavor framework. In section \ref{sec:2.2}, the extension of this formalism in the three-flavor landscape is described. In section \ref{sec:2.3}, we introduce matter-effect in our discussion, incorporating standard neutrino-matter interactions within both two-flavor and three-flavor scenarios. Section \ref{sec:2.4} provides an overview of several relevant neutrino oscillation experiments and their contributions to measuring oscillation parameters. In section \ref{sec:2.4}, we present the current status of the six oscillation parameters and the so-called eight-fold degeneracies, discussing their experimental determination. Finally, in section \ref{sec:2.5}, we provide a concluding overview of this chapter.\\

\section{The mathematical framework of two-flavor neutrino oscillation}
\label{sec:2.1}
There are several approaches to construct the neutrino oscillation framework. Neutrinos are fermions and occupy their places in the lepton family of the SM, in the form of $SU(2)$ doublets. So, ideally, the Dirac equation should lead us to the neutrino oscillation probability. However, neutrino oscillation does not manifest any spin nature of itself, and hence, for the ease of our calculations, we use the Schr\"{o}dinger equation considering neutrino as a free particle instead of assuming it as a wave packet. Neutrinos are produced or detected at their flavor eigenstates ($\nu_e$, $\nu_\mu$, or $\nu_\tau$) or weak eigenstates (or gauge states), but they are in the mass eigenstates or physical states ($\nu_1$ and $\nu_2$) while propagating through the vacuum or the matter. The mass eigenstates and flavor eigenstates are considered to form a complete orthonormal basis to conserve the flavor transition probability, and hence a flavor eigenstate, $\nu_\alpha$ is connected to the mass eigenstate $\nu_k$ through a unitary matrix $U$ in the following fashion.

\begin{align}
\ket{\nu_{\alpha}}=\sum_{k=1}^{2}U_{\alpha k}\ket{\nu_{k}},
\end{align}
where $\alpha$ is the flavor index [$\alpha$ = \{$e$, $\mu$\}, or \{$e, \tau$\}, or \{$\mu, \tau$\}] and k is the mass index, $U_{\alpha k}$ are the elements of the unitary matrix connecting the bridge between this two orthonormal basis, ensuring the probability conservation. The completeness of the basis indicates that $\sum_{\alpha} \ket{\nu_\alpha}\bra{\nu_\alpha} = \sum_{k=1,2} \ket{\nu_k}\bra{\nu_k}=\mathds{1}$, whereas the orthonormality ensures that $ \langle \nu_\alpha | \nu_\beta \rangle = \delta_{\alpha \beta}$  and 
 $ \langle \nu_i | \nu_j \rangle=  \delta_{ij}$. The crucial point is to be noted here that the three mass eigenstates are non-degenerate ($i.e.$, the mass of each eigenstate must not be equal, $m_1\neq m_2$, where $m_k$ is the mass of the eigenstate $\nu_k$). These mass eigenstates evolve with time while neutrinos propagate through the vacuum or the matter, and hence, the flavor eigenstates are also affected by time evolution. As the mass eigenstates are the eigenstates of the Hamiltonian, we can write the time evolution of the mass eigenstates in the following way.\\
\par Let us consider a system where a neutrino beam of flavor state $\ket{\nu_{\alpha}(0,0)}$ is generated at space-time point $(x,t)=(0,0)$.\\
We know while the neutrino will propagate through the vacuum they are in mass eigenstates and its time evolution will be governed by the time-dependent Schr\"{o}dinger equation, $i.e.$,\\
\begin{align}
i\frac{\partial}{\partial t}\ket{\nu_{k}(x,t)}&=E\ket{\nu_{k}(x,t)}\\
&=\frac{-1}{2m_{k}}\frac{\partial^2}{\partial^2 x}\ket{\nu_{i}(x,t)},
\end{align}
	where, $k=1,2$. Here, we have used the natural units ($i.e.$, $\hbar$=1).\\
	Now, 
	\begin{align}
	i\frac{\partial}{\partial t}\ket{\nu_{k}(x,t)}&=E\ket{\nu_{k}(x,t)}\\
	\Rightarrow&\frac{\partial\ket{\nu_{k}(x,t)}}{\ket{\nu_{k}(x,t)}}=-iE\partial t	.
	\end{align}\\
	Performing partial integration, we get \\
	\begin{align}
\ket{\nu_{k}(x,t)}=e^{-iE_{k}t}\ket{\nu_{i}(0,0)},
\end{align}
and the trial solution $\ket{\nu_{k}(x,t)}=e^{ip_{k} x}$ satisfies the equation\\
\begin{align}
\frac{-1}{2m_{k}}\frac{\partial^2}{\partial^2 x}\ket{\nu_{k}(x,t)}=E\ket{\nu_{k}(x,t)}.
\end{align}
So finally,\\
\begin{align}
\ket{\nu_{k}(x,t)}&=e^{-i(E_{k}t-p_{k}x)}\ket{\nu_{k}(0,0)}\\
&=e^{-i\phi_{k}}\ket{\nu_{k}(0,0)}, 
\end{align}
where,\\
\begin{equation}
 \phi_{k}=E_{k}t-p_{k}x.
 \label{equn:2.10}
 \end{equation}
At some later space-time point (x,t), the time evolved flavor state $\ket{\nu}_\alpha$ will be\\
\begin{align}
\ket{\nu_{\alpha}(x,t)} &=\sum_{k=1}^{2} U_{\alpha k}\ket{\nu_{k}(x,t)} \\
&=\sum_{k=1}^{2} U_{\alpha k} e^{-i\phi_{k}}\ket{\nu_{k}(0,0)} .
\label{eqn:2.12}
\end{align}
So,\\
\begin{align}
\ket{\nu_{k}(0,0)}=\sum_{\gamma}U^*_{\gamma k}\ket{\nu_{\gamma}(0,0)}. 
\label{eqn:2.13}
\end{align}
Here $\gamma$ is a dummy index of the unitary matrix while being inverted.\\
We have to find the evolution of flavor states, but from the Schr\"{o}dinger equation, we got the time evolution of mass eigenstates as they are the eigenstates of the governing Hamiltonian.\\
So, substituting equation \ref{eqn:2.13} into equation \ref{eqn:2.12}, we get,\\
\begin{align}
	\ket{\nu_{\alpha}(x,t)}&=\sum_{k=1}^{2}U_{\alpha k}e^{-i\phi_{k}}\sum_{\gamma=1}^{2}U^*_{\gamma k}\ket{\nu_{\gamma}(0,0)}\\
	&=\sum_{\gamma}\sum_{k}U^*_{\gamma k}e^{-i\phi_{k}}U_{\alpha k}\ket{\nu_{\gamma}(0,0)}.
\end{align}
Transition amplitude for detecting a neutrino of flavor $\beta$ at space-time point (t,x) when generation of a neutrino of flavor $\alpha$ at space-time point (0,0) is\\
\begin{align}
A(\ket{\nu_{\alpha}(0,0)}\rightarrow\nonumber \ket{\nu_{\beta}(x,t)})&=\braket{\nu_{\beta}(x,t)|\nu_{\alpha}(0,0)}\\\nonumber
 &=\sum_{\gamma}\sum_{k}U_{\gamma k}e^{i\phi_{k}}U^*_{\beta k}\braket{\nu_{\gamma}(0,0)|\nu_{\alpha}(0,0)}\\\nonumber
 &=\sum_{\gamma}\sum_{k}U_{\gamma k}e^{i\phi_{k}}U^*_{\beta k}\delta_{\gamma\alpha}\\
 &=\sum_{k}U_{\alpha k}e^{i\phi_{k}}U^*_{\beta k}.
\end{align}
So, the oscillation probability is,
\begin{align}
	P(\nu_{\beta}\rightarrow\nu_{\alpha})&=|A(\ket{\nu_{\beta}(0,0)}\rightarrow \nonumber\ket{\nu_{\alpha}(x,t)})|^2\\\nonumber
	&=|\sum_{k}U_{\alpha k}e^{i\phi_{k}}U^*_{\beta k}|^2\\ \nonumber
	&=\sum_{k}U_{\alpha k}e^{i\phi_{k}}U^*_{\beta k}\sum_{j}U^*_{\alpha j}e^{-i\phi_{j}}U_{\beta j}\\
	&=\sum_{j=1}^{2}\sum_{k=1}^{2}U_{\alpha k}U^*_{\beta k}U^*_{\alpha_j}U_{\beta j}e^{-i(\phi_{j}-\phi_{k})} .
    \label{equn:2.17}
\end{align}
For the two-flavor oscillation, the $(2\times2)$ unitary matrix is just a rotation matrix in 2d.\\
\begin{equation}
U= \begin{pmatrix}
 U_{\alpha_1} & U_{\alpha_2} \\
 U_{\beta_1} & U_{\beta_2} \\
\end{pmatrix}
=	\begin{pmatrix}
\cos\theta & \sin\theta\\
-\sin\theta & \cos\theta\\
	\end{pmatrix},
\end{equation}
where $\theta$ is a parameter called the mixing angle. For the two-flavor oscillation, it is the only mixing parameter as a $2\times2$ unitary matrix can be parameterized by only one parameter. We show the calculation of equation \ref{equn:2.17} term by term.\\
For \underline{\bf{$k=1,\,j=1;$}}\\
\begin{align}
P_1(\nu_{\beta}\rightarrow\nonumber\nu_{\alpha})&=U_{\alpha 1}U^*_{\beta 1}U^*_{\alpha_1}U_{\beta 1}e^{-i(\phi_{1}-\phi_{1})}\\
&=|U_{\beta 1}|^2|U_{\alpha_1}|^2.
\end{align}
For \underline{\bf{$k=2,\,j=2;$}}\\
\begin{align}
P_2(\nu_{\beta}\rightarrow\nu_{\alpha})=|U_{\beta 2}|^2|U_{\alpha_2}|^2.
\end{align}
For \underline{\bf{$k=1,\,j=2;$}}\\
\begin{align}
P_3(\nu_{\beta}\rightarrow\nu_{\alpha})=U_{\alpha 1}U^*_{\beta 1}U^*_{\alpha 2}U_{\beta 2}e^{-i(\phi_{2}-\phi_{1})}.
\end{align}
For \underline{\bf{$k=2,\,j=1;$}}\\
\begin{align}
P_4(\nu_{\beta}\rightarrow\nu_{\alpha})=U_{\alpha 2}U^*_{\beta 2}U^*_{\alpha 1}U_{\beta 1}e^{-i(\phi_{1}-\phi_{2})}.
\end{align}
Finally, \\
\begin{align}
P(\nu_{\beta}\rightarrow\nonumber\nu_{\alpha})&=|U_{\beta 1}|^2|U_{\alpha 1}|^2+|U_{\beta 2}|^2|U_{\alpha 2}|^2+U_{\alpha 1}U^*_{\beta 1}U_{\alpha 2}U^*_{\beta 2}\left(e^{i\left(\phi_{2}-\phi_{1}\right)}+e^{-i\left(\phi_{2}-\phi_{1}\right)}\right)\\\nonumber
&=(\sin^2\theta\cos^2\theta+\cos^2\theta\sin^2\theta)+2\,U_{\alpha 1}U^*_{\beta 1}U_{\alpha 2}U^*_{\beta 2}\cos(\phi_{2}-\phi_{1})\\\nonumber
&=2\sin^2\theta\cos^2\theta+2\cos\theta(-\sin\theta)\sin\theta\cos\theta\cos(\phi_{2}-\phi_{1})\\\nonumber
&=2\sin^2\theta\cos^2\theta\,[1-\cos(\phi_{2}-\phi_{1})]\\\nonumber
&=4\sin^2\theta\cos^2\theta\sin^2\left(\frac{\phi_{2}-\phi_{1}}{2}\right)\\\nonumber
&=\sin^2(2\theta)\sin^2\left(\frac{\phi_{2}-\phi_{1}}{2}\right).\\
\label{equn:2.23}
\end{align}
By equation \ref{equn:2.10},
\begin{align}
\phi_{k}=E_{k}t-p_{k}x.
\end{align}
So,
\begin{align}
\phi_{2}-\phi_{1}=(E_2-E_1)t-(p_2-p_1)x.
\end{align}
\textit{At this point, we make our first assumption that neutrinos are relativistic}, so $t=x=L$. Here, $L$ is the length of the baseline, and $E$ is the true neutrino energy. As we are in natural unit scale, $c=1$, and 
\begin{align}
p_i&=\nonumber\sqrt{E^2_i-m^2_i}\\\nonumber
&=E_i\sqrt{1-\frac{m^2_i}{E^2_i}}\\\nonumber
&\approx E_i(1-\frac{m^2_i}{2E^2_i})\\\nonumber
&=E_{i}-\frac{m^2_i}{2E_i},\\
\label{equn:2.26}
\end{align}
as neutrino masses are very small compared to their energy.\\
By equation \ref{equn:2.26},\\
\begin{align}
 \phi_{2}-\phi_{1}=\left(\frac{m^2_2}{2E_2}-\frac{m^2_1}{2E_1}\right)L.
 \end{align}
 \textit{Our next assumption posits that each neutrino flavor state is associated with a well-defined momentum p, implying that all mass eigenstates $\nu_i$ (i= 1,2) possess an identical momentum. This premise is grounded in the notion that all mass eigenstates emanate from a common source and traverse coherently along the same trajectory.} It is to be noted here that mass eigenstates are assumed as plane wave solutions. Henceforth, we will consider $E_1=E_2=E$ in our calculation.\\
 Now,\\
 \begin{align}
 \phi_{2}-\phi_{1}&=\nonumber\left(\frac{m^2_2}{2E}-\frac{m^2_1}{2E}\right)\,L\\
 &=\frac{\Delta m^2\,L}{2E},
 \end{align}
 where, $\Delta m^2=m^2_2-m^2_1$.\\
 By equation \ref{equn:2.23}, we can write the probability of the neutrino oscillation from the flavor $\alpha$ to the flavor $\beta$ is \\
 \begin{align}
 P(\nu_{\alpha}\rightarrow\nu_{\beta})=\sin^2\left(2\theta\right)\sin^2\left(\frac{\Delta m^{2}\,L}{4E}\right).
 \label{eq:2-flavor vacuum oscillation formula}
 \end{align}
 \par Here, the first term signifies the amplitude part of the oscillation (governed by the parameter ``mixing angle" $\theta$), and the second term signifies the frequency part (governed by the parameter ``mass-squared difference" $\Delta m^2$). They are fixed in Nature. Here, we have calculated the oscillation probability by imposing natural units ($i.e.$, $\hbar$=c=1).\\
 \par  Now, if we want this oscillation probability formula compatible with the data obtained from the experiments, we need to convert the units of the equation \ref{eq:2-flavor vacuum oscillation formula} $\left[\Delta m^2\, (\mathrm{in \,\,eV}^2), \, L \,(\mathrm{in\,\, eV}^{-1}),\, E\, (\mathrm{in\,\, eV})\right]$ to the units used by the experiments [$i.e.$, $\Delta m^2\, (\mathrm{in\,\,eV}^2), \, L \,(\mathrm{in \,\,km}),\, E\, (\mathrm{in\,\,GeV})$]. The quantity  $\left(\dfrac{\Delta m^2\, L}{4E}\right)$ is in the argument of a sinusoidal function and needs to be dimensionless. After implementing those conversion factors, the final form of the neutrino oscillation probability from a flavor $\alpha$ to be transited to a flavor $\beta$ through vacuum in the two-flavor landscape is,
 \begin{align}
P(\nu_{\alpha}\rightarrow\nu_{\beta})=\sin^2\left(2\theta\right)\sin^2\left(1.27\times\frac{\Delta m^{2}\,L}{E}\right),
\label{eq:2 flavor app oscillation probability}
 \end{align}
where $L$ is the length over which neutrino changes its flavor (in km), E is the true neutrino energy (in GeV), and $\Delta m^2_{21}$ is the mass-squared difference (in $\mathrm{eV}^2$), where $\Delta m^2= m^2_2-m^2_1$. It is noteworthy that, two flavor neutrino oscillation only suggests the non-trivial mixing and mass-induced flavor conversion in the neutrino sector but cannot predict anything regarding the CP violation. 
Similarly, the expression of the survival (or disappearance) probability in the two-flavor landscape can be written as,
\begin{align}
P(\nu_{\alpha}\rightarrow\nu_{\alpha})=1-\sin^2\left(2\theta\right)\sin^2\left(1.27\times\frac{\Delta m^{2}\,L}{E}\right).
\label{eq:2 flavor oscillation probability}
 \end{align}

\begin{figure}[htb!]
	\centering
	\includegraphics[width=\linewidth]{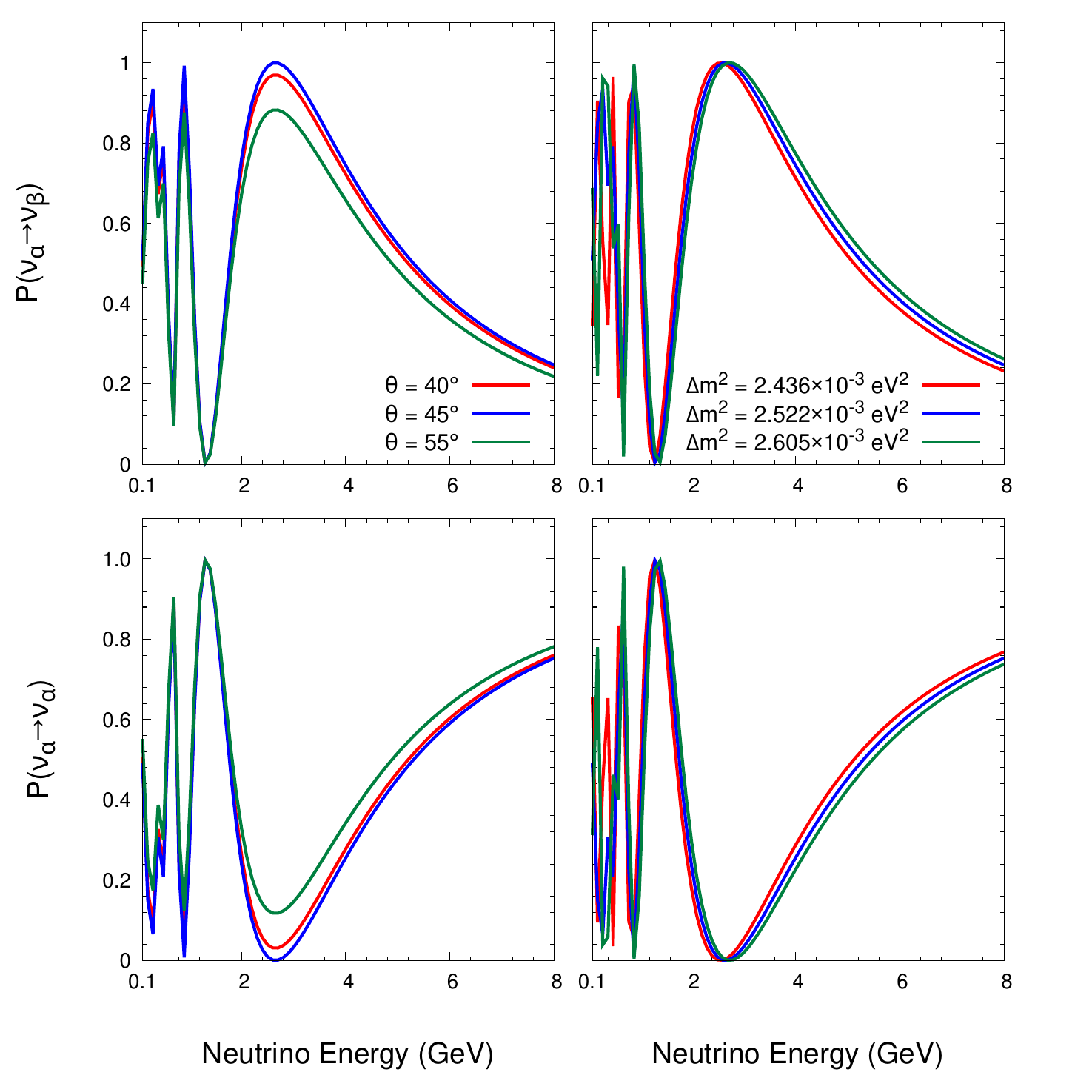}	
    \caption{\footnotesize{Neutrino oscillation probability in vacuum at two flavor case as the function of true neutrino energy (in GeV) is depicted here. The upper (lower) panel represents $\nu_\alpha\rightarrow\nu_\beta$ appearance ($\nu_\alpha\rightarrow\nu_\alpha$ disappearance) probability using Equation \ref{eq:2 flavor app oscillation probability}. In the left panel, we observe the oscillation probability for three choices of mixing angle $\theta\,\,i.e.,\,\,40^\circ,\,45^\circ,\,\mathrm{and}\,\,\, 55^\circ$ (shown in red, blue, and green lines), respectively, assuming $L= 1300$ km and $\Delta m^2=+2.522\times10^{-3}$ eV$^2$. In the right panel, we compute the same but for three different choices of $\Delta m^2,\,\,i.e.,\,2.436\times10^{-3}$ eV$^2,\, 2.522\times10^{-3}$ eV$^2, \mathrm{and}\,\,\, 2.605\times10^{-3}$ eV$^2$ (shown by red, blue, and green lines), respectively, assuming a given mixing angle $\theta=45^\circ$.}}
	\label{fig:oscillation_prob_vacuum_analytical}
\end{figure} 
 \par Equation \ref{eq:2 flavor app oscillation probability} resembles with the standard oscillation formula in the acoustic medium, where the first part signifies the amplitude part (governed by the mixing angle $\theta$), whereas the second part depicts the frequency part (governed by the mass-squared difference $\Delta m^2$). If $\theta$ becomes $45^\circ$, appearance probability (\ref{eq:2 flavor app oscillation probability}) becomes maximum. Hence the value $\theta=45^\circ$ is called the maximal mixing (MM) value \cite{Xing:2015fdg}. Any other values of $\theta$ show lesser appearance probability than the maximal mixing for a given set of values of oscillation parameters. On the other hand, the shift of the frequency can be tuned to different energy and baseline lengths by controlling the value of $\Delta m^2$. 

 Figure~\ref{fig:oscillation_prob_vacuum_analytical} illustrates that the variations in the mixing angle enable precise modulation of the amplitude of the neutrino oscillation probability, while adjustments to the mass-squared difference primarily govern the phase shift of the oscillation pattern. The mixing angle $\theta$ signifies the extent of flavor mixing, with maximal mixing corresponding to the highest transition probability or the deepest suppression in survival probability. Moreover, in the oscillation frequency domain, an increase in the value of $\Delta m^2$ induces a phase shift toward higher neutrino energies, resulting in a rightward displacement of the oscillation peak.\\
 \section{Formalism of neutrino oscillation probability in $3\nu$ paradigm}
 \label{sec:2.2}
 The discovery of the non-zero value of $\theta_{13}$ in the Daya Bay \cite{DayaBay:2012fng, DayaBay:2022orm} experiment connects the atmospheric neutrino sector and the solar neutrino sector, establishing the $3\nu$ paradigm on a strong footing. At the same time, the two-flavor neutrino oscillation delineates the values of the solar and atmospheric oscillation parameters, and the $3\nu$ paradigm with non-zero $\theta_{13}$ gives rise to the possibilities of observing CP violation in the neutrino sector. Now, we discuss the formalism of the three-flavor neutrino oscillation.
 \par Let, a neutrino of flavor $ \alpha$ is produced at $ x=0 $ with energy $E$, then the state of the neutrino at $x=0$ can be expressed as\\
 	\begin{align}
 		\ket{\nu_{\alpha}(x=0)}& =\nonumber\ket{\nu_\alpha}\\
 		&=\sum_{j=1}^{3}(U_{PMNS})^*_{\alpha j}\ket{\nu_j}.\\\nonumber
  	 	\end{align}
 At the position $x=L$, the neutrino state can be written as,\\
 \begin{align}
 	\ket{\nu_{\alpha}(x=L)}&=\nonumber\sum_{j=1}^{3}e^{ip_jL}(U_{PMNS})^*_{\alpha j}\ket{\nu_j}\\
 	&=e^{ip_kL}\sum_{j=1}^{3}e^{i(p_j-p_k)L}(U_{PMNS})^*_{\alpha j}\ket{\nu_j}.\\\nonumber
 \end{align}
   	
   	Assuming $m_j<<E$ we can approximate, \\
   	\begin{align}
   		p_j&=\nonumber\sqrt{E^2-m^2_j}\\
   		&=E-\frac{m^2_j}{2E}+...\\\nonumber.
   	\end{align}
   So,
   \begin{align}
   	p_j-p_k \approx -\frac{\Delta m^2_{jk}}{2E}\\\nonumber
   \end{align}
   where, $\Delta m^2_{jk}=m^2_j-m^2_k$ . \\
   	   	Hence, we can write
   	   	\begin{align}
   	   		\ket{\nu_{\alpha}(x=L)}&=e^{ip_kL}\sum_{j=1}^{3}exp\left(-i\frac{\Delta m^2_{jk}\,L}{2E}\right)(U_{PMNS})^*_{\alpha j} \ket{\nu_j}.\\\nonumber
   	   	\end{align}
        \vspace{-1.5mm}
   	So, the probability amplitude of observing the neutrino flavor $\beta$ at $x=L$ is given by (neglecting the overall phase as it will be cancelled out while multiplying with its complex conjugate while obtaining the probability),   
	\begin{align}
  	\textrm{A}_{\beta\alpha}&=\nonumber\braket{{\nu_\beta}|\nu_{\alpha}(x=L)}\\\nonumber
	&=\nonumber\left[\sum_{k=1}^{3}\bra{\nu_k}(U_{PMNS})_{\beta k}\right]\left[\sum_{j=1}^{3}\exp\left(-i\frac{\Delta m^2_{jk}\,L}{2E}\right)(U_{PMNS})^*_{\alpha j}\ket{\nu_j}\right]\\\nonumber
   	&=\nonumber\sum_{j=1}^{3}(U_{PMNS})_{\beta j}\exp\left(-i\frac{\Delta m^2_{jk}\,L}{2E}\right)(U_{PMNS})^*_{\alpha j}\\
   	&=\sum_{j=1}^{3}U_{\beta j}\exp\left(-i\frac{\Delta m^2_{jk}\,L}{2E}\right)U^*_{\alpha j}.\\\nonumber
  	\end{align}	
  	So, the probability of oscillation from $\ket{\nu_\alpha}$ to $\ket{\nu_\beta}$ with true neutrino energy $E$ and baseline $L$ is given by,\\
  	\begin{align}
  		P(\nu_\alpha \rightarrow \nu_\beta)&=\nonumber|A_{\beta\alpha}|^2\\\nonumber
  		&=\nonumber|\sum_{j=1}^{3}U_{\beta j}exp\left(-i\frac{\Delta m^2_{jk}\,L}{2E}\right)U^*_{\alpha j}|^2\\\nonumber
  		&=\nonumber \left[\sum_{j=1}^{3}U_{\beta j}exp\left(-i\frac{\Delta m^2_{jk}\,L}{2E}\right)U^*_{\alpha j}\right]\times\left[\sum_{i=1}^{3}U_{\alpha i}exp\left(i\frac{\Delta m^2_{ik}\,L}{2E}\right)U^*_{\beta i}\right]\\
  		&=\sum_{j=1}^{3}\sum_{i=1}^{3}exp\left[-i\frac{\left(\Delta m^2_{jk}-\Delta m^2_{ik}\right)\,L}{2E}\right]\,U_{\beta j}U^*_{\alpha j}U_{\alpha i}U^*_{\beta i}\\\nonumber
  		\label{Equation 8}\\
  		&=\sum_{i}\sum_{j}\left(A_{ij}+iB_{ij}\right)\times\left(C_{ij}+iD_{ij}\right);    (say),\\\nonumber
  	\end{align}
   	   	where we assume, $exp\left[-i\dfrac{\left(\Delta m^2_{jk}-\Delta m^2_{ik}\right)\,L}{2E}\right]=A_{ij}+iB_{ij}$ and $U_{\beta j}U^*_{\alpha j}U_{\alpha i}U^*_{\beta i}=C_{ij}+iD_{ij}$. As the oscillation probability is a measurable quantity, so it should be real. Hence, we can drop the imaginary component of equation \ref{Equation 8}.\\
        
   	   	We break the sum of equation \ref{Equation 8} in two parts, $i.e.,$\\
   	   	\[
\sum_{i=1}^{3} \sum_{j=1}^{3} = \sum_{\substack{i=j=1}}^{3} + \sum_{\substack{i \neq j \\ i=1,\, j=1}}^{3}.
\]        
   	   	\textbf{\underline{Case - 1 :}}\\
   	   	for $i=j$,\\
   	  \begin{align}
   	  	 exp\left[-i\frac{\Delta m^2_{ii}\,L}{2E}\right]U_{\beta i}U^*_{\alpha i}U_{\alpha i}U^*_{\beta i}	&= U_{\beta i}U^*_{\alpha i}U_{\alpha i}U^*_{\beta i}.\\\nonumber
   	  \end{align}
   	   	\textbf{\underline{Case - 2 :}}\\
   	   for $i\neq j$,\\
   	   \begin{align}
   	   Re\left(exp\left[-i\frac{\Delta m^2_{ji}\,L}{2E}\right]\right)&=\nonumber \cos\left(\frac{\Delta m^2_{ji}\,L}{2E}\right)\\
   	   &=\cos(\Delta_{ji});\,\,(say, A_{ij}),\\
   	   Im\left(exp\left[-i\frac{\Delta m^2_{ji}\,L}{2E}\right]\right)&=\nonumber -\sin\left(\frac{\Delta m^2_{ji}\,L}{2E}\right)\\
   	   &=-\sin(\Delta_{ji});\,\,(say, B_{ij}).\\\nonumber
   	   \end{align}	
   	   	So, from equation \ref{Equation 8}, considering the real part only, we can write\\
   	   	\begin{align}
    P (\nu_\alpha \rightarrow \nu_\beta) &= A_{ij}C_{ij} - B_{ij}D_{ij} \nonumber \\
    &= Re (U^*_{\alpha i} U_{\beta i}U_{\alpha j}U^*_{\beta j}) 
    - 2\,Re (U^*_{\alpha i} U_{\beta i}U_{\alpha j}U^*_{\beta j}) \sin^2\left(\frac{\Delta_{ji}}{2}\right) \nonumber \\
    &\quad + Im\,(U^*_{\alpha i} U_{\beta i}U_{\alpha j}U^*_{\beta j}) \sin(\Delta_{ji}). \label{Equation12}
\end{align}
	   	Here, we have used $\cos(\Delta_{ji})=1-2\sin^2\left(\dfrac{\Delta_{ji}}{2}\right)$. So, taking case - 1 ($i.e.$, $i=j$) and case - 2 ($i\neq j$) together, we can write,
  \begin{align}
    P(\nu_\alpha \rightarrow \nu_\beta) &= \sum_{i}\sum_{j} \Big[ U_{\beta i}U^*_{\alpha i}U_{\alpha i}U^*_{\beta i} 
    + Re\,(U^*_{\alpha i} U_{\beta i}U_{\alpha j}U^*_{\beta j}) \nonumber \\
    &\quad - 2Re\,(U^*_{\alpha i} U_{\beta i}U_{\alpha j}U^*_{\beta j}) \sin^2\left(\frac{\Delta_{ji}}{2}\right) \nonumber \\
    &\quad + Im\,(U^*_{\alpha i} U_{\beta i}U_{\alpha j}U^*_{\beta j}) \sin(\Delta_{ji}) \Big]. \label{Equation}
\end{align}
  Now,
\begin{align}
    \left(\sum_{i=1}^{3}U^*_{\alpha i}U_{\beta i}\right)^2 &= 
    \left(U^*_{\alpha 1}U_{\beta 1} + U^*_{\alpha 2}U_{\beta 2} + U^*_{\alpha 3}U_{\beta 3} \right)^2 \nonumber \\
    &= \sum_{i}(U^*_{\alpha i}U_{\beta i})^2 + \sum_{i \neq j} 2(U^*_{\alpha i} U_{\beta i}U_{\alpha j}U^*_{\beta j}). \label{Equation13}
\end{align}
   So, from the equation \ref{Equation13}, we can write the final generalized form of the equation of neutrino oscillation probability in $3\nu$ framework as \\
   \begin{align}
    P(\nu_\alpha \rightarrow \nu_\beta) &= \delta_{\alpha\beta} 
    - 4\sum_{i>j} Re\,(U^*_{\alpha i} U_{\beta i}U_{\alpha j}U^*_{\beta j}) \sin^2\left(\frac{\Delta_{ji}}{2}\right) \nonumber \\
    &\quad + 2\sum_{i>j} Im\,(U^*_{\alpha i} U_{\beta i}U_{\alpha j}U^*_{\beta j}) \sin(\Delta_{ji}). \label{Equation14}
\end{align}
If $i$ and $j$ take only values from 1 to 2, the above equation reduces to the two-flavor neutrino oscillation expression. The third term of equation \ref{Equation14} represents the CP violation. The PMNS matrix can be written as,
\begin{equation}
   	  U= 
   	   \begin{pmatrix}
	1 & 0 & 0 \\
	0 & c_{23} & s_{23} \\
	0 & -s_{23} & c_{23}
\end{pmatrix} 
\begin{pmatrix}
	c_{13} & 0 & s_{13}e^{-i\delta} \\
	0 & 1 & 0 \\
	-s_{13}e^{i\delta} & 0 & c_{13}
\end{pmatrix}
\begin{pmatrix}
	c_{12} & s_{12} & 0 \\
	-s_{12} & c_{12} & 0 \\
	0 & 0 & 1
\end{pmatrix} \nonumber
   \end{equation}
   \begin{align}
   & =
   	\begin{pmatrix}
   	   {c_{12}c_{13}} & {s_{12}c_{13}} & {s_{13}e^{-i\delta}}\\
   	  {-s_{12}c_{23}-c_{12}s_{13}s_{23}e^{i\delta}} & {c_{12}c_{23}-s_{12}s_{13}s_{23}e^{i\delta}} & {c_{13}s_{23}}\\
   	   {s_{12}s_{23}-c_{12}s_{13}c_{23}e^{i\delta}} & {-c_{12}s_{23}-s_{12}s_{13}c_{23}e^{i\delta}} & {c_{13}c_{23}}
   	   \end{pmatrix}
       \label{PMNS matrix}
   \end{align}
   \begin{align}
   & \equiv
   	\begin{pmatrix}
   	U_{\alpha 1} & U_{\alpha 2} & U_{\alpha 3}\\
   	U_{\beta 1} & U_{\beta 2} & U_{\beta 3}\\
   	U_{\gamma 1} & U_{\gamma 2} & U_{\gamma 3}
   	\end{pmatrix}.
   \end{align}
 Here, $c_{ij}$ and $s_{ij}$ stand for $\cos\theta_{ij}$ and $\sin\theta_{ij}$, respectively, where $i,j= 1,2,3; \,\,i\neq j$. We are ignoring the Majorana phases here as they do not affect the neutrino oscillation probability. The advantage of considering the PMNS matrix as a suitable unitary mixing matrix connecting the flavor basis and mass basis of neutrinos is that the ratios of some matrix elements of $U$ give us the values of the oscillation parameters distinctly. For example,\\
 \begin{equation}
     \tan^2\theta_{23} = \frac{|U_{\beta3}|^2}{|U_{\gamma3}|^2}, \\\nonumber
 \end{equation}
 and,  \begin{equation}
     \tan^2\theta_{12} = \frac{|U_{\alpha2}|^2}{|U_{\alpha1}|^2} \\\nonumber.
 \end{equation}
The 1-3 element of the PMNS matrix gives us the value of $\delta_{\mathrm{CP}}$ if the value of $\theta_{13}$ is well known. The ratios of other matrix elements of $U$ also give the value of the oscillation parameters, $e.g.$, 
\ \begin{equation}
     \cos^2\theta_{12} = \frac{\left| U_{\alpha1} \right|^2}{ 1 - \left|U_{\alpha3} \right|^2} \nonumber,\,\,\sin^2\theta_{12} = \frac{\left| U_{\alpha2} \right|^2}{ 1 - \left|U_{\alpha3} \right|^2} \nonumber, \\
   \cos^2\theta_{23} = \frac{\left| U_{\gamma3} \right|^2}{ 1 - \left|U_{\alpha3} \right|^2} \nonumber,\,\,\sin^2\theta_{23} = \frac{\left| U_{\beta3} \right|^2}{ 1 - \left|U_{\alpha3} \right|^2}. \nonumber \\  
\end{equation}
Now, substituting the values of the matrix elements from equation \ref{PMNS matrix} to equation \ref{Equation14}, we get the final expression of neutrino oscillation probability in vacuum. If we focus on appearance channels only ($i.e.,\, \alpha \neq \beta$ in equation \ref{Equation14}) and keep the terms upto second order of ($\alpha\cdot\sin(2\theta_{13})$), where $\alpha=\dfrac{\Delta m^2_{21}}{\Delta m^2_{31}}$, then equation \ref{Equation14} takes the following form \cite{Akhmedov:2004ny}.\\

\begin{align}
    P(\nu_{\mu} \rightarrow \nu_{e}) \simeq 
    &\ \bm{\sin^2\theta_{23}} \sin^2(2\theta_{13}) 
    \sin^2\Delta \nonumber \\
    &+ (\alpha\Delta) \sin(2\theta_{13})\sin(2\theta_{12})\sin(2\theta_{23}) \bm{\sin (\delta_{\mathrm{CP}})} \sin^2(\Delta) \nonumber \\
    &+ (\alpha\Delta) \sin(2\theta_{13})\sin(2\theta_{12})\sin(2\theta_{23}) \bm{\cos( \delta_{\mathrm{CP}})} \cos(\Delta)\sin(\Delta) \nonumber \\
    &+ (\alpha\Delta)^2 \cos^2\theta_{23} \sin^2(2\theta_{12}).
    \label{Four_term_app_prob}
\end{align}
\par The first term in the equation \ref{Four_term_app_prob} is the leading term as it does not depend on $\alpha$ (as $\alpha<<1$) and governed by the atmospheric mixing angle $\theta_{23}$. Here, $\Delta=\dfrac{\Delta m^2_{31}\,L}{4E}$. Note that, this term depends on $\sin^2\theta_{23}$ and sensitive to the octant of $\theta_{23}$, hence the $\nu_\mu\rightarrow\nu_e$ appearance channel helps to resolve the octant of $\theta_{23}$. Although, there is a sub-leading effect of the second term, it gives the signature of CP violation as $\sin(\delta_{\mathrm{CP}})$ changes sign as $\delta_{\mathrm{CP}}$ changes its sign when we go from neutrino to antineutrino, $i.e.,$ it is a CP-odd term. Hence, the $\nu_\mu\rightarrow\nu_e$ appearance channel helps to study CP violation (Dirac $\bm{\delta_{\mathrm{CP}}}$) in the neutrino sector. The third term is the CP-even term. The last term is the solar term and depends on the solar mixing angle $\theta_{12}$. But, its contribution in the appearance channel is very much suppressed due to its $\alpha^2$ dependence.\\
\par Similarly, we can study some features of the $\nu_\mu\rightarrow\nu_\mu$ disappearance probability in vacuum by writing its expression following the previous approximations,
\begin{align}
    P(\nu_{\mu} \rightarrow \nu_{\mu}) \simeq 
    &\ 1- \bm{\sin^22\theta_{23}} \sin^2(\Delta) \nonumber \\
    &+ 4\sin^2\theta_{13}\sin^2\theta_{23}\cos(2\theta_{23})\sin^2\Delta \nonumber\\
    &+ (\alpha\Delta) \sin2\Delta [\cos^2\theta_{12}\sin^22\theta_{23}-2\sin\theta_{13}\sin2\theta_{12}\sin^2\theta_{23}\sin2\theta_{23}\bm{\cos\delta_{\mathrm{CP}}}] \nonumber \\
    &- (\alpha\Delta)^2 [\sin^22\theta_{12}\cos^2\theta_{23}+\cos^2\theta_{12}\sin^22\theta_{23}(\cos2\Delta-\sin^2\theta_{12})].
    \label{Four_term_disapp_prob}
\end{align}
The first salient feature of the above expression is that the leading term is proportional to $\sin^22\theta_{23}$ and therefore it is not sensitive to the octant of $\theta_{23}$, but its strength is larger than the $\nu_\mu\rightarrow\nu_e$ appearance probability in equation \ref{Four_term_app_prob}, which is suppressed by $\sin^2(2\theta_{13})$. Hence, the disappearance channel is rich in statistics. Furthermore, this channel is not sensitive to the CP violation as it only contains $\cos\delta_{\mathrm{CP}}$ term. The motivations behind showcasing only these two channels throughout this thesis are i) the $\nu_\mu\rightarrow\nu_e$ appearance channel \cite{Huber:2002mx, Cervera:2000kp} is sensitive to two of the most important unresolved issues in the $3\nu$ paradigm, $i.e.,$ CP violation and exclusion of the wrong octant solution of $\theta_{23}$, ii) the $\nu_\mu\rightarrow\nu_\mu$ disappearance channel ensures to bring good precision in measuring the atmospheric oscillation parameters $\theta_{23}$ and $\Delta m^2_{31}$ owing to high statistics in this channel in long-baseline (LBL) experiments (for details, see the next chapter on long-baseline experiments). For antineutrinos, the sign of $\delta_{\mathrm{CP}}$ changes its sign ($i.e., P(\nu_\alpha\rightarrow\nu_\beta;\,\delta_{\mathrm{CP}})\rightarrow P(\bar{\nu}_\alpha\rightarrow\bar{\nu}_\beta;\,-\delta_{\mathrm{CP}})$).
\par The CP-odd asymmetry in neutrino oscillation can be written as,
\begin{align}
    \Delta P_{\alpha \beta} &= P(\nu_\alpha\rightarrow\nu_\beta) - P(\bar{\nu}_\alpha\rightarrow\bar{\nu}_\beta) \nonumber\\
    &= \pm 16J\sin\left(\frac{\Delta m^2_{21}L}{4E}\right) \sin\left(\frac{\Delta m^2_{31}L}{4E}\right) \sin\left(\frac{\Delta m^2_{32}L}{4E}\right).
    \label{Delta P}
\end{align}
Here, $\Delta P_{\alpha \beta}$ denotes the amount of CP-odd asymmetry in the neutrino sector, $J$ is the Jarlskog invariant ($J\equiv Im[U_{e1}U^*_{\mu 1}U^*_{e 2}U_{\mu 2}]$) \cite{Jarlskog:1985ht, Jarlskog:1985cw}, which is a constant quantity irrespective of the choice of basis, and can be expressed under the standard PMNS parameterization as,
\begin{equation}
    J = \frac{1}{8} \cos\theta_{13} \sin (2\theta_{12}) \sin (2\theta_{13}) \sin (2\theta_{23}) \sin \delta_{\mathrm{CP}}.\\
    \label{Jarlskog invariant}
\end{equation}
Equations \ref{Delta P} and \ref{Jarlskog invariant} denote the necessary conditions for CP violation in the neutrino sector \cite{Antusch:2001ck}, $i.e.,$%
\begin{itemize}[topsep=-0.5pt, itemsep=10pt, parsep=0pt, partopsep=0pt]
    \item the mixing angles $\neq 0^\circ, \, 90^\circ,$
    \item the value of $\delta_{\mathrm{CP}} \neq 0^\circ, \pm 180^\circ$,
    \item the value of $\Delta m^2_{ij} \neq 0$.
\end{itemize}

\section{Manifestation of matter effect in neutrino oscillation}
\label{sec:2.3}
Neutrinos interact with the ambient matter via weak interactions and it can modify the oscillation probabilities of neutrinos. The first experimental evidence of matter effects in neutrino oscillations came from the solar neutrino experiments, particularly the Sudbury Neutrino Observatory (SNO) \cite{SNO:2002hgz} and Super-Kamiokande \cite{Super-Kamiokande:2013mie}, around the early 2000s. These experiments confirmed the Mikheyev-Smirnov-Wolfenstein (MSW) effect \cite{Wolfenstein:1977ue, Mikheyev:1989dy, Mikheyev:1985zog}, which describes how the oscillation probability of neutrinos gets modified as they travel through dense matter.
\par The modified Hamiltonian incorporating the neutrino-matter interaction can be written as,
\begin{equation}
    \hat{H}_{eff} = \hat{H}_{vac}+ \hat{H}_{matt}.
    \label{eq:general hamiltonian matter effect}
\end{equation}
To have a close look at how the eigenvalues are modified under the influence of the neutrino-matter interaction, we take the help of the time-dependent Schr\"{o}dinger equation, $i.e.,$
\begin{equation}
    i \frac{d}{dt} \ket{\nu(t)} = \hat{H}_{eff} \ket{\nu(t)}.
    \label{Schrodinger eqn}
\end{equation}
This equation \ref{Schrodinger eqn} can be rewritten in the flavor eigenstate as,
\begin{equation}
    i \frac{d}{dt} \nu_{\alpha} (t) = \sum_{\beta} H_{\alpha\beta} \nu_{\beta} (t),
\end{equation}
where $\alpha,\beta$ denote the flavor indices and $H_{\alpha\beta}= \langle \nu_\alpha | H | \nu_\beta \rangle$ are the matrix elements. On the other hand, $\ket{\nu_k}$ denotes the mass eigenstate and hence the eigenstate of the free Hamiltonian $\hat{H}_{vac},\, i.e.,$
\begin{equation}
\hat{H}_{vac} \ket{\nu_k} = E_k\ket{\nu_k},
\label{vacuum equation}
\end{equation}
where, $E_k$ denotes the eigenvalue of the free Hamiltonian and can be written as $E_k=\sqrt{\vec{p}^2+m^2_k}$, $\vec{p}$ being the momentum of the mass eigenstate $\ket{\nu_k}$. Neutrinos interact with the electrons of the ordinary matter through charged-current (CC) interaction via the $W^\pm$ boson of the SM. This interaction induces a matter potential confronting which the neutrino changes its flavor while propagating through a medium. This potential can be depicted as,
\begin{align}
    V_{CC,\alpha} &= \sqrt{2} G_F N_e, \quad [\alpha = e] \\\nonumber
    &= 0, \quad [\alpha = \mu,\tau].
    \label{eq:V_cc}
\end{align}
Neutrinos mainly interact with the ambient electron of the matter while performing charged current interaction. There is another kind of interaction of neutrinos with the matter via the neutral mediator $Z^0$ boson of the SM, hence it is called neutral current (NC) interaction \cite{Linder:2005fc}, the matter potential of which can be expressed as,
\begin{align}
    V_{NC,\alpha} &= -\frac{G_F}{\sqrt{2}} N_n, \quad [\alpha = e, \mu, \tau], \\\nonumber,
\end{align}
where, $N_e$ and $N_n$ signify the electron number and neutron number densities of the matter, and $G_F$ denotes the Fermi coupling constant 
 ($G_F=1.166\times10^{-5}$ GeV$^{-2}$). Neutrinos interact with the matter via coherent and forward scattering \cite{COHERENT:2020iec}. Due to the scarcity of naturally produced muon and tau in the Earth matter, CC interaction only occurs with the electrons of the medium. Now, the CC interaction potential can be expressed in terms of some measurable parameters used in the experiments directly, $i.e.,$ line-averaged Earth matter density  $\rho_{\text{avg}}$ \cite{Kelly:2018kmb} and the above-mentioned number densities ($i.e., N_e \,\mathrm{and}\, N_n$) as the following fashion,
 \begin{equation}
     V_{CC}\approx 7.6\times\left(\frac{N_e}{N_p+N_n}\right)\times\left[\frac{\rho_{\text{avg}}}{10^{14} \,\mathrm{g/cm^{3}}}\right] \mathrm{eV}.
     \label{CC matter pot}
 \end{equation}
Considering further the medium of interaction is electrically neutral ($i.e., N_e=N_p$) and isoscaler ($i.e., N_p=N_n$), equation \ref{CC matter pot} can be rewritten as,
\begin{equation}
    V_{CC} = 7.6 \times 0.5 \times\left[\frac{\rho_{\text{avg}}}{10^{14}\, \mathrm{g/ cm^{3}}}\right] \mathrm{eV}.
\end{equation}
The total matter potential $V(\nu)_\alpha = V_{CC,\alpha}+V_{NC,\alpha}$ changes its polarity while considering antineutrinos instead of neutrinos $i.e., V(\nu)_\alpha = -V(\bar{\nu})_\alpha $. This $V_\alpha$ is the eigenvalue of the Hamiltonian $\hat{H}_{\mathrm{matt}}$ in weak basis, $i.e.,$
\begin{equation}
    \hat{H}_{\mathrm{matt}}\ket{\nu_\alpha}= V_\alpha\ket{\nu_\alpha}.
\end{equation}
 At tree level, neutral-current interactions mediated by the $Z^0$ boson are flavour universal, giving identical contributions to all active neutrinos, and hence it does not affect the oscillation probability. However, at the one-loop level, radiative corrections to the neutrino--$Z^0$ vertex involve virtual $W$ bosons and the corresponding charged lepton partner. Since these corrections depend on the charged lepton mass, the large difference between $m_\mu$ and $m_\tau$ breaks flavor universality, leading to a small but finite splitting between the effective matter potentials of $\nu_\mu$ and $\nu_\tau$~\cite{Botella:1986wy}. The difference between the matter potential confronted by $\nu_\mu$ and $\nu_\tau$ can be written as \cite{Mirizzi:2009td},
	\begin{align}
		\Delta V_{\tau\mu}=V_\tau-V_\mu=-\dfrac{3\sqrt{2}}{2\pi^2}G_Fm^2_\tau\bigg(ln \dfrac{m_W}{m_\tau} -\dfrac{1}{2}\bigg)\dfrac{N_e}{m^2_W},
		\label{eq:Higher_MSW_analysis}
	\end{align}
	where, $m_\tau$ and $m_W$ denote mass of tau and $W$ boson. For Earth matter, the relative contribution of this splitting with respect to the potential $V_{\text{CC,e}}$ is
	\begin{align}
		\dfrac{\Delta V_{\tau\mu}}{V_{\text{CC,e}}}\sim -2.5\times10^{-4}.
	\end{align}
	Although in standard interaction of active neutrinos, this effect is negligible, but from the context of dense environments like supernova, this slight difference between the interaction potentials becomes significant. It is the first non-zero splitting for the NC interaction of $\nu_\mu$ and $\nu_\tau$ (or in the 2-3 sector) which could affect neutrino oscillation in dense media.
	
 From equation \ref{vacuum equation}, it is evident that the energy of the neutrinos is the eigenvalues of the vacuum Hamiltonian in mass basis. To convert it into the physically measurable flavor basis, we need to diagonalize it through the unitary mixing matrix $U$ by the following equation
\begin{equation}
    \hat{H}^f_{vac}= U\hat{H}^m_{vac}U^\dagger.
\end{equation}. 
In the matrix form, it can be written as,
\begin{equation}
    H^f = \mathbf{U}
    \begin{bmatrix}
        E_1^m & 0 & 0 \\
        0 & E_2^m & 0 \\
        0 & 0 & E_3^m
    \end{bmatrix}
    \mathbf{U}^\dagger.
\end{equation}
After invoking the influence of the matter effect in our picture, this mixing matrix $U$ cannot diagonalize the effective Hamiltonian $\hat{H}_{eff}$, and it will not be diagonal even on the previous mass basis, considered so far. Hence, we need to construct a new type of mass basis (say, $\ket{\nu^m_i}$) where $\hat{H}_{vac}$ is diagonal. In this matter-modified mass basis $\ket{\nu^m_i}$, the diagonalized form of the effective Hamiltonian will be $\hat{H}^m=diag\, (E^m_1, E^m_2, E^m_3)= diag\, \frac{1}{2E}(\Tilde{m}^2_1, \Tilde{m}^2_2, \Tilde{m}^3_3)=\Tilde{U}\hat{H}_{eff} \Tilde{U}^\dagger$. Here, $\Tilde{U}$ is the matter-modified mixing matrix (kindly note that, $\ket{\nu^m_i}\neq\ket{\nu_i},\,\Tilde{U}^m\neq U$). Like the previous case, here also we assume equal momentum approximation ($i.e., E_1=E_2=E_3=E$) and $\Tilde{m}^2_i$ signifies the matter-modified mass squared term. So, the flavor basis can now be expressed in terms of the matter-modified mass basis through this relation $\ket{\nu_\alpha}=\Tilde{U}\ket{\nu^m_i}$. Adorned with this new mass basis and new mixing matrix, we can show the neutrino oscillation probability while neutrinos interact through matter with the two new plug-ins, $U_{\alpha i}\rightarrow \Tilde{U}^m_{\alpha i}$ and $E_i\rightarrow E^m_i$, where $E^m_i$ is the matter-modified energy eigenvalue of each mass state. Now, following the footprints of the same calculations mentioned in the section \ref{sec:2.2}, we can write down the expression of the neutrino oscillation probability under the influence of the matter effect as,
\begin{align}
    P(\nu_\alpha \rightarrow \nu_\beta) &= \delta_{\alpha\beta} 
    - 4\sum_{i>j} Re (\Tilde{U}^*_{\alpha i} \Tilde{U}_{\beta i}\Tilde{U}_{\alpha j}\Tilde{U}^*_{\beta j}) \sin^2\left(\frac{\Tilde{\Delta}_{ji}}{2}\right) \nonumber \\
    &\quad + 2\sum_{i>j} Im (\Tilde{U}^*_{\alpha i} \Tilde{U}_{\beta i}\Tilde{U}_{\alpha j}\Tilde{U}^*_{\beta j}) \sin(\Tilde{\Delta}_{ji}), \label{eq: matt modified oscillation probability}
\end{align}
where, $\Tilde{\Delta}_{ji} = \dfrac{\Delta\Tilde{m}^2_{ji}L}{4E}$ denotes the modified frequency term with two measurable quantities, baseline length ($L$ in km) and the neutrino energy ($E$ in GeV).

\subsection{Overview of the matter effect in two-flavor neutrino oscillation}
To understand the influences of the matter effect on the neutrino oscillation and the modifications of its expression, we consider the simple two-flavor oscillation first. Following the equation \ref{eq:general hamiltonian matter effect} and \ref{eq:V_cc}, we can rewrite the effective Hamiltonian in the matrix form as,
\begin{align}
\label{eq:2 flavor matter effect}
    \hat{H}_{\text{eff}} &= \mathbf{U}
    \begin{bmatrix}
         0 & 0 \\
         0 & \dfrac{\Delta m^2}{2E}
    \end{bmatrix}
    \mathbf{U}^\dagger
    + \begin{bmatrix}
         V_{CC} & 0 \\
         0 & 0
    \end{bmatrix} \\
    &= \begin{bmatrix}
         \cos\theta & \sin\theta \\
         -\sin\theta & \cos\theta
    \end{bmatrix}
    \begin{bmatrix}
         0 & 0 \\
         0 & \dfrac{\Delta m^2}{2E}
    \end{bmatrix}
    \begin{bmatrix}
         \cos\theta & -\sin\theta \\
         \sin\theta & \cos\theta
    \end{bmatrix}
    + \begin{bmatrix}
         V_{CC} & 0 \\
         0 & 0
    \end{bmatrix} \nonumber \\
    &= \dfrac{\Delta m^2}{4E}
    \begin{bmatrix}
         \left(1-\cos2\theta+\dfrac{4EV_{CC}}{\Delta m^2}\right) & \sin2\theta \\
         \sin2\theta & 1+\cos2\theta
    \end{bmatrix}. \nonumber
\end{align}
If we diagonalize the matrix mentioned in \ref{eq:2 flavor matter effect}, then the eigenvalues are,
\begin{align}
    E^m_1=\frac{V_{CC}}{2}+\frac{\Delta m^2}{4E} \left[1-\sqrt{\sin^22\theta+\left(\cos2\theta-\frac{2EV_{CC}}{\Delta m^2}\right)^2}\right]\\
   E^m_2=\frac{V_{CC}}{2}+\frac{\Delta m^2}{4E} \left[1+\sqrt{\sin^22\theta+\left(\cos2\theta-\frac{2EV_{CC}}{\Delta m^2}\right)^2}\right]\nonumber.
\end{align}
Now, we can parameterize the modified mixing matrix $\Tilde{U}^m$ with the matter-modified oscillation parameter $\theta^m$ and write the form of $\Tilde{U}^m$ as,
\begin{align}
    U^m &= 
    \begin{bmatrix}
         \cos\theta^m & \sin\theta^m \\
         -\sin\theta^m & \cos\theta^m
    \end{bmatrix}.
    \end{align}
The link between the matter-modified oscillation parameter ($\theta^m$) and the same in vacuum can be expressed as,
\begin{equation}
    \sin2\theta^m=\frac{\sin2\theta}{\sqrt{\sin^22\theta+\left[\cos2\theta-\dfrac{2EV_{CC}}{\Delta m^2}\right]^2}}.
    \label{resonance condition}
\end{equation}
If, in the denominator of the equation \ref{resonance condition}, $\cos2\theta=\dfrac{2EV_{CC}}{\Delta m^2}$ , $\theta^m$ will be $45^\circ$ (maximal mixing value). This phenomenon is called ``MSW (Mikhev-Smirnov-Wolfenstein) resonance" \cite{Smirnov:2019kto}, and the expression $\cos2\theta=\dfrac{2EV_{CC}}{\Delta m^2}$ is the criterion of this resonance. Neutrinos can show the MSW resonance if $\Delta m^2>0$, whereas antineutrinos can show it when $\Delta m^2<0$, as $V_{CC}$ changes its polarity for antineutrinos. 
\par Now, following the recipes used to derive the equation \ref{eq:2 flavor oscillation probability}, the revised expression of the probability of neutrino oscillation from a flavor $\alpha$ to a flavor $\beta$ is,
\begin{align}
P(\nu_{\alpha} \rightarrow \nu_{\beta}) &= \sin^2(2\theta^m) \sin^2\left( 1.27 \times \frac{\Delta \Tilde{m}^{2} L}{E} \right) \\
&= \sin^2(2\theta^m) \sin^2\left[\frac{\Delta m^2 L}{4E}\sqrt{\sin^22\theta+\left(\cos2\theta-\frac{2EV_{\mathrm{CC}}}{\Delta m^2}\right)^2}    \right].
\label{eq:2 flavor oscillation probability matter}
\end{align}
\par Some key takeaways from the equation \ref{eq:2 flavor oscillation probability matter} are,
\begin{itemize}
    \item If we change the polarity of $\Delta m^2$, the oscillation probability is affected due to the change of the $\theta^m$ and $\Delta \Tilde{m}^2$. So, the matter effect is a good portal to investigate the correct mass hierarchy in Nature.
    \item If we consider the extreme values of the matter potential V, then for $V=0$, the equation \ref{eq:2 flavor oscillation probability matter} reduces to the equation \ref{eq:2 flavor oscillation probability}. On the other hand, if $V\rightarrow \infty$, then the oscillation probability vanishes due to the lack of oscillation in the presence of an infinitely dense medium.
    \item Non-degenerate value of $m^2$ is crucially required for the neutrino oscillation either in vacuum or in matter.   
\end{itemize}

\subsection{The influence of matter effect in $3\nu$ framework}
By advancing our next step to the experimentally confirmed three-flavor case, we just expand the matrix form of the Hamiltonian [from equation \ref{eq:general hamiltonian matter effect}] in $3\times3$ dimension by introducing one additional mass-squared splitting, $i.e.$, atmospheric mass-splitting ($\Delta m^2_{31}$). The form of the effective Hamiltonian introducing matter effect in three flavors \cite{Agarwalla:2021zfr} is,
\begin{align}
    \hat{H}_{\text{eff}} &= \mathbf{U}
    \begin{bmatrix}
         0 & 0 & 0\\
         0 & \dfrac{\Delta m^2_{21}}{2E} & 0\\
         0 & 0 & \dfrac{\Delta m^2_{31}}{2E}
    \end{bmatrix}
    \mathbf{U}^\dagger
    + \begin{bmatrix}
         V_{CC} & 0 & 0\\
         0 & 0 & 0\\
         0 & 0 & 0\\
    \end{bmatrix}. 
    \label{eq:3 flavor matter effect}
\end{align}
Note that, here, $U$ is the ($3\times3$) unitary mixing matrix and is considered as the usual PMNS matrix for convenience. Now, we have to follow the same way we show in the previous section. But, to construct a new basis where the matter-modified Hamiltonian will be diagonalized in $3\nu$ framework is a bit tideous job. However, after diagonalizing the Hamiltonian, we land up to the following expression of matter-modified neutrino oscillation probability in $\nu_\mu\rightarrow\nu_e$ appearance channel by taking the approximation of considering only upto the 2nd order of ($\alpha\cdot\sin(2\theta_{13})$) \cite{Akhmedov:2004ny}.
\begin{align}
    P(\nu_{\mu} \rightarrow \nu_{e}) \simeq 
    &\ \bm{\sin^2\theta_{23}} \sin^2(2\theta_{13}) 
   \frac{\sin^2[(A-1)\Delta]}{(A-1)^2} \nonumber \\
    &+ \alpha \sin(2\theta_{13})\sin(2\theta_{12})\sin(2\theta_{23}) \bm{\sin (\delta_{\mathrm{CP}})} \sin(\Delta) \frac{\sin(A\Delta)}{A} \bm{\frac{\sin[(A-1)\Delta]}{(A-1)}}\nonumber \\
    &+ \alpha \sin(2\theta_{13})\sin(2\theta_{12})\sin(2\theta_{23}) \cos (\delta_{\mathrm{CP}}) \cos(\Delta) \frac{\sin(A\Delta)}{A} \frac{\sin[(A-1)\Delta]}{(A-1)} \nonumber \\
    &+ \alpha^2 \cos^2\theta_{23} \sin^2(2\theta_{12}) \frac{\sin^2(A\Delta)}{A^2},
    \label{Four_term_app_prob_matter}
\end{align}
where, $A=\pm\dfrac{2\sqrt{2}G_FN_eE}{\Delta m^2_{31}}$ stands for the dimensionless normalized matter potential [positive (negative) sign is for neutrino (antineutrino)] and $\Delta = \dfrac{\Delta m^2_{31}L}{4E}$. 
The same characteristic features of this equation through $\nu_\mu\rightarrow\nu_e$ appearance channel mentioned in \ref{Four_term_app_prob} hold good in the equation \ref{Four_term_app_prob_matter} also. Additionally, we highlight one more fact crucially contributed only by the matter effect in this channel. In the second term ($i.e.$, CP-violating term) of equation \ref{Four_term_app_prob_matter}, if we consider the interactions of antineutrinos with the matter, then the sign of the matter potential will be changed, \textit{i.e.,} $A\rightarrow -A $ and $\delta_{\mathrm{CP}}\rightarrow -\delta_{\mathrm{CP}}$. Consequently, the value of the term $\dfrac{\sin[(A-1)\Delta]}{(A-1)}$ will be changed. If the value of $\delta_{\mathrm{CP}}$ becomes zero, the matter term $\dfrac{\sin[(A-1)\Delta]}{(A-1)}$ will still affect the value of the appearance probability, producing a matter-induced (fake) CP violation. In this case, even if we do not consider the change of the sign of intrinsic $\delta_{\mathrm{CP}}$ ($i.e.,\,\sin(\delta_{\mathrm{CP}})$), the value of the last term ($i.e.,\,\dfrac{\sin[(A-1)\Delta]}{(A-1)}$) will be changed.\\ Thus, it is difficult to understand if the value of appearance probability changes due to intrinsic $\delta_{\mathrm{CP}}\,(i.e., \sin(\delta_{\mathrm{CP}}))$ term or due to the contribution of matter effect ($i.e., \dfrac{\sin[(A-1)\Delta]}{(A-1)}$). Therefore, the matter effect mimics a CP-violating signal in this channel, creating ambiguity in identifying whether the observed change in appearance probability originates from intrinsic $\delta_{\mathrm{CP}}$ or from matter-induced effects. This confusion is enhanced if the strength of the matter potential ($A$) becomes relatively high due to the higher line-averaged density of the Earth matter. Hence, we focus on an important point from the above expression. The $\nu_\mu\rightarrow\nu_e$ appearance channel is contaminated with fake CP violation \cite{deGouvea:2016pom, Agarwalla:2022xdo}. The amount of contamination increases with the increment of the value of the line-averaged Earth matter density. So, this channel solely is not a good probe to measure the intrinsic $\delta_{\mathrm{CP}}$ in Nature. 
\par Similarly for the $\nu_\mu\rightarrow\nu_\mu$ disappearance channel, we can write the oscillation probability under the influence of matter effect as,
\begin{samepage}
\begin{flalign}
     P(\nu_\mu \to \nu_\mu) \nonumber\\
    &= 1 - \sin^2 2\theta_{23} \sin^2 \Delta  - 4 s_{13}^2 s_{23}^2 \frac{\sin^2 (A-1)\Delta}{(A-1)^2}\nonumber \\
    &+ \alpha^2 \sin^2 2\theta_{12} c_{23}^2 \frac{\sin^2 A \Delta}{A^2} - (\alpha \Delta)^2 c_{12}^4 \sin^2 2\theta_{23} \cos 2\Delta  \nonumber \\
    &+ \frac{1}{2A} \alpha^2 \sin^2 2\theta_{12} \sin^2 2\theta_{23} \left( \sin \frac{\sin A \Delta}{A} \cos (A-1)\Delta - \frac{\Delta}{2} \sin 2\Delta \right) \nonumber \\
    &- \frac{2}{A-1} s_{13}^2 \sin^2 2\theta_{23} \left( \sin A \Delta \cos A \Delta \frac{\sin (A-1)\Delta}{(A-1)} - \frac{A\Delta}{2} \sin 2\Delta \right) \nonumber \\
    &- 2\alpha s_{13} \sin 2\theta_{12} \sin 2\theta_{23} \cos \delta_{\text{CP}} \cos \Delta \frac{\sin A \Delta}{A} \frac{\sin (A-1)\Delta}{(A-1)} \nonumber \\
    &+ \frac{1}{A-1} \alpha s_{13} \sin 2\theta_{12} \sin 4\theta_{23}\bm{\cos \delta_{\text{CP}}} \sin \Delta\left[A \sin \Delta - \frac{\sin A \Delta}{A} \cos (A-1)\Delta)\right], &&
    \label{eq:disapp oscillation prob matter}
\end{flalign}
\end{samepage}
where, $c_{ij}\, (s_{ij})$ stands for $\cos\theta_{ij}\,(\sin\theta_{ij})$. The polarities of $A$ and intrinsic $\delta_{\mathrm{CP}}$ are changed if we switch from neutrino to antineutrino. Similarly, considering neutrinos also, if we shift from the normal mass ordering (NMO, \textit{i.e.,} $\Delta m^2_{31}>0$) to the inverted mass ordering (IMO, \textit{i.e.,} $\Delta m^2_{31}<0$), there will be a sign flip of $A$ as it depends on the atmospheric mass-splitting. From equation \ref{eq:disapp oscillation prob matter}, it is evident that, $\nu_\mu\rightarrow\nu_\mu$ disappearance channel is statistically enriched, and hence is utilized to investigate any phenomenology demanding the statistical dominance (e.g., bringing improved precision on measuring the oscillation parameters by reducing their uncertainties). The cons of this channel are that the probability is not free from the degeneracy of $\theta_{23}$ and hence is not helpful to exclude the $\theta_{23}$ octant. It is also not helpful to probe intrinsic $\delta_{\mathrm{CP}}$ in Nature due to containing even terms of $\delta_{\mathrm{CP}}$.

\section{A keen insight into different neutrino oscillation parameters, their current status, the eight-fold degeneracy, and some pending puzzles in $3\nu$ framework}
\label{sec:2.4}
The cornerstone of three-flavor oscillation was laid by the LEP experiment \cite{ALEPH:2005ab} in 1990 through the invisible decay of $Z$ boson and is firmly established by the discovery of non-zero $\theta_{13}$ in the Daya Bay experiment in 2012 \cite{DayaBay:2012fng, DayaBay:2022orm} with adequate statistical significance. Throughout this thesis, we maintain the unitarity property of the mixing matrix between flavor basis and mass basis. For a $N\times N$ unitary mixing matrix, there should be $N(N-1)$ independent parameters; $i.e., \dfrac{N(N-1)}{2}$ number of mixing angles, $\dfrac{(N-1)(N-2)}{2}$ number of Dirac CP-phases, and $(N-1)$ number of Majorana CP-phases. After standing on the ground of $3\nu$ landscape ($i.e., N=3$),  there should be three mixing angles ($\theta_{12}, \theta_{13}$, and $\theta_{23}$), one Dirac $\delta_{\mathrm{CP}}$ phase, and two Majorana CP-phases \cite{Zhukovsky:2016bee}. As Majorana phases do not affect the neutrino oscillation probability, we do not consider them in our discussion. We see that non-degenerate mass states \cite{Raychaudhuri:2000zn} are pivotal for the mass-induced neutrino oscillation. Hence, two more oscillation parameters are important to consider, solar mass-splitting ($\Delta m^2_{21}$) and atmospheric mass-splitting ($\Delta m^2_{31}$). These six independent neutrino oscillation parameters construct the backbone of the oscillation perspective of active neutrinos in the three-flavor framework. In this section, we give an overview on these neutrino oscillation parameters along with their experimental confirmations. Unlike the precisely measured values of the oscillation parameters in the quark oscillations, neutrino oscillation parameters bear a wide range of uncertainties. The most uncertain neutrino oscillation parameter is Dirac $\delta_{\mathrm{CP}}$ followed by $\theta_{23}$ \cite{Capozzi:2021fjo}. However, due to the advancement of several present and future experimental setups, it becomes possible to reduce this uncertainty significantly, and hence oscillation phenomenology has stood in the domain of the precision era \cite{Abada:2025jpk}. There are some remaining puzzles in the three-flavor landscape. We still do not know, in Nature, whether lower octant (LO) or higher octant (HO) of $\theta_{23}$ reigns; this is called ``octant ambiguity" \cite{Agarwalla:2013ju}. We also do not know precisely, if the mass states of the active neutrinos are ordered according to the normal mass ordering or the inverted mass ordering; this is called ``mass hierarchy ambiguity" \cite{Petcov:2005rv, Blennow:2013oma}. We still do not know about the existence of CP violation in Nature, to establish the matter-antimatter asymmetry in the leptonic sector \cite{Branco:2011zb, Singh:2024nvt}. 
\par Although the six neutrino oscillation parameters ($i.e., \theta_{12}, \theta_{13}, \theta_{23}, \Delta m^2_{21}, \Delta m^2_{31}, \delta_{\mathrm{CP}}$) are independent quantities, they are however correlated among each other. The expression of oscillation probabilities ($i.e.,$ equation \ref{Four_term_app_prob_matter} and \ref{eq:disapp oscillation prob matter}) reflect that for some sets of the values of some specific oscillation parameters ($i.e., \theta_{23}, \delta_{\mathrm{CP}}$, and the sign of $\Delta m^2_{31}$) represent the same oscillation probability, which is called $degeneracy$ \cite{Kajita:2006bt}. The first discussion of degeneracies in neutrino oscillations dates back to the late 1990s and early 2000s when long-baseline experiments such as K2K, MINOS, and T2K started probing the $\nu_\mu\rightarrow\nu_e$ appearance. Barger \cite{Barger:2001yr}, Giunti, and Marfatia (2001) demonstrated that the neutrino oscillation probability exhibits intrinsic degeneracy, wherein distinct sets of values for the CP-violating phase ($\delta_{\mathrm{CP}}$) and the reactor mixing angle $\theta_{13}$ yield nearly identical transition probabilities, making their independent determination ambiguous. This is called \textit{intrinsic degeneracy}, and it is two-fold. However, after the remarkable precision measurement of $\theta_{13}$ by Daya Bay \cite{DayaBay:2012fng, DayaBay:2022orm}, this degeneracy is resolved \cite{Kajita:2006bt}. Burguet-Castell et al. (2002) extended this analysis by examining the effects of \textit{mass hierarchy degeneracy} \cite{Burguet-Castell:2002ald}. They showed that oscillation probabilities remain symmetric when the sign of $\Delta m^2_{31}$ is reversed, leading to a two-fold degeneracy between the normal mass ordering (NMO) and the inverted mass ordering (IMO). Minakata and Nunokawa (2003) proposed the idea of another two-fold $octant\, degeneracy$ \cite{Nunokawa:2006mc}, pointing out that since oscillation probabilities depend on \(\sin^2\theta_{23}\), they cannot differentiate between the lower octant (LO: \(\theta_{23} < 45^\circ\)) and the higher octant (HO: \(\theta_{23} > 45^\circ\)). So, there exists a total $2\times2\times2=8$ fold degeneracies as all three degeneracies mentioned are independent of each other. Phenomenological studies of any parameters mentioned here affect to the uncertainty of the other conjugate parameters through these degeneracies. Hence, a degeneracy-free study is one of the main goals of oscillation phenomenology through different methodologies. Due to the different interactions of neutrinos and antineutrinos with matter, long-baseline experiments like NO$\nu$A, T2K, and DUNE leverage this asymmetry to determine the true mass hierarchy. Exceptional measurement of the value of $\theta_{13}$ by Daya Bay \cite{DayaBay:2012fng, DayaBay:2022orm} helps to break the intrinsic $(\theta_{13}-\delta_{\mathrm{CP}})$ degeneracy. The spectral analysis also opens an avenue to tackle several degeneracies in the upcoming long-baseline experiments like DUNE due to its enriched energy resolutions \cite{Agarwalla:2021bzs}. The synergistic approach of different experiments also helps to reduce these degeneracies \cite{Agarwalla:2024kti}.
\subsection{Oscillation parameters in solar (1-2) sector} 
Photons coming out of the Sun's outer surface undergo multiple scattering while coming from the center to the surface of the Sun. On the other hand, neutrinos can penetrate all the inner layers of the Sun and come to the Sun's surface almost confronting no interactions. Hence, solar neutrinos are important as a very fast messenger of the information at Sun's core. Solar neutrinos ($i.e., \nu_e$) is generated in the Sun's core through the fusion process, called pp-chain \cite{Redchuk:2020hjv} and CNO cycle \cite{Bahcall:1964gx}.
The main reaction of pp-chain helping to neutrino synthesis is,\\
\begin{equation}
p + p \rightarrow {}^2\text{H} + e^+ + \nu_e.
\end{equation}
This reaction predominantly contributes to the production of solar neutrinos, with a branching ratio of approximately 0.86, yielding a high flux of \(10^{10} \, \text{cm}^{-2} \text{s}^{-1}\) and neutrino energies extending up to 420 keV. Neutrinos originating from other nuclear reactions within the Sun typically exhibit energies ranging from a few keV to several MeV. The first indication of neutrino mixing emerged from the solar neutrino anomaly, a discrepancy revealed by experiments specifically designed to detect solar neutrinos. The Homestake Experiment (1968) \cite{Cleveland:1998nv, Davis:1968cp}, led by Ray Davis et al., utilized chlorine-based detectors to measure the flux of solar electron neutrinos (\(\nu_e\)). Remarkably, the observed flux was only 30\% of the expected value predicted by the Standard Solar Model (SSM) \cite{10.1007/978-94-009-0541-2_3}, highlighting a significant anomaly in solar neutrino detection. In the 1980s, the Super-Kamiokande experiment confirmed this neutrino deficit using a water Cherenkov detector, while in the 1990s, gallium-based detectors further substantiated the discrepancy. A major breakthrough occurred in 2001 when the Sudbury Neutrino Observatory (SNO) observed that the electron neutrino flux measured via charged-current interactions was notably lower than the SSM prediction \cite{Aharmim:2009gd, Aharmim:2008kc}. SNO provided the first direct evidence that this deficit could be explained by flavor oscillations, where \(\nu_e\) undergoes oscillation into \(\nu_\mu\) and \(\nu_\tau\). The Super-Kamiokande experiment \cite{Hu:2022ufh} subsequently determined the first precise value of the solar mixing angle as \(\sin^2\theta_{12} \approx 0.3\). Additionally, the KamLAND reactor neutrino experiment \cite{KamLAND:2002uet} provided an independent measurement of the mixing parameters, yielding \(\theta_{12} \approx 34^\circ\) and \(\Delta m^2_{21} \approx 7.5 \times 10^{-5} \,\, \text{eV}^2\). Future experiments JUNO (Jiangmen Underground Neutrino Observatory) \cite{JUNO:2023ete} and Hyper-Kamiokande (HK) \cite{Martinez-Mirave:2021cvh} have the prospect to bring more precision on the measurements of $\theta_{12}$ and $\Delta m^2_{21}$.

\subsection{Overview of atmospheric oscillation parameters (2-3 sector)}
The prime source of the atmospheric neutrinos ($\nu_\mu$) is the decay of the short-lived pions ($\sim 10^{-8}$ sec) generated through the interactions between primary cosmic showers and the atmospheric air \cite{Hamilton:1943dne, Reines:1965qk}. These kaons and short-lived pions further decay to produce atmospheric neutrinos via the following process.
\begin{equation}
\begin{aligned}
    \pi^+, K^+ &\rightarrow \nu_{\mu} \mu^+ \rightarrow \nu_{\mu} e^+ \nu_e \bar{\nu}_{\mu}, \\
    \pi^-, K^- &\rightarrow \bar{\nu}_{\mu} \mu^- \rightarrow \bar{\nu}_{\mu} e^- \bar{\nu}_e {\nu}_{\mu}.
\end{aligned}
\end{equation}
Since each charged pion decay produces one $\nu_\mu$ and each muon decay produces one $\nu_e$, the naive expectation for the flux ratio is:
\begin{equation}
\dfrac{\nu_\mu+\bar{\nu}_\mu}{\nu_e+\bar{\nu}_e}\approx2.
\end{equation}
The \textit{Super-Kamiokande} experiment \cite{Super-Kamiokande:2015qek} observed a significant deficit in the flux of \textit{upward-going muon neutrinos} — those traversing long distances through the Earth — compared to theoretical expectations, while the \textit{electron neutrino flux} remained consistent with predictions. This discrepancy, known as the \textit{atmospheric neutrino anomaly}, provided the first strong indication of neutrino oscillations \cite{Kajita:2010zz}. The \textit{branching ratio} for kaon decays producing $\nu_\mu$ is lower ($\sim 64\%$) than that of pion decays since a substantial fraction of kaons undergo hadronic decays that do not yield neutrinos ($\sim 99\%$). At \textit{low and moderate energies} ($< 100$ GeV), pions dominate the atmospheric neutrino flux due to their \textit{greater abundance and shorter lifetime}, leading to rapid decay. However, at \textit{higher energies} ($> 100$ GeV), kaons contribute more significantly, as their longer decay length allows them to decay before interacting with the atmosphere. Consequently, pion decays play the dominant role in atmospheric neutrino production. In \textit{1998}, Super-Kamiokande provided a breakthrough by resolving the anomaly through neutrino oscillations \cite{Super-Kamiokande:1998kpq}, demonstrating that \textit{approximately 50\% of $\nu_\mu$} traveling over distances comparable to Earth's diameter \textit{undergo flavor conversion into $\nu_\tau$}. At that time, they gave their best-fit of atmospheric mass-squared splitting as $\Delta m^2_{31} = 2.2 \times 10^{-3}$ eV$^2$ inside the physical region $0<\sin^22\theta<1$, marking a major milestone in neutrino physics.\\
The dominant oscillation channels for atmospheric neutrinos are 
$\nu_\mu \to \nu_\mu$ (survival) and $\nu_\mu \to \nu_\tau$ (transition). However, due to the significantly larger rest-mass energy of the tau lepton, detecting $\nu_\tau$ remains challenging. The typical detectable energy range of atmospheric neutrinos spans $0.1$–$100$ GeV, while high-energy neutrino telescopes such as \textit{IceCube} \cite{IceCube:2019cia} extend sensitivity to the $10$–$1000$ GeV range. The distance traveled by these neutrinos is determined by the \textit{zenith angle}, which depends on the relative positioning of the detector and the source of neutrinos. Thus, \textit{energy and zenith angle} serve as the two primary control parameters in this sector. The propagation length varies from $\sim10$ km for downward-going neutrinos to $\sim12,700$ km for those traversing the Earth's diameter. As they travel through matter over long distances, atmospheric neutrinos provide a unique opportunity to probe \textit{oscillation parameters} such as $\theta_{23}$ and $\Delta m^2_{31}$ with high precision. In India, the \textit{Kolar Gold Field (KGF)} experiment \cite{Krishnaswamy:1971be} was among the earliest efforts to detect atmospheric neutrinos, while the \textit{Indian Neutrino Observatory (INO)} \cite{ICAL:2015stm} has future plans to contribute significantly to studying their properties. On the international front, the \textit{KM3NeT/ORCA} experiment \cite{Drakopoulou:2023nkh}, located in the Mediterranean Sea, is advancing research on the \textit{neutrino mass hierarchy} and improving the precision of atmospheric mass-splitting measurements. Their latest analysis, based on $355$ days of data, provides updated constraints on $\sin^2\theta_{23}$ and $\Delta m^2_{31}$ \cite{Schumann:2023kil}.

\subsection{Reactor Neutrinos (1-3 sector): the bridge connecting the other two oscillation sectors as a hallmark of $3\nu$ paradigm}
The 1-3 (reactor) sector of the PMNS matrix is crucially pivotal in giving the first signature of the three flavor framework connecting the solar sector and the atmospheric sector. Not only that, reactor antineutrinos ($\bar{\nu_e}$) were detected first in the history of the discovery of the ``neutrinos" in 1956 by Clyde Cowan and Frederick Reines \cite{Cowan:1956rrn} at the Savannah River Plant (nuclear reactor) in South Carolina, USA, where the reactor produced a high flux of electron antineutrinos from beta decay of fission products. The electron antineutrinos produced there via the inverse beta decay (IBD) process,
\begin{equation}
\bar{\nu}_e + p \rightarrow e^+ + n.
\end{equation}
The typical energy of the reactor antineutrinos are of 10 MeV, while the energy of antineutrinos taking part in the IBD is $\sim 2$ MeV \cite{Kopeikin:2004cn, Ma:2012bm}. Most of the energy is occupied by the positron and the neutron gets captured by the H or other doping nucleus of the scintillator. This neutron capturing ensures a delayed signal signifying an event in the reactor. The main oscillation channel investigated in the reactor sector is $\bar{\nu}_e\rightarrow\bar{\nu}_e$ disappearance.\\ 
\par \textbf{KamLAND} (Kamioka Liquid Scintillator Antineutrino Detector) \cite{KamLAND:2002uet} \cite{KamLAND:2004mhv} was the first experiment to establish an upper limit on the mixing angle $\theta_{13}$ in 2010, constraining it to $\sin^2 2\theta_{13} < 0.2$ at a 90\% confidence level. From 2011 to 2013, KamLAND provided evidence for a non-zero $\theta_{13}$ \cite{KamLAND:2013rgu} and improved its precision, thereby reducing the uncertainty in its measurement. Additionally, the combined analysis of KamLAND and SNO data \cite{Balantekin:2004hi} significantly refined the determination of the solar neutrino oscillation parameters, yielding $\tan^2\theta_{12} = 0.47^{+0.06}_{-0.05}$ and $\Delta m^2_{21} = 7.59^{+0.21}_{-0.21} \times 10^{-5} \text{ eV}^2$. KamLAND, located at an average baseline of 80 km, observes reactor antineutrinos from 55 nuclear reactors in Japan, making it a pivotal long-baseline reactor neutrino experiment. It also achieved a groundbreaking milestone as the first experiment to detect geo-neutrinos, providing direct evidence of antineutrinos from radioactive decays within the Earth's crust and mantle. Its extension, KamLAND-Zen (operational since 2011) \cite{KamLAND-Zen:2024eml, Ozaki:2019uyd}, is dedicated to the search for neutrinoless double-beta decay ($0\nu\beta\beta$), aiming to probe the majorana nature of neutrinos and the possible violation of lepton number conservation.\\
\par During the early 2010s, three additional experiments among the eight originally proposed began data collection: Double Chooz \cite{DoubleChooz:2019qbj}, Daya Bay \cite{DayaBay:2012fng}, and RENO \cite{Kim:2013sza}. Double Chooz (France) is an extension of the original Chooz experiment, while RENO (Reactor Experiment for Neutrino Oscillation) in South Korea and Double Chooz both employ liquid scintillator detectors for reactor antineutrino detection. RENO consists of two detector modules, whereas Daya Bay features a more extensive setup with eight detector modules strategically placed around six nuclear reactors at a baseline of 1.9 km. The first indication of a nonzero $\theta_{13}$ came from Double Chooz, while Daya Bay provided the first definitive measurement in 2012 with a statistical significance of $5.2\sigma$ C.L., firmly establishing the three-flavor neutrino oscillation framework. The Daya Bay result marked a major milestone in neutrino physics, as it confirmed that $\theta_{13}$ is not only non-zero but also relatively large. This discovery was pivotal, as it enabled the experimental study of CP violation in the lepton sector, with profound implications for understanding the matter-antimatter asymmetry of the Universe.\\
\par \textbf{JUNO} (Jiangmen Underground Neutrino Observatory) \cite{JUNO:2024jaw} is a next-generation medium-baseline reactor neutrino experiment under construction in China. The central detector consists of 20 kilotons of linear alkylbenzene-based liquid scintillator, making it one of the largest scintillator-based neutrino detectors in the world. Located at a baseline of $\sim 53$ km from the Yangjiang and Taishan nuclear power plants, it will observe an intense flux of electron antineutrinos ($\bar{\nu}_e$). A distinctive feature of JUNO is its excellent energy resolution ($3\%$ at 1 MeV), which is crucial for resolving the neutrino mass ordering. Additionally, its medium-baseline distance allows for significant matter effects in neutrino oscillations. JUNO is expected to provide precise constraints on the solar mixing parameters ($\theta_{12}, \Delta m^2_{21}$) and the atmospheric mass-squared splitting ($\Delta m^2_{31}$), though, like all reactor neutrino experiments, it is not sensitive to the atmospheric mixing angle ($\theta_{23}$), as reactor $\bar{\nu}_e$ do not participate in $\nu_\mu \rightarrow \nu_\tau$ oscillations.\\
\section{Summary}
\label{sec:2.5}
Several key sources of neutrinos have been discussed above, providing an overview of the discovery of the four neutrino oscillation parameters of the PMNS matrix ($\theta_{12}, \theta_{13}, \theta_{23},$ and $\delta_{\mathrm{CP}}$). In addition to these well-known sources, there exist other fundamental neutrino sources that play a crucial role in advancing the frontiers of neutrino physics.\\
\textbf{Astrophysical neutrinos} serve as a powerful probe for addressing several fundamental open questions in physics and cosmology, such as the nature of dark matter, the flavor composition of neutrinos, neutrino interactions at ultra-high energies, the internal structure of the Sun, and the matter-antimatter asymmetry of the Universe. These neutrinos span a wide energy range, from a few MeV to several TeV. Some of the leading neutrino telescopes dedicated to detecting astrophysical neutrinos include IceCube \cite{IceCube:2013low}, Baikal-GVD \cite{Baikal-GVD:2020xgh}, KM3NeT/ARCA \cite{KM3Net:2016zxf, KM3NeT:2024paj}, and P-ONE \cite{Henningsen:2024nmb}.\\
\textbf{Geo-neutrinos} are primarily electron antineutrinos ($\bar{\nu}_e$) produced by the natural radioactive decays occurring within Earth's interior \cite{Fiorentini:2007te, Mantovani:2003yd}, predominantly in the crust and mantle. These neutrinos originate from beta decays of long-lived radioactive isotopes, such as $^{238}\mathrm{U}$, $^{232}\mathrm{Th}$, and $^{40}\mathrm{K}$. KamLAND \cite{Bellini:2021sow} was the first experiment to detect geo-neutrinos, and leading detectors such as Borexino \cite{Borexino:2019gps}, SNO$^+$ \cite{Semenec:2023mjj}, and JUNO continue to advance geo-neutrino observations. Their measurements, when combined with seismic wave studies, provide valuable insights into Earth's internal structure and its radiogenic heat contribution.\\
Among the diverse sources of neutrinos, artificially produced neutrinos offer a unique advantage—their energy and propagation distance can be precisely controlled. This capability is harnessed in \textbf{Long-Baseline (LBL) experiments} \cite{Agarwalla:2014fva}, which play a pivotal role in the precision era of neutrino oscillation physics. Leading LBL experiments such as T2K \cite{T2K:2011qtm}, NO$\nu$A \cite{NOvA:2004blv}, and MINOS \cite{MINOS:2011amj} have significantly contributed to our understanding of neutrino phenomenology. Looking ahead, next-generation high-precision experiments like DUNE \cite{DUNE:2021tad} and T2HK \cite{Hyper-Kamiokande:2018ofw} hold immense potential to unravel the remaining puzzles in the $3\nu$ paradigm. The next chapter is dedicated to the long-baseline experiments completely, which is one of the main content of this thesis.\\ 
To get a quick overview of the experiments contributed in measuring the specific neutrino oscillation parameter, the following chart may be useful.
\begin{itemize}
    \item \textbf{Solar Parameters:}
    \begin{itemize}
        \item $\Delta m^2_{21}$: Strongest constraint from \textbf{KamLAND}, also constrained by other Solar experiments, e.g.- \textbf{GALLEX}.
        \item $\theta_{12}$: Best measured by \textbf{Solar experiments}, with additional input from \textbf{KamLAND}.
    \end{itemize}
    
    \item \textbf{Atmospheric and Long-Baseline Parameters:}
    \begin{itemize}
        \item $|\Delta m^2_{31}|$: Best constrained by \textbf{Long-Baseline (LBL) experiments}, with additional input from \textbf{Atmospheric + Reactor experiments}.
        \item $\theta_{23}$: Primarily determined by \textbf{Atmospheric + LBL experiments}.
    \end{itemize}
    
    \item \textbf{Reactor and Long-Baseline Constraints:}
    \begin{itemize}
        \item $\theta_{13}$: Most precisely measured by \textbf{Reactor experiments}, with additional constraints from \textbf{Solar + KamLAND} and \textbf{Atmospheric + LBL} experiments.
        \item $\delta_{\mathrm{CP}}$: Constrained mainly by \textbf{LBL experiments}, with additional input from \textbf{Atmospheric experiments}.
        \item Reactor experiments are blind in measuring $\theta_{23}$.
    \end{itemize}
    
    \item \textbf{Mass Ordering:}
    \begin{itemize}
        \item Determined through a combination of \textbf{LBL + Reactor} and \textbf{Atmospheric} experiments.
        \item Complementary constraints come from \textbf{Cosmology} and \textbf{Neutrinoless Double Beta Decay ($0\nu\beta\beta$)}.
    \end{itemize}
\end{itemize}


\chapter{A Review on Long-baseline Neutrino Oscillation Experiments}
\label{C3} 
Neutrinos, originating from various sources, differ not only in type but also in energy and the distance that they travel. As mentioned earlier, they are observed across an energy spectrum ranging from eV to PeV, depending on their sources and production mechanisms. For certain sources, both the energy of neutrinos and their propagation distance span a wide range (e.g., atmospheric and astrophysical neutrinos). Due to matter effects, neutrino mass-mixing parameters get evolved and the oscillation probabilities get modified in a non-trivial fashion as a function of energy and path length. Since these oscillation parameters $L$ and $E$ are inherently correlated, achieving high precision for a single parameter is challenging without addressing the uncertainties in the others. Therefore, a well-controlled, high-precision neutrino beam can be useful to tackle the uncertainties in neutrino energy and flux. This motivated the development of artificial setups designed to generate approximately mono-energetic neutrino beams, with a well-determined path from the source to the detector \cite{Agarwalla:2014fva}. \\
\par In 1960, Melvin Schwartz \cite{Schwartz:1960hg} proposed that an intense neutrino beam could be created using an accelerator, which would allow controlled laboratory-based studies of neutrino interactions. This idea was realized in the famous Brookhaven Neutrino Experiment (BNL E-734, 1962-1963) at the Alternating Gradient Synchrotron (AGS) \cite{Shiltsev:2019rfl}. The concept of producing a neutrino beam using accelerated protons became the foundation for modern long-baseline neutrino experiments. The experiment at BNL successfully demonstrated the existence of muon neutrino ($\nu_\mu$) as distinct from the electron neutrino ($\nu_e$). High-energy protons were directed onto a beryllium target, producing a shower of secondary particles, mainly pions and some kaons ($\pi^\pm$ and $K^\pm$). The pions and kaons decayed in a 200-meter-long decay tunnel, producing a beam of neutrinos via the decay process,
\begin{align}
    \pi^+ \rightarrow \mu^++\nu_\mu,\nonumber\\
    K^+ \rightarrow \mu^++\nu_\mu.
\end{align}
Since neutrinos feebly interact with matter, they pass through a thick steel shield ($\sim$13.5 m of steel), which absorbs all remaining hadrons and muons, leaving only a pure beam of muon neutrinos. A 10-ton aluminum spark chamber detector was placed 21 meters from the target to observe neutrino interactions. The neutrinos were expected to interact with nuclei in the aluminum via charged-current interaction,
\begin{align}
    \nu_\mu+n \rightarrow \mu^-+p.
\end{align}
The experiment only observed muons, proving that the beam contained a new neutrino species — the muon neutrino ($\nu_\mu$), distinct from the electron neutrino ($\nu_e$). This discovery was crucial in establishing the two-neutrino model, and in 1988, the Nobel Prize was awarded to Leon Lederman, Melvin Schwartz, and Jack Steinberger for this work \cite{Nobel1988}. In this chapter, we explore the key features of long-baseline experiments, outlining their general setups, primary objectives in advancing neutrino oscillation research, and significant contributions from past, present, and future \cite{Agarwalla:2024aee} LBL experiments worldwide.\\
\section{General setup of LBL experiments and their beam specifications}
In long-baseline experiments, neutrino beams are produced with well-defined energies and a minimal uncertainty in neutrino flux. These beams typically span from a few hundred MeV to a few GeV, while neutrinos traverse hundreds of kilometers from their source to the detector. The process begins with the production and acceleration of high-energy proton beams, reaching tens to hundreds of GeV, which are then directed onto a solid graphite target. As these energetic protons strike the target, they come to a halt, triggering a cascade of pions and kaons, which play a crucial role in neutrino production \cite{Feldman:2012jdx, Diwan:2003bp}.
\par Before decaying, pions ($\pi^\pm$) and kaons ($K^\pm$) pass through magnetic horns (conductors carrying strong pulsed electric currents) \cite{Ichikawa:2012ey, Kahn:2003zz, Baussan:2011yg}. These horns use the Lorentz force, generated by both electric and magnetic fields, to selectively focus either positively or negatively charged particles into a well-defined beam, while deflecting the unwanted ones. For example, when $\pi^+$ and $K^+$ are focused, they produce neutrinos, whereas focusing $\pi^-$ and $K^-$ results in antineutrinos. While the selected particles move forward, others are steered away by the electromagnetic force (Lorentz force) inside the horn. Finally, the focused particles (mostly $\sim99\%$ pions) enter into a decay pipe — an evacuated tunnel about 50 to 200 meters long — where they undergo further decay as per the following decay channels.
\begin{align}
    \pi^+ \rightarrow \mu^++\nu_\mu,\nonumber\\
   \pi^- \rightarrow \mu^-+\bar{\nu}_\mu.
   \label{eq:nu_production_LBL}
\end{align}
The branching ratio of pions decaying into muons is $>99\%$. While pions can also decay through other channels, such as $\pi^+ \rightarrow e^+ + \nu_e$, these occur with a negligible branching ratio of about $10^{-4}$. The low branching ratio for the decay channel of $\pi^+ \rightarrow e^+ + \nu_e$ is due to the helicity suppression. To ensure a pure, nearly mono-energetic neutrino ($\nu_\mu$) beam, a beam dump is used to absorb any remaining particles, including muons, as well as leftover pions and kaons.\\
\subsection{Concept of near detector and far detector in LBL setup}
Long-baseline neutrino oscillation experiments investigate how neutrinos change their flavors as they propagate over distances of hundreds to thousands of kilometers. Because neutrino oscillation probabilities depend on both energy and baseline, these experiments utilize two detectors: a near detector (ND) situated close to the neutrino source and a far detector (FD) placed at a distant location. Comparing measurements from these near and far detectors enables us to determine the fundamental neutrino oscillation parameters while mitigating the systematic uncertainties.\\
\par \textbf{Near detectors} (ND) are kept close to the source of the neutrinos ($\sim$ few hundred meters). Sometimes it consists of several detector components to study different types of neutrino interactions. Near detectors (NDs) \cite{DUNE:2021tad, Emberger:2018pgr, Burgman:2021csn} play a crucial role in characterizing the neutrino beam before oscillations occur. These detectors are designed with specific features to measure the neutrino flux, energy spectra, and interaction cross-sections as precisely as possible. It is smaller in size than far detectors, but large enough to collect significant event statistics. The energy resolution of the near detector is very high ($<100$ MeV), reducing the uncertainty in the measured neutrino spectra. It is also enriched with fine-grained tracking to reconstruct interaction vertices and outgoing particles. It also utilizes shielding and timing cuts to eliminate the beam-induced backgrounds. For T2K, there are two near detectors: ND280 (off-axis, plastic scintillator, magnetized) \cite{Yevarouskaya:2023yth, T2K:2022atj} and INGRID (on-axis, iron scintillator, non-magnetized) \cite{Abe:2011xv}. ND280 measures the neutrino flux and cross-section at the same off-axis angle as Super-Kamiokande, whereas INGRID monitors the neutrino beam direction and intensity. For NO$\nu$A, there is a 300-ton liquid scintillator kept at around 1 km away from the neutrino source as a near detector \cite{ NOvA:2024rov} with good energy resolution and particle-tracking capability. In MINOS, at a distance of 1 km away from the source, a 980-ton magnetized ($\sim$ 0.5 Tesla) iron-scintillator tracking calorimeter was placed as a near detector \cite{MINOS:2008hdf}. In DUNE, there will be a near detector (575 m away from the source) having multiple specifications \cite{DUNE:2021tad}. The DUNE ND system is planned to consist of three major sub-detectors: ND-LAr (Liquid Argon Time Projection Chamber - LArTPC), ND-GAr (High-Pressure Gaseous Argon TPC - HPgTPC, also called ``SAND (System for on-Axis Neutrino Detection)", magnetized), and ND-LAr Off-Axis System (associated with DUNE PRISM (Precision Reaction Independent Spectrum Measurement)) \cite{Hasnip:2023ygr, Hasnip:2025gyi}. It is one of the most advanced near detectors. The ND-LAr Off-Axis System will be able to move off-axis up to $\pm~30$ meters to sample different neutrino fluxes and also can move sideways relative to the beam. It will measure neutrino interactions at different beam angles, providing a way to study the neutrino flux in different energy regions. It will also help to constrain systematic uncertainties in flux modeling. \\
\par \textbf{Far detectors} (FDs) are pivotal in long-baseline neutrino oscillation experiments, positioned hundreds to thousands of kilometers away from the neutrino source, typically an accelerator-based beamline. Their primary aim is to observe the beam neutrinos after oscillation, providing key insights into measuring neutrino mixing parameters, mass ordering, and CP violation. To enhance the interaction rates, far detectors are significantly larger (on the kiloton scale) than near detectors, as neutrinos interact weakly with matter. They are placed deep underground to reduce cosmic background interference, ensuring cleaner signal detection. Additionally, their design enables precise identification of different neutrino flavors, key to identify neutrino oscillation. We elaborate the specifications of far detectors of some past, present, and future long-baseline experiments in the upcoming subsections with more details.\\
\subsection{Beam specifications of LBL experiments}
Neutrino beams are categorized as wide-band beams (WBB) or narrow-band beams (NBB) based on their energy spectra, which are intrinsically linked to the focusing and subsequent decay of secondary pions and kaons generated within the target. Likewise, their classification as on-axis or off-axis beams pertains to the trajectory of the neutrinos relative to the decay pipe's alignment.\\
\par A \textbf{wide-band neutrino beam} exhibits a broad and continuous energy distribution shaped by the momentum spectra of secondary pions and kaons \cite{Barger:2006vy, Papadimitriou:2016ksv}. The peak energy of such a beam is experiment-dependent, but generally falls within a few GeV to approximately 10 GeV. This broad energy spectrum facilitates the detection of multiple oscillation peaks (or a range of \( L/E \) values), enhancing sensitivity to important searches in neutrino oscillation frontiers such as the neutrino mass ordering and CP violation. Additionally, the statistical uncertainty in the beam is comparatively lower. The DUNE experiment plans to utilize a wide-band neutrino beam spanning in the range of 0.1 to 10 GeV, with a peak energy around 2.5 GeV. Similarly, the MINOS experiment utilized a wide-band neutrino beam with a peak energy around 3 GeV.\\
\par A \textbf{narrow-Band neutrino beam} is a beam characterized by a narrow energy range, achieved by selecting pions and kaons within a specific momentum range before they undergo decay into neutrinos. This selection results in a beam where most neutrinos possess similar energies, thereby reducing the uncertainties associated with the energy-dependent effects in neutrino interactions, as it effectively probes a fixed \( L/E \) ratio. Such a beam minimizes the background from deep-inelastic neutrino scatterings, thereby enhancing signal purity in oscillation and cross-section measurements \cite{Steinberger:1988zd, NuTeV:1998aks}. Additionally, it significantly mitigates the energy-smearing effects in cross-section measurements. A narrow-band neutrino beam is generated by precisely controlling the momentum of the parent hadrons (primarily pions and kaons) before they decay into neutrinos. A magnetic horn system is employed to select hadrons within a definite momentum window (e.g., 4–6 GeV/c for a \(\sim\) 2 GeV neutrino beam). Further refinement is achieved through hadron spectrometers and collimators, ensuring that only hadrons with the desired momentum reach the decay pipe. As a result, the produced neutrino spectrum remains narrow and well-defined, peaking at a specific energy. The energy of the neutrino beam and that of the parent pions (for both wide-band and narrow-band beams) are related as follows,
\begin{align}
    E_\nu \approx \frac{0.43\times p_\pi}{1+\gamma^2\theta^2},
    \label{eqn:3.4}
\end{align}
where \( p_\pi \) denotes the momentum of the pion, \( \gamma \) represents its Lorentz boost factor, and \( \theta \) denotes the angle between the trajectory of the emitted neutrino and the decay pipe (or the direction of the parent pion). The energy spread in a narrow-band beam is relatively low (\(\sim 10\text{–}20\%\)), but the statistics is significantly low. The currently running LBL experiments T2K and NO$\nu$A utilize this type of neutrino beam and the upcoming Hyper-Kamiokande experiment will also use this kind of neutrino beam from accelerators.\\
\par Neutrino experiments often use high-intensity neutrino beams produced at accelerator facilities. These beams are categorized into on-axis and off-axis configurations depending on the angle at which the neutrinos travel relative to the center of the beamline. The distinction between these two configurations has significant implications for neutrino physics, particularly in long-baseline neutrino oscillation experiments. \\
\par An \textbf{on-axis neutrino beam} consists of neutrinos propagating along the central axis of the hadron decay pipe, meaning their trajectories are collinear with those of the parent mesons \cite{Angelico:2019gyi}. The resulting neutrino energy spectrum is broad, exhibiting a substantial high-energy tail. Since neutrinos originate from the full phase space of the decaying mesons, the neutrino flux is maximized. It is helpful for experiments studying a wide range of neutrino energies, such as cross-section measurements. DUNE and MINOS experiment use the on-axis neutrino beam.\\
\par An \textbf{off-axis neutrino beam} consists of neutrinos that emerge at a small angle \( \theta \) relative to the central axis of the decay pipe \cite{McDonald:2001mc, Karpova:2025lvb}. This is achieved by positioning the detector slightly off (non-collinear) the beam axis. Compared to an on-axis beam, the neutrino energy spectrum is narrower and peaks at a lower energy. Additionally, the suppression of the high-energy tail reduces backgrounds from deep inelastic scattering. The mean neutrino energy is determined by the energy of the parent mesons and their decay kinematics, following the relation in equation \ref{eqn:3.4}. This beam configuration allows for better control over systematic uncertainties in oscillation experiments and helps mitigate backgrounds from neutral-current interactions. Off-axis beams are particularly advantageous in long-baseline neutrino oscillation experiments, as they concentrate the neutrino flux within a narrow energy range where the oscillation probability is maximized, thereby reducing unwanted backgrounds. Most modern long-baseline oscillation experiments, such as T2K (with an off-axis angle of \(2.5^\circ\)) \cite{T2K:2012qoq} and NO$\nu$A (with an off-axis angle of \(0.8^\circ\)) \cite{Lackey:2019niq}, employ an off-axis configuration to enhance the signal over background ratio. By utilizing an off-axis beam, experiments can fine-tune the neutrino energy spectrum to enhance the flux and maximize the oscillation probability.\\
\section{A Panorama of LBL Neutrino Experiments}
\subsection{K2K (KEK to Kamioka) [1999 - 2004]}
The K2K (KEK to Kamioka) experiment \cite{K2K:2004iot} was the first long-baseline neutrino oscillation experiment, conducted between 1999 and 2004, aimed at verifying the atmospheric neutrino oscillation results observed by Super-Kamiokande (Super-K) using a controlled neutrino beam. With a 250 km baseline \cite{K2K:2002icj}, K2K employed the 12 GeV Proton Synchrotron (PS) at KEK in Tsukuba, Japan, as its neutrino source. The experimental setup comprised of a near detector at KEK and a far detector at Super-Kamiokande in Kamioka, Japan. A wide-band neutrino beam peaking at 1.3 GeV was generated by bombarding an aluminum target with $6.0\times10^{12}$ protons per ejection, producing pions and kaons, which subsequently decayed into neutrinos. The far detector, a 50 kt water Cherenkov detector, recorded neutrino interactions, while the near detector, with a 1 kt volume, characterized the initial neutrino flux. K2K \cite{Nishikawa:2004qy} successfully observed the disappearance of muon neutrinos, reinforcing the evidence for neutrino oscillations, and measured the mass-squared difference as $\Delta m^2_{32} = (2.8 \pm 0.3) \times 10^{-3}$ eV², while establishing a lower bound on the atmospheric mixing angle, $\sin^2 2\theta_{23} > 0.9$.\\
\subsection{MINOS (Main Injector Neutrino Oscillation Search) [2005 - 2012]/ MINOS+ [2013 - 2016]}
MINOS was a pioneering long-baseline neutrino oscillation experiment with a baseline of 735 km, linking Fermilab in Illinois to the Soudan Mine in Minnesota \cite{MINOS:2014rjg}. It harnessed the high-intensity NuMI (Neutrinos at the Main Injector) beam, a formidable muon neutrino source generated at Fermilab \cite{MINOS:2020llm}. In this setup, 120 GeV protons from the Main Injector bombarded on a graphite target, initiating the production of pions and kaons, which subsequently decayed mostly into muon neutrinos. The experiment’s 5.4 kt far detector, positioned 710 m underground, was meticulously engineered with alternating layers of 2.54 cm thick steel plates and 1 cm thick plastic scintillator strips, enabling precise tracking of neutrino interactions and flavor transitions. MINOS played a pivotal role in refining the atmospheric oscillation parameters and established a stringent lower bound of $\sin^2(2\theta_{23}) > 0.95$.  Over its operational tenure, it collected $1.07 \times 10^{21}$ protons on target (P.O.T) in $\nu_{\mu}$ mode and $0.33 \times 10^{21}$ P.O.T in $\bar{\nu}_{\mu}$ mode. In 2013, MINOS evolved into its successor, MINOS+, which leveraged an upgraded proton beam, accumulating $0.58 \times 10^{21}$ POT in $\nu_{\mu}$ mode within its first two years of operation \cite{Evans:2017brt}. Beyond its precision measurements of atmospheric oscillation parameters, MINOS also offered compelling evidence supporting the non-zero value of $\theta_{13}$ \cite{MINOS:2011amj}.\\
\subsection{OPERA (Oscillation Project with Emulsion-tRacking Apparatus) [2006 - 2018]}
The OPERA experiment \cite{OPERA:2015wbl} was one of the first generation long-baseline neutrino experiments designed to detect the third-generation neutrino, $\nu_\tau$, within a $\nu_\mu$ beam, making $\nu_\mu \rightarrow \nu_\tau$ its primary oscillation channel. Due to its relatively large rest mass energy ($\sim$ 1.7 GeV), the tau lepton ($\tau$) - and consequently $\nu_\tau$ - is challenging to observe in most of the neutrino experiments. Unlike some long-baseline experiments such as MINOS or NO$\nu$A, OPERA primarily relied on a far detector at LNGS without a dedicated near detector \cite{OPERA:2018nar}. However, the CERN Neutrinos to Gran Sasso (CNGS) beamline included monitoring systems to characterize the neutrino beam at its source, effectively serving as a near-detector substitute by analyzing the extracted 400 GeV proton beam from the CERN Super Proton Synchrotron (SPS) before its journey to Gran Sasso. The far detector, located deep underground at the Gran Sasso National Laboratory (LNGS) in Italy, was a hybrid setup combining electronic trackers with emulsion film technology to precisely identify $\nu_\tau$ interactions. With a 730 km baseline and a far detector with a 1.25 kt volume, OPERA employed scintillator strips for event localization. A key component of the detector was the Emulsion Cloud Chamber bricks, which consisted of stacked lead plates interleaved with nuclear emulsion films, enabling high-precision vertex reconstruction. Additionally, resistive plate chambers (RPCs) provided crucial information on tracking and time-of-flight measurements. OPERA utilized a $\sim$17 GeV neutrino beam to effectively detect tau neutrinos, making it a significant contribution in neutrino physics.\\
\subsection{T2K (Tokai-to-Kamioka) [2010 - till date]}
T2K is one of the pioneering second-generation long-baseline experiments featuring both near and far detectors. It employs two near detectors positioned at 280 meters from the neutrino production target at J-PARC: ND280 \cite{Yevarouskaya:2023yth, T2K:2022atj}, which is $2.5^\circ$ off-axis, and INGRID (Interactive Neutrino GRID) \cite{Abe:2011xv}, which is aligned on-axis. ND280 comprises of multiple sub-systems designed for various measurements, including neutral current background identification, high-precision momentum determination, and particle identification using energy loss gradients (dE/dX) through time projection chambers (TPCs). Additionally, it incorporates a fine-grained detector made of polystyrene scintillator bars serving as an active neutrino interaction target and electromagnetic calorimeters for detecting electrons and photons. On the other hand, INGRID consists of 16 iron-scintillator modules arranged in a cross-shaped configuration, facilitating monitoring of beam stability, direction, and intensity. There are ongoing plans to reduce INGRID's energy threshold from 450 MeV/c to 300 MeV/c \cite{Roth:2024mwe}, which would enhance neutron resolution to $(15 - 30)\%$ \cite{Munteanu:2019llq}. T2K utilizes a beam of muon neutrinos (or antineutrinos) generated at the Japan Proton Accelerator Research Complex (J-PARC) in Tokai and detected at the Super-Kamiokande detector, situated 295 km away in Kamioka. A 515 kW beam of 30 GeV protons is directed onto a graphite target, resulting in the production of charged mesons, which subsequently decay in flight, yielding neutrinos. This process produces a narrow-band neutrino flux with a peak energy around 0.6 GeV. The focusing horns can be tuned to select positively or negatively charged particles, thereby determining whether the far detector receives a neutrino-dominant or antineutrino-dominant beam. Then the beam traverses 295 km baseline to reach Super-Kamiokande. Super-Kamiokande, the far detector of T2K, is a massive water Cherenkov detector located 1,000 meters underground in the Mozumi Mine. It consists of a 50 kt volume of ultra-pure water enclosed within a cylindrical stainless-steel tank measuring 39.3 m in diameter and 41.4 m in height (fiducial volume is 22.5 kt). Inside, 13,000 photomultiplier tubes capture Cherenkov radiation emitted when charged particles exceed the speed of light in water. The resulting Cherenkov ring patterns, which differ for various particle types such as muons and electrons, enable particle identification. The T2K experiment primarily investigates the $\nu_\mu\rightarrow\nu_e$ appearance and $\nu_\mu\rightarrow\nu_\mu$ disappearance channels to explore CP violation in the neutrino sector, resolve the $\theta_{23}$ octant degeneracy, and achieve precise measurements of the 2-3 oscillation parameters. Since its commencement in 2010, T2K has accumulated a total exposure of $3.6\times 10^{21}$ protons-on-target (P.O.T.) by 2020, with (1.97 and $1.63) \times 10^{21}$ P.O.T. recorded in antineutrino and neutrino modes, respectively, at the far detector. Future projections estimate a total exposure of $7.8 \times 10^{21}$ P.O.T. with a 750 kW beam power, evenly distributed between neutrino and antineutrino modes, in accordance with ref. \cite{T2K:2014xyt}. For both appearance and disappearance channels, we consider an uncorrelated $5\%$ systematic uncertainty for signal events and $10\%$ for the background events. For both appearance and disappearance channels, we consider $5\%\,(10\%)$ systematic uncertainty for the signal (background) events. The latest results of T2K can be found in ref. \cite{T2K:2023mcm}.\\
\subsection{NO$\nu$A (NuMI Off-Axis $\nu_e$ Appearance) [2014 - till date]}
NO$\nu$A is a prominent second-generation long-baseline neutrino experiment currently operating in the U.S. It utilizes the neutrino beam from the NuMI (Neutrinos at the Main Injector) \cite{Adamson:2015dkw, NOvA:2004blv, NOvA:2019cyt} facility at Fermilab, which runs at a beam power of 700 kW. The experiment's far detector is positioned at 810 km from the source (in Ash River, Minnesota) at an off-axis angle of $0.8^\circ$, with the neutrino beam having an average energy of approximately 2 GeV. The far detector is a 14 kt fiducial mass tracking calorimeter filled with the liquid scintillator. Additionally, NO$\nu$A features a near detector located at 1 km from the source, sharing the same off-axis angle as the far detector. The primary oscillation channels under investigation in this experiment include $\nu_\mu\rightarrow\nu_e$ and $\bar{\nu}_\mu\rightarrow\bar{\nu}_e$. Like T2K, one of NO$\nu$A’s key objectives is to explore CP violation in the neutrino sector and determine the neutrino mass ordering. Furthermore, it plays a crucial role in improving the precision of neutrino oscillation parameter measurements \cite{NOvA:2021nfi}. In our analysis, we take into account the full projected exposure of NO$\nu$A, totaling $3.6\times10^{21}$ P.O.T, distributed equally between neutrino and antineutrino modes, as outlined in ref. \cite{NOvA:2007rmc}. For both appearance and disappearance channels, we assume systematic uncertainties of $5\%$ for signal events and $10\%$ for the background events.\\
\subsection{T2HK (Tokai-to-Hyper-Kamiokande) experiment in \\JAPAN}
T2HK is an upcoming third-generation long-baseline neutrino experiment designed to advance our understanding of neutrino oscillation parameters. This experiment features a 187 kt water Cherenkov far detector (Hyper-Kamiokande) and will receive beam from the J-PARC facility in Tokai covering a baseline of 295 km. T2HK will employ two near detectors: ND280 (Near Detector 280) and IWCD (Intermediate Water Cherenkov Detector). ND280, positioned approximately 280 m from the beam source, is responsible for measuring the initial neutrino flux and energy spectra before oscillation. It includes improved tracking and calorimetry compared to T2K’s ND280, which helps in reducing systematic uncertainties. IWCD \cite{Akutsu2021IWCD, IWCD_TDR_2022}, a 1 kt detector located at 1 km from the source, enables neutrino flux measurements at varying off-axis angles. The experiment utilizes a 30 GeV proton beam with a beam power of 1.3 MW directed onto a graphite target, generating an artificial, narrow-band neutrino beam with a $2.5^\circ$ off-axis angle. The neutrino beam will achieve its first oscillation maximum at 0.6 GeV and will have an energy range of 100 MeV to 3 GeV.
Unlike DUNE, T2HK operates with an asymmetric runtime distribution: 2.5 years in neutrino mode and 7.5 years in antineutrino mode. This corresponds to the total exposure of $2.7\times10^{22}$ protons-on-target, leading to a total of 2431 kt$\cdot$MW$\cdot$year. The experiment primarily investigates four oscillation processes: $\nu_\mu\rightarrow\nu_e$ appearance, $\bar{\nu}_\mu\rightarrow\bar{\nu}_e$ appearance, $\nu_\mu\rightarrow\nu_\mu$ disappearance, and $\bar{\nu}_\mu\rightarrow\bar{\nu}_\mu$ disappearance. It assumes systematic uncertainties of $5\%$ for appearance channels (with an improved uncertainty of $2.7\%$) and $3.5\%$ for disappearance channels \cite{Hyper-Kamiokande:2018ofw}. T2HK is expected to provide a robust signal of CP violation, resolve the octant degeneracy of $\theta_{23}$.\\

\par The T2HKK (Tokai-to-Hyper-Kamiokande to Korea) experimental setup comprises of two far detectors: one situated in Japan, termed as the Japanese Detector (JD), and the other proposed in Korea, referred to as the Korean Detector (KD). The KD is designed to receive the same neutrino beam as JD, but at an extended baseline of 1100 km, optimizing its sensitivity at the second oscillation maximum. Like JD, the proposed KD will also be a 187 kt water Cherenkov detector, ensuring sufficient statistics at the second oscillation maximum. The expected systematic uncertainties for the JD+KD setup are 5\% for the appearance channel and 3.5\% for the disappearance channel \cite{Panda:2022vdw}.\\
\subsection{DUNE (Deep Underground Neutrino Experiment) in U.S.A}
The upcoming international DUNE project hosted by Fermilab represents a major step forward in the roadmap of neutrino physics. It is another important next-generation long-baseline experiment planned to begin data collection around 2030. One of its main goals is to study CP violation in the neutrino sector, which could help to explain why the Universe has more matter than antimatter. DUNE is expected to resolve the octant ambiguity of $\theta_{23}$ and determine the correct order of neutrino masses due to significant Earth's matter effect that neutrinos feel while traveling from Fermilab to Homestake. The experiment \cite{DUNE:2018tke} has a long 1285 km baseline, stretching from Fermilab to the Sanford Underground Research Facility (SURF) in South Dakota. The far detector is a 40 kt liquid argon time projection chamber (LArTPC). The experiment is planned to be upgraded gradually over time. The neutrino beam is produced by the Main Injector at the Long-Baseline Neutrino Facility (LBNF), where a 1.2 MW, 120 GeV proton beam strikes a graphite target, creating charged mesons that decay into neutrinos. This results in a wide-band neutrino flux, spanning from a few hundred MeV to tens of GeV, peaking at around 2.5 GeV, with most of the neutrinos in the 1 to 5 GeV range. The near detector complex consists of three main parts: ND-LAr (Liquid Argon Near Detector) – ArgonCube, ND-GAr (Gaseous Argon Detector) – High-Pressure Gaseous TPC, and SAND (System for on-Axis Neutrino Detection). These components help in calibrating the neutrino flux, reducing uncertainties in cross-section measurements, and checking the stability of the beam over the time. According to the DUNE Technical Design Report \cite{DUNE:2020jqi}, the experiment will run for a total of 10 years, with equal time spent in neutrino and antineutrino modes (5 years each). This corresponds to an annual exposure of $1.1\times10^{21}$ P.O.T, leading to a total exposure of 480 kt$\cdot$MW$\cdot$year. DUNE will mainly study four key neutrino oscillation channels: $\nu_\mu\rightarrow\nu_e$ appearance, $\bar{\nu}_\mu\rightarrow\bar{\nu}_e$ appearance, $\nu_\mu\rightarrow\nu_\mu$ disappearance, and $\bar{\nu}_\mu\rightarrow\bar{\nu}_\mu$ disappearance, assuming systematic uncertainties of $2\%$ for appearance channels and $5\%$ for disappearance channels.


\chapter{A close look on the departure of 2-3 mixing angle from maximal mixing with DUNE in the perspective of present neutrino oscillation data}
\label{C4} 
Recent global fit analyses of 3$\nu$ oscillation data show a preference for normal mass ordering (NMO) at 2.5$\sigma$ 
and provide 1.6$\sigma$ indications for lower $\theta_{23}$ octant 
($\sin^2\theta_{23} < 0.5$) and leptonic CP violation ($\sin\delta_{\mathrm{CP}} < 0$). In this work, we study in detail the capabilities of DUNE to establish the deviation from maximal $\theta_{23}$ and to resolve its octant in light of the 
current data. Introducing for the first time, a bi-events plot in the plane of total $\nu$ and $\bar\nu$ disappearance events, we discuss the impact of $\sin^2\theta_{23}$ - $\Delta m^2_{31}$ degeneracy in establishing non-maximal $\theta_{23}$ and show how this degeneracy can be resolved with the help of spectral analysis. A 3$\sigma$ (5$\sigma$) determination of non-maximal $\theta_{23}$ is possible in DUNE with an exposure of 336 kt$\cdot$MW$\cdot$years if the true value of $\sin^2\theta_{23} \lesssim 0.465~(0.450)$ or $\sin^2\theta_{23} \gtrsim 0.554~(0.572)$. We study the role of appearance and disappearance channels, systematic uncertainties, marginalization over oscillation parameters, and the importance of spectral analysis in establishing non-maximal $\theta_{23}$. We observe that both $\nu$ and $\bar\nu$ data are essential to settle the $\theta_{23}$ octant at a high confidence level. DUNE can resolve the octant of $\theta_{23}$ at 4.2$\sigma$ (5$\sigma$) using 336 (480) kt$\cdot$MW$\cdot$years of exposure for the present best-fit values of oscillation parameters. DUNE can improve the current relative 1$\sigma$ precision on $\sin^2\theta_{23}$ ($\Delta m^2_{31}$) by a factor of  4.4 (2.8) using 336 kt$\cdot$MW$\cdot$years of exposure.\\
\par A deeply-relevant and much-awaited result concerning neutrinos in recent times is the {\it hint} for violation of the CP symmetry in the leptonic sector. The T2K collaboration~\cite{T2K:2011qtm} in their 2019 results~\cite{T2K:2019bcf} have shown that their neutrino and antineutrino appearance data point towards CP being near-maximally violated $i.e.$ $\vert\sin\delta_{\rm CP}\vert$ is close to 1. They obtain a best-fit value of $\delta_{\rm CP}$ at $252^\circ$ while the CP-conserving values of $\delta_{\rm CP} = 0^\circ\,, \,\pm\, 180^\circ$ are ruled out at $95\%$ confidence level (C.L.). They also report a preference for the normal mass ordering over the inverted mass ordering at nearly $68\%$ confidence level. NMO and $\delta_{\rm CP} = 270^\circ$ is in fact one of the most favorable parameter combinations for which early hints regarding mass ordering and CP violation can be expected from the currently running long-baseline accelerator experiments~\cite{Prakash:2012az, Agarwalla:2012bv}. The same set of measurements are also being carried out by the NO$\nu$A experiment~\cite{NOvA:2007rmc, Ayres:2002ws, NOvA:2004blv} which operates at a longer baseline with more energetic neutrinos. The recent results from NO$\nu$A~\cite{NOvA:2021nfi} also show a preference for NMO, but their best-fit to $\delta_{\rm CP}$ is not in conjunction with T2K. NO$\nu$A's $\delta_{\rm CP}$ best-fit value of $148^\circ$ is $2.5\sigma$ away from T2K's best-fit. However, the two experiments agree on $\delta_{\rm CP}$ measurements when they assume IMO to be true -- each reporting a best-fit value around $270^\circ$. The tension between these two data sets is not yet at a statistically significant level and we need to wait for further data from T2K and NO$\nu$A to see if this tension persists. In any case, a $5\sigma$ {\it discovery} of any of the current unknowns in neutrino oscillation physics does not seem to be within the reach of either of these experiments~\cite{Agarwalla:2012bv}. Nonetheless, these results are quite important and play an important role in the global fit studies.

\begin{figure}[htb!]
	\centering
	\includegraphics[width=\linewidth]{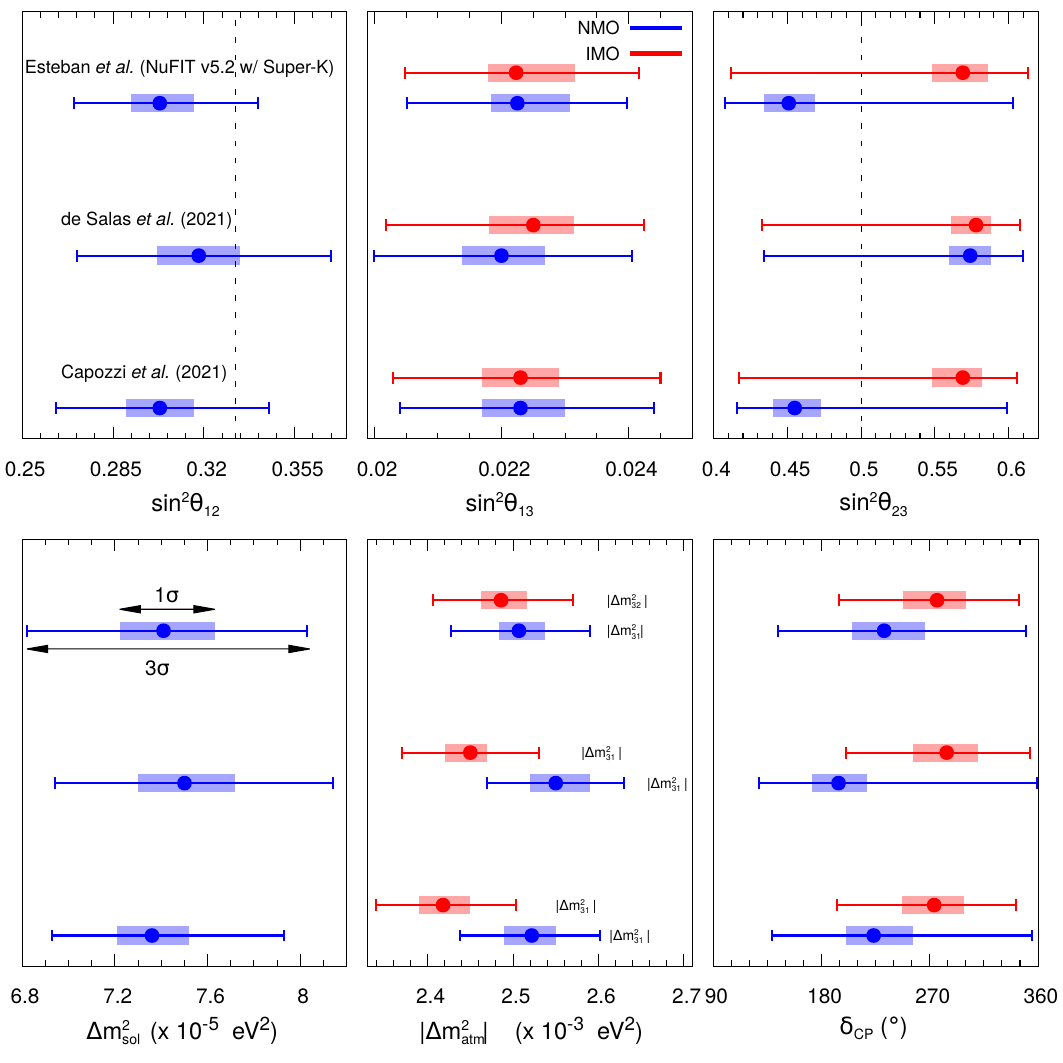}
	\caption{\footnotesize{Current $1\sigma$ (see rectangular boxes)  and $3\sigma$ (see horizontal lines) allowed ranges of the neutrino oscillation parameters obtained from the global fit studies performed by Esteban $et~ al.$~\cite{NuFIT}, de Salas $et~ al.$~\cite{deSalas:2020pgw}, and Capozzi $et~ al.$~\cite{Capozzi:2021fjo}. The blue (red) lines and boxes represent the values for NMO (IMO). In each panel, the best-fit value of the respective oscillation parameter is shown by blue (red) dots for NMO (IMO).
Vertical black dashed lines in the panels related to $\sin^2\theta_{12}$ and $\sin^2\theta_{23}$ show their corresponding values in the tri-bimaximal mixing scheme. Note that the measurements of $\sin^{2}\theta_{12}$ and $\Delta m^{2}_{\mathrm{sol}}\, (\equiv \Delta m^{2}_{21})$ are not sensitive to the choice of mass ordering. }}
	\label{fig:1}
\end{figure}

Fig.~\ref{fig:1} summarizes our current understanding of the six neutrino oscillation parameters in the standard three-neutrino framework. It confirms that we have already attained a remarkable precision on solar oscillation parameters ($\Delta m^2_{21}$ and $\sin^2\theta_{12}$), atmospheric mass-splitting ($\Delta m^2_{31}$), and reactor mixing angle ($\theta_{13}$). In this figure, we compare the $1\sigma$ (shown with colored rectangular blocks) and $3\sigma$ allowed (shown with horizontal lines) regions of the oscillation parameters that have been calculated by doing a combined analyses of the existing global oscillation data~\cite{NuFIT,deSalas:2020pgw, Capozzi:2021fjo, Esteban:2020cvm}. These works take into account the data from the Solar (Gallex and GNO~\cite{Cleveland_1998}, SAGE~\cite{SAGE:2009eeu}, the four phases of Super-K (SK I-IV)~\cite{Super-Kamiokande:2005wtt, Super-Kamiokande:2008ecj, Super-Kamiokande:2010tar}, SNO~\cite{SNO:2011hxd}, and Borexino I-III~\cite{Bellini:2011rx, Borexino:2008fkj, BOREXINO:2014pcl}), atmospheric (IceCube/DeepCore~\cite{IceCube:2014flw, IceCube} and the four phases of Super-K (SK I-IV)~\cite{Super-Kamiokande:2017yvm, data-superk}), reactor (KamLAND~\cite{KamLAND:2013rgu}, Daya Bay~\cite{DayaBay:2018yms}, and RENO~\cite{RENO:2018dro, reno}), and the accelerator experiments (MINOS~\cite{MINOS:2013utc, MINOS:2013xrl}, T2K~\cite{t2k}, and NO$\nu$A~\cite{nova}). All the three studies find that the earlier tension between Solar and KamLAND data has been reduced considerably after incorporating the recent results from Super-K Phase IV 2970 days of solar data (energy spectra and day-night asymmetry)~\cite{sk}. Additionally, both Esteban $et~ al.$~\cite{Esteban:2020cvm, NuFIT} and Capozzi $et ~al.$~\cite{Capozzi:2021fjo} also consider the recent Super-K Phase IV atmospheric data~\cite{sk}.

In Fig.~\ref{fig:1}, the blue (red) regions are obtained assuming NMO (IMO). Note that in the case of the solar oscillation parameters $i.e.$ $\sin^2\theta_{12}$ and $\Delta m^2_{21}$, the IMO and NMO regions are identical. Vertical black dashed lines in the panels related to $\sin^2\theta_{12}$ and $\sin^2\theta_{23}$ depict their corresponding values in the tri-bimaximal mixing scheme~\cite{Harrison:2002er,Harrison:2002kp,King:2014nza}. 
Note that while Esteban $et~ al.$ and de Salas $et~ al.$ quote the values of atmospheric mass-splitting in terms of $\Delta m^{2}_{31}$, Capozzi $et~ al.$ express it in terms of $\Delta m^{2} = m^{2}_{3}- (m^{2}_{1} + m_{2}^{2})/2$, where $\Delta m^{2}_{31}$ = $\Delta m^{2} + \Delta m^{2}_{21}/2$ for both NMO and IMO.

A novel aspect of Fig.~\ref{fig:1} is that all three global fits now rule out $\delta_{\rm CP} \in \left[ 0, \sim135^\circ \right]$ at $3\sigma$ confidence level and $\delta_{\rm CP} \in \left[ 0, \sim180^\circ \right]$ at $1\sigma$ confidence level, while predicting the best-fit value to lie somewhere in the range $\left[ 200^\circ,~230^\circ \right]$. The constraint in $\delta_{\rm CP}$ is  essentially due to the data from T2K and NO$\nu$A as discussed earlier. As far as the neutrino mass ordering is concerned, all three global fits show preference for NMO, ruling out IMO at close to $2.5\sigma$~\cite{Esteban:2020cvm,NuFIT,Capozzi:2021fjo,deSalas:2020pgw}. Therefore, for the sake of simplicity, in this work, we show our results assuming NMO both in data and fit. We observe that the results do not change much for IMO.

Another important feature that emerges from Fig.~\ref{fig:1} is that the current $3\sigma$ allowed range in $\sin^2\theta_{23}$ is $\sim 0.4~ \rm{to}~ 0.6$. This range is still relatively large as compared to the current uncertainties on $\theta_{12}$ and $\theta_{13}$ and it spans on either sides of $\sin^2\theta_{23}  = 0.5$. The value $\sin^2\theta_{23}  = 0.5$ (or equivalently $\sin^22\theta_{23}  = 1$) corresponds to the case of maximal mixing (henceforth, referred to as {\it maximality}) between the $\nu_{2}, \, \nu_{3}$ and $\nu_{\mu}, \, \nu_{\tau}$ eigenstates, which can in principle allows for a complete flavor transition between $\nu_{\mu}$ and $\nu_{\tau}$. However, the recent global fit studies suggest that $\sin^2\theta_{23}$ $\neq$ 0.5 (see Fig.~\ref{fig:1}) or $\sin^22\theta_{23}  \neq 1$. This leads to the so-called octant degeneracy of $\theta_{23}$ $i.e.$ a lack of knowledge regarding whether $\theta_{23}$ is less than $45^\circ$ (denoted as lower octant, LO) or greater than $45^\circ$ (labelled as higher octant, HO)~\cite{Fogli:1996pv, Barger:2001yr, Minakata:2002qi, Minakata:2004pg, Hiraide:2006vh}. But before the question of octant of $\theta_{23}$ arises, it is vital to establish the exclusion of maximality at a high significance, which is the main thrust of this work. In Ref.~\cite{Capozzi:2021fjo}, the authors find a preference at 1.6$\sigma$ for $\theta_{23}$ in the LO with respect to the secondary best-fit in HO. They obtain a best-fit value of $\sin^{2}\theta_{23}$ = 0.455 in the LO assuming NMO and disfavor maximal $\theta_{23}$ mixing at $\sim$1.8$\sigma$. However, there is a slight disagreement between the three global fit studies as far as the measurement of $\theta_{23}$ is concerned (see top right panel in Fig.~\ref{fig:1}). In Ref. \cite{deSalas:2020pgw}, de Salas $et~ al.$ find a best-fit in the HO around $\sin^2\theta_{23} \sim 0.57$ assuming NMO, while Capozzi $et~ al.$~\cite{Capozzi:2021fjo} and Esteban $et~ al.$~\cite{Esteban:2020cvm} obtain the best-fit around $\sin^{2}\theta_{23} \sim 0.45$ in the LO. This difference in the best-fit value of $\sin^2\theta_{23}$ is probably due to the recent Super-K Phase I-IV 364.8 kt$\cdot$yrs of atmospheric data~\cite{sk} that only Capozzi $et~ al.$ and Esteban $et~ al.$ consider in their latest analyses.\\
\par The issue of non-maximal $\theta_{23}$ and the resolution of its octant (if $\sin^{2}2\theta_{23} \neq 1$) have far-reaching consequences as far as the models explaining neutrino masses and mixings are concerned~\cite{Mohapatra:2006gs, Albright:2006cw, Albright:2010ap, King:2013eh, King:2014nza}. Some examples of such models are quark-lepton complementarity~\cite{Raidal:2004iw, Minakata:2004xt,Ferrandis:2004vp, Antusch:2005ca}, $A_{4}$ flavor symmetry~\cite{Ma:2002ge,Ma:2001dn, Babu:2002dz,Grimus:2005mu, Ma:2005mw}, and $\mu$-$\tau$ permutation symmetry~\cite{Fukuyama:1997ky, Mohapatra:1998ka,Lam:2001fb, Harrison:2002et, Kitabayashi:2002jd, Grimus:2003kq, Ghosal:2003mq,Koide:2003rx, Mohapatra:2005yu}. The $\mu$-$\tau$ permutation symmetry is of particular interest since the current oscillation data strongly indicates that this symmetry is not exact in Nature. A high-precision measurement of 2-3 mixing angle and the measurement of its octant are inevitable to disclose the pattern of deviations from the above-mentioned symmetries, which in turn will help us to explain tiny neutrino masses and one small and two large mixing angles in the lepton sector~\cite{Xing:2014zka, Xing:2015fdg}. It has also been shown that without an accurate measurement of $\theta_{23}$, a precise measurement of $\delta_{\rm CP}$ will not be possible~\cite{Minakata:2013eoa}.

There are several studies in the literature addressing the issues related to the 2-3 mixing angle in the context of various neutrino oscillation experiments. For example, see Refs.~\cite{Antusch:2004yx, Minakata:2004pg, GonzalezGarcia:2004cu, Choudhury:2004sv, Choubey:2005zy, Indumathi:2006gr, Kajita:2006bt, Hagiwara:2006nn, Samanta:2010xm, Agarwalla:2013hma, Minakata:2013eoa, Ge:2013zua, Ge:2013ffa, Chatterjee:2013qus, Choubey:2013xqa, Bass:2013vcg, Coloma:2014kca, Bora:2014zwa, Das:2014fja,Nath:2015kjg, Ghosh:2015ena, Agarwalla:2016xlg, Ballett:2016daj}. In this work, we analyze in detail the sensitivities of the next generation, high-precision long-baseline neutrino oscillation experiment DUNE (Deep Underground Neutrino Experiment)~\cite{DUNE:2015lol, DUNE:2020lwj, DUNE:2020ypp, DUNE:2020jqi, DUNE:2021cuw, DUNE:2021mtg} to establish the deviation from maximal $\theta_{23}$ and to resolve its octant at high confidence level in light of the current neutrino oscillation data. While estimating DUNE's capability for the discovery of non-maximal $\theta_{23}$, we shed light on some relevant issues such as: (i) the individual contributions from appearance and disappearance channels, (ii) the impact of systematic uncertainties and marginalization over oscillation parameters, and (iii) importance of spectral analysis and data from both neutrino and antineutrino runs. We also study how much improvement DUNE can offer in the precision measurements of $\sin^{2}\theta_{23}$ and $\Delta m^2_{31}$ as compared to their current precision. While estimating the achievable precision on these parameters in DUNE, we also quantify the contribution from individual appearance and disappearance channels and demonstrate the importance of having both neutrino and antineutrino data.

The layout of this chapter is as follows. In Sec.~\ref{probability}, we discuss the potential of DUNE's baseline and energy in establishing deviation from maximal $\theta_{23}$ at the level of probabilities. Next, in Sec.~\ref{events}, we describe the key features of DUNE which are relevant for our numerical simulation and discuss the impact of possible correlations and degeneracies among $\sin^{2}\theta_{23}$, $\Delta m^{2}_{31}$, and $\delta_{\mathrm{CP}}$ at the level of total event rates, bi-events, and event spectra. In Sec.~\ref{results}, we quantify the performance of DUNE to establish non-maximal $\theta_{23}$. We also address several issues which are relevant to achieve the above-mentioned goals. In Sec.~\ref{conclusion}, we summarize our findings and make concluding remarks.\\
\section{Discussion at the level of probabilities}
\label{probability}

\begin{table}[htb!]
		\label{tableprecision}

		\centering
		\resizebox{\columnwidth}{!}{%
			\begin{tabular}{|c|c|c|c|c|c|}
				\hline \hline
				\multirow{2}{*}{\textbf{Parameter}} & \multirow{2}{*}{\textbf{Best-fit}} & \multirow{2}{*}{\textbf{1$\sigma$ range}} & \multirow{2}{*}{\textbf{2$\sigma$ range}} & \multirow{2}{*}{\textbf{3$\sigma$ range}} & \textbf{Relative 1$\sigma$}\\
				& & & & &\textbf{Precision (\%)}\\
				\hline \hline
				$\Delta m^2_{21}/10^{-5}$ $\mathrm{eV^{2}}$ & 7.36 & 7.21 - 7.52 & 7.06 - 7.71 & 6.93 - 7.93 & 2.3\\
				\hline
				$\sin^{2}\theta_{12}/10^{-1}$ & 3.03 & 2.90 - 3.16 & 2.77 - 3.30 & 2.63 - 3.45 & 4.5\\
				\hline
				$\sin^{2}\theta_{13}/10^{-2}$ & 2.23 & 2.17 - 2.30 & 2.11 - 2.37 & 2.04 - 2.44 &  3.0\\
				\hline
				$\sin^2\theta_{23}/10^{-1}$ & 4.55 & 4.40 - 4.73 & 4.27 - 5.81 & 4.16 - 5.99 & 6.7 \\
				\hline
				$\Delta m^2_{31}/10^{-3}$ $\mathrm{eV^2}$ & 2.522 & 2.490 - 2.545 & 2.462 - 2.575 & 2.436 - 2.605 & 1.1\\
				\hline
				$\delta_{\text{CP}}$/$^\circ$ & 223  & 200 - 256 & 169 - 313 & 139 - 355 & 16\\
				\hline \hline
			\end{tabular}
			}
			\caption{\footnotesize{The benchmark values of the oscilation parameters and their corresponding ranges that we consider in our study assuming NMO. In second column, we mention the best-fit values as given in Ref.~\cite{Capozzi:2021fjo}. The third, fourth, and fifth columns depict the current 1$\sigma$, 2$\sigma$, and 3$\sigma$ allowed ranges, respectively under NMO scheme. The sixth column depicts the present relative 1$\sigma$ precision on various oscillation parameters as given in Ref.~\cite{Capozzi:2021fjo}.}}
			\label{table:one}
	\end{table}

In the three-neutrino framework, the flavor eigenstates $\vert \nu_{\alpha}\rangle~\left(\alpha = e, \mu, \tau \right)$ and the mass eigenstates $\vert \nu_{i} \rangle ~\left(i=1,2,3\right)$ are connected by the $3\times3$ unitary Pontecorvo-Maki-Nakagawa-Sakata (PMNS) matrix $U$:
\begin{equation}
\vert \nu_{\alpha}\rangle  = \sum_{i} U^\ast_{\alpha i}\vert\nu_{i} \rangle ~~~ {\rm and} ~~~
\vert \bar{\nu}_{\alpha} \rangle = \sum_{i} U_{\alpha i}\vert \bar{\nu}_{i} \rangle\, .
\end{equation}
Following the standard Particle Data Group convention~\cite{Zyla:2020zbs}, the vacuum PMNS matrix $U$ is parametrized in terms of the three mixing angles ($\theta_{23}$, $\theta_{13}$, $\theta_{12}$) and one Dirac-type CP phase ($\delta_{\mathrm{CP}}$). The probability that a neutrino, with flavor $\alpha$ and energy $E$, after traveling a distance $L$, can be detected as a neutrino with flavor $\beta$ is given by
\begin{equation}
P_{\alpha\beta} = \delta_{\alpha\beta} - 4 \sum_{j>i} \mathcal{R} \left( U^\ast_{\alpha j}U_{\beta j} U_{\alpha i} U^\ast_{\beta i}\right)\sin^2\frac{\Delta m^2_{ji} L}{4E} + 2 \sum_{j>i} \mathcal{I} \left( U^\ast_{\alpha j}U_{\beta j} U_{\alpha i} U^\ast_{\beta i}\right)\sin\frac{\Delta m^2_{ji} L}{2E}\, ,
\end{equation}
where, $\Delta m^2_{ji} = m^2_{j} - m^2_{i}$.
Approximate analytical expressions for oscillation probabilities including matter effect have been derived in Ref.~\cite{Akhmedov:2004ny}, retaining terms only up to second order in the small parameters $\sin^2\theta_{13}$ and $\alpha \left(\equiv \Delta m^2_{21}/\Delta m^2_{31}\right)$. The analytical expression for muon neutrino survival probability ($P_{\mu \mu}$) under the constant matter density approximation is given in Eq. 33 of Ref.~\cite{Akhmedov:2004ny}. Considering the current best-fit values of oscillation parameters (see second column in Table~\ref{table:one}), we have $\sin^2\theta_{13}\approx 0.02$, $\alpha \approx 0.03$, $\alpha\sin\theta_{13}\approx 0.004$, and $\alpha^2\approx 0.0008$. Therefore, ignoring the sub-leading terms which are of the order $\alpha^2$ and approximating $\cos \theta_{13}$ equal to 1, Eq. 33 of Ref. \cite{Akhmedov:2004ny} simplifies to
\begin{eqnarray}
\label{pmumushort} 
  P_{\mu\mu} &\approx & 1 - M \sin^22\theta_{23} - N \sin^2\theta_{23} - R \sin 2\theta_{23} + T \sin 4\theta_{23} \, ,
\end{eqnarray}
\newpage
where,
\begin{align}
	\label{eqM}
	M =\;& \sin^2\Delta 
	- \alpha \cos^2\theta_{12}\,\Delta\sin2\Delta \nonumber\\
	&+ \frac{2}{\hat A -1}\sin^2\theta_{13}
	\bigg(
	\sin\Delta\,\cos(\hat A\Delta)\,
	\frac{\sin[(\hat A - 1)\Delta]}{\hat A - 1}
	- \frac{\hat A}{2}\,\Delta\sin2\Delta
	\bigg)\,,
\end{align}

%
\begin{equation}
\label{eqN}
N = 4\sin^2\theta_{13}\frac{\sin^2(\hat A - 1) \Delta}{(\hat A-1)^2}\, ,
\end{equation}
\begin{equation}
\label{eqR}
R = 2\alpha\sin\theta_{13}\sin2\theta_{12}\cos\delta_{\rm CP} \cos\Delta\frac{\sin\hat A \Delta}{\hat A}\frac{\sin(\hat A - 1)\Delta}{\hat A - 1}\, ,
\end{equation}

and
\begin{equation}
\label{eqT}
T = \frac{1}{\hat{A}-1}\alpha\sin\theta_{13}\sin2\theta_{12}\cos\delta_{\rm CP}\sin\Delta\bigg(\hat A \sin\Delta - \frac{\sin \hat A \Delta}{\hat A}\cos(\hat A - 1)\Delta \bigg)\, .
\end{equation}
In the above equations, $\Delta \equiv \Delta m^{2}_{31}L/4E$ and $\hat{A} \equiv A/\Delta m^{2}_{31}$. The Wolfenstein matter term, $A = 2\sqrt{2}G_{F}N_{e}E = 7.6\times10^{-5} \times \rho~(\mathrm{g/cm^{3}}) \times E$ (GeV), where $G_{F}$ is the Fermi coupling constant, $N_{e}$ is the ambient electron density, $E$ is the energy of neutrino, and $\rho$ is the constant matter density through which neutrino propagates. In Eq.~\ref{pmumushort}, all the terms containing $\theta_{23}$ (see Eqs.~\ref{eqM} to~\ref{eqT}) provide crucial information to establish non-maximal $\theta_{23}$ and contribute towards the precision measurement of $\theta_{23}$, which are the focus of this work. The first term in Eq.~\ref{eqM}, which is proportional to $\sin^{2} {\Delta}$, is the leading term in muon neutrino survival channel and contributes the most to address the above-mentioned physics issues. The term in Eq.~\ref{eqN}, which is the leading term in $\nu_{\mu} \rightarrow \nu_{e}$ appearance channel, is suppressed by the small quantity $\sin^{2}\theta_{13}$, and the terms in Eqs.~\ref{eqR} and~\ref{eqT} are proportional to the quantitity $\alpha \sin \theta_{13}$ which is around $\sim 0.004$. Therefore, these terms provide sub-leading contributions towards establishing deviation from maximal mixing and to precisely measure the value of $\theta_{23}$. Note that the terms in Eqs.~\ref{eqR} and~\ref{eqT} are proportional to $\cos \delta_{\mathrm{CP}}$. These terms may help to measure the value of $\delta_{\mathrm{CP}}$, but they are blind to CP asymmetry. On the other hand, the terms in Eq.~\ref{eqN} and~\ref{eqT}, which are proportional to $\sin^{2}\theta_{23}$ and $\sin 4\theta_{23}$, respectively provide information on the octant of $\theta_{23}$. 

The main sensitivity to settle the octant of $\theta_{23}$ stems from $\nu_{\mu} \rightarrow \nu_{e}$ appearance channel ($P_{\mu e}$), which when expressed up to first order in $\alpha\sin\theta_{13}$ is given by (ignoring the term $\propto$ $\alpha^{2}$ and $\cos\theta_{13}\approx 1$)
\begin{equation}
\label{eqpmue}
P_{\mu e} \approx N \sin^2\theta_{23} + O \sin2\theta_{23}\cos\left(\Delta + \delta_{\mathrm {CP}}\right)\, ,
\end{equation}
where, 
\begin{equation}
\label{eqO}
O = 2\alpha\sin\theta_{13}\sin2\theta_{12}\frac{\sin\hat A \Delta }{\hat A}\frac{\sin(\hat A -1)\Delta }{\hat A-1}\, .
\end{equation}
%
Note that the first term in Eq.~\ref{eqpmue} is sensitive to octant of $\theta_{23}$, while the second term is sensitive to CP phase $\delta_{\mathrm {CP}}$. This leads to an octant\,-\,$\delta_{\mathrm {CP}}$ degeneracy in the measurements made via appearance channel. However, this degeneracy can be resolved with the help of balanced neutrino and antineutrino data in appearance mode as discussed for the first time in Ref.~\cite{Agarwalla:2013ju}. Since, both the terms in $P_{\mu e}$ contain information on $\theta_{23}$ (see Eq.~\ref{eqpmue}), they contribute towards establishing deviation from maximal $\theta_{23}$ (see discussion in Sec.~\ref{appdisapp} and Fig.~\ref{fig:7}) and to precisely measure the value of $\sin^{2}\theta_{23}$. 

	\begin{figure}[htb!]
	\centering
	\includegraphics[width=0.49\linewidth]{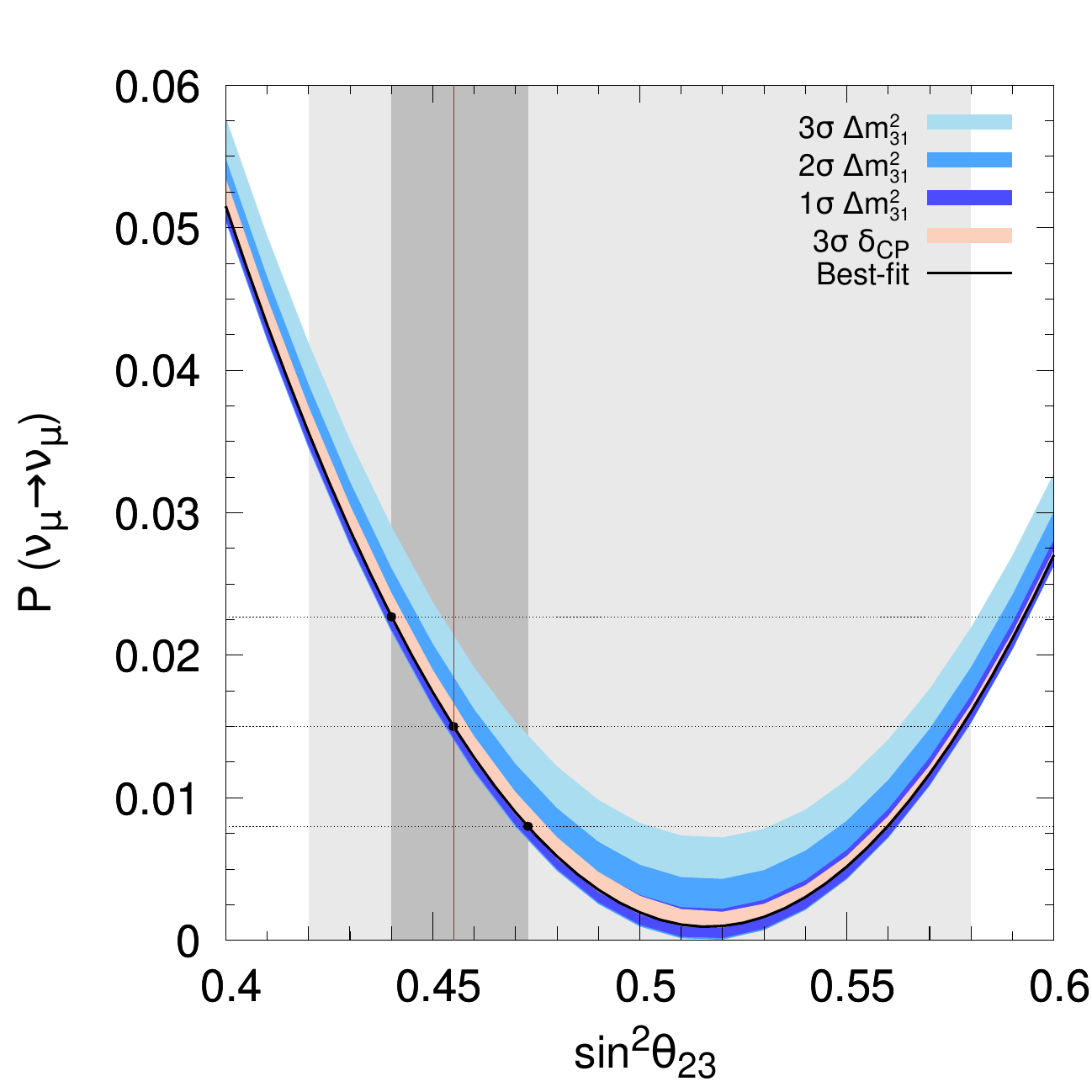}
	\includegraphics[width=0.49\linewidth]{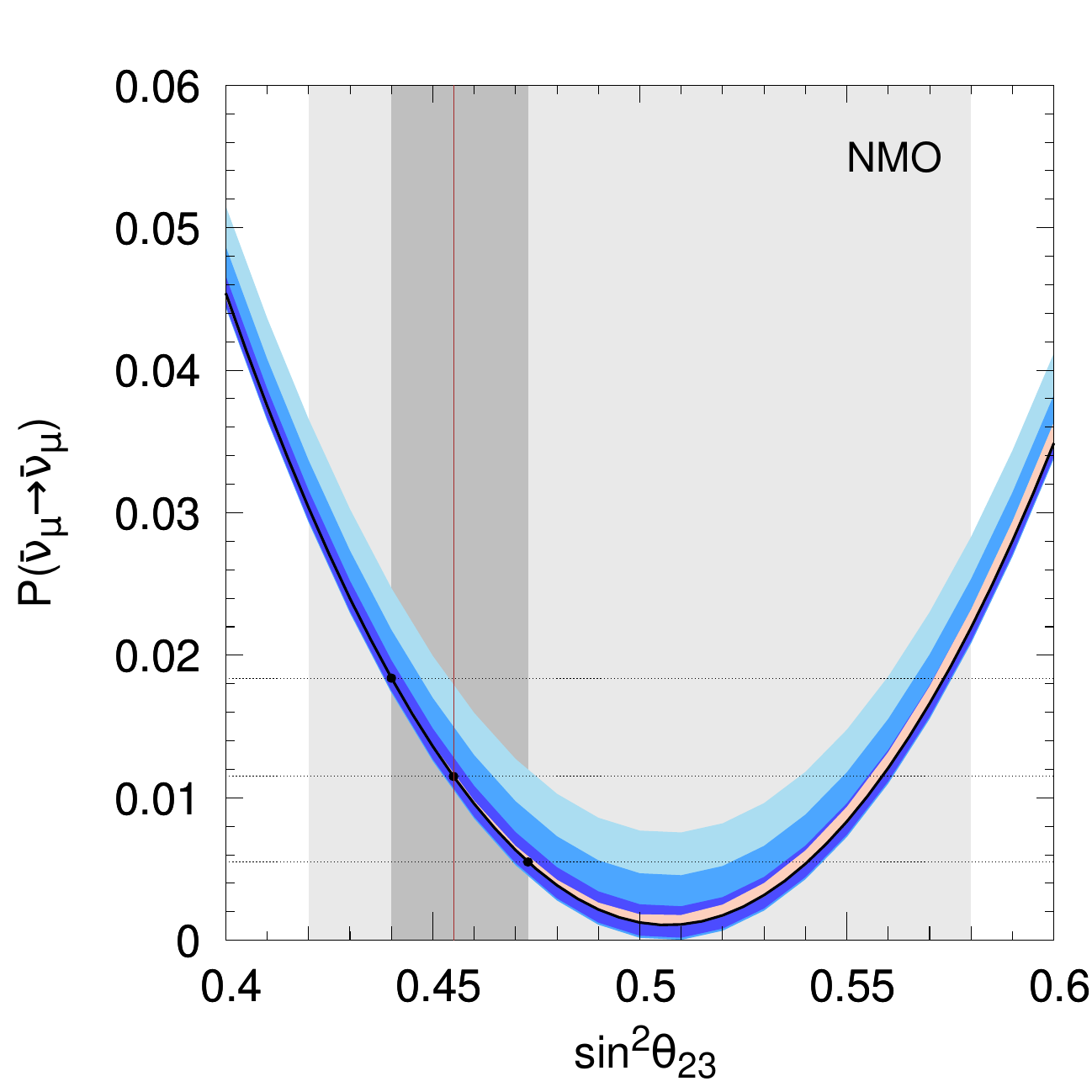}
	\includegraphics[width=0.49\linewidth]{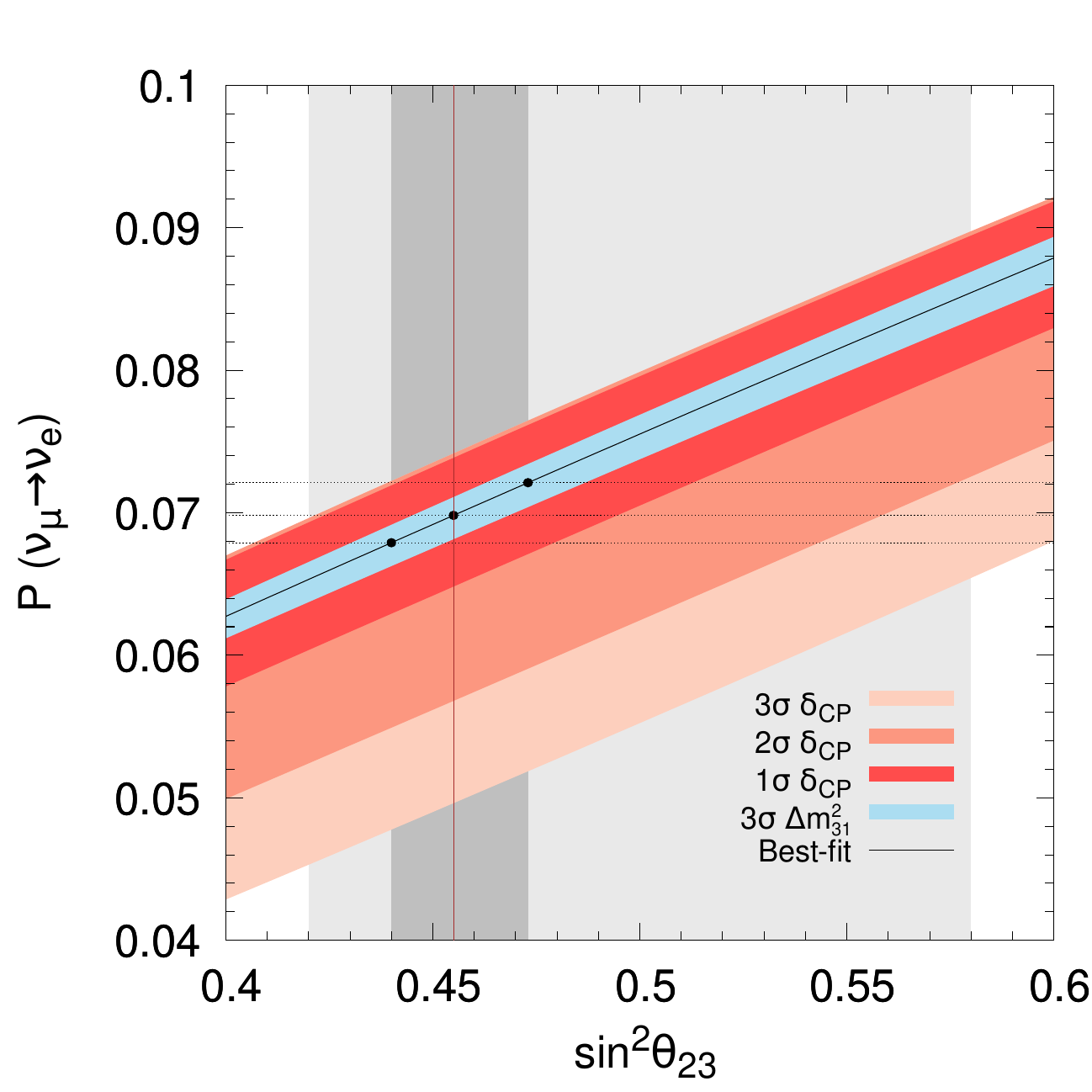}
	\includegraphics[width=0.49\linewidth]{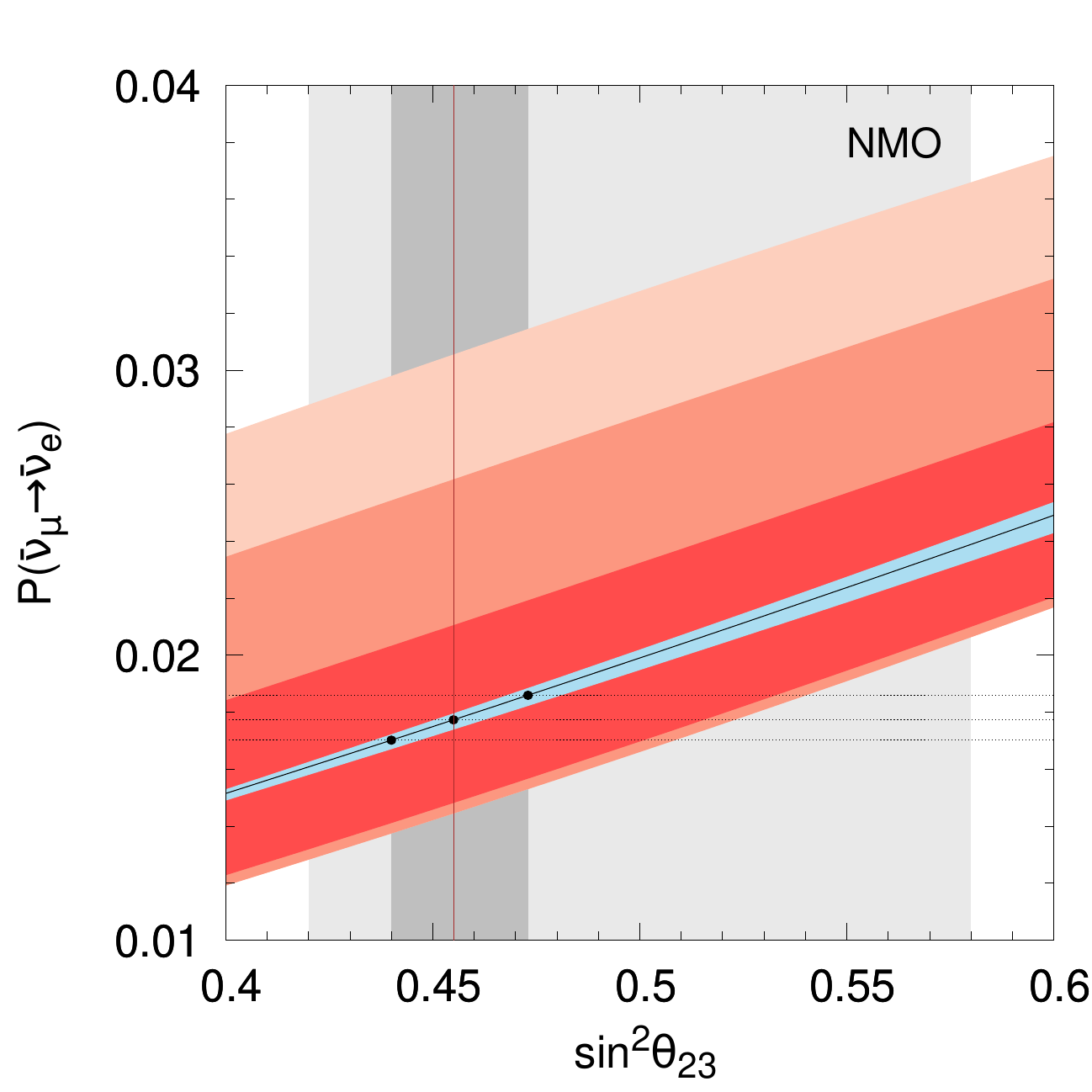}
	\caption{\footnotesize{Probability as a function of $\sin^2\theta_{23}$ for $E = 2.5$ GeV, $ L=1284.9$ km, and $\rho=2.848~ \rm{g/cm}^{3}$ assuming NMO. The top (bottom) panels are for the disappearance (appearance) channel. The left (right) panels are for neutrino (antineutrino). The black solid curves show the probability considering the best-fit values of oscillation parameters as given in Table~\ref{table:one}. The three shaded blue (red) regions depict the variations in probability due to the present $1\sigma,~2\sigma,~\rm{and}~3\sigma$ allowed ranges in $\Delta m^2_{31}$ ($\delta_{\rm CP}$). The dark (light)-shaded grey area shows the currently allowed $1\sigma~ (2\sigma)$ region in $\sin^{2}\theta_{23}$ with the best-fit value of $\sin^{2}\theta_{23} = 0.455$ as shown by the vertical brown line. See Table~\ref{table:one} for details. Note that y-ranges are different in the bottom two panels.}}
	\label{fig:2}
\end{figure} 
To understand the role of oscillation channels $P_{\mu \mu}$ and $P_{\mu e}$ in distinguishing a non-maximal $\sin^{2}\theta_{23}$ from maximal mixing $i.e.$ $\sin^{2}\theta_{23} = 0.5$, we draw Fig.~\ref{fig:2}. In this figure, we show oscillation probabilities as a function of $\sin^2\theta_{23}$. To generate this figure, we use our benchmark values of the oscillation parameters and their corresponding ranges from Table~\ref{table:one}. The top panels are for $P_{\mu\mu}$ while the bottom panels are for $P_{\mu e}$. In the left (right) panels, we show the probabilities for neutrino (antineutrino).\\
The solid black curves in each of these figures show the probability corresponding to the best-fit values of $\Delta m^2_{31}$ and $\delta_{\rm CP}$ for each $\sin^2\theta_{23}$ whereas the different shades of blue and red bands correspond to the variation in probability due to $1,~2,$ and $3\sigma$ variations in $\Delta m^2_{31}$ and $\delta_{\rm CP}$, respectively. To generate these probabilities, we choose $ E = 2.5~\rm GeV$ and $ L=1284.9$ km which correspond to the DUNE's baseline and peak-energy fluxes. $E \sim 2.5~\rm GeV$ also corresponds to the first oscillation maximum (minimum) in $P_{\mu e}$ ($P_{\mu\mu}$). From Fig.~\ref{fig:2}, we make the following observations.\\
\begin{itemize}
\item 
$P_{\mu\mu}$ varies a lot as $\Delta m^2_{31}$ is varied while it changes only marginally with respect to $\delta_{\rm CP}$\footnote{$P_{\mu\mu}$ depends only on $\cos\delta_{\rm CP}$ at the sub-leading interference term proportional to $\alpha\sin^2\theta_{13}$. See Eq. 33 of Ref. \cite{Akhmedov:2004ny}.}.
The opposite behavior is seen in the case of $P_{\mu e}$ \cite{Coloma:2014kca, Minakata:2013eoa}. Therefore, we see significant degeneracies among the oscillation parameters $\Delta m^2_{31}$ and $\sin^2\theta_{23}$ in $P_{\mu\mu}$ channel. For $P_{\mu e}$ channel, the degeneracies are observed among $\delta_{\rm CP}$ and $\sin^2\theta_{23}$. 
\item 
For values of $\sin^2\theta_{23}$ in the HO which are very close to $\sin^2\theta_{23}=0.5$ $i.e.$ $\sin^2\theta_{23} \in [0.5, 0.53]$, $P_{\mu\mu}$ shows a flat behavior, which means that the slope of $P_{\mu\mu}$ is nearly 0. Also, note that the minimum of $P_{\mu\mu}$ occurs in HO, slightly away from $\sin^2\theta_{23} = 0.5$. It happens due to finite $\theta_{13}$ corrections~\cite{Raut:2012dm}. At the same time, for values of $\sin^2\theta_{23}$ in the LO which are very close to $\sin^2\theta_{23}=0.5$, $P_{\mu\mu}$ is steep. However, for values of $\sin^2\theta_{23}$ which are adequately far from $\sin^2\theta_{23}=0.5$, $P_{\mu\mu}$ is very steep in both LO and HO. We observe these features in both neutrino and antineutrino probabilities.

We can understand the above mentioned features in the following fashion. Let us consider only the first three terms on the R.H.S in Eq.~\ref{pmumushort}, neglecting the last two terms which are $\alpha \sin \theta_{13}$ suppressed. Now, it is easy to show that 

\begin{equation}
\frac{dP_{\mu\mu}}{d\sin^2\theta_{23}} \approx -\,4M\left(1-2\sin^2\theta_{23}\right) - N\, . 
\end{equation} 

Thus, $dP_{\mu\mu}/d\sin^2\theta_{23} \propto 1-2\sin^2\theta_{23}$ and therefore, $P_{\mu\mu}$ is steep when $\sin^2\theta_{23}$ is far from 0.5 and $P_{\mu\mu}$ is expected to be flat when $\sin^2\theta_{23}$ is close to 0.5. However, the minimum of $P_{\mu\mu}$ occurs when $dP_{\mu\mu}/d\sin^2\theta_{23} = 0 $, $i.e.$ at

\begin{equation}
\sin^2\theta_{23} \approx \frac{1}{2}\left(1+\frac{N}{4M}\right) \, . 
\end{equation}

Given that N and M are positive quantities, the minimum of $P_{\mu\mu}$ occurs in HO at a value larger than $\sin^2\theta_{23}$ = 0.5, but not at $\sin^2\theta_{23}$ = 0.5. This shift in the location of minimium of $P_{\mu\mu}$ is mainly governed by the N term, which is proportional to $\sin^{2}\theta_{13}$~\cite{Raut:2012dm}. Therefore, we observe a flat behavior in $P_{\mu\mu}$ around this shifted minimum in HO.

\item 
$P_{\mu e}$ shows a monotonic increase with respect to $\sin^2\theta_{23}$. This is true for both LO and HO, and in case of both neutrinos and antineutrinos. 
\end{itemize} 

Thus, based on the observations made above, we expect the results to have the following features: 
\begin{itemize}
\item Since, we expect the combination of $P_{\mu\mu}$ and $P_{\mu e}$ channels to resolve the degeneracies that are present in each of them individually, we do not expect the sensitivity to exclude non-maximal $\theta_{23}$, be too much affected by the choice of $\Delta m^2_{31}$ and $\delta_{\rm CP}$ within the given $3\sigma$ range. 
\item For values of $\sin^2\theta_{23}$ in the HO and very close to 0.5, we expect that the sensitivity to establish deviation from maximality will come mainly from the appearance channel. However, for $\sin^2\theta_{23}$ values farther away from 0.5, the disappearance channel will contribute significantly. In the LO, we expect the main sensitivity to come from the disappearance channel even for $\sin^2\theta_{23}$ values very close to 0.5. 
\end{itemize} 
While the above arguments have been made using the probabilities calculated with a particular choice of $E = 2.5~\rm GeV$, we will see in the results section that these features hold in general.   

\section{Discussion at the level of events}
\label{events} 

We start this section by mentioning the salient features of DUNE which are crucial for our numerical simulations. Then, we show the total appearance and disappearance event rates in neutrino and antineutrino modes as a function of $\sin^{2}\theta_{23}$, $\Delta m^{2}_{31}$, and $\delta_{\mathrm{CP}}$ to establish some physics issues which are necessary to understand our main results. We also exhibit the bi-events plot in the plane of neutrino - antineutrino disappearance events and display their event spectra.

\subsection{Salient features of DUNE}
\label{experiment}

In order to calculate the expected event rates in DUNE and to estimate its sensitivity towards various physics issues, we use the publicaly available software GLoBES (General Long Baseline Experiment Simulator)~\cite{Huber:2004ka, Huber:2007ji}. We consider the simulation details as described in Ref.~\cite{DUNE:2021cuw}. DUNE will look for $\nu_{\mu}\rightarrow \nu_{\mu}$ (disappearance) and $\nu_{\mu}\rightarrow \nu_{e}$ (appearance) oscillations in both neutrino and antineutrino modes. Neutrinos are produced at the LBNF's Main Injector in Fermilab, Illinois, Chicago where protons of energy 120 GeV and power 1.2 MW are bombarded on a graphite target. This leads to the production of charged mesons which then decay in flight producing the neutrinos. Using the desired polarity in the horn-focusing system, neutrino or antineutrino mode can be selected. The neutrino flux at DUNE is wide-band with energies ranging from few hundreds of MeV to few tens of GeV, but the flux peaks at around $~$2.5 GeV with the majority of the flux lying in $1$ GeV to $5$ GeV region. 
These neutrinos first see a near detector (ND) placed 574 m downstream from the source and a far detector (FD) located roughly 5000 ft below the Earth's surface at the Sanford Underground Research Facility (SURF) in Lead, South Dakota, USA. The main purpose of ND is to precisely measure the unoscillated neutrino flux so as to reduce the systematic uncertainties related to fluxes. The distance between the source of production of neutrinos in Fermilab and the FD is 1284.9 km and the neutrinos traverse through Earth matter of roughly constant density of around $2.848~\rm g/cm ^{3}$. The FD is a 40 kt liquid argon time projection chamber (LArTPC) and is placed underground in order to minimize cosmogenic and atmospheric backgrounds. 
We consider a total run-time of 7 years equally divided in neutrino and antineutrino modes with a total of 1.1$ \times$ 10$^{21}$ protons on target (P.O.T.). This corresponds to a total exposure of 336 kt$\cdot$MW$\cdot$years equally shared in neutrino and antineutrino modes. 
The energy resolution of the FD in $\left(0.5 - 5\right)$ GeV range is around $\left(15 - 20\right)\%$. 
Our assumptions on systematic uncertainties are based on the material provided in Ref.~\cite{DUNE:2021cuw}. The errors are bin-to-bin correlated and are same for both neutrinos and antineutrinos. In the appearance channel $i.e.$ for the electron events, the normalization error is $2\%$ while for the disappearance channel $i.e.$ for the muon events, the normalization error is $5\%$. For the background events, the error varies from $5\%$ to $20\%$. 

\subsection{Appearance and disappearance event rates as a function of $\sin^{2}\theta_{23},\ \Delta m^{2}_{31},$ and $\delta_{\rm CP}$}

In Table~\ref{table:two}, we show the total neutrino and antineutrino event rates for DUNE as a function of the oscillation parameters for both disappearance and appearance channels. The three columns correspond to LO ($\sin^2\theta_{23} = 0.455$) on the left, MM ($\sin^2\theta_{23} = 0.5$) in the center and HO ($\sin^2\theta_{23} = 0.599$) on the right. The central number in each cell, shown in boldface corresponds to the total number of events for the best-fit values of $\delta_{\rm CP} = 223^\circ$ and $\Delta m^2_{31} = 2.522\times 10^{-3}~\rm eV^2$, while considering $\sin^2\theta_{23}$ in LO, MM, and HO in second, third, and fourth columns, respectively. Rest of the oscillation parameters are kept at their respective best-fit values (see Table~\ref{table:one} for details). We determine other two numbers in the left and right (top and bottom) of central value by varying $\delta_{\rm CP}$ ($\Delta m^2_{31}$) from its central best-fit value to $3\sigma$ lower and upper bounds, respectively (as explained in the schematic diagram above Table~\ref{table:two}).

\begin{table}[htb!]
	\begin{center}
		\begin{small}
			\begin{tabular}{||c|c|c|c|c||}
			\hline \hline
			    \multicolumn{5}{||c||}{} \\
			    \multicolumn{5}{||c||}{$\mathcal{N}\left(223, 2.436\right)$} \\
			    \multicolumn{5}{||c||}{$\uparrow$} \\
			    \multicolumn{5}{||c||}{$\mathcal{N}\left(139, 2.522\right) \leftarrow \mathcal{N}\left(223, 2.522\right) \rightarrow \mathcal{N}\left(355, 2.522\right)$} \\
			    \multicolumn{5}{||c||}{$\downarrow$} \\
			    \multicolumn{5}{||c||}{$\mathcal{N}\left(223, 2.605\right)$} \\
			    \multicolumn{5}{||c||}{} \\
			    \multicolumn{5}{||c||}{$\mathcal{N}\left(x, y\right)$ where  $\mathcal{N}$: total number of events, $x$: $\delta_{\rm CP}$ in degrees, $y$: $\Delta m^2_{31}$ in $10^{-3}~\rm eV^2$ }\\
			    \multicolumn{5}{||c||}{} \\
			    \hline \hline
			    \multicolumn{5}{c}{} \\
			
				\hline\hline
				\multicolumn{2}{||c|}{Channel} & \text{LO}  & \text{MM} & \text{HO} \\
				\hline
				 &  & 1104  & 1193 & 1383 \\
				
				& & \boldsymbol{$\uparrow$} & \boldsymbol{$\uparrow$} & \boldsymbol{$\uparrow$}\\
				
			\multirow{5}{*}{\rotatebox[origin=c]{90}{Appearance}}& \text{$\nu$} &
				{820 $\leftarrow$$\boldsymbol{1121}$$\rightarrow$ 969}  & 				{908 $\leftarrow$$\boldsymbol{1211}$$\rightarrow$ 1058}  & 
								{1107 $\leftarrow$$\boldsymbol{1403}$$\rightarrow$ 1254}   \\
				
				& & \boldsymbol{$\downarrow$}& \boldsymbol{$\downarrow$}&\boldsymbol{$\downarrow$}\\
				
				& & 1135  & 1226 &
				1421 \\
				\cline{2-5}
				&  & 206  & 227 & 277 \\
				
				& & \boldsymbol{$\uparrow$}& \boldsymbol{$\uparrow$}&\boldsymbol{$\uparrow$}\\

				& \text{$\bar\nu$} & {267 $\leftarrow$$\boldsymbol{208}$$\rightarrow$ 258}  & {289 $\leftarrow$$\boldsymbol{230}$$\rightarrow$ 280}  & {338 $\leftarrow$$\boldsymbol{279}$$\rightarrow$ 329}   \\
				
				& & \boldsymbol{$\downarrow$}&\boldsymbol{$\downarrow$}&\boldsymbol{$\downarrow$}\\

				& & 210  & 232 & 281 \\
				\hline \hline
				 &  & 11018  & 10797 & 11249 \\
				
				& & \boldsymbol{$\uparrow$}& \boldsymbol{$\uparrow$}& \boldsymbol{$\uparrow$}\\
				
				\multirow{5}{*}{\rotatebox[origin=c]{90}{Disappearance}} & \text{$\nu$} & {10870 $\leftarrow$$\boldsymbol{10870}$$\rightarrow$ 10896}  & {10646 $\leftarrow$$\boldsymbol{10646}$$\rightarrow$ 10663}  & {11100 $\leftarrow$$\boldsymbol{11100}$$\rightarrow$ 11095} \\
				
				& & \boldsymbol{$\downarrow$}& \boldsymbol{$\downarrow$}&\boldsymbol{$\downarrow$}\\

				& & 10758  & 10532 & 10986 \\
				\cline{2-5}
				&  & 6397  & 6310 & 6565 \\
				
				& & \boldsymbol{$\uparrow$}&\boldsymbol{$\uparrow$}&\boldsymbol{$\uparrow$}\\

				&\text{$\bar\nu$} & {6306 $\leftarrow$$\boldsymbol{6306}$$\rightarrow$ 6280}  & {6219 $\leftarrow$$\boldsymbol{6219}$$\rightarrow$ 6193}  & {6477 $\leftarrow$$\boldsymbol{6477}$$\rightarrow$ 6452} \\
				
				& & \boldsymbol{$\downarrow$}&\boldsymbol{$\downarrow$}&\boldsymbol{$\downarrow$}\\

				& & 6234  & 6146 & 6406 \\
				\hline\hline
				
			\end{tabular}
		\end{small}
	\end{center}
	\caption{\footnotesize{Total appearance and diappearance event rates in $\nu$ and $\bar{\nu}$ mode. We assume 3.5 years of $\nu$ run and 3.5 years of $\bar{\nu}$ run and estimate the event rates for three different choices of $\sin^{2}\theta_{23}$: 0.455 (LO), 0.5 (MM), and 0.599 (HO). The central number in each cell corresponds to the current best-fit values of $\delta_{\rm CP}$ = 223$^{\circ}$ and $\Delta m^2_{31}$ = 2.522 $\times$ 10$^{-3}$ eV$^{2}$ assuming NMO. The other four numbers in each cell show the number of events corresponding to the present $3\sigma$ lower and upper bounds in $\Delta m^2_{31}$ (up and down arrows) and $\delta_{\rm CP}$ (left and right arrows). For clarity, see the schematic diagram given above this table.}}
	\label{table:two}
\end{table}

\begin{figure}[htb!]
	\centering
	\includegraphics[width=0.49\linewidth]{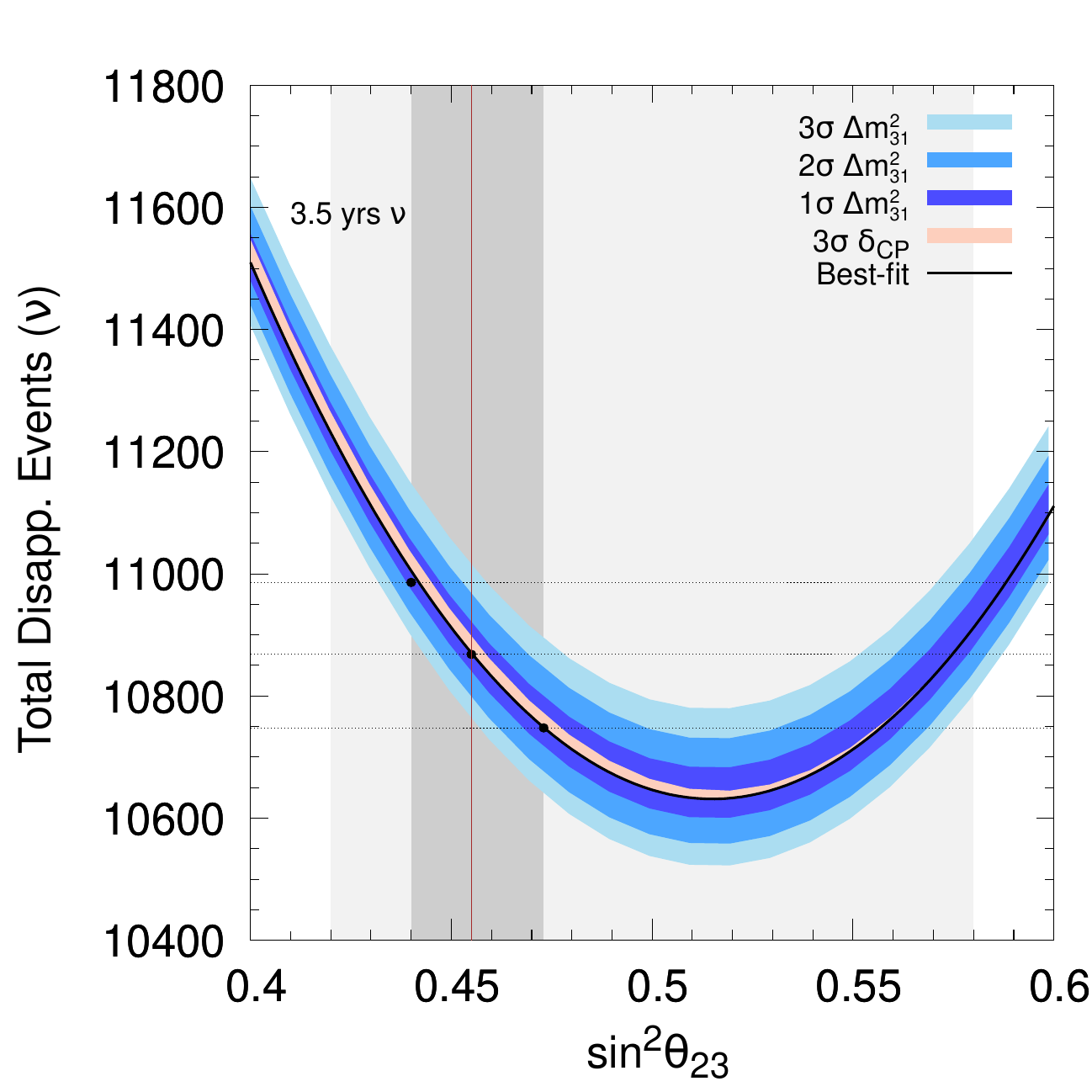}
	\includegraphics[width=0.49\linewidth]{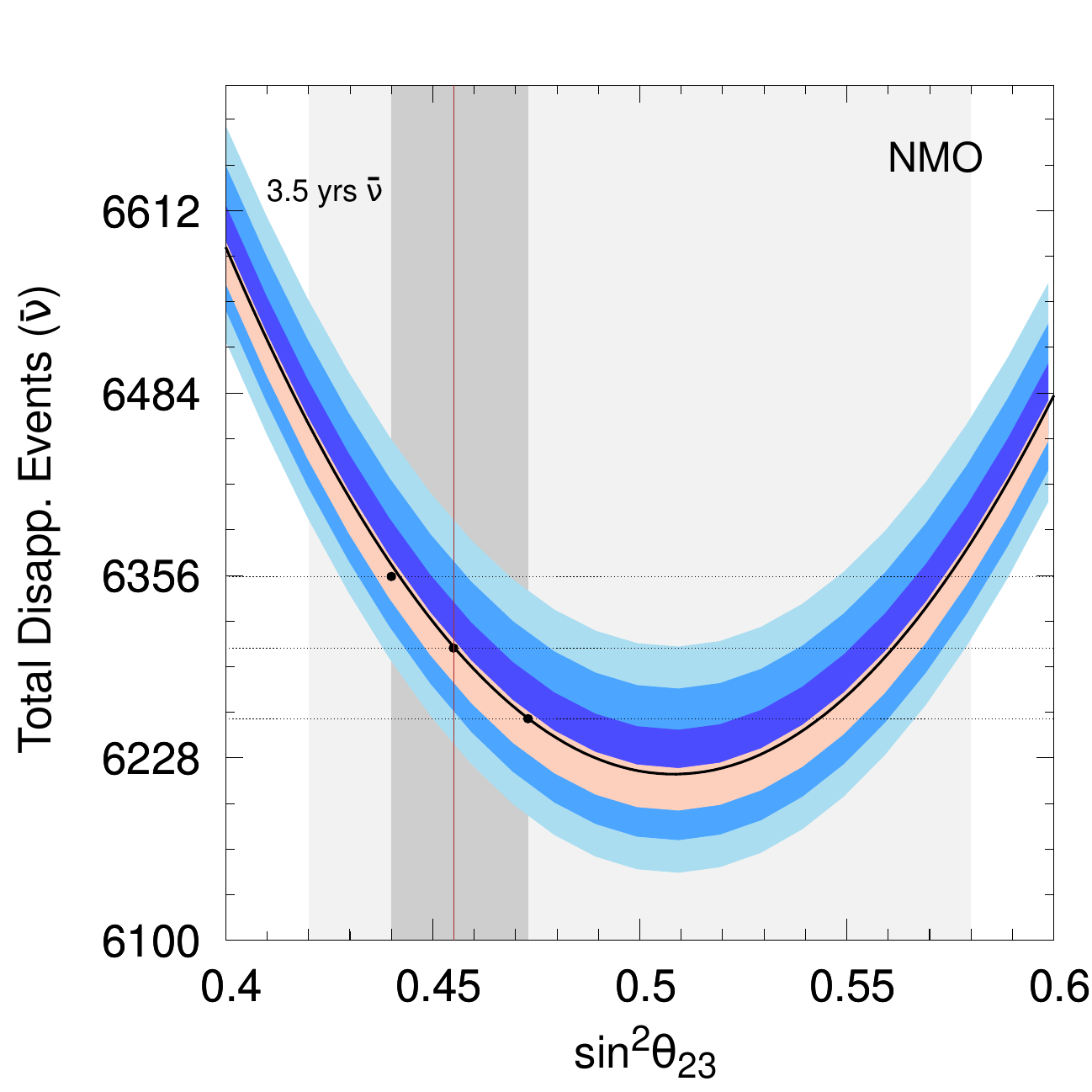}
	\includegraphics[width=0.49\linewidth]{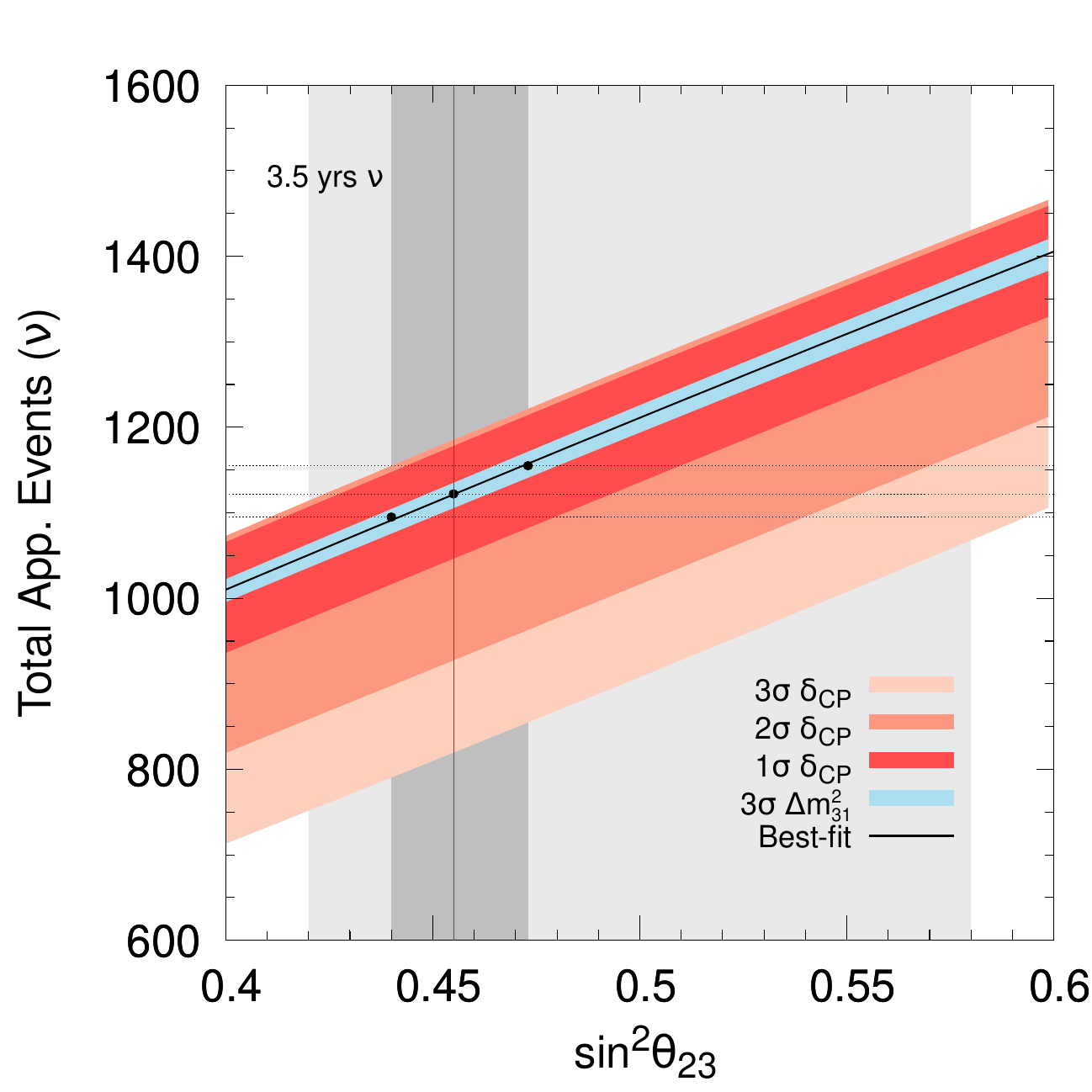}
	\includegraphics[width=0.49\linewidth]{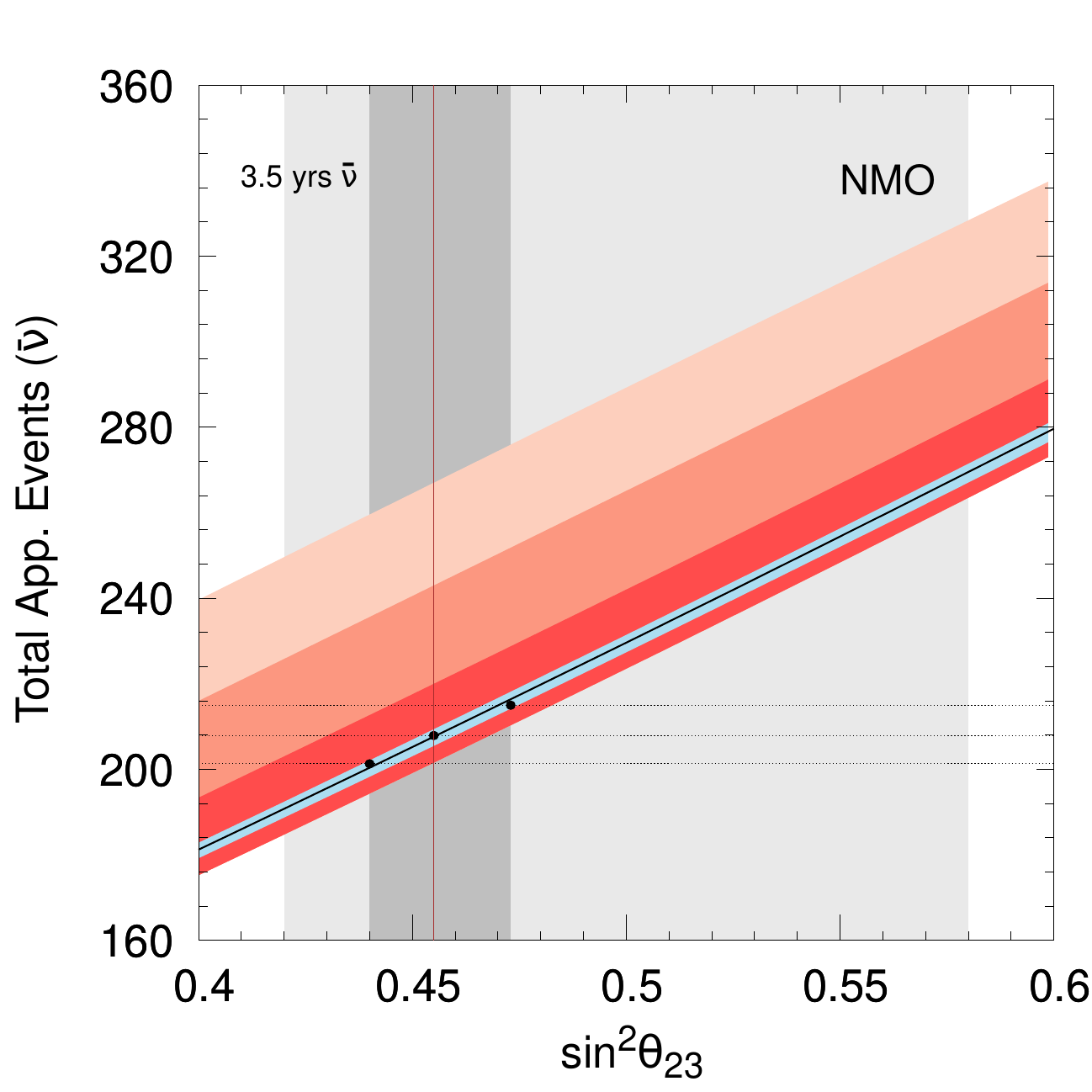}
	\caption{\footnotesize{Total event rates as a function of $\sin^2\theta_{23}$ for DUNE assuming NMO. The top (bottom) panels are for disappearance (appearance) channel. The left (right) panels are for neutrino (antineutrino) assuming 3.5 years of run. The black solid curves show the event rates considering the best-fit values of oscillation parameters as given in Table~\ref{table:one}. The three shaded blue (red) regions show the variations in events due to present $1\sigma,~2\sigma,~ \rm{and}~3\sigma$ allowed ranges in $\Delta m^2_{31}$ ($\delta_{\rm CP}$). The dark (light)-shaded grey area shows the currently allowed $1\sigma~ (2\sigma)$ region in $\sin^{2}\theta_{23}$ with the best-fit value of $\sin^{2}\theta_{23} = 0.455$ as shown by the vertical brown line. See Table~\ref{table:one} and related text for details. Note that y-ranges are different in all the four panels. }}
	\label{fig:3}
\end{figure}
From Table~\ref{table:two}, we note the following: 
\begin{itemize}
\item The appearance events change by a lot when $\sin^2\theta_{23}$ or $\delta_{\rm CP}$ is varied. This is observed in the case of both neutrino and antineutrino events. Thus, there are distinct $\theta_{23}$ - $\delta_{\rm CP}$ pairs which give same number of total events.  
\item The change in appearance events due to variation in $\Delta m^2_{31}$ is very small and the number of events in the three cases (best-fit, $3\sigma$ upper bounds, and 3$\sigma$ lower bounds) are almost degenerate. 
\item In the case of disappearance events, the central number for LO and HO are close to one another, but they are different from MM. The numbers also change significantly with respect to $\Delta m^2_{31}$. As far as $\delta_{\rm CP}$ variation is concerned, the event numbers show almost no change. Thus, in the case of disappearance events, there appears a $\sin^2\theta_{23}$ - $\Delta m^2_{31}$ degeneracy at the level of total rates.
\end{itemize} 
The observations made above are in line with the physics discussion done before in Sec.~\ref{probability} based on probabilities. As an example, we note that for 
$\nu$ appearance events, the numbers can vary between 820 to 969 for LO and 908 to 1058 for MM as $\delta_{\rm CP}$ is varied in the current $3\sigma$ range. Thus, there is a significant overlap in $\nu$ appearance events for LO and MM due to unknown $\delta_{\mathrm{CP}}$. However, for $\nu$ disappearance channel, the same oscillation parameter sets give number of events in the range 10870 to 10896 and 10646 to 10663, respectively, which may help to reduce the degenearcy as observed in appearance channel. The reverse argument can also be made where, for $\nu$ disappearance events, the numbers vary between 10758 to 11018 for LO and 10532 to 10797 for MM as $\Delta m^2_{31}$ is varied in its current $3\sigma$ range. But in the case of neutrino appearance channel, the corresponding events are lie in the range of 1104 to 1135 for LO and 1193 to 1226 for MM. Thus, the degeneracy that exists in the disappearance channel is partially resolved through the measurements made using appearance channel.

In Fig.~\ref{fig:3}, we show the disappearance (top panels) and appearance (bottom panels) event rates for various choices of $\sin^2\theta_{23}$ in the range $[0.4, 0.6]$. To generate this figure, we use the benchmark values of the oscillation parameters and their corresponding ranges as given in Table~\ref{table:one}. We see that the total event rates follow the same behavior as previously seen in Fig.~\ref{fig:2}, where we show $P_{\mu\mu}$ and $P_{\mu e}$ as a function of $\sin^2\theta_{23}$ assuming $E= 2.5~\rm GeV$. Note that though in Sec.~\ref{probability}, we discuss the main physics issues assuming a particular value of $E=2.5$ GeV, similar features are retained at the total event rates level as well in Fig.~\ref{fig:3}.

\begin{figure}[htb!]
\centering
\includegraphics[width=0.63\linewidth]{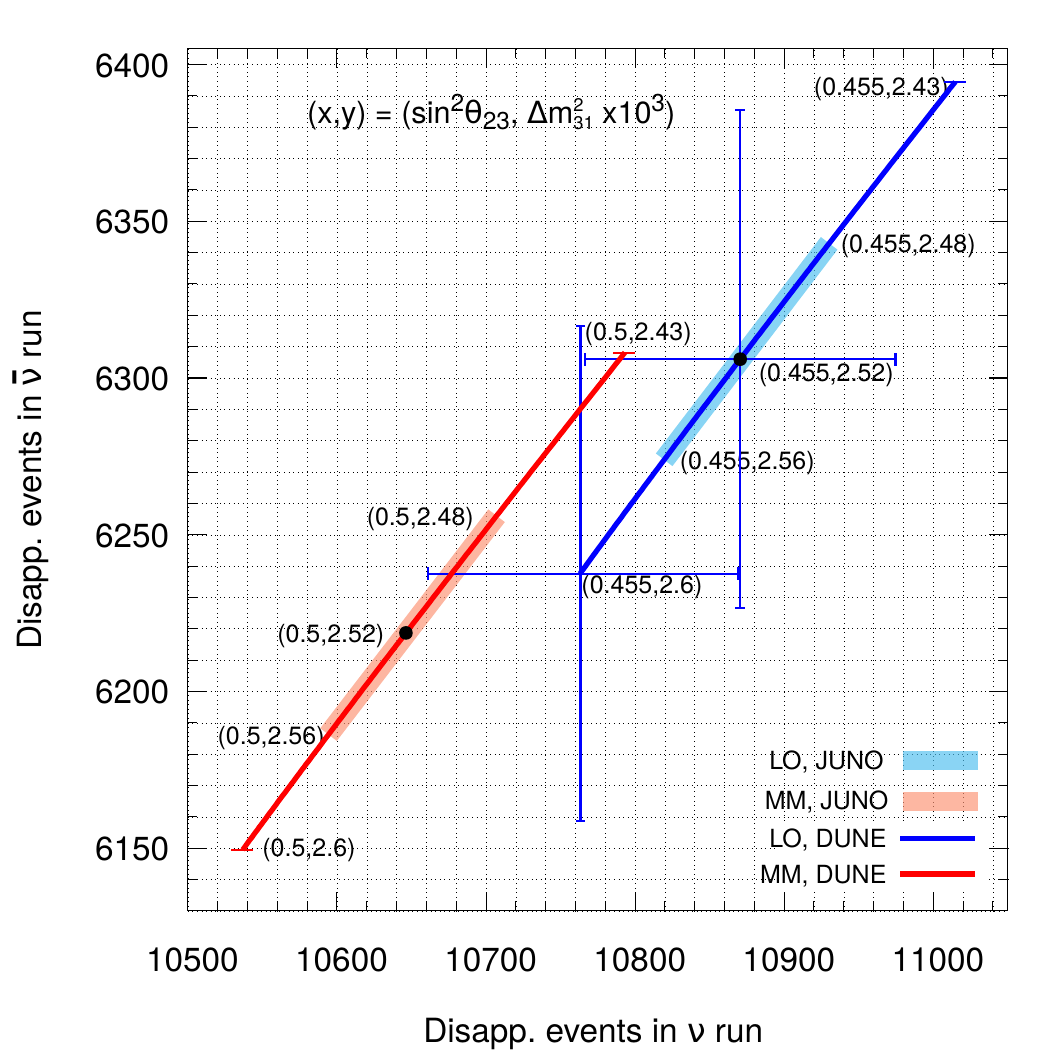}
\caption{\footnotesize{Bi-events plot for DUNE in the plane of neutrino - antineutrino disappearance events assuming 336 kt$\cdot$MW$\cdot$years of exposure equally divided in neutrino and antineutrino modes. The blue line is obtained by varying $\Delta m^{2}_{31}$ in its 3$\sigma$ range of $[2.43 : 2.6] \times 10^{-3}~\rm{eV}^{2}$ with $\sin^2\theta_{23}=0.455$ (LO). The red line depicts the same with $\sin^2\theta_{23}=0.5$ (MM). The black dot on each line shows the disappearance events corresponding to the best-fit value of $\Delta m^2_{31} = 2.52 \times 10^{-3}$ eV$^{2}$. The values of other oscillation parameters are taken from Table~\ref{table:one} assuming NMO. The blue (red) rectangular region on blue (red) line portrays the variation in event rates due to 3$\sigma$ range in $\Delta m^{2}_{31}$ expected from JUNO~\cite{EPS-HEP-Conference2021, JUNO:2015zny}. The horizontal (vertical) error bars for the points [($\sin^{2}\theta_{23} = 0.455, \Delta m^{2}_{31} = 2.52 \times 10^{-3} \rm{eV}^{2}$) and ($\sin^{2}\theta_{23} = 0.455, \Delta m^{2}_{31} = 2.6 \times 10^{-3} \rm{eV}^{2}$)] show the 1$\sigma$ statistical uncertainties which are obtained by taking the square root of the $\nu \,(\bar{\nu})$ disappearance events.}} 
\label{fig:4}
\end{figure}

In Fig.~\ref{fig:4}, we show the dependence of disappearance events on the oscillation parameters $\Delta m^2_{31}$ and $\sin^2\theta_{23}$ through the bi-events plot where we display the total neutrino (antineutrino) disappearance events on the x-axis (y-axis) assuming 3.5 years of run. We obtain blue curve by varying $\Delta m^{2}_{31}$ in its current 3$\sigma$ range of $[2.43 : 2.6] \times 10^{-3}~\rm{eV}^{2}$ assuming the current best-fit of $\sin^2\theta_{23}=0.455$ (see LO in legends). The red curve portrays the same with $\sin^2\theta_{23}=0.5$ (see MM in legends). The black dot on each line shows the disappearance events corresponding to the best-fit value of $\Delta m^2_{31} = 2.52 \times 10^{-3}$ eV$^{2}$. The values of other oscillation parameters are taken from Table~\ref{table:one} assuming NMO and $\delta_{\rm CP}$ = 223$^{\circ}$. The blue (red) rectangular region on blue (red) curve shows the variation in event rates due to allowed 3$\sigma$ range in $\Delta m^{2}_{31}$ as expected from JUNO~\cite{EPS-HEP-Conference2021, JUNO:2015zny}. The horizontal (vertical) error bars for the points [($\sin^{2}\theta_{23} = 0.455, \, \Delta m^{2}_{31} = 2.52 \times 10^{-3} \rm{eV}^{2}$) and ($\sin^{2}\theta_{23} = 0.455, \, \Delta m^{2}_{31} = 2.6 \times 10^{-3} \rm{eV}^{2}$)] show the 1$\sigma$ statistical uncertainties which are obtained by taking the square root of the neutrino (antineutrino) disappearance events.

The 1$\sigma$ statistical uncertainty in neutrino disappearance events corresponding to the benchmark oscillation parameters $\sin^2\theta_{23} = 0.455$ and $\Delta m^2_{31} = 2.522 \times 10^{-3}~\rm eV^2$ (see horizontal error bar around the black dot on blue line) has some overlap with the neutrino events on the red line. The same is true for antineutrino disappearance events (see the vertical error bar around the black dot on blue line). These overlapping regions due to 1$\sigma$ statistical fluctuations get reduced when we consider the variation in event rates for the allowed 3$\sigma$ range in $\Delta m^{2}_{31}$  (2.48 $\times$ 10$^{-3}$ eV$^2$ to 2.56$ \times$ 10$^{-3}$ eV$^2$, see rectangular red region) as expected from JUNO~\cite{EPS-HEP-Conference2021, JUNO:2015zny}. Therefore, we can conclude that based on only total event rates, a definitive exclusion of MM is not possible at high confidence level for the current best-fit values of oscillation parameters. We demonstrate later that the $\sin^2\theta_{23}$ - $\Delta m^2_{31}$ degeneracy that is present at the total event rates level can be resolved by including the spectral shape information along with total event rates and we can establish the deviation from maximal $\theta_{23}$ at high confidence level in DUNE. If we consider the event rates and their 1$\sigma$ statistical uncertainties corresponding to the oscillation parameters $\sin^2\theta_{23} = 0.455$ (current best-fit) and $\Delta m^2_{31} = 2.6 \times 10^{-3}~\rm eV^2$ (current 1$\sigma$ upper bound) on the blue line then we see more overlap with the event rates on the red line corresponding to MM solution. The overlap is less for the benchmark oscillation parameters $\sin^2\theta_{23} = 0.455$ (current best-fit) and $\Delta m^2_{31} = 2.43 \times 10^{-3}~\rm eV^2$ (current 1$\sigma$ lower bound) on the blue line.

\subsection{Disappearance event spectra to resolve $\sin^{2} \theta_{23}-\Delta m^{2}_{31}$ degeneracy}

\begin{figure}[htb!]
	\centering
	\includegraphics[width=0.49\linewidth]{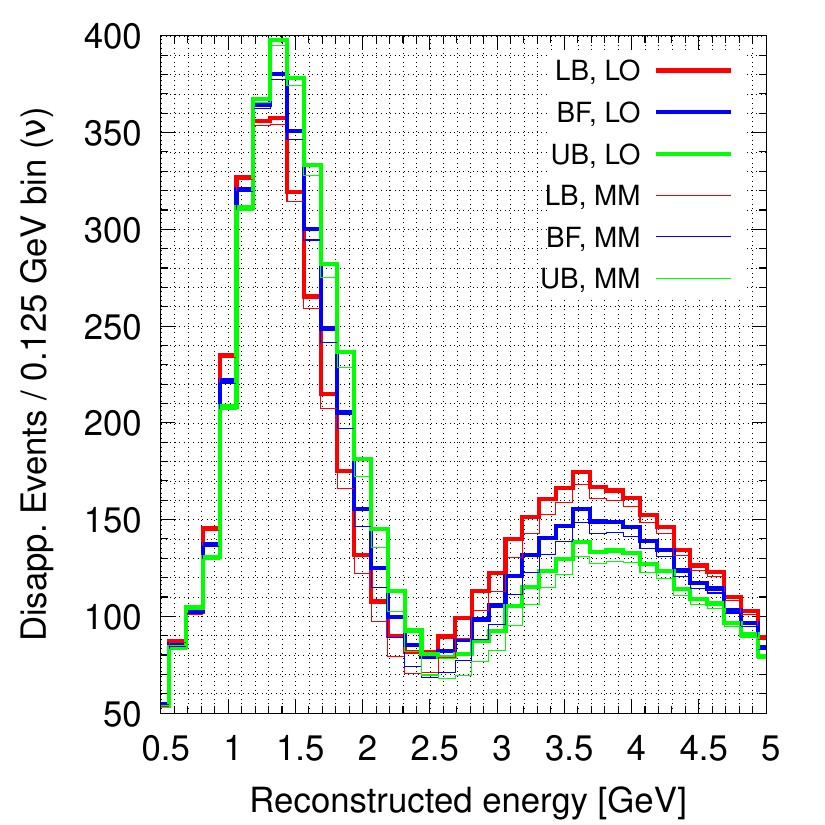}
	\includegraphics[width=0.49\linewidth]{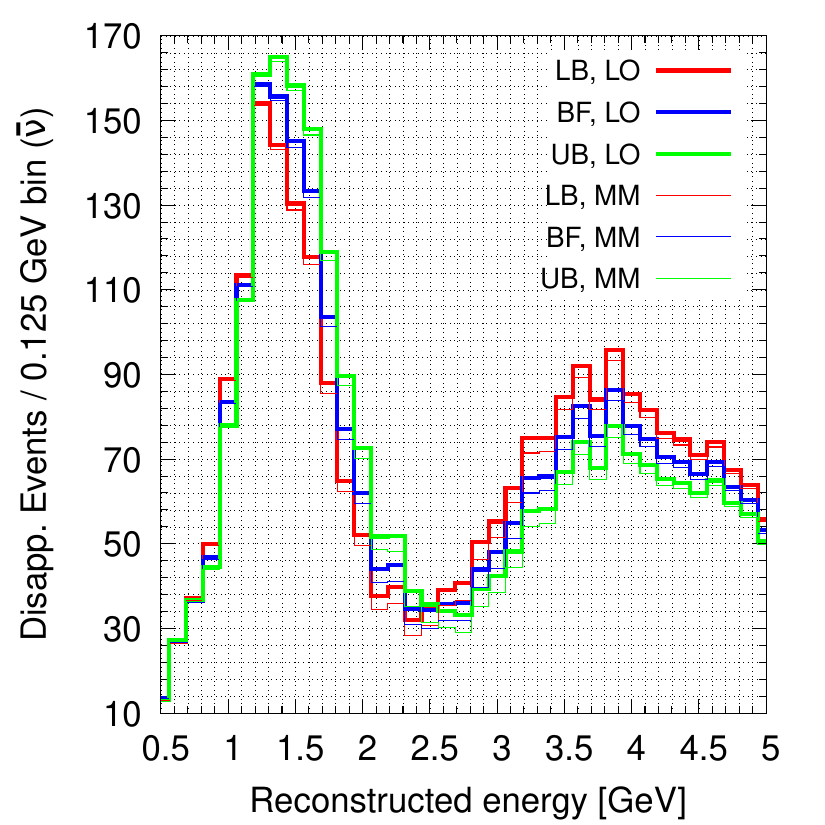}
	\includegraphics[width=0.49\linewidth]{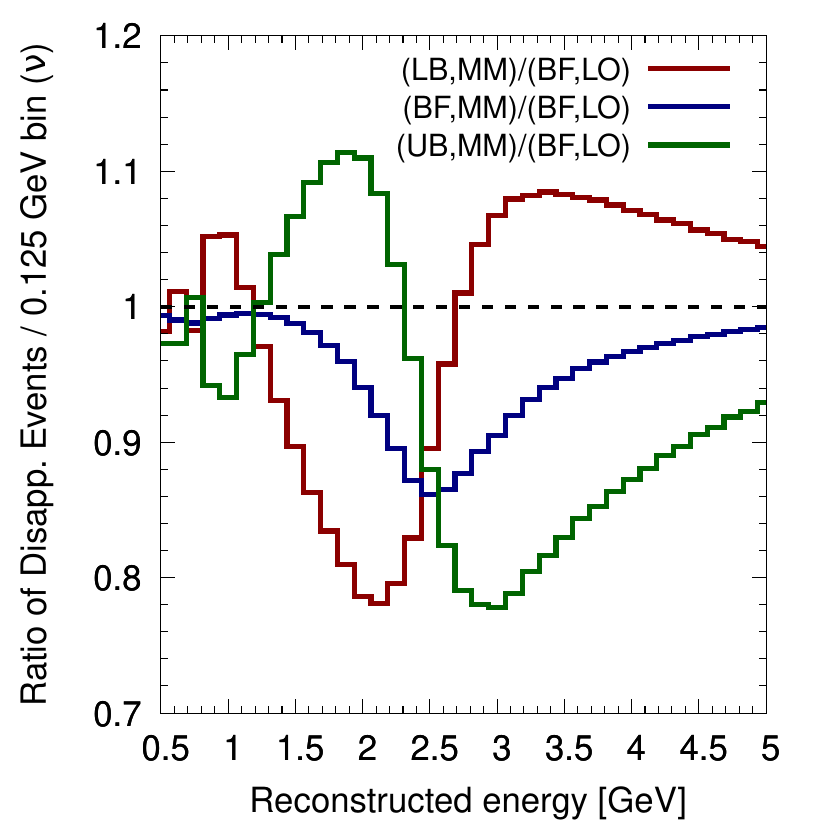}
	\includegraphics[width=0.49\linewidth]{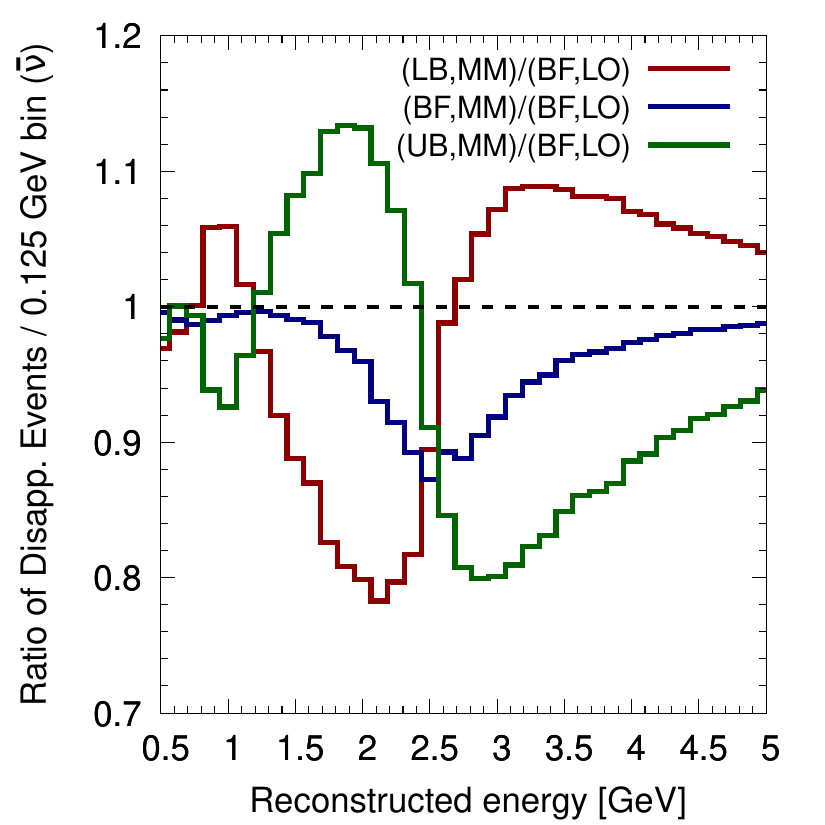}
	\caption{\footnotesize{In the top left (right) panel, we show the expected neutrino (antineutrino) disappearance event spectra as a function of reconstructed neutrino (antineutrino) energy assuming 3.5 years of $\nu$ ($\bar{\nu}$) run. The thick (thin) colored histograms correspond to a LO (MM) value of $\sin^{2}\theta_{23}$ = 0.455 (0.5). For a given value of $\sin^{2}\theta_{23}$, we depict the event spectra for three different choices of $\Delta m^2_{31}$: 2.522$\times 10^{-3}$ eV$^2$ [current best-fit (BF), blue lines], 2.436$\times 10^{-3}$ eV$^2$ [current 3$\sigma$ lower bound (LB), red lines], and 2.605$\times 10^{-3}$ eV$^2$ [current 3$\sigma$ upper bound (UB), green lines]. In the bottom left panel, we show the ratio of neutrino disappearance events in each energy bin as a function of reconstructed neutrino energy assuming 3.5 years of $\nu$ run. We present the same for antineutrino in the bottom right panel. The brown, blue, and green curves show the ratio of $N_{1}/{N}$, $N_{2}/{N}$, and $N_{3}/{N}$, respectively, where $N$: events in a given energy bin for $\left(\Delta m^2_{31} = 2.522\times 10^{-3}~\rm eV^2, \sin^2\theta_{23}=0.455\right)$, $N_{1}$: events in a given energy bin for $\left(\Delta m^2_{31} = 2.436\times 10^{-3}~\rm eV^2, \sin^2\theta_{23}=0.5\right)$, $N_{2}$: events in a given energy bin for $\left(\Delta m^2_{31} = 2.522\times 10^{-3}~\rm eV^2, \sin^2\theta_{23}=0.5\right)$, and $N_{3}$: events in a given energy bin for $\left(\Delta m^2_{31} = 2.605\times 10^{-3}~\rm eV^2, \sin^2\theta_{23}=0.5\right)$.
}}
	\label{fig:5}
\end{figure}

In Fig.~\ref{fig:5}, we show the event spectra for $\nu_{\mu}\rightarrow\nu_{\mu}$ disappearance channel as a function of the reconstructed neutrino energy. The top left (top right) panel is for neutrino (antineutrino) disappearance events. In the top panel, the thick and thin colored histograms correspond to a LO and MM value of $\sin^{2}\theta_{23}$, respectively. For a given value of $\sin^{2}\theta_{23}$, we exhibit the event spectra for three different choices of $\Delta m^2_{31}$: 2.522 $\times 10^{-3}$ eV$^2$ (BF, see blue lines), 2.436 $\times 10^{-3}$ eV$^2$ (LB, see red lines), and 2.605 $\times 10^{-3}$ eV$^2$ (UB, see green lines). Looking at Fig.~\ref{fig:5}, it seems that there is significant degeneracy between LO and MM when a full $3\sigma$ variation in $\Delta m^2_{31}$ is considered. However, on observing closely, it can be seen that the events for energy bins on either side of the oscillation minimum (maximum in the case of $P_{\mu e}$) at $ E = 2.5~GeV$ behave oppositely when $\Delta m^2_{31}$ is varied. This is made more evident in the lower panel of Fig.~\ref{fig:5} where the left (right) figure corresponds to neutrino (antineutrino). The three set of curves correspond to ratio of events in each reconstructed energy bin - $N_{i}/N$ (for $i =1,2,3$) where N is the number of events when $\sin^2\theta_{23} = 0.455$ and $\Delta m^2_{31} = 2.522 \, \times  10^{-3}~\rm eV^2$. In the ratio, $N_{1}, N_{2}, N_{3}$ are number of events when $\sin^2\theta_{23} = 0.5$ and $\Delta m^2_{31} = 2.438 \, \times 10^{-3}~\rm eV^2$, $2.522 \times 10^{-3}~\rm eV^2$, and $2.602 \times 10^{-3}~\rm eV^2$, respectively. It can be seen from the lower panels in Fig.~\ref{fig:5}, that the ratio of events approach 1 (reduction in sensitivity towards exclusion of maximality) on one side of the oscillation minimum while moving farther away from 1 compared to the blue line on the other side of the oscillation minimum. This explains that, while some of the energy bins decrease the sensitivity in deviation from the maximal choice of $\theta_{23}$, the other energy bins help to increase.  Therefore, we conclude that doing a spectral analysis further breaks the $\sin^2\theta_{23}$ - $\Delta m^2_{31}$ degeneracy seen in total event rates and therefore current 3$\sigma$ uncertainty in $\Delta m^2_{31}$ will not affect the sensitivity of DUNE in establishing non-maximal mixing of $\theta_{23}$.

\section{Our findings}
\label{results}

In this section, we demonstrate the capability of DUNE to address three important issues related to atmospheric oscillation parameter: (i) possible deviation of $\theta_{23}$ from maximal mixing (45$^{\circ}$), (ii) the correct octant of $\theta_{23}$ if it turns out to be non-maximal in Nature, and (iii) the achievable precision on the atmospheric oscillation parameters $\sin^2\theta_{23}$ and $\Delta m^2_{31}$ in light of current neutrino oscillation data. To estimate the median sensitivities in the frequentist approach~\cite{Blennow:2013oma}, we use the following definition of Poissonian $\chi^2$

\begin{equation}
\chi^2 (\vec{\omega}, \, \kappa_{s}, \, \kappa_{b})= \underset{( \vec{\lambda}, \, \kappa_{s}, \,\kappa_{b})}{\mathrm{min}}\left\{  2\sum_{i=1}^{n}(\tilde{y_i}-x_i-x_i\mathrm{ln}\frac{\tilde{y_i}}{x_i})+\kappa^2_{s}+ \kappa^2_{b}\right\}\, , 
\label{chi2} 
\end{equation}

where, n is the total number of reconstructed energy bins and 

\begin{equation}
\tilde{y_i}\,(\vec{\omega},\{\kappa_{s},\kappa_{b}\}) = N^{th}_i(\vec{\omega})[1+\pi^s\kappa_{s}]+N^b_i(\vec{\omega})[1+\pi^b\kappa_{b}]\, .
\label{chi}
\end{equation}
In the above equation, $N^{th}_i\,(\vec{\omega})$ denotes the predicted number of signal events in the $i$-th energy bin for a set of oscillation parameters $\vec{\omega} = \left\lbrace \theta_{23},\theta_{13},\theta_{12},\Delta \mathrm{m}^{2}_{21}, \Delta \mathrm{m}^{2}_{31},\delta_{\mathrm{CP}}\right\rbrace$ and $\vec{\lambda}$ is the set of oscillation parameters which we have marginalized in the fit. For an instance, when we address the issue of deviation from maximality, $\vec{\lambda} = \{\Delta m_{31}^2,~\delta_{\mathrm{CP}}\}$. $N^{b}_i\,(\vec{\omega})$ represents the number of background events in the $i$-th energy bin where the neutral (charged) current backgrounds are independent (dependent) on $\vec{\omega}$. The quantity $\pi^s$ ($\pi^b$) is the normalization uncertainty on signal (background). The quantities $\kappa_{s}$ and $\kappa_{b}$ are the systematic pulls~\cite{Huber:2002mx,Fogli:2002pt,Gonzalez-Garcia:2004pka} on signal and background, respectively. We incorporate the prospective data in Eq.~\ref{chi2} using the variable $x_{i}= N^{ex}_i\, +\, N^b_i$, where $N^{ex}_i$ denotes the observed charged current signal events in the $i$-th energy bin and $N^b_i$  represents the background as mentioned earlier.

\subsection{Deviation from maximal $\theta_{23}$}

As discussed previously in Fig.~\ref{fig:1}, the three global analyses of the oscillation data do not agree on the octant in which the best-fit value of $\theta_{23}$ lies. Further, they all find $\sin^2\theta_{23}=0.5$ to be allowed at $3\sigma$ confidence level.
Therefore, before we address the issue of resolving the octant of $\theta_{23}$, it is imperative to question at what confidence level maximal 2-3 mixing can be ruled out. We define $\Delta\chi^2$ for deviation from maximal $\theta_{23}$ as follows: 
\begin{equation}
\Delta \chi^2_{\rm DM}= \underset{(\vec{\lambda},~ \kappa_{s},\kappa_{b})}{\mathrm{min}}\left\{ \chi^2\left(\sin^2\theta_{23}^{\mathrm{true}} \in [0.4,0.6]\right) - \chi^2\left(\sin^2\theta_{23}^{\mathrm{test}} = 0.5\right)\right\},  
\end{equation} 
Here, $\vec{\lambda} = \{\delta_{\mathrm{CP}}, \, \Delta m^{2}_{31}\}$ are the oscillation parameters over which the $\Delta\chi^2$ has been marginalized, while $\kappa_{s}~\text{and}~ \kappa_{b}$ are the systematic pulls on signal and background, respectively.

\begin{figure}[htb!]
	\centering
	\includegraphics[width=0.49\linewidth]{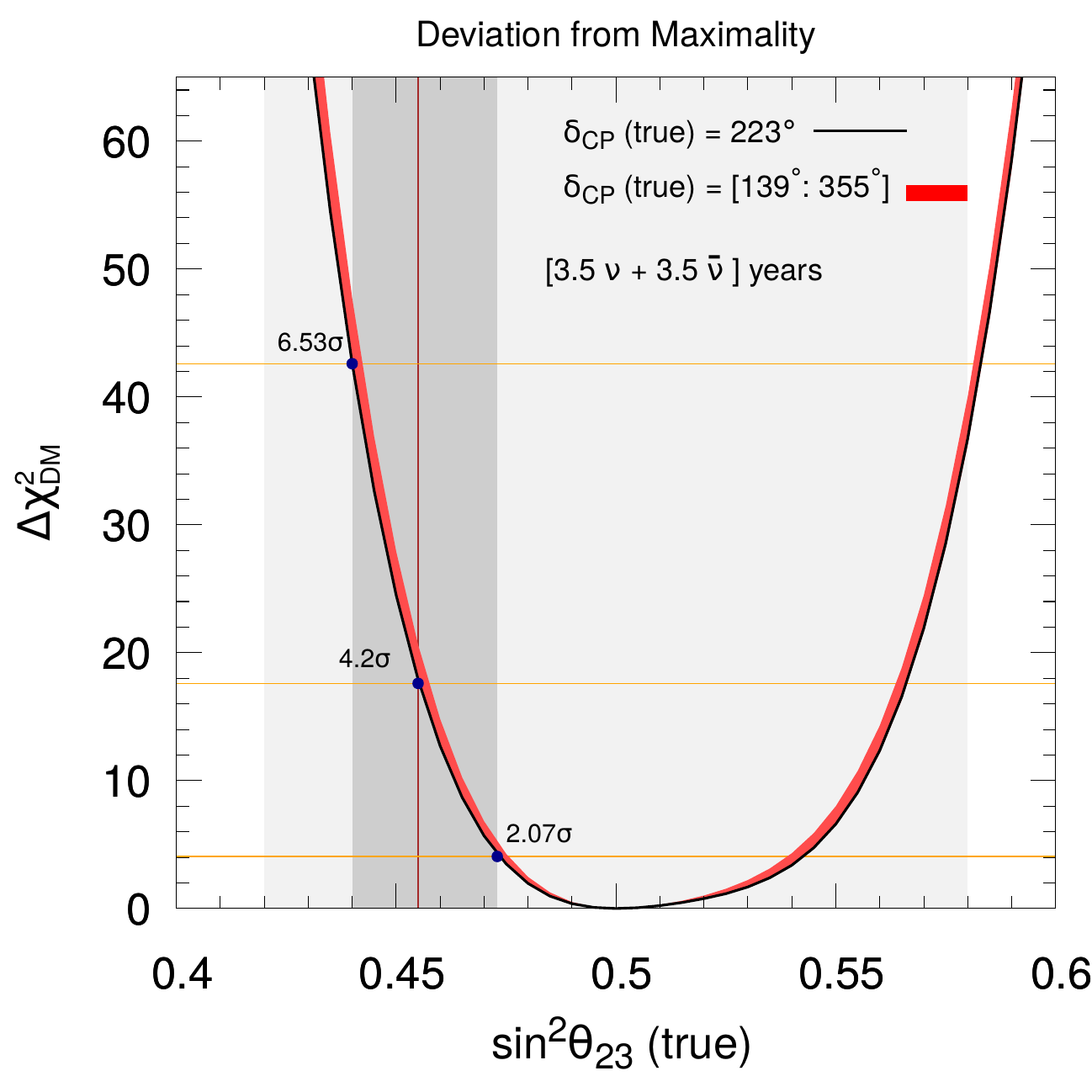}
	\includegraphics[width=0.49\linewidth]{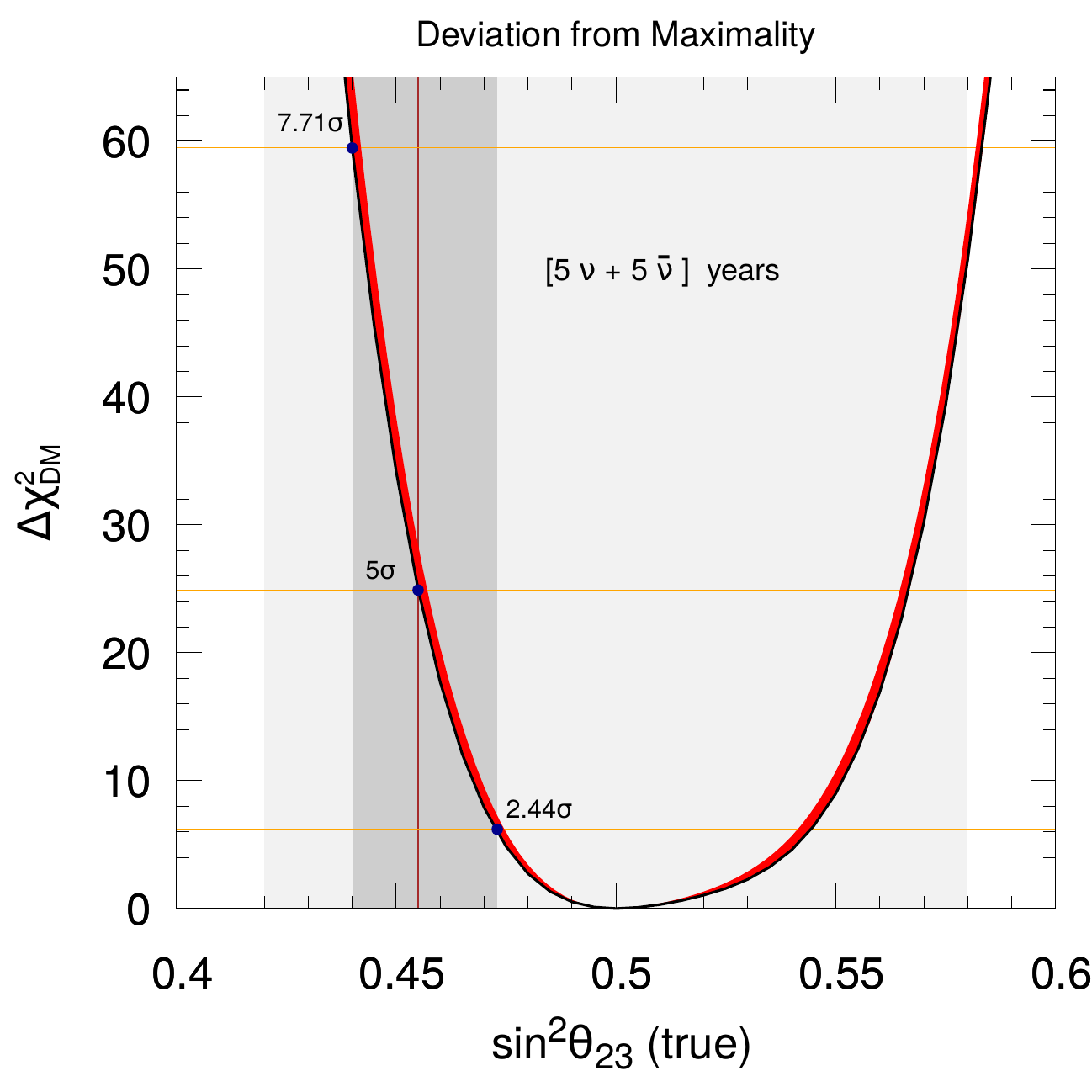}
	\caption{\footnotesize{The black curve in the left (right) panel shows the potential of DUNE to establish the deviation from maximal $\theta_{23}$ as a function of true $\sin^2\theta_{23}$ assuming true NMO and $\delta_{\mathrm{CP}}$ (true) = 223$^{\circ}$ with 7 (10) years of total exposure equally divided in $\nu$ and $\bar{\nu}$ modes. The red bands portray the same for true $\delta_{\mathrm{CP}}$ in the range of 139$^{\circ}$ to 355$^{\circ}$. In the fit, we marginalize over the current 3$\sigma$ range of $\Delta m^{2}_{31} $ and $\delta_{\rm CP}$, while keeping rest of the oscillation parameters fixed at their present best-fit values as shown in Table~\ref{table:one}. The dark (light)-shaded grey area shows the currently allowed $1\sigma$ $(2\sigma)$ region in $\sin^{2}\theta_{23}$ as obtained in the global fit study~\cite{Capozzi:2021fjo} with the best-fit value of $\sin^{2}\theta_{23} = 0.455$ as shown by vertical brown line. The horizontal orange lines show the sensitivity (experessed in $\sigma = \sqrt{\Delta\chi^2_{\rm DM}}$) for the current best-fit and 1$\sigma$ upper and lower bounds of $\sin^2\theta_{23}$.}}
	\label{fig:6}
\end{figure}

In Fig.~\ref{fig:6}, we show the potential of DUNE to establish deviation from maximal $\theta_{23}$ as a function of the true $\sin^2\theta_{23}$ in the range of 0.4 to 0.6. The black lines in both left and right panels display the ability of DUNE in establishing deviation from maximal $\theta_{23}$ assuming true NMO and $\delta_{\rm CP}$ (true) = 223$^{\circ}$. In the left panel, we show the results with nominal neutrino and antineutrino runs of 3.5 years each, while in the right panel we show results with 5 years of running in each mode. The red bands in Fig.~\ref{fig:6} portray the variation in $\Delta \chi^2_{\rm DM}$ for true $\delta_{\rm CP}$  in its current 3$\sigma$ allowed range of 139$^{\circ}$ to 355$^{\circ}$ (see Table~\ref{table:one}). Left panel reveals that for $\sin^2\theta_{23}$ (true) = 0.47 (current 1$\sigma$ upper bound), 0.455 (current best-fit), and 0.44 (current 1$\sigma$ lower bound), DUNE can exclude maximal mixing solution at 2.07$\sigma$, 4.2$\sigma$, and 6.5$\sigma$, respectively assuming true NMO and with a total 7 years of run (corresponds to a total exposure of 336 kt$\cdot$MW$\cdot$years) equally divided in neutrino and antineutrino modes. For a total 10 years of run (corresponds to a total exposure of 480 kt$\cdot$MW$\cdot$years), the above sensitivities get enhanced to 2.44$\sigma$, 5$\sigma$, and 7.71$\sigma$, respectively (see right panel). We observe that a 3$\sigma$ (5$\sigma$) determination of non-maximal $\theta_{23}$ is possible in DUNE with a total exposure of 7 years if the true value of $\sin^2\theta_{23} \lesssim 0.465~(0.450)$ or $\sin^2\theta_{23} \gtrsim 0.554~(0.572)$ for any value of true $\delta_{\mathrm{CP}}$ in its present 3$\sigma$ range and true NMO (see left panel). When we increase the total exposure from 7 years to 10 years, we see a marginal enhancement in the sensitivity (see right panel).

\subsubsection{Contributions from appearance and disappearance channels and role of systematics}
\label{appdisapp}

\begin{figure}[htb!]
	\centering
	\includegraphics[width=0.65\linewidth]{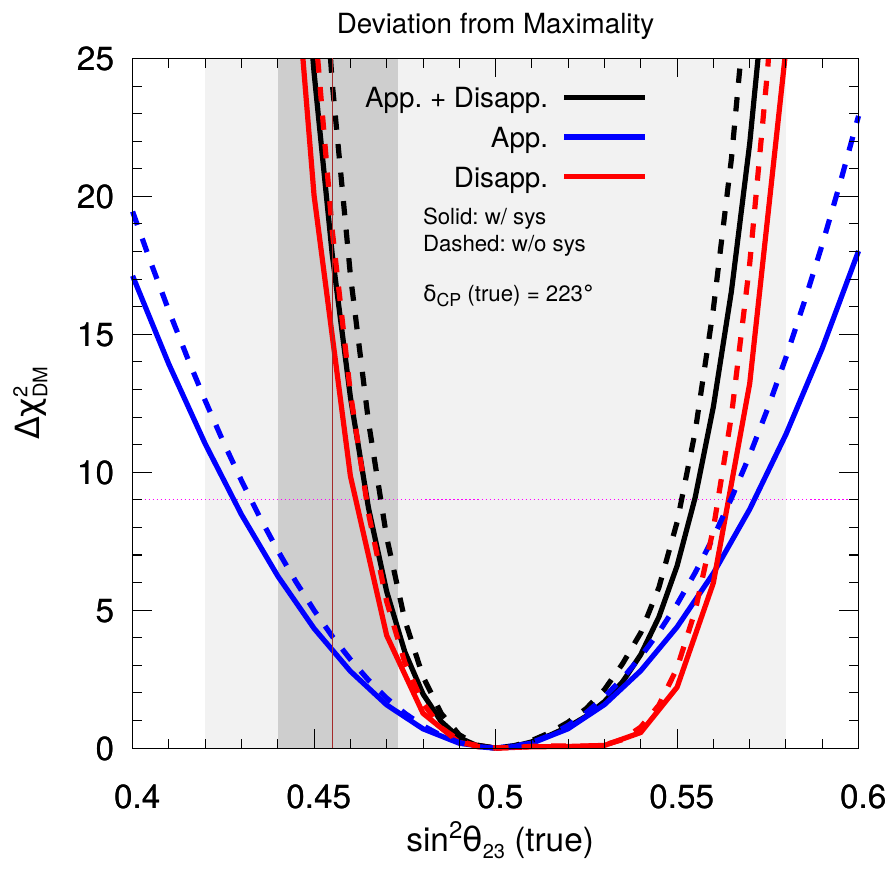}
	\caption{\footnotesize{Performance of DUNE to establish deviation from maximality as a function of true $\sin^2\theta_{23}$ assuming true NMO and $\delta_{\rm CP}$ (true) = 223$^\circ$ with 7 years of exposure equally divided in $\nu$ and $\bar{\nu}$ modes. Red, blue, and black curves show the sensitivity obtained using disappearance channel, appearance channel and their combinations, respectively. The solid (dashed) lines depict the results with (without) systematic uncertainties. In the fit, we marginalize over the current 3$\sigma$ range of $\Delta m^{2}_{31} $ and $\delta_{\rm CP}$, while keeping rest of the oscillation parameters fixed at their present best-fit values as shown in Table~\ref{table:one}. The dark (light)-shaded grey area shows the currently allowed $1\sigma$ $(2\sigma)$ region in $\sin^{2}\theta_{23}$ as obtained in the global fit study~\cite{Capozzi:2021fjo} assuming NMO with the best-fit value of $\sin^{2}\theta_{23} = 0.455$ as shown by vertical brown line. The capability of DUNE to establish non-maximal $\theta_{23}$ at 3$\sigma$ ($\Delta \chi^{2}_{\mathrm{DM}} = 9$) confidence level is shown by horizontal pink dotted line.}}
	\label{fig:7}
\end{figure}

We now explore how the appearance and disappearance channels individually contribute towards the exclusion of MM. In Fig.~\ref{fig:7}, we show the $\Delta \chi^2_{\rm DM}$ as a function of true $\sin^2\theta_{23}$ for appearance (in solid blue), disappearance (in solid red) and combined (in solid black). It is interesting to note that for true  values of $\sin^2\theta_{23}$ in HO that are very close to MM, the appearance channel provides better sensitivity towards the exclusion of MM. However, for $\sin^2\theta_{23} \gtrsim 0.56$, the $\Delta \chi^2$ increases very rapidly. In the case of LO, we see that it is mainly the disappearance channel that contributes to the exclusion of MM.  
In order to understand such a behavior, we refer to Section~\ref{probability}, where we discuss that, for $\sin^2\theta_{23} \gtrsim 0.5$, $P_{\mu\mu}$ shows a flat behavior while $P_{\mu e}$ increases linearly. It is only when $\sin^2\theta_{23}$ is a little away from $0.5$ that $P_{\mu\mu}$ increases steeply. On the other hand, in LO, $P_{\mu\mu}$ is very steep even for values which are close to $0.5$. In this figure, we also discuss the role that systematic uncertainties play in deteriorating the sensitivity of DUNE towards exclusion of MM. We consider two scenarios here which are shown in Fig.~\ref{fig:7}. In the first case, we consider an ideal experimental setup with no systematic uncertainties (shown with dashed curves in Fig.~\ref{fig:7}). In the second case, we consider the DUNE's nominal systematic uncertainties (shown by solid curves in Fig.~\ref{fig:7}) described in Ref. \cite{DUNE:2021cuw}. Looking at Fig.~\ref{fig:7}, it appears that both appearance and disappearance channels are affected by the systematics especially when going from a no systematics ideal experimental setup to the realistic situation. For example, at true $\sin^2\theta_{23} = 0.455$, systematic uncertainties deteriorate the MM-exclusion from $\Delta\chi^2_{\rm DM} = 25$ to $\Delta\chi^2 = 20$. In order to explore this point further, we generate results for three more choices of systematic uncertainties. The results are shown in Table~\ref{table:three}.

\begin{table}[htb!]
	\begin{center}
		\begin{small}
			\begin{tabular}{|c|c|c|c|c|c|c|}
				\hline\hline
                 {True $\sin^2\theta_{23}$} & Channels & 2\%, 5\% & 0\%, 0\% & 5\%, 5\% & 5\%, 10\% & 10\%, 10\%\\
 				\hline
				 & App.+Disapp.&
				$\boldsymbol{17.64}$ &
				$\boldsymbol{24.13}$  & 
				$\boldsymbol{16.88}$ &
				$\boldsymbol{16.74}$ &
				$\boldsymbol{15.42}$\\
				
				$\boldsymbol{0.455}$ & App.&	 $\boldsymbol{3.52}$ & $\boldsymbol{4.05}$ & $\boldsymbol{2.33}$  & $\boldsymbol{2.33}$ & $\boldsymbol{1.05}$ \\
				
			(Best-fit)	& 	Disapp.& $\boldsymbol{14.31}$ & $\boldsymbol{18.79}$ & $\boldsymbol{14.31}$  & $\boldsymbol{14.16}$ & $\boldsymbol{14.16}$ \\

				\hline \hline
				 & App.+Disapp.&
				$\boldsymbol{4.28}$ &
				$\boldsymbol{5.72}$ & 
				$\boldsymbol{3.88}$ &
				$\boldsymbol{3.84}$ &
				$\boldsymbol{3.42}$\\
				
				$\boldsymbol{0.473}$ & 	App.& $\boldsymbol{1.27}$ & $\boldsymbol{1.47}$ & $\boldsymbol{0.84}$  & $\boldsymbol{0.84}$ & $\boldsymbol{0.38}$ \\
				
				($1\sigma$ upper bound) & 	Disapp.& $\boldsymbol{2.99}$ & $\boldsymbol{3.88}$ &
				$\boldsymbol{2.99}$  & 
				$\boldsymbol{2.97}$ & 
				$\boldsymbol{2.97}$ \\
				
				\hline\hline
				
			\end{tabular}
		\end{small}
	\end{center}
	\caption{\footnotesize{Impact of systematics uncertainties on the determination of non-maximal $\sin^{2}\theta_{23}$. We show results for $\sin^2\theta_{23}$ (true) = 0.455 (current best-fit) and  $\sin^2\theta_{23}$ (true) = 0.473 (current $1\sigma$ upper bound) assuming true NMO and $\delta_{\rm CP}$ (true) = 223$^\circ$ with 7 years of exposure equally shared in $\nu$ and $\bar{\nu}$ modes. We estimate the sensitivity for different choices of normalization uncertainties on appearance and disappearance events [(2\%, 5\%), (0\%, 0\%), (5\%, 5\%), (5\%, 10\%), and (10\%, 10\%)], where (2\%, 5\%) is the benchmark choice~\cite{DUNE:2021cuw}. We keep the normalization uncertainties on various backgrounds fixed at their default values as given in Ref.~\cite{DUNE:2021cuw}. Results are given for the appearance channel, disappearance channel, and their combination. In the fit, we marginalize over the current 3$\sigma$ range of $\Delta m^{2}_{31}$ and $\delta_{\rm CP}$, keeping rest of the oscillation parameters fixed at their present best-fit values as shown in Table~\ref{table:one}.}}
	\label{table:three}
\end{table}

We show the results for two values of $\sin^2\theta_{23}$ corresponding to the current best-fit of 0.455 and the present $1\sigma$ upper bound of 0.473 (see Table~\ref{table:one}). The three rows in Table~\ref{table:three} correspond to combined data from appearance and disappearance, only appearance data and only disappearance data. The different columns correspond to various choice of the systematic errors denoted as $(x\%,y\%)$ where $x\%$ denotes the normalization error in the measurement of electron-like events (due to appearance) while $y\%$ denotes the normalization error in the measurement of muon-like events (due to disappearance). We do not change the systematic uncertainties for background events and consider the same systematic uncertainty values for both neutrino and antineutrino channels. Our results show that while the sensitivity certainly deteriorates as we go from the ideal case of $(0\%, 0\%)$ to the nominal values of $(2\%, 5\%)$, there is a negligible decrease in sensitivity due to only disappearance channel as the errors are increased further. The appearance channels are affected more because of systematic uncertainties, but since their contribution to the overall sensitivity to establish deviation from maximal $\theta_{23}$ is marginal, it does not affect much. Therefore, we conclude that DUNE's sensitivity to MM exclusion will not be systematics dominated and similar performance can be expected even with somewhat worse systematics. 

\subsubsection{Advantage due to spectral analysis 
and impact of marginalization over oscillation parameters}  

\begin{figure}[htb!]
	\centering
	\includegraphics[width=0.7\textwidth]{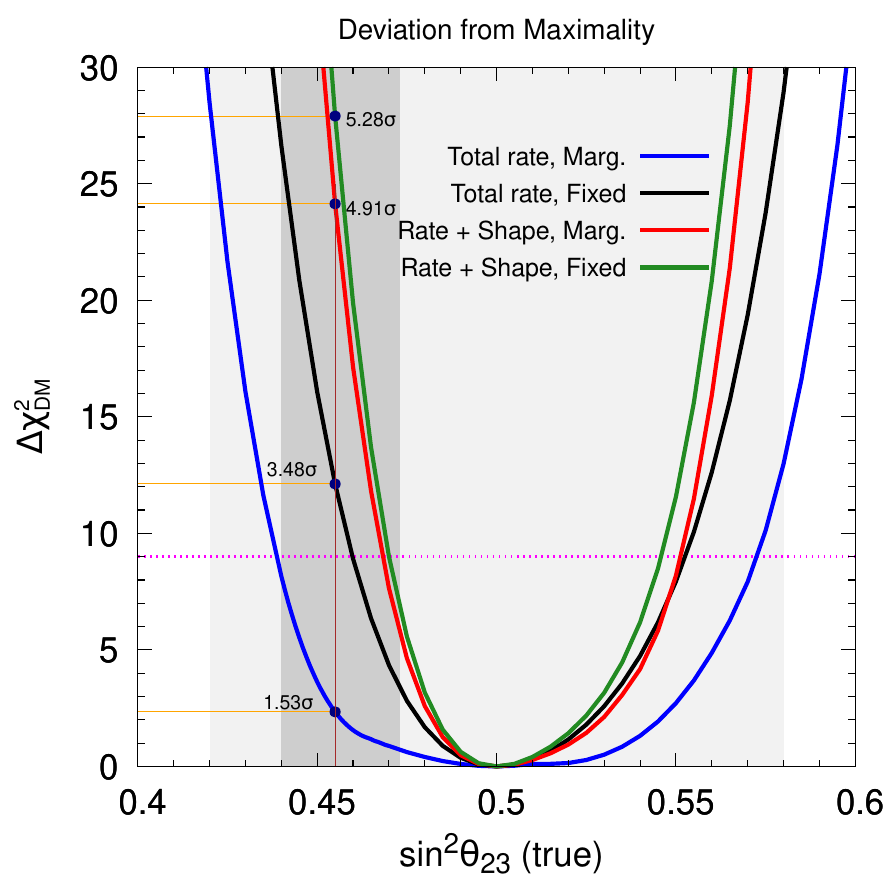}
	\caption{\footnotesize{Potential of DUNE to establish deviation from maximality as a function of true $\sin^2\theta_{23}$ assuming true NMO and $\delta_{\rm CP}$ (true) = 223$^\circ$ with the combined 3.5 years $\nu$ + 3.5 years $\bar{\nu}$ run. Blue (Black) curve shows the performance based on total event rates when $\Delta m^{2}_{31}$ and $\delta_{\mathrm{CP}}$ are marginalized (fixed) in the fit. Red (green) curve depicts the sensitivity based on rate + shape analysis when $\Delta m^{2}_{31}$ and $\delta_{\mathrm{CP}}$ are marginalized (fixed) in the fit. See text for details. The dark (light)-shaded grey area shows the currently allowed $1\sigma$ $(2\sigma)$ region in $\sin^{2}\theta_{23}$ as obtained in the global fit study~\cite{Capozzi:2021fjo} with the best-fit value of $\sin^{2}\theta_{23} = 0.455$ as shown by vertical brown line. The horizontal orange lines show the sensitivity (experessed in $\sigma = \sqrt{\Delta\chi^2_{\rm DM}}$) due to individual runs for the current best-fit value of $\sin^2\theta_{23}$. The capability of DUNE to establish non-maximal $\theta_{23}$ at 3$\sigma$ ($\Delta \chi^{2}_{\mathrm{DM}} = 9$) confidence level is shown by horizontal pink dotted line. For simplicity, we do not consider systematic uncertainties in this figure.}}
	\label{fig:8}
\end{figure}

We now explore the benefit of spectral shape information in DUNE on top of total event rates in establishing deviation from maximality. In Fig.~\ref{fig:8}, we show $\Delta\chi^2_{\rm DM}$ as a function of true $\sin^2\theta_{23}$ for four cases. The blue and black curves are obtained based on total event rates, while the red and green curves show the sensitivity when we include the spectral shape information along with total event rates. For both total rate and rate + shape analyses, we estimate the sensitivities in the fixed-parameter and marginalized scenarios. In the fixed-parameter case, we keep all the oscillation parameters fixed at their best-fit values (see second column in Table~\ref{table:one}) in both data and fit, while in the marginalized case, we minimize over $\Delta m^2_{31}$ and $\delta_{\rm CP}$ in their current 3$\sigma$ ranges. This comparison between fixed-parameter (see black and green lines) and marginalized (see blue and red lines) scenarios enable us to see how much the sensitivity gets deteriorated due to the uncertainties on $\Delta m^2_{31}$ and $\delta_{\rm CP}$. While establishing non-maximal $\theta_{23}$, the bulk of the sensitivity stems from the disappearance channel (see Fig.~\ref{fig:7}) and the uncertainty on $\Delta m^{2}_{31}$ affects this channel more than $\delta_{\rm CP}$. This can be seen from the top panels of Figs.~\ref{fig:2} and~\ref{fig:3}, and Fig.~\ref{fig:6} also confirms that the impact of $\delta_{\rm CP}$ is minimal in establishing the deviation from maximality. At the same time, we expect that the upcoming medium-baseline reactor experiment JUNO will measure $\Delta m^{2}_{31}$ with utmost precision~\cite{EPS-HEP-Conference2021, JUNO:2015zny} before DUNE will start taking data. Therefore, it makes complete sense to analyze the potential of DUNE to establish non-maximal $\theta_{23}$ in the fixed-parameter scenario (see black and green lines). However, Fig.~\ref{fig:8} also reveals that the impact of uncertainty on $\Delta m^{2}_{31}$ in the marginalized case is substantialy reduced when we exploit the spectral shape information (see red line) $-$ thanks to the intense wide-band beam resulting into high-statitstics in disappearance mode and excellent energy resolution of LArTPC detector in DUNE~\cite{Friedland:2018vry, DeRomeri:2016qwo}. 

We see from Fig.~\ref{fig:8} that the ability of DUNE to exclude maximal mixing solution in the fit for $\sin^{2}\theta_{23}$ (true) = 0.455, gets significantly enhanced from 1.53$\sigma$ (see blue line) to 4.91$\sigma$ (see red line) when we include spectral shape information in the analysis. We see this improvement in the sensitivity because the impact of $\Delta m^2_{31} - \sin^2\theta_{23}$ degeneracy gets reduced substantially when we perform rate + shape analysis instead of using only total rates. As we already demonstrate before using Fig.~\ref{fig:5} in Sec.~\ref{events} that we can reduce the impact of this degeneracy because of the fact that energy bins on either side of the oscillation minimum in disappearance events show different behavior with respect to a change in the value of $\Delta m^2_{31}$. For this reason, in the fit, the test value of $\Delta m^2_{31}$ does not get deviate much from its central best-fit value. Nevertheless, we observe that even in the case of rate + shape analysis, the uncertainty in $\Delta m^{2}_{31}$ reduces the potential of DUNE to establish deviation from maximality for $\sin^{2}\theta_{23}$ (true) = 0.455 from 5.28$\sigma$ to 4.91$\sigma$ while going from fixed-parameter case to marginalized scenario. So, an ultra-precise measurement of $\Delta m^{2}_{31}$ in future will undoubtedly enhance DUNE's capability to establish deviation from maximal $\theta_{23}$. For the sake of simplicity, while addressing the advantage due to spectral analysis and the impact of marginalization over oscillation parameters in Fig.~\ref{fig:8},  we do not take into account the systematic uncertainties in the analysis.

\subsubsection{Individual contributions from  neutrino and antineutrino runs}

\begin{figure}[htb!]
	\centering
	\includegraphics[width = 0.49\linewidth,height=0.51\linewidth]{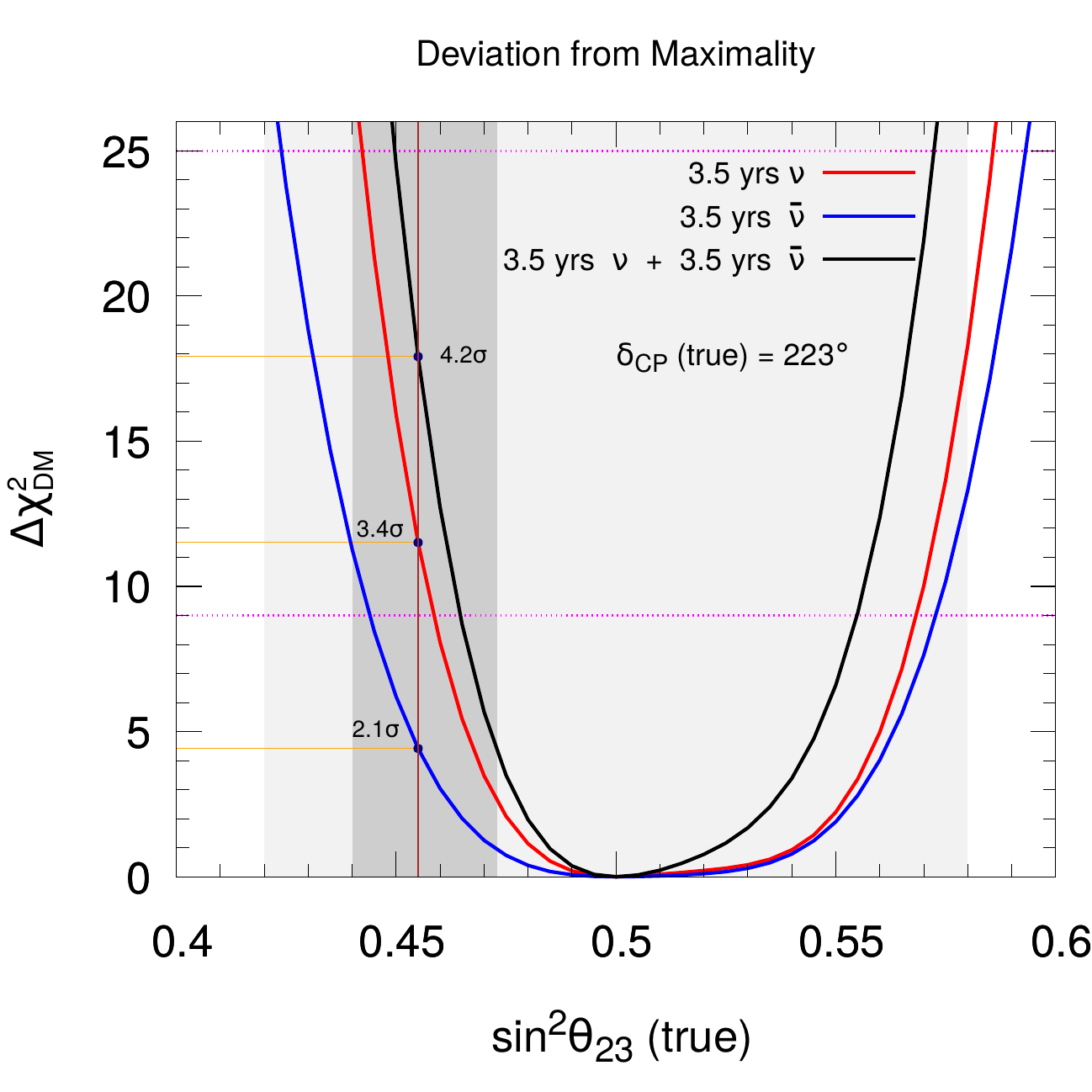}
	\includegraphics[width = 0.49\linewidth,height=0.51\linewidth]{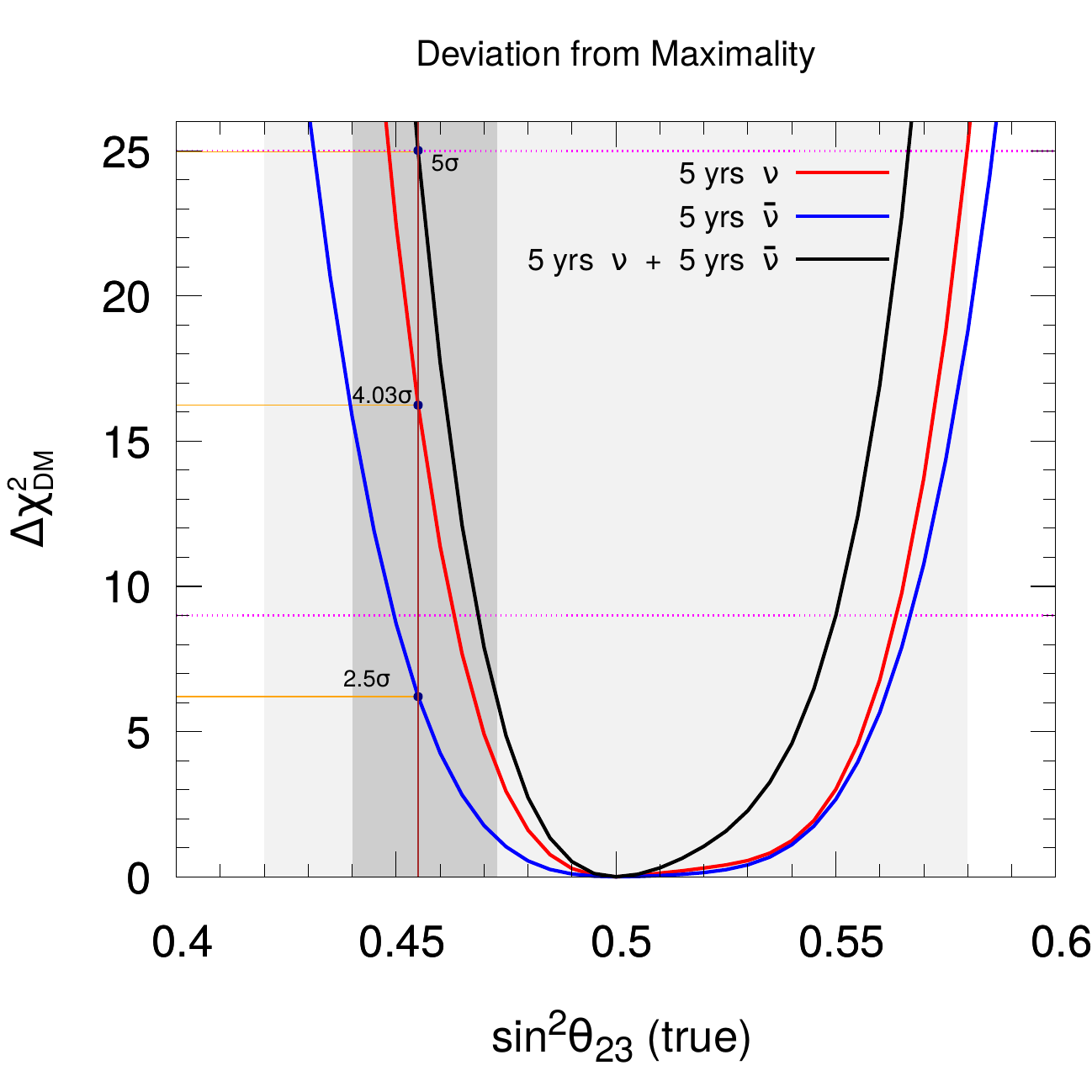}
	\caption{\footnotesize{Potential of DUNE to establish the deviation from maximal $\theta_{23}$ as a function of true $\sin^2\theta_{23}$ assuming true NMO and $\delta_{\rm CP}\, \mathrm{(true)} = 223^\circ$. The red, blue, and black curves in the left (right) panel are drawn assuming 3.5 (5) years of neutrino run, 3.5 (5) years of antineutrino run, and the combined 7 (10) years of $\nu + \bar{\nu}$ run, respectively. In the fit, we marginalize over the current 3$\sigma$ range of $\Delta m^{2}_{31}$ and $\delta_{\rm CP}$, while keeping rest of the oscillation parameters fixed at their present best-fit values as shown in Table~\ref{table:one}. The dark (light)-shaded grey area shows the currently allowed $1\sigma$ $(2\sigma)$ region in $\sin^{2}\theta_{23}$ as obtained in the global fit study~\cite{Capozzi:2021fjo} assuming NMO with the best-fit value of $\sin^{2}\theta_{23} = 0.455$ as shown by vertical brown line. The horizontal orange lines show the sensitivity (experessed in $\sigma = \sqrt{\Delta\chi^2_{\rm DM}}$) due to individual runs for the current best-fit value of $\sin^2\theta_{23}$. The capability of DUNE to establish non-maximal $\theta_{23}$ at 3$\sigma$ ($\Delta \chi^{2}_{\mathrm{DM}} = 9$) and 5$\sigma$ ($\Delta \chi^{2}_{\mathrm{DM}} = 25$) confidence levels are shown by horizontal pink dotted lines.
	}}
	\label{fig:9}
\end{figure} 

In Fig.~\ref{fig:9}, we demonstrate the capability of DUNE to establish non-maximal $\theta_{23}$ assuming true NMO and $\delta_{\rm CP}\, \mathrm{(true)} = 223^\circ$. The red, blue, and black curves in the left (right) panel are drawn assuming 3.5 (5) years of neutrino run, 3.5 (5) years of antineutrino run, and the combined 7 (10) years of $\nu + \bar{\nu}$ run, respectively. We observe that the sensitivity of DUNE to exclude maximal $\theta_{23}$ gets improved significantly when we combine the data from both neutrino and antineutrino modes (see black curves) as compared to the stand-alone neutrino (see red curves) or antineutrino (see blue curves) run. Mostly, the data from neutrino run contributes in the combined analysis due to their superior statistics. We notice from the left panel that a 2.1$\sigma$, 3.4$\sigma$, and 4.2$\sigma$ determination of non-maximal $\theta_{23}$ is possible in DUNE considering 3.5 years of $\bar{\nu}$ run, 3.5 years of $\nu$ run, and the combined 3.5 years $\nu$ + 3.5 years $\bar{\nu}$ run, respectively assuming the present best-fit values of $\sin^{2}\theta_{23}$ (0.455) and $\delta_{\rm CP}$ (223$^\circ$) as their true choices and with true NMO. In the right panel, the sensitivities get improved to 2.5$\sigma$, 4$\sigma$, and 5$\sigma$ with 5 years of $\bar{\nu}$ run, 5 years of $\nu$ run, and the combined 5 years $\nu$ + 5 years $\bar{\nu}$ run, respectively.

\subsubsection{Performance as a function of exposure}

\begin{figure}[htb!]
	\centering
	\includegraphics[width = 0.7\linewidth]{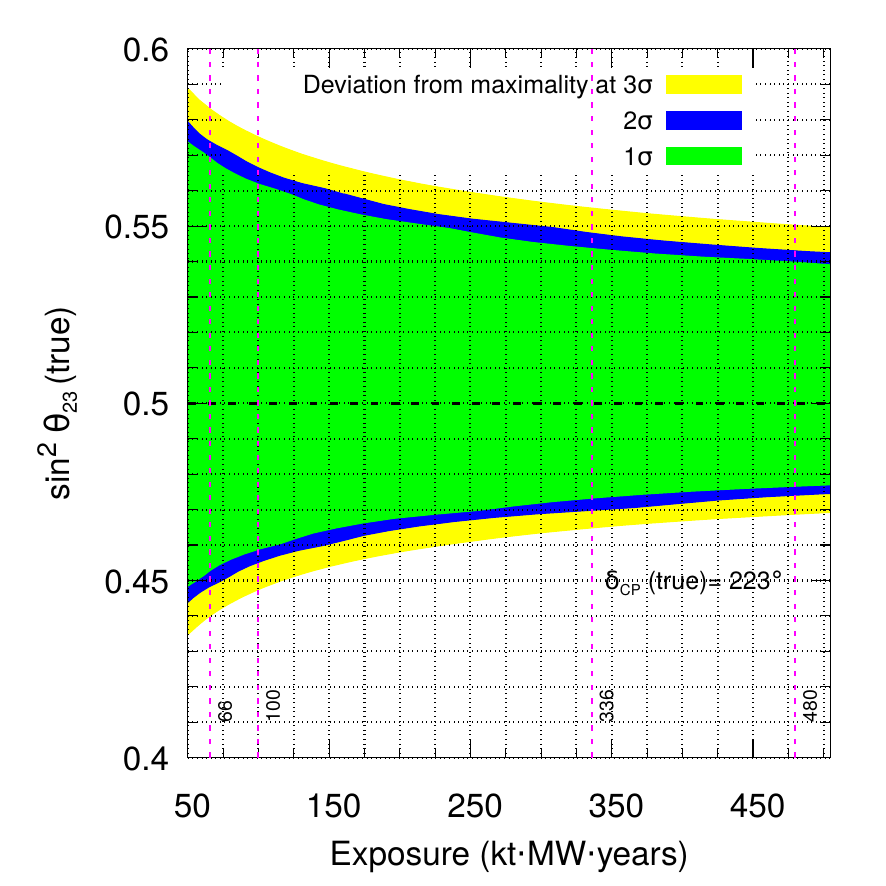}
	\caption{\footnotesize{The discovery of true values of non-maximal $\sin^{2}\theta_{23}$ as a function of exposure (kt$\cdot$MW$\cdot$years) at $3\sigma$ (yellow curves), $2\sigma$ (blue curves), and $1\sigma$ (green curves) confidence levels. For a given exposure, we assume equal run-time in both $\nu$ and $\bar{\nu}$ modes. We consider true NMO and $\delta_{\mathrm{CP}}$ (true) = 223$^{\circ}$. We marginalize over $\delta_{\mathrm{CP}}$ and $\Delta m^{2}_{31}$ in the fit in their allowed 3$\sigma$ ranges as given in Table~\ref{table:one}.}
	}
	\label{fig:10}
\end{figure}

 The DUNE collaboration is planning to adopt an incremental approach where they will gradually increase the exposure by adding the new detector modules to their setup and will also upgrade the beam power from 1.2 MW to 2.4 MW after 6 years \cite{DUNE:2020jqi, DUNE:2021mtg} 
. This staging approach is well justified in light of the challenges that appear while operating a high-power superbeam and in constructing a massive underground 40 kt liquid argon detector. A nominal deployment plan is discussed in Ref. \cite{DUNE:2020jqi}, where the collaboration plans to start the experiment with two far detector (FD) modules having a total fiducial mass of 20 kt and with a beam power of 1.2 MW. After one year, they plan to add one more FD module of 10 kt fiducial mass and after two more years, they will add another 10 kt FD module to have the total fiducial mass of 40 kt. After operating the experiment for six years with a beam power of 1.2 MW, there is also a plan to upgrade the beam power to 2.4 MW \cite{DUNE:2020jqi}.

In Fig.~\ref{fig:10}, we exhibit the performance of DUNE to establish the possible deviation of true values of $\sin^{2}\theta_{23}$ from MM choice ($\sin^{2}\theta_{23}^{\mathrm{test}}$ = 0.5) in the fit as a function of exposure expressed in the units of kt$\cdot$MW$\cdot$years. We show the results at 3$\sigma$ (see yellow curves), 2$\sigma$(see blue curves), and 1$\sigma$ (see green curves) confidence levels assuming $\delta_{\mathrm{CP}}$ (true) = 223$^{\circ}$, and true NMO. While obtaining the results, we marginalize over $\delta_{\mathrm{CP}}$ and $\Delta m^{2}_{31}$ in their presently allowed 3$\sigma$ ranges as given in Table~\ref{table:one}. We see a significant improvement in the discovery of a non-maximal $\theta_{23}$ while increasing the exposure from 50 kt$\cdot$MW$\cdot$years to 100 kt$\cdot$MW$\cdot$years. In Fig.~\ref{fig:10} we show for the first time the true values of $\sin^{2}\theta_{23}$ that DUNE can distinguish from $\sin^{2}\theta_{23}^{\mathrm{test}}= 0.5$ using these exposures at various confidence levels (see dashed vertical lines). While further increasing the exposure from 100 kt$\cdot$MW$\cdot$years to 336kt$\cdot$MW$\cdot$years  which is our benchmark choice, we see a marginal increment in the performance. Note that we hardly see any improvement in the sensitivity if we increase the exposure further which suggest that the statistics is not a limiting factor anymore and any possible reduction in the systematic uncertainties may enhance the results further.

\section{Summary and conclusions}
\label{conclusion}

We have achieved remarkable precision on the solar ($\theta_{12}, \, \Delta m^{2}_{21}$) and atmospheric ($\theta_{23}, \, \Delta m^{2}_{31}$) oscillation parameters over the last few years. According to Ref.~\cite{Capozzi:2021fjo}, the current relative 1$\sigma$ errors on $\sin^{2}\theta_{12}, \, \Delta m^{2}_{21}, \, \sin^{2}\theta_{23}$, and $\Delta m^{2}_{31}$ are 4.5\%, 2.3\%, 6.7\%, and 1.1\%, respectively. The recent hints for normal mass ordering (at $\sim$ 2.5$\sigma$), as well as for lower octant $\theta_{23}$ ($\theta_{23} <$ 45$^{\circ}$) and for $\delta_{\mathrm{CP}}$ in the lower half-plane ($\sin \delta_{\mathrm{CP}} < 0$) signify major developments in the three-flavor neutrino oscillation paradigm. The high-precision measurement of $\theta_{13}$ from the Daya Bay reactor experiment and the possible complementarities between the recent Super-K Phase I-IV atmospheric data and the appearance and disappearance data from the ongoing long-baseline oscillation experiments - NO$\nu$A and T2K, play an important role in providing these crucial hints. An accurate measurement of $\theta_{23}$ and resolution of its octant (if $\theta_{23}$ turns out to be non-maximal) are crucial to transform these preliminary hints into 5$\sigma$ discoveries. A discovery of non-maximal $\theta_{23}$ at high confidence level will serve as a crucial input to the theories of neutrino masses and mixings and it will certainly be a major breakthrough in addressing the the age-old flavor problem. In this paper, we explore in detail the sensitivities of the upcoming high-precision long-baseline experiment DUNE to establish the possible deviation from maximal $\theta_{23}$ and to resolve its octant at high confidence level in light of the recent neutrino oscillation data. 

We start the paper by showing the possible correlations and degeneracies among the oscillation parameters $\sin^{2}\theta_{23}, \, \Delta
m^{2}_{31}$, and $\delta_{\mathrm{CP}}$ in the context of $\nu_{\mu} \rightarrow \nu_{\mu}$ disappearance channel and $\nu_{\mu} \rightarrow \nu_{e}$ appearance channel at the probability and event levels. We introduce for the first time, a bi-events plot in the plane of total neutrino and antineutrino disappearance events to demonstrate the impact of $\sin^{2}\theta_{23} - \Delta m^{2}_{31}$ degeneracy in establishing deviation from maximality. Next, we show how the spectral shape information in neutrino and antineutrino disappearance events can play an important role to resolve this degeneracy.

Using the latest simulation details of DUNE~\cite{DUNE:2021cuw}, we observe that a 3$\sigma$ (5$\sigma$) determination of non-maximal $\theta_{23}$ is possible in DUNE with an exposure of 336 kt$\cdot$MW$\cdot$years if the true value of $\sin^2\theta_{23} \lesssim 0.465~(0.450)$ or $\sin^2\theta_{23} \gtrsim 0.554~(0.572)$ for any value of true $\delta_{\mathrm{CP}}$ in the present 3$\sigma$ range and true NMO. DUNE can exclude the maximal mixing solution of $\theta_{23}$ at 4.2$\sigma$ (5$\sigma$) with a total 7 (10) years of run (equally divided in neutrino and antineutrino modes) assuming the present best-fit values of $\sin^{2}\theta_{23}$ (0.455) and $\delta_{\mathrm{CP}}$ (223$^{\circ}$) as their true choices with true NMO. The same can be enhanced to 6.5$\sigma$ (7.7$\sigma$) if we assume $\sin^{2}\theta_{23}$ (true) = 0.44, which is the current 1$\sigma$ lower bound. On the other hand, the sensitivity can be reduced to 2.07$\sigma$ (2.44$\sigma$) if $\sin^{2}\theta_{23}$ (true) turns out to be 0.473, which is the current 1$\sigma$ upper bound.

We study the role that systematic uncertainties play in establishing  deviation from maximality by varying the normalization errors in both appearance and disappearance channels. We explore the contribution that each oscillation channel has on the sensitivity and show how performing a spectral analysis alleviates the possible reduction in sensitivity due to the marginalization over $\Delta m^2_{31}$ that is present when only total event rates are considered. We also explore the effect of exposure and the individual contributions from neutrino and antineutrino modes. We hope that this study serves as an important addition to several fundamental physics issues that can be explored by the high-precision long-baseline experiment DUNE and provides a boost to the physics reach of DUNE. For more details, see the reference \cite{Agarwalla:2021bzs}.



\chapter{Improved precision on 2-3 oscillation parameters using the synergy between DUNE and T2HK}
\label{C5} 
\begin{figure}[t!]
	\centering
	\includegraphics[width=\linewidth]{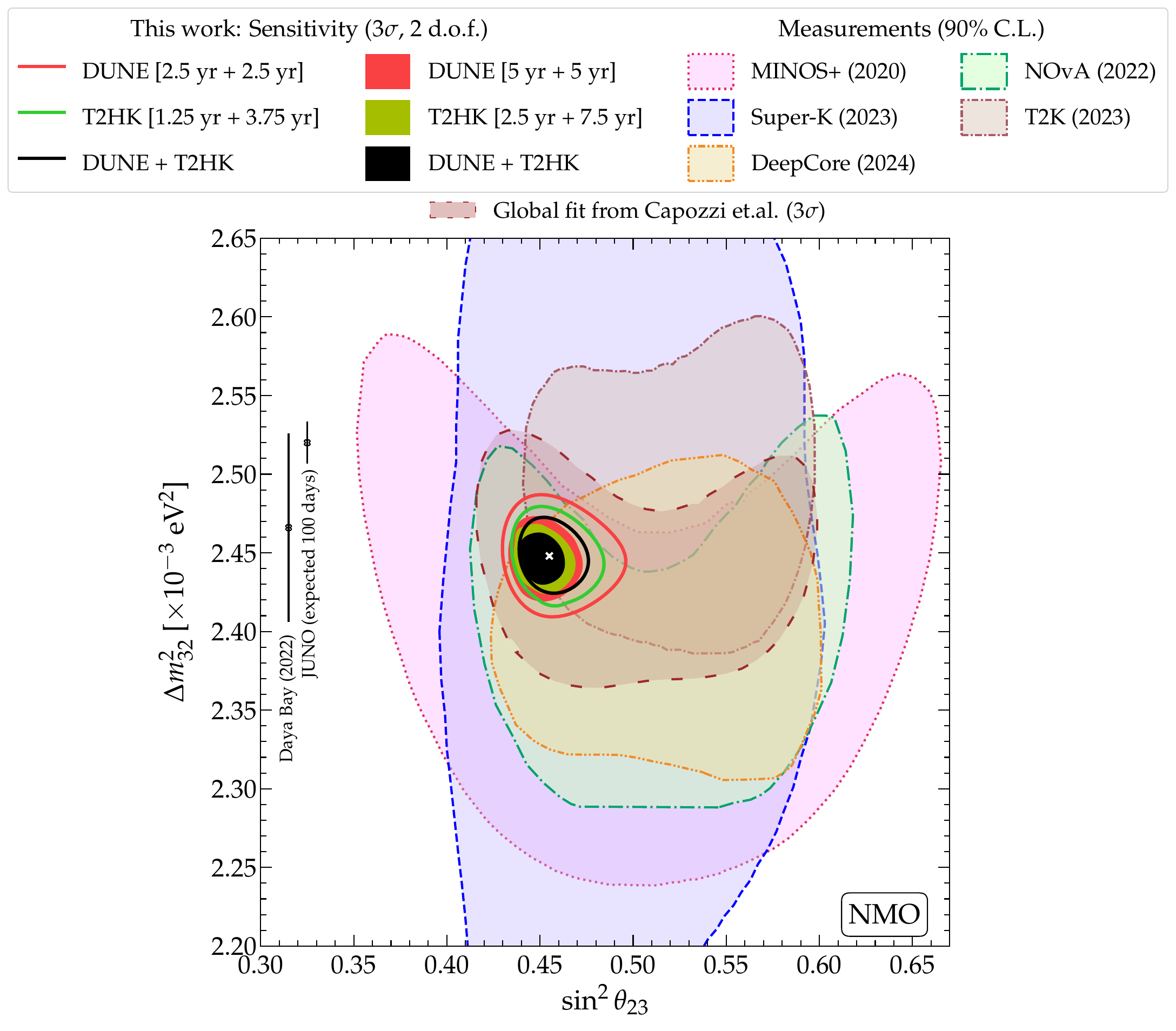}
	\caption{\footnotesize{Allowed ranges at 3$\sigma$ (2 d.o.f.) in the atmospheric mixing parameters, $\sin^{2}\theta_{23}$ and $\Delta m^{2}_{32}$, using DUNE, T2HK, and DUNE + T2HK. DUNE is expected to have an exposure of 480 kt$\cdot$MW$\cdot$yr and T2HK, an exposure of 2431 kt$\cdot$MW$\cdot$yr. We assume DUNE (T2HK) running for 5 (2.5) yr in $\nu$ and 5 (7.5) yr in $\bar{\nu}$ mode while estimating nominal exposure. We also depict the allowed ranges for the same when only half of the total projected exposure is considered (DUNE: [2.5 yr in $\nu$ + 2.5 yr in $\bar{\nu}$], T2HK: [1.25 yr in $\nu$ + 3.75 yr in $\bar{\nu}$]). Existing allowed ranges are from: Super-Kamiokande~\cite{Super-Kamiokande:2023ahc}, Tokai-to-Kamioka (T2K)~\cite{T2K:2023smv}, NuMI Off-Axis $\nu_{e}$ Appearance (NO$\nu$A)~\cite{NOvA:2021nfi}, Main injector neutrino oscillation search (MINOS+)~\cite{MINOS:2020llm}, and IceCube DeepCore~\cite{IceCube:2024xjj,Yu:2023tmw} at 90\% C.L. We also show existing (expected) bounds on $\Delta m^{2}_{32}$ from Daya Bay~\cite{DayaBay:2022orm} 
 (Jiangmen Underground Neutrino Observatory; JUNO ~\cite{JUNO:2022mxj}).  To have a complete picture, we also show the present global fit, following Ref.~\cite{Capozzi:2021fjo} at 3$\sigma$ (1dof). \textit{Our projected allowed ranges improve from the existing bounds by manifold.} See section~\ref{appendix1} for a detailed discussion.Further, figure~\ref{fig:allowed-ranges-nmo} elaborates on the importance of both neutrino and antineutrino modes in this sensitivity separately. Also, see figure~\ref{fig:appendix-allowed-ranges-nmo} for a few additional sensitivity curves.}} 
 \label{fig:moneyplot}
 \end{figure}

A high-precision measurement of $\Delta m^2_{31}$ and $\theta_{23}$ is inevitable to estimate the Earth's matter effect in long-baseline experiments which in turn plays an important role in addressing the issue of neutrino mass ordering and to measure the value of CP phase in $3\nu$ framework. After reviewing the results from the past and present experiments, and discussing the near-future sensitivities from the IceCube Upgrade and KM3NeT/ORCA, we study the expected improvements in the precision of 2-3 oscillation parameters that the next-generation long-baseline experiments, DUNE and T2HK, can bring either in isolation or combination. We highlight the relevance of the possible complementarities between these two experiments in obtaining the improved sensitivities in determining the deviation from maximal mixing of $\theta_{23}$, excluding the wrong-octant solution of $\theta_{23}$, and obtaining high precision on 2-3 oscillation parameters, as compared to their individual performances. We observe that for the current best-fit values of the oscillation parameters and assuming normal mass ordering (NMO), DUNE + T2HK can establish the non-maximal $\theta_{23}$ and exclude the wrong octant solution of $\theta_{23}$ at around 7$\sigma$ C.L. with their nominal exposures. We find that DUNE + T2HK can improve the current relative 1$\sigma$ precision on $\sin^{2}\theta_{23}~(\Delta m^{2}_{31})$  by a factor of 7 (5) assuming NMO. Also, we notice that with less than half of their nominal exposures, the combination of DUNE and T2HK can achieve the sensitivities that are expected from these individual experiments using their full exposures. We also portray how the synergy between DUNE and T2HK can provide better constraints on ($\sin^2\theta_{23}$ - $\delta_{\mathrm{CP}}$) plane as compared to their individual reach.
Current oscillation experiments in the three-neutrino paradigm depict potent complementarity. The Solar experiment's precision measurements of solar mixing angle have been combined with KamLAND's ability to determine solar mass splitting well, enabling their synergy to be utilized in the neutrino community for a very long time. The combination of these data and those from long-baseline (LBL) accelerator and atmospheric neutrinos represents the minimal dataset for the characterization of oscillation parameters. The unprecedented precision obtained on the reactor mixing angle ($\theta_{13}$) from short-baseline reactor data like Daya Bay~\cite{DayaBay:2022orm} has further alleviated the uncertainty in the measurements of other unknowns like $\theta_{23},\, \delta_{\mathrm{CP}}$, and the sign of $\Delta m^{2}_{31}$ indirectly by reducing the correlations. The atmospheric neutrino data, when complemented with the long-baseline data, allows us to gain valuable insight into the atmospheric parameters. For example, Super-K, with its extensive statistics, can impose strict constraints on the measurements of $\theta_{23}$, while MINOS/MINOS$+$, benefiting from a precisely known $L/E$ ratio, can offer more accurate measurements of the atmospheric mass splitting, as illustrated in figure~\ref{fig:moneyplot}. In this context, recently, the Super-K experiment has enhanced its precision on atmospheric neutrino oscillation parameters by using the number of tagged neutrons to enhance the separation between neutrino and antineutrino events, improving the efficiency to classify the multi-ring events using a boosted decision tree algorithm, and adding 48\% exposure by analyzing events from an expanded fiducial volume and from 1186 additional live-days, including the data which were collected after a major refurbishment of the detector in 2018~\cite{Super-Kamiokande:2023ahc}.

The presently running accelerators~\cite{Ali:2022mrp,Catano-Mur:2022kyq} and atmospheric experiments~\cite{sk,IceCube:2019dyb} along with the high-precision measurements from reactors~\cite{DayaBay:2018yms,Seo:2019shs} have helped to achieve the current relative 1$\sigma$ precision of about 1.1\% and 6.7\% in the atmospheric parameters: $\Delta m^{2}_{31}$ and $\sin^2\theta_{23}$, respectively~\cite{Capozzi:2021fjo}. The upcoming medium-baseline reactor oscillation experiment JUNO~\cite{JUNO:2022mxj} is expected to achieve considerable improvement in the precision of atmospheric mass-squared difference $\Delta m^2_{32}$ as compared to the current precision from Daya Bay~\cite{DayaBay:2022orm} reactor experiment, as can be seen from figure~\ref{fig:moneyplot}. The DeepCore array consisting of 8 dedicated strings with denser spacing in the central region of IceCube has enabled the detection and reconstruction of atmospheric neutrinos with energies as low as a few GeV, providing high-precision measurements of 2-3 oscillation parameters. Using convolutional neural networks with 9.3 years of data, the IceCube DeepCore has provided a new high-precision measurements of $\Delta m^{2}_{32} = 2.40^{+0.05}_{-0.04} \times 10^{-3}$  eV$^{2}$ and $\sin^{2}\theta_{23}= 0.54^{+0.04}_{-0.03}$~\cite{IceCube:2024xjj} assuming normal mass ordering (NMO), which are compatible and complementary with the existing measurements from the long-baseline experiments. A new extension of IceCube, namely the IceCube Upgrade to be deployed in the polar session of 2025/26 with seven new strings in the central region of DeepCore detector and an energy threshold of around 1 GeV is expected to improve the precision of 2-3 oscillation parameters by (20-30)\%~\cite{IceCube:2023ins}. In ref.~\cite{IceCube-Gen2:2019fet}, the combined sensitivity of the future JUNO, IceCube Upgrade, and PINGU data was estimated to resolve the pressing issue of neutrino mass ordering. The under-construction water Cherenkov neutrino detector KM3NeT/ORCA also has the potential to shed light on neutrino mass ordering and 2-3 oscillation parameters~\cite{KM3NeT:2021ozk}. In fact, recently, they announced their measurements of 2-3 oscillation parameters using an initial configuration with 6-detection units of photo-sensors corresponding to an exposure of 433 kt$\cdot$yr, collected in 510 days of data taking~\cite{KM3NeT:2024ecf}. They reveal a best-fit of $\sin^{2}\theta_{23}= 0.51$ and $\Delta m^{2}_{31} = 2.14 \times 10^{-3}$  eV$^{2}$ with their initial configuration, referred to as ORCA6. Further, in the Neutrino 2024 conference, the KM3NeT/ORCA collaboration showed slightly improved measurements of 2-3 oscillation parameters using an updated exposure of 715 kt$\cdot$yr~\cite{coelho_orca}. In ref.~\cite{KM3NeT:2021rkn}, a combined analysis of the prospective JUNO and KM3NeT/ORCA data was performed to determine the correct neutrino mass ordering.

The present hints of non-maximal $\theta_{23}$ from the global oscillation data~\cite{Capozzi:2021fjo,deSalas:2020pgw,Esteban:2020cvm,NuFIT} give rise to two probable solutions: $\theta_{23} < 45^{\circ}$ or the lower octant solutions (LO) and $\theta_{23} > 45^{\circ}$ or the higher octant solutions (HO)~\cite{Fogli:1996pv, Barger:2001yr, Minakata:2002qi, Minakata:2004pg, Hiraide:2006vh}. Prior to delving into the determination of the correct octant, it is imperative to eliminate the possibility of maximal mixing with a high level of confidence. This investigation was undertaken earlier, taking into account the forthcoming DUNE (Deep Underground Neutrino Experiment)~\cite{Agarwalla:2021bzs}. In this study, we explore the synergies between the two prominent upcoming neutrino experiments: DUNE~\cite{DUNE:2020jqi, DUNE:2020ypp}, designed to receive a wide-band on-axis neutrino beam traversing a distance of around 1300 km with substantial Earth matter effect, and T2HK (Tokai to Hyper-Kamiokande)~\cite{Hyper-KamiokandeProto-:2015xww, Hyper-Kamiokande:2018ofw}, which proposes the use of a narrow-band off-axis neutrino beam with a baseline of 295 km having minimal Earth matter effect. We investigate how the collaborative synergy of these setups enhances their individual sensitivities. Additionally, we assess the expected sensitivities of currently running long-baseline experiments: T2K (Tokai to Kamioka)~\cite{T2K:2011qtm} and NO$\nu$A (NuMI Off-Axis $\nu_{e}$ Appearance)~\cite{NOvA:2007rmc}, considering their full projected exposures. Our findings indicate that the combined DUNE + T2HK configuration can detect a departure from maximal $\theta_{23}$ with exceptional significance. Moreover, their complementarity turns out to be essential for achieving degeneracy-free and significantly precise measurements of both $\sin^2\theta_{23}$ and $\Delta m^{2}_{31}$.

As mentioned before, figure~\ref{fig:moneyplot} shows a part of our main results exhibiting the expected precision in 2-3 oscillation parameters: $\sin^{2}\theta_{23}$ and $\Delta m^{2}_{32}$ at 3$\sigma$ confidence level (C.L.) using the projected full and half exposures of standalone DUNE and T2HK, and their combination (DUNE + T2HK). To have a better comparison, we also include the currently allowed regions from the ongoing (T2K, NO$\nu$A, IceCube DeepCore, Super-Kamiokande), completed (Daya Bay and MINOS+), and upcoming (JUNO) experiments at 90\% C.L. assuming NMO. The details of these experiments (such as exposures, runtime, and current status) and the best-fit values of the oscillation parameters obtained from them are given in table~\ref{tab:other-experiments}. In figure~\ref{fig:moneyplot}, we also show the allowed region from the global fit of all the available oscillation data following Ref.~\cite{Capozzi:2021fjo} at 3$\sigma$, assuming NMO. More details related to this figure can be found in section~\ref{appendix2}. Note that in figure~\ref{fig:moneyplot}, we compare the sensitivities of various experiments in terms of $\Delta m^{2}_{32}$ (see the y-axis). However, we present all our results in terms of $\Delta m^{2}_{31}$ in the present manuscript. 
\begin{table}[htb!]
 \begin{tabular}{ | c | c | c | c | c | c|}
 \hline
 \multirow{3}{*}{Experiment}
 & \multicolumn{2}{c|}{Best-fit} &  \multirow{3}{*}{Exposure} &  \multirow{3}{*}{Span} &  \multirow{3}{*}{Status}\\ 
 \cline{2-3}
 &  \multirow{2}{*}{$\sin^{2}\theta_{23}$} & $\Delta m^{2}_{32}$ &    &  &\\
  & & $[10^{-3}] ~\text{eV}^2$ &   & &\\
  \hline
\multirow{3}{*}{T2K~\cite{T2K:2023smv}} & \multirow{3}{*}{0.56} & \multirow{3}{*}{2.49} &  P.O.T. & \multirow{3}{*}{2010 - 2020} & \multirow{3}{*}{Ongoing}\\
& & & $16.3 \times 10^{20}$ ($\nu$)  & & \\
& & & $19.7 \times 10^{20}$ ($\bar{\nu}$)  & & \\
\hline
\multirow{3}{*}{NO$\nu$A~\cite{NOvA:2021nfi}} & \multirow{3}{*}{0.57} & \multirow{3}{*}{2.41} & P.O.T. &  & \multirow{3}{*}{Ongoing} \\
& & & $13.6 \times 10^{20}$ ($\nu$) & 2016 - 2019 & \\
& & & $12.5 \times 10^{20}$ ($\bar{\nu}$) & 2014 - 2020 & \\
\hline
MINOS/ & \multirow{2}{*}{0.43} & \multirow{2}{*}{2.4} & $23.76 \times 10^{20}$ P.O.T. & 2005 - 2016 & \multirow{2}{*}{Completed}\\
MINOS$+$~\cite{MINOS:2020llm}& & & 60.75 kt$\cdot$yr & 2011 - 2016 & \\
\hline
Super-K & \multirow{2}{*}{0.49} & \multirow{2}{*}{2.4} & \multirow{2}{*}{484.2 kt$\cdot$yr}  & \multirow{2}{*}{1996 - 2020} & \multirow{2}{*}{Ongoing}\\ 
(I - V)~\cite{Super-Kamiokande:2023ahc} & & & & & \\
\hline
IceCube- & \multirow{2}{*}{0.54} & \multirow{2}{*}{2.4} & \multirow{2}{*}{9.3 yr} & \multirow{2}{*}{2012 - 2021} & \multirow{2}{*}{Ongoing}\\
DeepCore~\cite{IceCube:2024xjj} & & & & & \\
\hline
Daya Bay~\cite{DayaBay:2022orm} & - & 2.466 & 3158 live days & 2011 - 2020 & Completed\\
\hline
JUNO~\cite{JUNO:2022mxj} & - & 2.52 & 6 yr & - & Upcoming\\
\hline
 \end{tabular}
 \caption{\footnotesize{Existing and expected best-fit values, collected exposure, runtime, and the present status of atmospheric parameters using long-baseline, atmospheric, and reactor experiments. Collected exposure is expressed in protons on target (P.O.T.) for long-baseline experiments and in kt$\cdot$yr for atmospheric experiments. For certain cases, we give the number of years of data collection. Using these details, currently allowed regions in the $\sin^{2}\theta_{23}$ and $\Delta m^{2}_{32}$ plane is shown in figure~\ref{fig:moneyplot}. 
 }}
 \label{tab:other-experiments}
\end{table}

The other main results include sensitivity towards deviation from maximal $\sin^{2}\theta_{23}$, exclusion of wrong octant solutions of $\sin^{2}\theta_{23}$, and precision on atmospheric parameters, studied as a function of exposure. We find that the \textit{complementarity between DUNE and T2HK plays a crucial role in reducing the dependency on large projected exposures of standalone experiments by manifold. Furthermore, the synergy between them helps in removing the degeneracies introduced by the individual setups, if any. } \\

The roadmap of this chapter is laid out as follows. In section \ref{sec:events}, we summarize the characteristic features of DUNE and T2HK and discuss the effect of wrong-sign contaminations and variation of 2-3 oscillation parameters on the total event statistics and event spectra, respectively. Next, section \ref{sec:results}, elaborates on our results and discussions. We compute the sensitivities in establishing deviation from maximal $\sin^2\theta_{23}$, exclusion of wrong octant solutions of $\sin^2\theta_{23}$, and the precision of $\sin^2\theta_{23}$ and $\Delta m^2_{31}$, using DUNE and T2HK in both isolation and combination. We also analyze the effect of scaled exposure on the above-mentioned sensitivity studies. Then, section \ref{sec:contour} shows projected allowed ranges in $(\sin^2\theta_{23}-\delta_{\mathrm{CP}})$ plane, using half and full exposures in DUNE, T2HK, and their combination. Finally, in section \ref{conclusion}, we summarize our findings and provide concluding remarks. While section \ref{appendix1} comprehends the individual roles of $\nu$ and $\bar{\nu}$ modes, using  DUNE, T2HK, and DUNE + T2HK in $(\sin^2\theta_{23}-\delta_{\mathrm{CP}})$ plane; section \ref{appendix2} depicts past, present, and upcoming projected sensitivities in $(\sin^2\theta_{23}-\delta_{\mathrm{CP}})$ plane.

\section{Experimental details and total event rates}
\label{sec:events}

We initiate our discussion by comparing and contrasting the two upcoming long-baseline experiments under consideration: DUNE and T2HK. Following this, we compute the expected total appearance and disappearance event rates in both the $\nu$ and the $\bar{\nu}$ modes for the presently allowed 3$\sigma$ ranges in $\theta_{23}$ and $\Delta m^{2}_{31}$~\cite{Capozzi:2021fjo} using GLoBES~\cite{Huber:2004ka, Huber:2007ji}. Since the far detectors in both DUNE and T2HK are unable to differentiate between neutrinos and antineutrinos, we also discuss the effect of ``wrong-sign'' contamination, which is considered a part of the signal in both the experiments.
%
\subsection{Complementarities between DUNE and T2HK}
\label{sec:2a}
 
DUNE  and T2HK  are two promising long-baseline experiments expected to achieve significant aspects of physics beyond the three-neutrino oscillations. We consider a single-phase state-of-the-art 40 kt Liquid Argon Time Projection Chamber (LArTPC) far detector in DUNE and a 187 kt Water Cherenkov far detector in T2HK as referred in their cumulative design reports, respectively~\cite{DUNE:2021tad, Hyper-Kamiokande:2016srs}. The neutrino flux in DUNE is expected to be wide-band on-axis, ranging from a few hundreds of MeV to a few tens of GeV, peaking at 2.5 GeV. This wide-band nature enables DUNE access to an envelope of various $L/E$ ratios, where $L$ corresponds to the distance that neutrino travels from source to detector and $E$ refers to the neutrino beam energy. Contrastingly, T2HK is expected to use a 2.5$^{\circ}$ off-axis J-PARC neutrino beam, the flux expected to peak at 0.6 GeV. The higher baseline in DUNE (1285 km; from Fermilab to South Dakota) ensures sufficient matter effect, while the relatively shorter baseline in T2HK (295 km; from J-PARC proton synchrotron facility to Hyper-Kamiokande) secures better precision in measurements of the intrinsic CP phase. The line-averaged constant Earth matter density ($\rho_{\mathrm{avg}}$) in DUNE is considered to be 2.848 g/cm$^3$, while in T2HK, it is taken as 2.8 g/cm$^3$. In DUNE, we consider 2\% detector systematic uncertainties in the appearance and  5\% in disappearance signal events following ref.~\cite{DUNE:2021cuw}. The binned events in T2HK have been matched with ref.~\cite{Hyper-Kamiokande:2016srs}, considering 5\% in appearance and 3.5\% in disappearance systematic uncertainties in signal events. Recently, apart from considering 5\% of conservative systematic uncertainties in appearance events, the T2HK collaboration is expecting to improve this uncertainty to about 2.7\% by the time they start taking their data in real-time~\cite{Munteanu:2022zla}. Thus we compare our results with both of these choices in figure~\ref{fig:octant exclusion}. For the runtime in DUNE, they expect to witness a balanced run between neutrino and antineutrino modes following [5 years in $\nu$ + 5 years in $\bar{\nu}$], T2HK aims instead to have an almost equal number of events in both the modes, thus following the 1:3 ratio of [2.5 years in $\nu$  + 7.5 years in $\bar{\nu}$]. Owing to the much higher detector fiducial mass, T2HK expects to accumulate about $2.7 \times 10^{22}$ P.O.T. in total ten years, providing a benchmark exposure of 2431 kt$\cdot$MW$\cdot$yr, while DUNE envisions a P.O.T. of around $1.1 \times 10^{21}$ per year with a benchmark exposure of 480 kt$\cdot$MW$\cdot$yr. In table~\ref{table:one}, we mention our assumed benchmark values and the ranges, which are taken from ref.~\cite{Capozzi:2021fjo}.

As the far detector deployment schedule and beam power scenarios are both subject to change, the results shown in this work are consistently given in terms of exposure in units of kt$\cdot$MW$\cdot$yr, which is agnostic to the exact staging scenario but can easily be expressed in terms of experiment years for any desired scenario. For having a complete summary, we present the sensitivity studies of DUNE and T2HK along with the full potential of ongoing long-baseline experiments: T2K and NO$\nu$A. We present our findings using the entire projected exposures of $84.4$ kt$\cdot$MW$\cdot$yr, generating $7.8 \times 10^{21}$ P.O.T. with a 750 kW beam power, evenly distributed between neutrino and antineutrino modes, as outlined in the ongoing long-baseline experiment T2K~\cite{T2K:2014xyt}. Additionally, we conduct simulations for the full projected exposure of NO$\nu$A, amounting to 58.8 kt$\cdot$MW$\cdot$yr and producing $3.6 \times 10^{21}$ P.O.T. with a 700 kW beam power, equally divided between neutrino and antineutrino modes, in accordance with ref.~\cite{NOvA:2007rmc, Patterson:2012zs}. In both experiments, we assume uncorrelated 5\% and 10\% systematic errors on signal and background events for both appearance and disappearance event rates.
%
\begin{table}[t!]
\centering
\resizebox{\columnwidth}{!}{%
\begin{tabular}{|c|c|c|c|c|c|c|}
\hline 
\multirow{2}{*}{\textbf{Parameter}} & $\Delta m^2_{21}/10^{-5}$ & \multirow{2}{*}{$\sin^{2}\theta_{12}/10^{-1}$} & \multirow{2}{*}{$\sin^{2}\theta_{13}/10^{-2}$} & \multirow{2}{*}{$\sin^2\theta_{23}/10^{-1}$} & $\Delta m^2_{31}/10^{-3}$  & $\delta_{\text{CP}}$ \\
				&($\mathrm{eV^{2}}$) & & & &($\mathrm{eV^{2}}$)&($^\circ$)\\
				\hline 
				{\textbf{Benchmark}} & $7.36$ & $3.03$ & $2.23$ & $4.55$ & $2.522$ & $223$\\
    \hline
				\textbf{$3\sigma$ range}& - & - & - & 4.16 - 5.99 & 2.436 - 2.605 & 139 - 355\\
					\hline 
			\end{tabular}
			}
			\caption{\footnotesize{The benchmark values of the oscillation parameters and their corresponding 3$\sigma$ allowed ranges considered in our study assuming normal mass ordering (NMO) following the ref.~\cite{Capozzi:2021fjo}.}}
			\label{table:one}
	\end{table}

\subsection{Events due to wrong-sign contamination }
\label{sec:2c}
\begin{table}[t!]
 \centering
 \begin{tabular}{ | c  *{6}{>{\centering\arraybackslash}p{2.2cm} |}}
 \hline
 \multicolumn{2}{|c|}{\multirow{3}{*}{Experiment}}
 & \multicolumn{4}{c|}{Number of events (NMO)} \\
 \cline{3-6}
          & & \multicolumn{2}{c|}{Appearance} 
          & \multicolumn{2}{c|}{Disappearance} \\
          \cline{3-6}
          & & $\nu$ mode & $\bar{\nu}$ mode & $\nu$ mode & $\bar{\nu}$ mode \\
 \hline
  \multirow{4}{*}{DUNE} & w/  & \multirow{2}{*}{1592} & \multirow{2}{*}{294} & \multirow{2}{*}{14598} & \multirow{2}{*}{8270} \\
                         & wrong-sign    &  &  &  & \\            
                      & w/o    & \multirow{2}{*}{1576} & \multirow{2}{*}{186} & \multirow{2}{*}{13413} & \multirow{2}{*}{4360}   \\
                 & wrong-sign &  &  & &  \\
 \hline
 \multirow{4}{*}{T2HK} & w/  & \multirow{2}{*}{1598} & \multirow{2}{*}{919} & \multirow{2}{*}{10064} & \multirow{2}{*}{13949} \\
                 &  wrong-sign   & &  & & \\
                       & w/o    & \multirow{2}{*}{1588}  & \multirow{2}{*}{755} & \multirow{2}{*}{9487} & \multirow{2}{*}{8985}   \\
                       &  wrong-sign    &  & & &  \\
 \hline
 \end{tabular}
 \caption{\footnotesize{Total (Signal) appearance and disappearance event rates in DUNE and T2HK assuming 480 kt$\cdot$MW$\cdot$yr and 2431 kt$\cdot$MW$\cdot$yr of exposure, respectively.  We fix the values of the standard mixing parameters to their benchmark values from~\cite{Capozzi:2021fjo}; see table~\ref{table:one}. }}
 \label{table:two}
\end{table}
%

\begin{figure}[ht!]
	\centering
	\includegraphics[width=\linewidth]{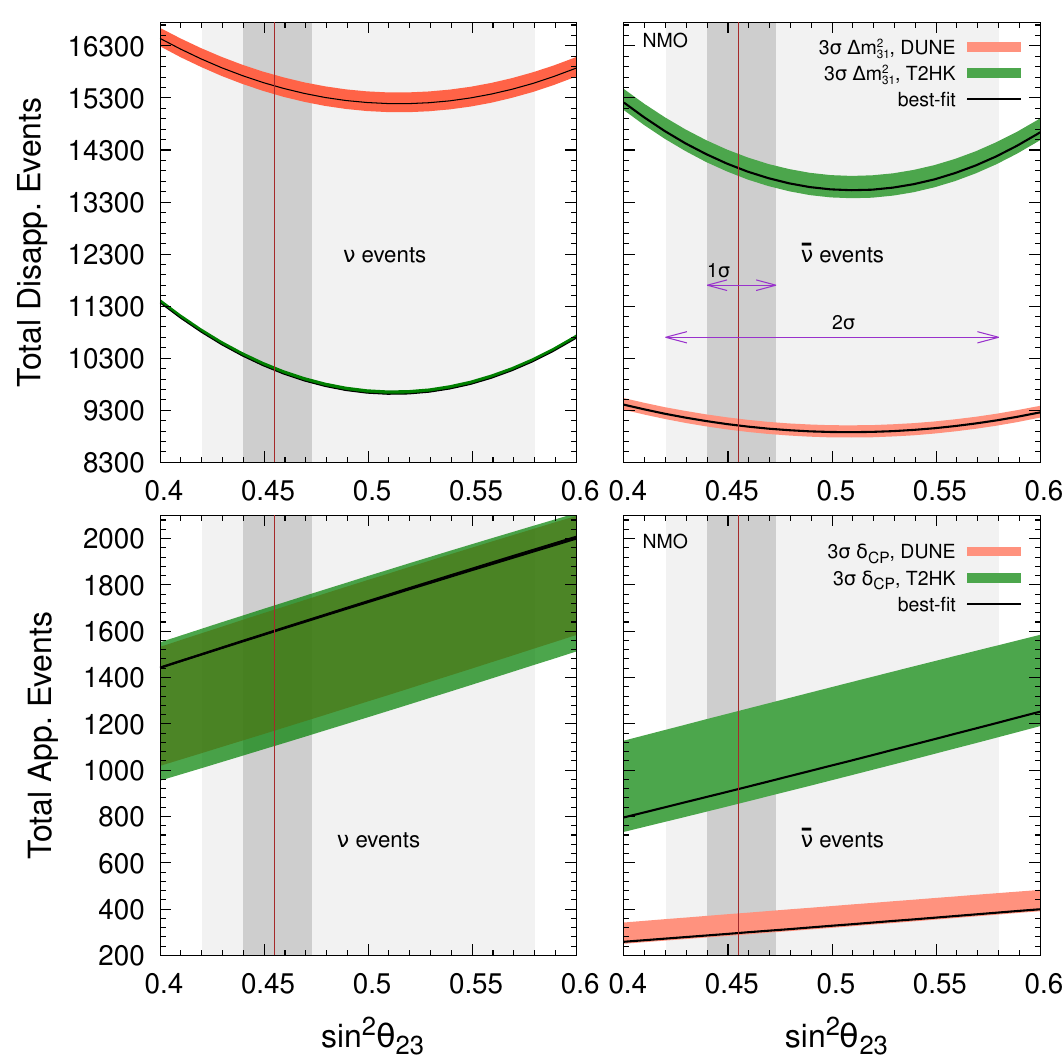}
	\caption{\footnotesize{Total (only signal) disappearance and appearance event rates as a function of $\sin^2\theta_{23}$ in DUNE and T2HK assuming NMO. The left and right panel depicts the same in neutrino and antineutrino modes for 480 kt$\cdot$MW$\cdot$yr and 2431 kt$\cdot$MW$\cdot$yr of exposure in DUNE and T2HK, respectively. Band depicts $~3\sigma$ uncertainty in $\Delta m^2_{31}$ (upper panel) and $\delta_{\rm CP}$ (lower panel). 
 }}
	\label{fig:1}
\end{figure}
In principle, reconstructing the charge of muons (to identify and segregate neutrinos from antineutrinos), event-by-event, is not feasible in DUNE and T2HK. There is always the occurrence of ``wrong-sign'' $ \bar{\nu}_{\mu}\, (\nu_{\mu})$ charged-current (CC) events when the primary beam is $\nu_{\mu} (\bar{\nu}_{\mu})$. Similarly, there is the contamination of ``wrong-sign'' $ \bar{\nu}_{e}\, (\nu_{e})$ charged-current (CC) events in neutrino (antineutrino) modes. \\
For the energy range under consideration, there is suppression in both the antineutrino cross-section and flux than a neutrino, which leads to a relatively more contribution of wrong-sign $\nu_{\mu}$ CC events in antineutrino mode than its counterpart. Apart from this, in Nature, positive mesons are more abundant than their negative counterpart as they are produced following the $pp$ or the $pn$ collisions~\cite{Dore:2018ldz}. Therefore, the neutrino beam is more intense than the antineutrino beam, and hence, the contamination of wrong-sign neutrinos in the antineutrino beam is higher. Both DUNE and T2HK follow horn current terminology, where the neutrino-enhanced beam is coined as forward horn current (FHC), and the antineutrino-enriched beam is the reverse horn current (RHC). In FHC, the wrong-sign flux is concentrated in the high-energy tail of the flux spectrum, where leptons are more likely to be forward and energetic due to the kinematics of neutrino and antineutrino scattering, while in RHC, they are concentrated around the low reconstructed energies~\cite{Katori:2016yel, DUNE:2020jqi}. Therefore, in general, we expect that the number of $\bar{\nu}$ events in $\nu$ beam to be way lesser than contamination of ${\nu}$ beam with $\bar{\nu}$.

In table \ref{table:two}, we compute the total event rates (signal) in $\nu_\mu\rightarrow\nu_\mu$ disappearance channel and $\nu_\mu\rightarrow\nu_{e}$ appearance channel for both neutrino and antineutrino modes in DUNE and T2HK, using benchmark oscillation parameters (refer to table~\ref{table:one}) with and without the inclusion of wrong-sign events. From the illustrative events shown in table~\ref{table:two} under two scenarios, we observe that the contribution from wrong-sign events is more in the RHC ($\bar{\nu}$ mode) than in FHC ($\nu$ mode). Moreover, in RHC, this contamination is more for the disappearance event rates (~50\% in DUNE, 35\% in T2HK) than appearance rates (~36\% in DUNE and ~17\% in T2HK). Consequently, given its larger size, T2HK is likely to exhibit greater effects from this contamination than DUNE. Following the general convention, in all our analyses henceforth, we have considered the wrong-sign contributions in both FHC and RHC signal events for both DUNE and T2HK.

\subsection{Total appearance and disappearance event rates}
\label{sec:2d}

Figure~\ref{fig:1} illustrates the total disappearance and appearance event rates in DUNE and T2HK. While the disappearance rates are affected mostly by uncertainty in $\Delta m^{2}_{31}$ and $\sin^{2}\theta_{23}$, the uncertainty in $\delta_{\rm CP}$ and $\sin^{2}\theta_{23}$ affects appearance rates predominantly, therefore we show bands of currently allowed 3$\sigma$ in $\Delta m^2_{31}$ for disappearance and $\delta_{\rm CP}$ in appearance event rates. The disappearance rates follow a U-shaped distribution when studied as a function of $\sin^{2}\theta_{23}$. This is because the disappearance rates $\propto (1-\sin^{2}\theta_{23})$. This points towards multiple combinations of $\sin^{2}\theta_{23}-\Delta m^2_{31}$ with same number of events~\cite{Agarwalla:2021bzs, Agarwalla:2022xdo}. Further, we observe that for values of $\sin^{2}\theta_{23}$ in the HO but close to 0.5, the curves show a flat behavior in both DUNE and T2HK. This hints that for these values of $\sin^{2}\theta_{23}$, sensitivity towards deviation from maximality will mostly come from the appearance rates, while disappearance rates may dominate later. Also, the minimum is not exactly at 0.5; instead, it is seen slightly shifted towards HO due to finite $\theta_{13}$ correction~\cite{Raut:2012dm}. Following higher runtime, more expected flux in neutrino mode, and substantial matter effect, we expect higher neutrino statistics in DUNE than T2HK, keeping NMO fixed. Similarly, higher runtime in antineutrino mode for T2HK implies higher antineutrino statistics. Having access to a wide-band beam makes DUNE capable of analyzing several $L/E$ ratios and more susceptible to a change in the value of $\Delta m^{2}_{31}$, unlike T2HK. This explains the higher neutrino disappearance statistics and a wider band when we vary $\Delta m^{2}_{31}$ in DUNE relative to T2HK. Further, DUNE's access to both the first and second oscillation maxima assures high disappearance rates in neutrino mode~\cite{DUNE:2020jqi} than T2HK, which has relatively fewer events at the second oscillation maximum~\cite{Hyper-Kamiokande:2018ofw}. In antineutrino mode, a higher runtime overcomes cross-section suppression in T2HK, which is not observed in the case of DUNE. Considering the appearance events in neutrino mode, DUNE has a higher runtime but lesser exposure, while T2HK has a lesser runtime but higher exposure; therefore, both experiments have almost similar event rates. In contrast, T2HK has higher statistics in $\bar{\nu}$ mode due to more runtime than DUNE.

Below, we show how the above contrasting features grant DUNE and T2HK the capability to probe atmospheric parameters organically, complementing each other.

\section{Projected sensitivities and its variation with total exposure}
\label{sec:results}
We project the expected sensitivities in DUNE, T2HK, and their combination to study the atmospheric parameters based on the detailed computation of event rates discussed above. Our results and analyses are based upon the current scenario in 3$\nu$ paradigm from the global oscillation data answering three crucial questions: (i) establishing deviation from maximal $\theta_{23}$, (ii) precision measurements in the 2-3 sector; $\sin^2\theta_{23}$ and $\Delta m^2_{31}$\,, and (iii) rejecting the wrong octant solutions of $\sin^2\theta_{23}$. Following the definition of Poissonian $\chi^2$~\cite{Baker:1983tu}, we estimate the median sensitivity~\cite{Cowan:2010js} of a given experiment in the frequentist approach~\cite{Blennow:2013oma} as  
\begin{equation}
\chi^2 (\vec{\omega}, \, \kappa_{s}, \, \kappa_{b,l})= \underset{( \vec{\lambda}, \, \kappa_{s}, \,\kappa_{b,l})}{\mathrm{min}}\left\{  2\sum_{i=1}^{n}(\tilde{y_i}-x_i-x_i\mathrm{ln}\frac{\tilde{y_i}}{x_i})+\kappa^2_{s}+ \sum_{l}\kappa^2_{b,l}\right\}\, , 
\label{eq:chi2} 
\end{equation}
where $n$ is the total number of reconstructed energy bins and $\vec{\lambda}$ is the set of oscillation parameters that are marginalized in the fit. The choice of set $\vec{\lambda}$ is discussed later in every subsection. Further,
\begin{equation}
\tilde{y_i}\,(\vec{\omega},\{\kappa_{s},\kappa_{b,l}\}) = N^{th}_i(\vec{\omega})[1+\pi^s\kappa_{s}] + \sum_{l}N^b_{i,l}(\vec{\omega})[1+\pi^b\kappa_{b,l}]\, .
\label{chi}
\end{equation}
Here, $N^{th}_i\,(\vec{\omega})$ is the number of signal events predicted in the $i$-th energy bin for a given set of oscillation parameters $\vec{\omega} = \left\lbrace \theta_{23}\,,\theta_{13}\,,\theta_{12}\,,\Delta \mathrm{m}^{2}_{21}\,, \Delta \mathrm{m}^{2}_{31}\,,\delta_{\mathrm{CP}}\right\rbrace$. $N^{b}_{i,l}\,(\vec{\omega})$ denotes the number of background events in the $i$-th energy bin where the neutral current (NC) backgrounds are independent of the oscillation parameter $\vec{\omega}$, while the charged current (CC) backgrounds is dependent on the oscillation parameters. $\pi^s$ is the pull term for systematic uncertainty on signal events. $\pi_{b,l}$ is the pull term for the systematic uncertainties on the $l$-th background contribution for any given channel in signal. These pull terms are uncorrelated with one another and have the same values in neutrino and antineutrino modes.  We incorporate the corresponding data in Eq.~\ref{eq:chi2} using the variable $x_{i}= N^{ex}_i\, +\, N^b_{i,l}$, where $N^{ex}_i$ indicates the observed CC signal events in the $i$-th energy bin via $\nu_{\mu}\rightarrow \nu_{\mu}$ and $\bar{\nu}_{\mu}\rightarrow \bar{\nu}_{\mu}$ disappearance channels and $\nu_{\mu}\rightarrow \nu_{e}$ and $\bar{\nu}_{\mu}\rightarrow \bar{\nu}_{e}$ appearance channels. Here, $N^b_{i,l}$  represents the $l$-th background contribution for a given channel. Throughout the simulation, we use publicly available software GLoBES~\cite{Huber:2002mx, Fogli:2002pt, Gonzalez-Garcia:2004pka}. We fix the mass ordering to NMO while generating data, as there are weak hints from global oscillation data favoring NMO at $\sim 2.5\sigma$~\cite{Capozzi:2021fjo, NuFIT}. In the fit, we marginalize over the allowed regions in oscillation parameters as mentioned in table~\ref{table:one}. We do not include any correlations among them as by the time these future experiments start taking data, these correlations will likely weaken~\cite{Song:2020nfh}. We consider the benchmark choices for $\theta_{12}$ and $\theta_{13}$ fixed~\cite{Capozzi:2021fjo}, as we do not expect the precision (2.8\%) achieved by Daya Bay to improve in the coming years~\cite{DayaBay:2022orm}. Although the present-day uncertainty in $\theta_{12}$ is 4.5\% ~\cite{Capozzi:2021fjo}, we do not expect the sensitivity in our study to get affected by it. We also fix the mass ordering in the fit as in the next decade, DUNE is expected to determine the mass ordering within initial years of data.

\subsection{Establishing deviation from maximal $\sin^{2}\theta_{23}$}
\begin{figure}[htb!]
	\centering
	\includegraphics[width=0.8\linewidth]{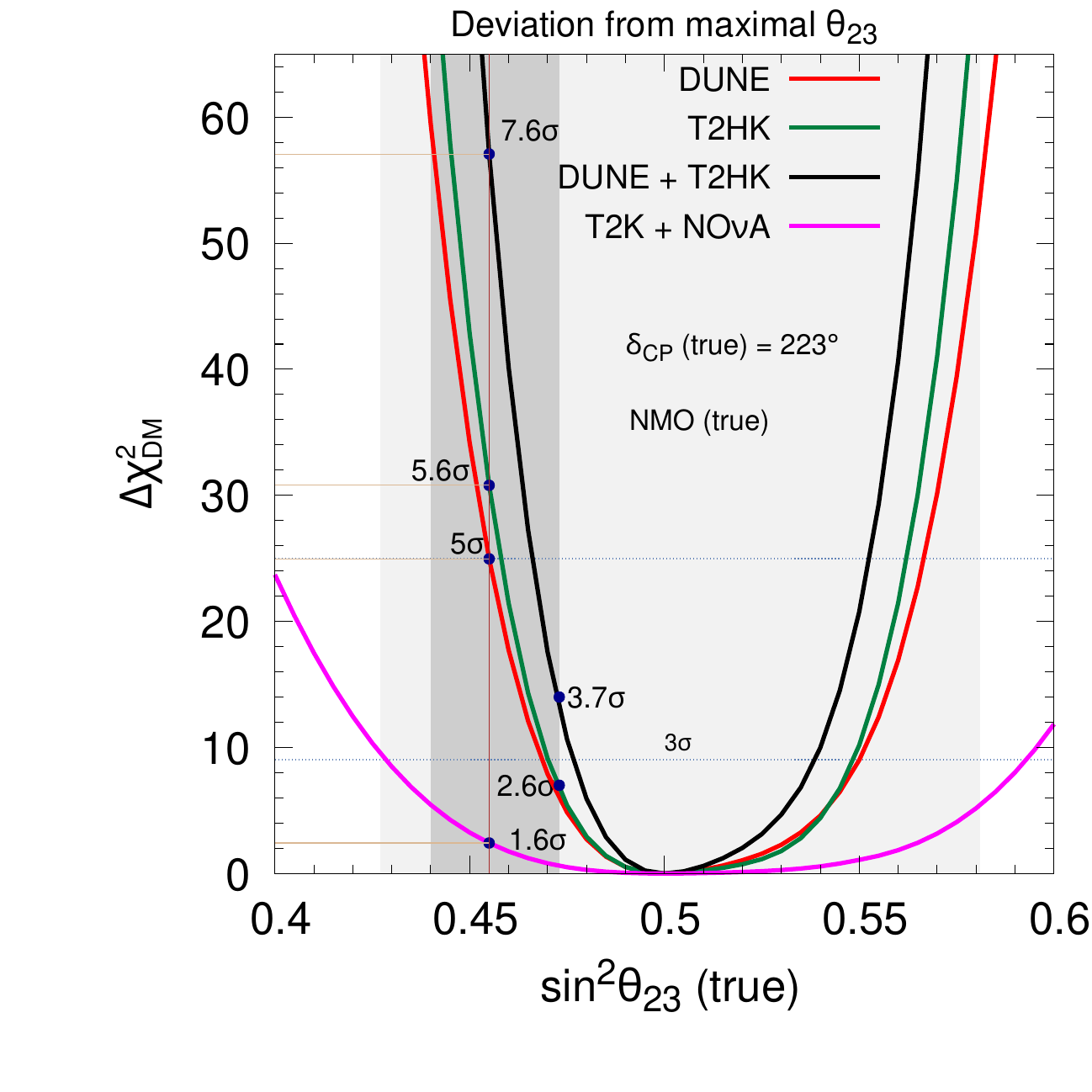}
	\caption{\footnotesize{ Sensitivity of DUNE, T2HK, DUNE + T2HK, and T2K [2.5 yr ($\nu$) + 2.5 yr ($\bar{\nu}$)] + NO$\nu$A [3 yr ($\nu$) + 3 yr ($\bar{\nu}$)] to exclude non-maximal solutions of $\sin^{2}\theta_{23}$ as a function of $\sin^2\theta_{23}$ in the data. For benchmark choices of oscillation parameters and the allowed ranges in $\Delta m^{2}_{31} $ and $\delta_{\rm CP}$ over which we marginalize in the fit, refer to table~\ref{table:one}. We assume exposures of 480 kt$\cdot$MW$\cdot$yr in DUNE, 2431 kt$\cdot$MW$\cdot$yr in T2HK, $84.4$ kt$\cdot$MW$\cdot$yr in T2K, and 58.8 kt$\cdot$MW$\cdot$yr in NO$\nu$A. For illustrative purpose, the benchmark choices of $\sin^{2}\theta_{23}$ is shown by a vertical brown line, projecting out the statistical confidence at the intersection with each curve. \textit{If in Nature, $\sin^{2}\theta_{23}$ is around the lower value of current 1$\sigma$ uncertainty ($\sim 0.473$), then the combination is the only solution to achieve $3\sigma$ with current benchmark values.} 
 }} 
	\label{fig:DM}
\end{figure}
We compute the statistical confidence with which DUNE, T2HK, and DUNE + T2HK can establish a deviation from maximal $\sin^{2}\theta_{23}$ by following 
\begin{equation}
\Delta \chi^2_{\text{DM}}= \underset{\delta_{\mathrm{CP}}\,,\,\Delta m^{2}_{31}}{\mathrm{min}}\left\{ \chi^2\left(\sin^2\theta_{23}^{\mathrm{test}} = 0.5\right)-\chi^2\left(\sin^2\theta_{23}^{\mathrm{true}} \in [0.4,0.6]\right) \right\},
\label{eq:deviation-from-maximality-chi2}
\end{equation}
where, $\vec{\lambda} = \{\delta_{\mathrm{CP}}, \, \Delta m^{2}_{31}\}$ is the set of oscillation parameters over which $\Delta\chi^2_{\rm DM}$ gets marginalized in the fit. So we generate the data for allowed uncertainty in $\sin^{2}\theta_{23}$ (refer to table~\ref{table:one}), while fixing it to 0.5 in the fit. There have been previous studies along this direction in ref.~\cite{Ballett:2016daj}\,, however here we consider the current best-fit values from ref.~\cite{Capozzi:2021fjo} which are similar to other global oscillation studies~\cite{NuFIT, Esteban:2020cvm, deSalas:2020pgw}. We also incorporate the latest collaboration estimates and ancillary files while using GLoBES.
\begin{figure}[t!]
	\centering
	\includegraphics[width=\linewidth]{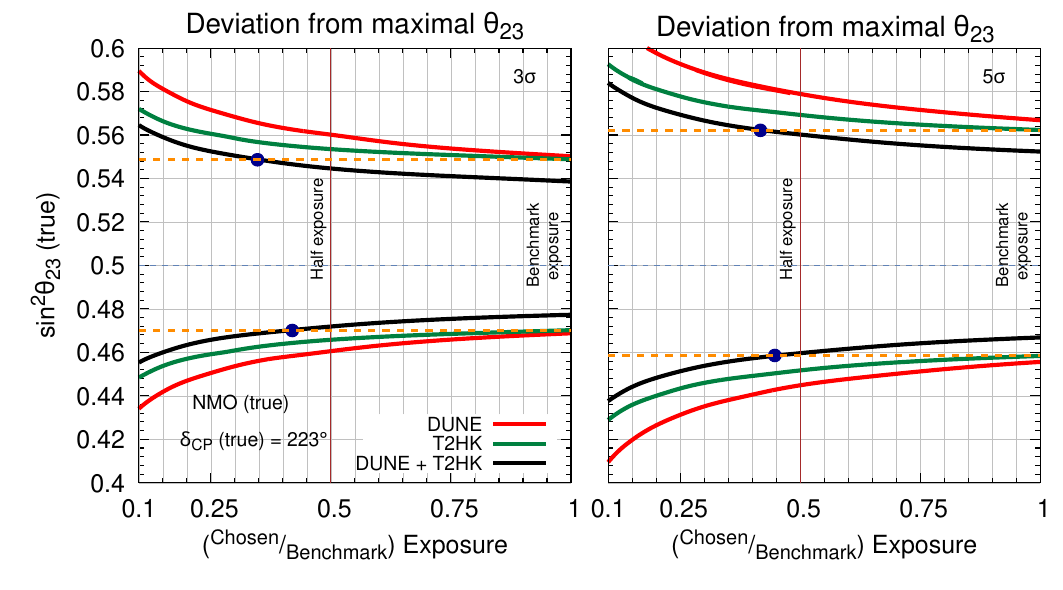}
	\caption{\footnotesize{3$\sigma$  (5$\sigma$) sensitivity of non-maximal $\sin^{2}\theta_{23}$ as a function of scaled (ratio of the chosen to the benchmark value of) exposure, assuming true NMO in the left (right) panel. The ratio reaches 1 at the benchmark choices of the corresponding experimental setups. We generate data by fixing all other oscillation parameters except $\sin^{2}\theta_{23}$ to their best-fit value (see table~\ref{table:one}). We marginalize over $\delta_{\mathrm{CP}}$ and $\Delta m^2_{31}$ in the test-statistics using the ranges of marginalization in table~\ref{table:one}. Dashed orange lines and solid blue circles are used to project and compare the maximum sensitivity attainable by standalone DUNE and T2HK using their nominal exposures with the exposure needed by DUNE + T2HK to achieve the same sensitivity. \textit{We find that using approximately 40\% of their exposures, DUNE + T2HK together achieves comparable sensitivity to each experiment independently running at their nominal exposures.}
  }}
	\label{fig:Exposure_DM}
\end{figure}

Figure~\ref{fig:DM} depicts the sensitivity in establishing deviation from maximality ($\Delta \chi^2_{\rm DM}$ in Eq.~\ref{eq:deviation-from-maximality-chi2}) as a function of true $\sin^{2}\theta_{23}$ for DUNE, T2HK, and their combination. As expected, it is smallest when the true and test equals for $\sin^{2}\theta_{23}$, increasing on both sides as we go away, implying the major contribution from $\sin^{2}2\theta_{23}$ in the leading term of disappearance channel (refer to an elaborate discussion in ref.~\cite{Agarwalla:2021bzs}). However, the U-shape around the $\sin^{2}\theta_{23} = 0.5$ is not symmetric because of the non-zero value of $\theta_{13}$. We observe that even after using the projected full exposure in the present-day long-baseline experiments: T2K and NO$\nu$A; they have lesser sensitivities. One of the major drawbacks in the present-generation experiments is much higher systematic uncertainties in both $\nu_{\mu}\rightarrow \nu_{\mu}$ and $\bar{\nu}_{\mu}\rightarrow \bar{\nu}_{\mu}$ disappearance rates (refer to last paragraph in Sec.~\ref{sec:2a} for corresponding values). The spread of curve around $\sin^{2}\theta_{23} = 0.5$ in all the experimental setups is less when $\sin^{2}\theta_{23} < 0.5$ than $\sin^{2}\theta_{23} > 0.5$. This can be explained due to the finite $\theta_{13}$ corrections~\cite{deGouvea:2005hk, Coloma:2012ut, Raut:2012dm} (Also, refer to discussion around fig.~\ref{fig:1} in Sec.~\ref{sec:2d}). Establishing sensitivity to deviations from maximality primarily depends on the disappearance statistic. In ref.~\cite{Agarwalla:2021bzs}, we also analyze and infer that uncertainty in the values of $\delta_{\mathrm{CP}}$ have minimal impact on this sensitivity. While T2HK, owing to huge disappearance statistics and corresponding lesser expected systematic uncertainties (refer to Sec.~\ref{sec:2a}), is able to achieve better sensitivity than DUNE, DUNE provides better measurements of $\Delta m^{2}_{31}$. Their combination (DUNE + T2HK) makes use of the complementary features among them and achieves a nearly 8$\sigma$ statistical confidence in establishing non-maximal $\sin^{2}\theta_{23}$, considering full exposures in the two experiments and the benchmark values. Furthermore,  \textit{if in Nature, $\sin^{2}\theta_{23}$ is around the upper value of current 1$\sigma$ uncertainty ($\sim 0.473$), then the combination is the only solution to achieve $3\sigma$ with current benchmark values.}

 The product of runtime, fiducial detector mass, and beam power provides the expected experimental exposure. The quantity of exposure is often considered interchangeably with runtime in phenomenological studies of neutrino oscillation. Currently, the DUNE collaboration also envisions a staged approach instead of undertaking the mammoth task of setting up a full-fledged DUNE detector of 40 kt fiducial mass~\cite{DUNE:2020jqi}. Therefore, it becomes imperative to discuss sensitivity study as a function of exposure. In figure~\ref{fig:Exposure_DM}, we study the nature of $\Delta \chi^2_{\rm DM}$ at 3$\sigma$ and 5$\sigma$ in Eq.~\ref{eq:deviation-from-maximality-chi2} as a function of scaled exposure for the standalone DUNE, T2HK, and their combination. We observe that initially, with an increase in exposure, the sensitivity to establish the deviation from maximal $\theta_{23}$ increases. However, the sensitivity after reaching half of their individual benchmark exposures reaches almost saturation. Horizontal illustrative dashed orange lines are drawn to depict the values of true $\sin^2\theta_{23}$ beyond which T2HK cannot differentiate between MM and the true values of $\sin^{2}\theta_{23}$ at its standard exposure, while the blue dots represent the intersection of the dashed orange line with the projected sensitivity curve of DUNE + T2HK. This implies that given the current benchmark choices of oscillation parameters in table~\ref{table:one}\,, the range of true values of $\sin^{2}\theta_{23}$ that can be differentiated from MM choices, by DUNE + T2HK with just $\sim 0.5$ of their nominal exposures, cannot be achieved by either of the individual experiments even with their respective projected exposures. When statistics are less, systematics become crucial. Therefore, at lower exposure, T2HK always performs better than DUNE to establish non-maximal $\sin^{2}\theta_{23}$, irrespective of the lower or higher octant of true choices of $\sin^{2}\theta_{23}$, because of better systematic uncertainties in disappearance rates. However, with the increment in exposure, the disappearance statistics in both DUNE and T2HK become similar. The complementary between DUNE + T2HK is essential to achieve a significant sensitivity at 5$\sigma$, even with high exposure. While the high precision measurements of DUNE in $\Delta m^{2}_{31}$ (due to substantial matter effect), complements the sensitivity of DUNE + T2HK at lower exposure, figure~\ref{fig:Exposure_DM} clearly shows that after a certain exposure, this study is no longer statistics-driven for achieving the sensitivity at 3$\sigma$. Nevertheless, a higher confidence level (5$\sigma$) is predominantly statistics-driven.

\subsection{Exclusion of wrong octant solutions of $\sin^{2}\theta_{23}$} 
\begin{figure}[htb!]
	\centering
	\includegraphics[width=0.8\linewidth]{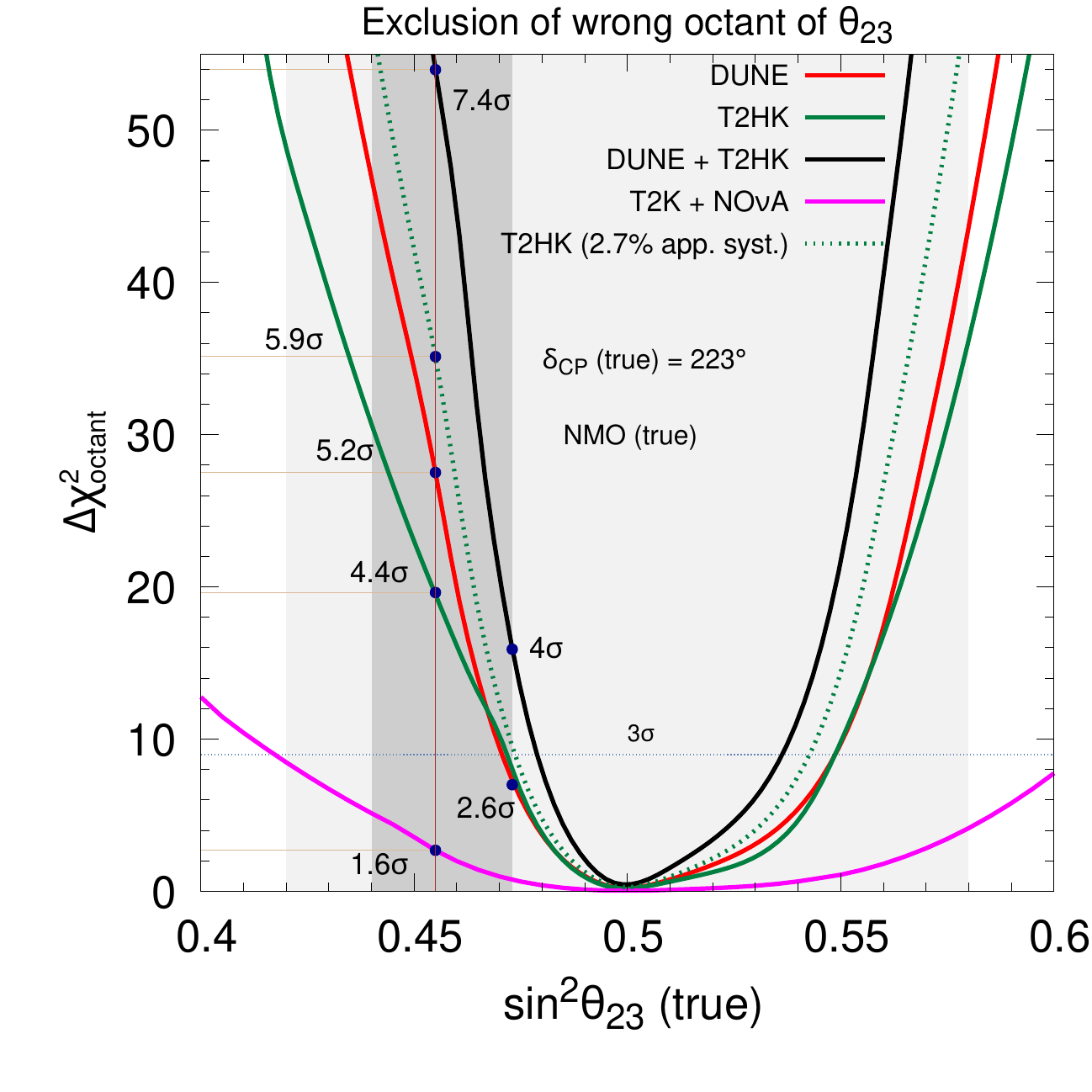}
	\caption{\footnotesize{ Sensitivity towards exclusion of wrong octant solutions in DUNE, T2HK, their combination, and T2K [2.5 yr $(\nu)$ + 2.5 yr $(\bar{\nu})$] + NO$\nu$A [3 yr $(\nu)$ + 3 yr $(\bar{\nu})$] as a function of $\sin^{2}\theta_{23}$ in data. The sensitivity of T2HK with an estimated improvement of nominal appearance systematic uncertainties to 2.7\% is also shown. In the fit, we marginalize over $\Delta m^{2}_{31}$ and $\delta_{\rm CP}$, keeping others fixed at their best-fit values (refer to table~\ref{table:one}). For illustrative purposes, we project out the statistical confidence for each curve, when $\sin^{2}\theta_{23} = 0.455$ in data.
 }}
	\label{fig:octant exclusion}
\end{figure}
\begin{figure}[htb!]
	\centering
	\includegraphics[width=\linewidth]{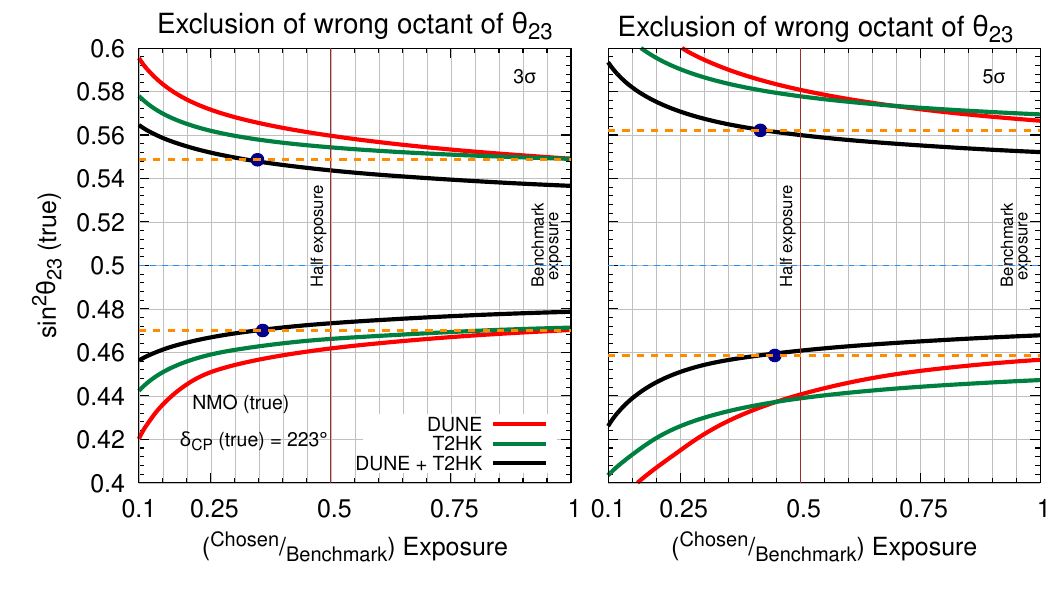}
	\caption{\footnotesize{3$\sigma$  (5$\sigma$) exclusion of wrong octant of $\sin^{2}\theta_{23}$ as a function of scaled exposure is shown in left (right) panel. 1 depicts the benchmark exposure. We marginalize over the allowed regions of $\delta_{\mathrm{CP}}$ and $\Delta m^2_{31}$ in the fit. Refer to table~\ref{table:one}. \textit{ At lower exposure, the complementarity between DUNE + T2HK is the only solution to attain a 5$\sigma$ discovery in ruling out the wrong octant solutions for a significant range of $\sin^2 \theta_{23}$ in Nature. 
 } }}
 \label{fig:octant-eclusion-vs-scaled-exposure}
 \end{figure}

Following the discussion of deviation from maximality, we study in this section the efficiency in establishing the octant of $\sin^{2}\theta_{23}$ by rejecting the hypothesis of wrong octant solutions. For this, we define 
\begin{eqnarray}
    \Delta \chi^{2}_{\text{octant}} &=& \underset{(\vec{\lambda})}{\mathrm{min}}\left\{ \chi^2\left(\sin^2\theta_{23}^{\mathrm{test}} \right) - \chi^2\left(\sin^2\theta_{23}^{\mathrm{true}}\right)\right\}\,,
\label{eq:octant-exclusion-chi2}
\end{eqnarray}
where $\sin^{2}\theta_{23}^{\mathrm{true}}$ refers to one octant, say LO; then, we generate data with [0.4, 0.5), while in the fit, we use the opposite octant, which in this case is HO $\in (0.5,0.6]$. Similar changes can be made by generating data with HO and excluding the LO hypothesis in the theory. In Eq.~\ref{eq:octant-exclusion-chi2}\,, $\vec{\lambda} = \{ \Delta m^{2}_{31}, \delta_{\rm CP}\}$. We use the corresponding allowed ranges from table~\ref{table:one}. Figure~\ref{fig:octant exclusion} depicts $\Delta \chi^{2}_{\text{octant}}$ as a function of $\sin^{2}\theta_{23}$ with which we generate data. It shows that alone DUNE and T2HK have similar sensitivity. T2HK is more stringent at lower significance, and DUNE is stronger at higher confidence. Large appearance systematic uncertainties in T2HK do not deteriorate the sensitivity, at least for lower significance due to comparable neutrino and antineutrino statistics~\cite{Agarwalla:2021bzs}. However, for attaining a higher significance, better systematic uncertainties in appearance rates are essential, which is a characteristic feature in DUNE (refer to section~\ref{sec:events}). The complementarity between DUNE and T2HK improves the standalone experiments' performance by almost $\sim 1.5$ times for the benchmark values from table~\ref{table:one}. The effect of improved appearance systematic uncertainties is distinctly visible when we consider the expected 2.7\% in T2HK~\cite{Munteanu:2022zla} instead of the nominal 5\%. Once the systematics are improved, T2HK performs better than DUNE irrespective of the true values of $\sin^{2}\theta_{23}$ in Nature. Sensitivity towards the exclusion of the wrong octant is dependent on both disappearance and appearance statistics, with the latter being dominant. For consistency, we have also checked the octant exclusion sensitivity using the best fit values and their corresponding allowed 3$\sigma$ ranges for minimization in the test-statistics from ref.~\cite{NuFIT:current}. We find that the results align closely with those shown in figure~\ref{fig:octant exclusion}.

In figure~\ref{fig:octant-eclusion-vs-scaled-exposure}\,, we study the efficacy of experiments in isolation and combination in ruling out the wrong octant of $\sin^{2}\theta_{23}$ as a function of scaled exposure. We observe that with 0.25 of their nominal exposures, DUNE alone will be able to differentiate about $\sim 45\%$ of $\sin^{2}\theta_{23}$ from wrong octant solutions; this improves to $\sim 50\%$ in T2HK, while their combination, DUNE + T2HK can differentiate $\sim 60\%$. So with just 0.25 of their individual exposures, it is possible in the combined setup to exclude the wrong octant for more than half of currently allowed $\sin^{2}\theta_{23}$ (refer to table~\ref{table:one}) at 3$\sigma$. Increasing beyond half of the nominal exposure does not help much, as the exclusion of the wrong octant solutions no longer remains statistics-driven. Further improvement in the allowed ranges of $\delta_{\mathrm{CP}}$ from the ongoing long-baseline experiments: T2K~\cite{T2K:2023smv} and NO$\nu$A~\cite{Carceller:2023kdz} may help to remove $(\sin^{2}\theta_{23} - \delta_{\mathrm{CP}})$ degeneracy and thus improve the sensitivity in wrong octant exclusion. We also notice that at 3$\sigma$ about $\sim 73\%$ of $\sin^{2}\theta_{23} \in [0.4, 0.6]$ can differentiate between correct and wrong octant solutions using the combined DUNE + T2HK setup, given the current benchmark values and projected exposure holds. For a higher confidence level (5$\sigma$), DUNE + T2HK is the only solution to attain sensitivity towards the exclusion of wrong octant solutions. We also observe that the discovery potential of DUNE + T2HK in excluding wrong octant solutions of $\sin^{2}\theta_{23}$, achievable with just $\sim$0.45 times of their individual exposures, is comparable to the sensitivity attained by standalone experiments using their nominal exposures.


\subsection{Precision measurements of $\sin^{2}\theta_{23}$ and $\Delta m^{2}_{31}$}
\label{precisionapp}

\begin{figure}[htb!]
\includegraphics[width=\linewidth]{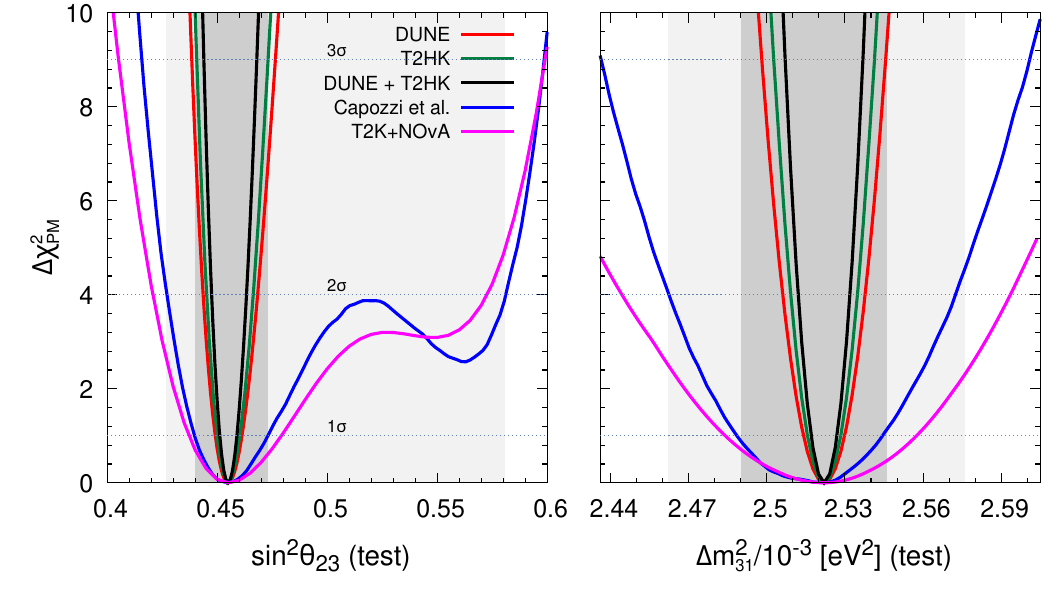}
\caption{\footnotesize{Expected achievable precision on $\sin^2\theta_{23}$ and $\Delta m^2_{31}$ around the respective benchmark values (refer to table~\ref{table:one}) using DUNE, T2HK, DUNE + T2HK, T2K [2.5 yr ($\nu$) + 2.5 yr ($\bar{\nu}$)] + NO$\nu$A [3 yr ($\nu$) + 3 yr ($\bar{\nu}$)], and global-fit from ref.~\cite{Capozzi:2021fjo}. In the fit, we marginalize over allowed region in $\delta_{\rm CP}$ and $\Delta m^{2}_{31}$ while determining precision on $\sin^2\theta_{23}$. Similarly, we perform marginalization over $\delta_{\rm CP}$ and $\sin^2\theta_{23}$ while determining precision measurements in $\Delta m^{2}_{31}$. Refer to table~\ref{table:four} for the computed values of the relative 1$\sigma$ precision, following Eq.~\ref{precision1}.  
}}
\label{fig:6}
\end{figure}
\begin{table}[t!]
	\centering
	\resizebox{\columnwidth}{!}{%
		\begin{tabular}{|c|c|c|c|c|c|c|}
			\hline
			\multirow{3}{*}{Parameter} & \multicolumn{6}{c|}{Relative 1$\sigma$ precision (\%)}\\
			\cline{2-7}
			& T2HK& DUNE & T2HK+DUNE & T2K+NO$\nu$A &Capozzi $et~al.$  & JUNO \\
			\hline
			$\sin^{2}\theta_{23}$ & 1.18 & 1.40 & 0.88 & 7.10 & 6.72 & ---\\
			\hline
			$\Delta m^{2}_{31}$ & 0.25& 0.31 & 0.20 & 0.99 & 1.09 & 0.2\\
			\hline
		\end{tabular}
	}
\caption{\footnotesize{Relative 1$\sigma$ precision computed from figure~\ref{fig:6}, following Eq.~\ref{precision1}. In addition, we also give present-day global-fit precision from ref.~\cite{Capozzi:2021fjo} and expected relative 1$\sigma$ precision using JUNO with an estimated 6 years of run~\cite{NavasNicolas:2023fza}. }}
\label{table:four}
\end{table}
\begin{figure}[t!]
	\centering
	\includegraphics[width=\linewidth]{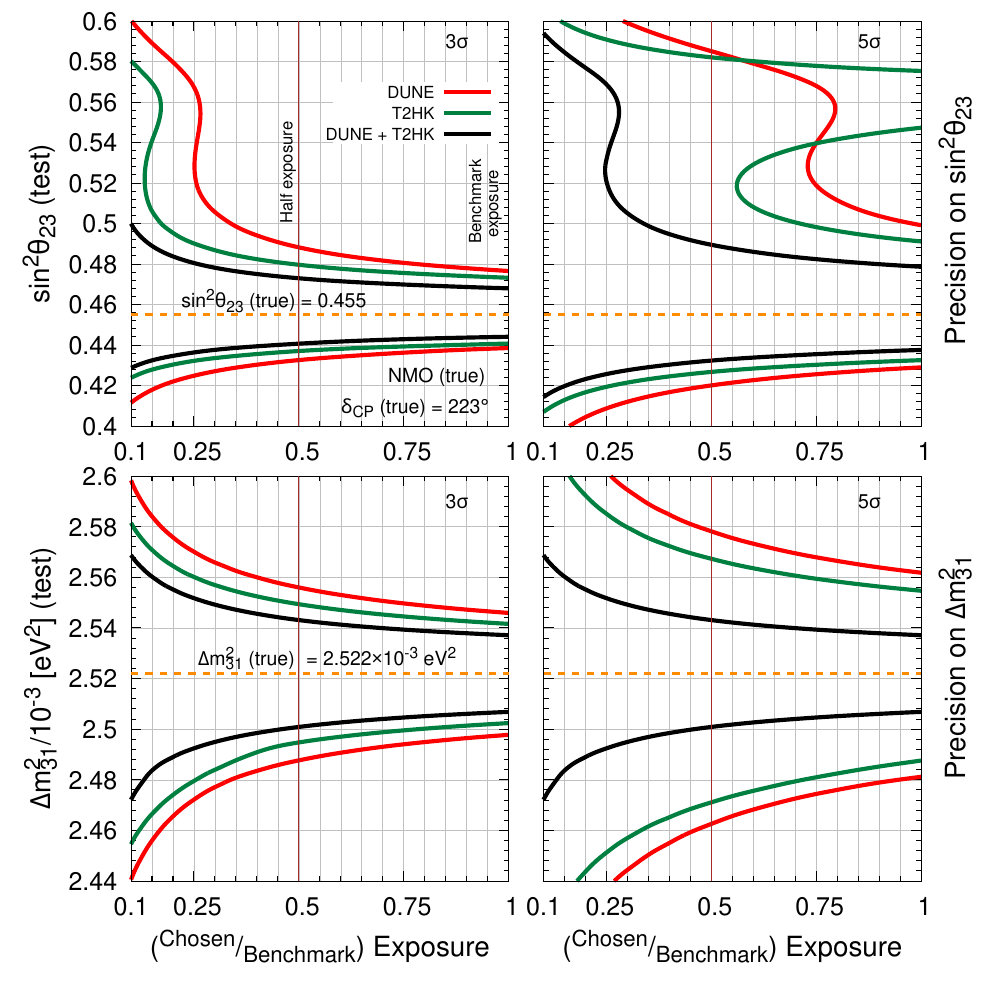}
	\caption{\footnotesize{Precision on $\sin^2\theta_{23}$ and $\Delta m^2_{31}$ around their benchmark values (refer to table~\ref{table:one}) as a function of scaled exposure is shown. The upper (lower) panel depicts the precision on $\sin^2\theta_{23}$ ($\Delta m^2_{31}$) at two different C.L. In the fit, we marginalize over the allowed ranges in  $\delta_{\mathrm{CP}}$ and $\Delta m^2_{31}$ ($\sin^2\theta_{23}$) while producing upper (lower) panel. \textit{Considering the benchmark choices, high statistical confidence (5$\sigma$) precision on both $\sin^2\theta_{23}$ and $\Delta m^2_{31}$ that can be achieved by DUNE + T2HK with just 0.25 of their individual exposures cannot be attained by standalone experiments even with their full exposures.} }}
	\label{figure:Exposure_precision}
\end{figure}
Following the exclusion of wrong octant solutions, it is imperative to question the precision in determining the value of atmospheric parameters: $\sin^2\theta_{23}$ and $\Delta m^2_{31}$. We compute the  statistical confidence for determining precision measurements on $\sin^{2}\theta_{23}$ by defining
\begin{equation}
\Delta \chi^2_{\text{PM}} = \underset{(\vec{\lambda})}{\mathrm{min}}\left\{ \chi^2\left(\sin^{2}\theta_{23}^{\mathrm{test}}\in [0.4,0.6] \right) - \chi^2\left(\sin^{2}\theta_{23}^{\mathrm{true}} = 0.455 \right)\right\}\,,
\label{precision-chisq-th23}
\end{equation}
while precision on $\Delta m_{31}^{2}$ is evaluated using
\begin{equation}
\Delta \chi^2_{\text{PM}} = \underset{(\vec{\lambda})}{\mathrm{min}}\left\{ \chi^2\left(\Delta m_{31}^{2,\mathrm{test}} \in [2.436,2.605]\times 10^{-3} \right) - \chi^2\left(\Delta m_{31}^{2,\mathrm{true}} = 2.522 \times 10^{-3} \right)\right\}\;.
\label{precision-chisq-m31}
\end{equation} 
 Here, test choices represent the corresponding allowed ranges of values, while the true choice is kept fixed at the benchmark choices (refer to table~\ref{table:one}). $\vec{\lambda}$ defines the set of oscillation parameters over which we perform marginalization in the fit given by $\vec{\lambda} = \{ \delta_{\rm CP}, \Delta m^{2}_{31}\}$ in Eq.~\ref{precision-chisq-th23} and $\vec{\lambda} = \{\delta_{\rm CP}, \sin^{2}\theta_{23}\}$ in Eq.~\ref{precision-chisq-m31}, respectively. For ease of quantifying, table~\ref{table:four} computes relative $1\sigma$-precision, defined as,
\begin{equation}
p(\zeta)\, =\, \frac{\zeta^{\rm max} - \zeta^{\rm min}}{6.0 \,\times \, \zeta^{\rm true}}\, \times \, 100\%\, .
\label{precision1}
\end{equation} 
Here, $\zeta^{\rm max}$ and $\zeta^{\rm min}$ depict the allowed upper and lower test values of each curve in the corresponding parameters (refer to figure~\ref{fig:6}) at $3\sigma$, respectively. We also quote the expected relative $1\sigma$-precision on $\Delta m^{2}_{31}$ from the upcoming reactor experiment, JUNO~\cite{NavasNicolas:2023fza} with the projected 6 years of runtime.

From figure~\ref{fig:6}\,, we observe that precision measurements in $\sin^2\theta_{23}$ allow a weak ($\sim 1.2\sigma$) clone solution in higher octant with the present global fit of oscillation data~\cite{Capozzi:2021fjo}. Although, with the full exposures of T2K + NO$\nu$A, we expect subtle improvement in it. However, the precision around the benchmark value of $\sin^2\theta_{23}$ is better using the present global fit oscillation data than the full projected exposures of present long-baseline experiments. This is because of huge disappearance statistics from other ongoing atmospheric experiments like Super-K and IceCube DeepCore. Similarly, we observe that the precision on $\Delta m^{2}_{31}$ using the global fit of oscillation data has already surpassed the expected precision using the full projected exposure of T2K + NO$\nu$A. This is mostly because of the input from reactor experiments like Daya Bay in the global fit. Additionally, we observe that due to the high energy resolution and large statistics in DUNE and T2HK, respectively, the standalone experiments can rule out the clone solution in $\sin^2\theta_{23}$. Comparatively, T2HK outperforms DUNE in precision measurements because of its extensive disappearance statistics and superior disappearance systematic uncertainties. A longer runtime in antineutrino mode also benefits T2HK, as both neutrino and antineutrino modes are crucial for achieving better precision measurements~\cite{Agarwalla:2013ju}. Table~\ref{table:four} helps in quantifying this improvement, showing the benefit of the interplay between DUNE and T2HK. Combining them improves the present-day~\cite{Capozzi:2021fjo} achievable precision on $\sin^2\theta_{23}$ and $\Delta m^2_{31}$ by a factor of $\sim$7 and $\sim$5, respectively. Further, we also make a comparison with the upcoming reactor experiment, JUNO~\cite{NavasNicolas:2023fza}. The achievable precision on $\sin^{2}\theta_{23}$ due to these next-generation experiments is truly remarkable.\\
In figure~\ref{figure:Exposure_precision}\,, we study precision on both the atmospheric parameters as a function of scaled exposure. While the appearance channel is dominated by $\sin^{2}\theta_{23}$, disappearance channel is mostly influenced by $\sin^{2}2\theta_{23}$~\cite{Mikheyev:1985zog,Wolfenstein:1977ue,Cervera:2000kp,Freund:2001ui}. Hence, in resolving the issue of the wrong octant, appearance events play a crucial role, while for achieving better precision around the correct octant, disappearance events are essential. While standalone DUNE and T2HK bearing low exposures ($\sim 0.25$ times nominal exposure) cannot rule out clone solutions in $\sin^{2}\theta_{23}$ at 3$\sigma$, the combined DUNE + T2HK provide degeneracy-free measurements. Furthermore for achieving a discovery potential (5$\sigma$), standalone DUNE is unable to rule out the clone solutions in $\sin^{2}\theta_{23}$ even after achieving the projected exposure, T2HK needs $\sim 0.8$ of nominal exposure for excluding the wrong octant solutions. However, \textit{DUNE + T2HK can provide degeneracy-free precision on $\sin^{2}\theta_{23}$ at 5$\sigma$ by considering only $\sim 0.3$ times of their individual benchmark exposures}. In the standalone setup, DUNE performs better than T2HK because of the higher systematic uncertainties assumed in the appearance events of T2HK (5\% refer section~\ref{sec:2a}) than DUNE (2\%). As discussed earlier, appearance events are responsible for fixing the correct octant of $\sin^2\theta_{23}$ and thus removing any clone solutions. The combined precision displays the benefit of synergy between DUNE and T2HK, which can accomplish 5$\sigma$ precision around the correct octant. Furthermore, we observe that reaching 3$\sigma$ precision becomes saturated after a while; thus, it is no longer statistics-dominated. However, a degeneracy-free precision can be achieved by DUNE + T2HK at 5$\sigma$ even at lesser exposures if $\sin^{2}\theta_{23}$ turns out to be in LO in Nature. Similarly, in the case of $\Delta m^{2}_{31}$, \textit{using approximately 20\% of the individual exposures of DUNE and T2HK together can achieve an impressive relative 1$\sigma$ precision of 0.25\%. However, at the same exposure level, standalone DUNE and T2HK cannot distinguish between the benchmark values of $\Delta m^{2}_{31}$ and the currently allowed ranges from Table~\ref{table:one} when analyzed at 5$\sigma$.}

\section{Allowed regions in ($\sin^{2}\theta_{23}- \delta_{\mathrm{CP}}$) plane}
\label{sec:contour}
\begin{figure}[htb!]
	\centering
	\includegraphics[width=\linewidth]{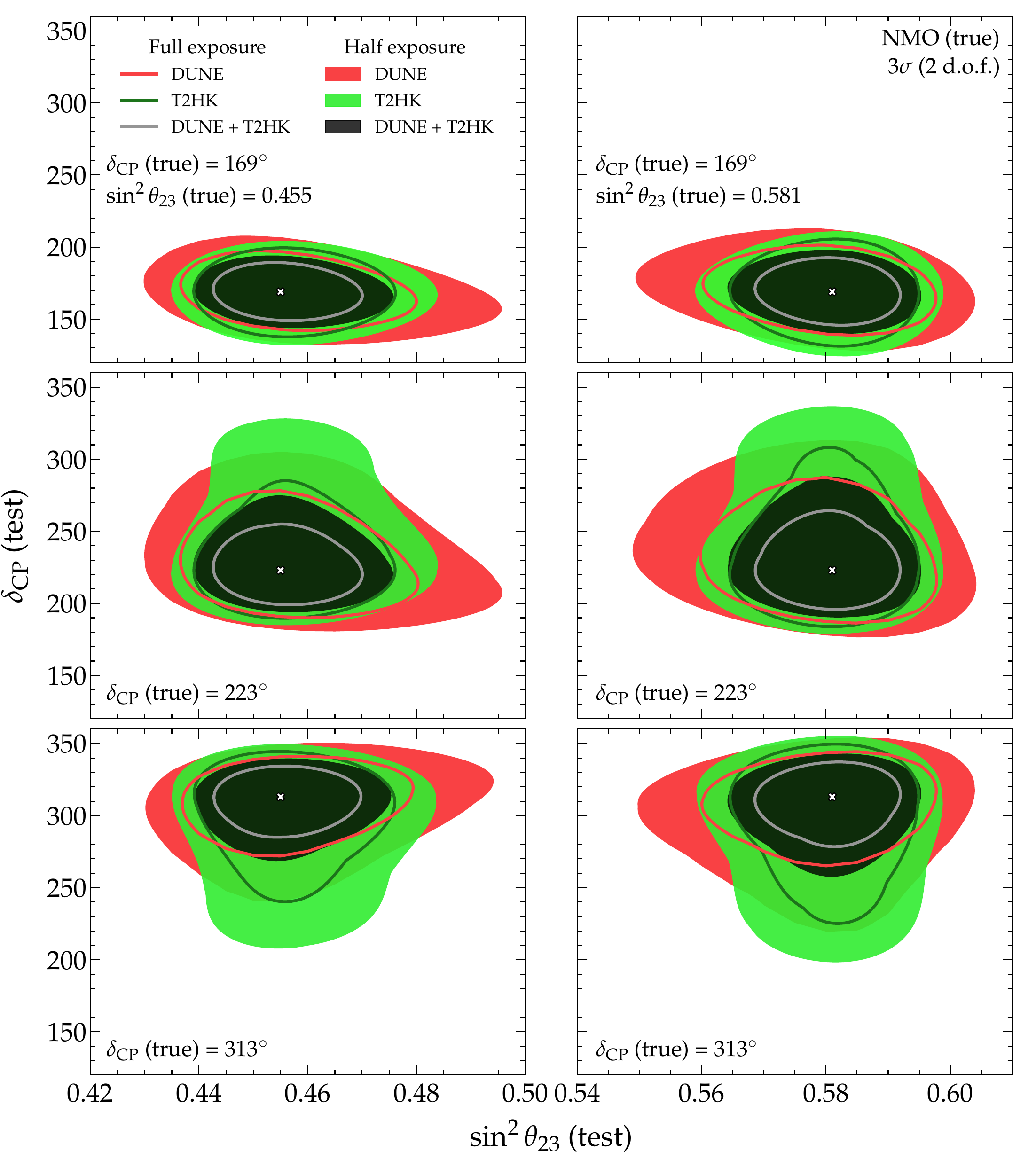}
	\caption{\footnotesize{Allowed regions of the test atmospheric mixing angle, $\sin^2\theta_{23}$  and CP phase, $\delta_\mathrm{CP}$. The true values correspond to the benchmark values and other illustrative choices following ref.~\cite{Capozzi:2021fjo} (see section~\ref{sec:contour} for detail). The test-statistics is scanned over $\sin^{2}\theta_{23}$  and $\delta_{\mathrm{CP}}$ (refer to Eq.~\ref{eq:chi-allowed-ranges}). \textit{Combination is the only solution to exclude clone solutions hinted by the standalone experiments at half exposures.}
 }}
	\label{fig:th23-dcp-allowed-ranges}
\end{figure}
As discussed previously, appearance events are necessary for extracting the correct octant of $\sin^2\theta_{23}$, while disappearance events are essential in obtaining a high-precision around the correct octant of $\sin^2\theta_{23}$. In this section therefore, we study the correlation between $\sin^{2}\theta_{23}$ and $ \delta_{\mathrm{CP}}$ in light of the current allowed oscillation parameter space. For this, we follow
\begin{eqnarray}
    \Delta \chi^{2} &=& \underset{(\vec{\lambda},\, \Delta m^{2}_{31})}{\mathrm{min}}\left\{ \chi^2\left(\sin^{2}\theta_{23}^{\mathrm{test}}\in [0.4,0.6], \delta_{\mathrm{CP}}^{\mathrm{test}} \in [0^{\circ}, 360^{\circ}] \right) - \chi^2 (\sin^{2}\theta_{23}^{\mathrm{true}},\delta_{\mathrm{CP}}^{\mathrm{true}} )\right\},
    \label{eq:chi-allowed-ranges}
\end{eqnarray}
where we use the benchmark values, $\pm 2\sigma \text{ in } \delta_{\mathrm{CP}} = 169^{\circ}$ and 313$^{\circ}$, and + 2$\sigma$ in $\sin^{2}\theta_{23} = 0.581$, following ref.~\cite{Capozzi:2021fjo} to generate data in each panel while scanning over the mentioned ranges in Eq.~\ref{eq:chi-allowed-ranges} of $\sin^2\theta_{23}$ and $\delta_{\mathrm{CP}}$. We marginalize over allowed ranges in $\Delta m^{2}_{31}$ (refer table~\ref{table:one}) in the fit.

In figure~\ref{fig:th23-dcp-allowed-ranges}\,, the benefit of exploiting the complementarity between DUNE and T2HK is clearly visible. DUNE with wide-beam is able to analyze various $L/E$ ratios. It also has access to the second oscillation maximum and 2\% appearance systematic uncertainties due to the magnificent LArTPC detector. These attribute to a better precision around the CP phase. Moreover, the large matter effect due to the long baseline helps in better precision measurement of $\Delta m^{2}_{31}$. However, more matter effect also induces extrinsic CP, which deteriorates the precision measurements in CP phase. This can be resolved by T2HK; with less matter effect, it provides better access to the intrinsic CP phase. Further, the huge disappearance statistics help in obtaining a better precision on $\sin^{2}\theta_{23}$. For comparison, the efficacy of DUNE + T2HK is visible in excluding the clone solutions hinted by the standalone experiments (in middle and lower panels), even with just half exposure. Further increasing the exposure from half to full in DUNE + T2HK leads to a quantitative decrement in the allowed region. For the upper panel, we observe that the allowed region is more or less similar in DUNE, T2HK, and the combination for the CP phase but varies with exposure for $\sin^{2}\theta_{23}$. This is because both DUNE and T2HK have better precision on $\delta_{\mathrm{CP}}$ when it is away from the CP-violating phases~\cite{Coloma:2012wq,Nath:2015kjg,Machado:2015vwa}. However, for achieving better precision on $\sin^{2}\theta_{23}$, large disappearance statistics are needed (therefore, T2HK consistently performs better than DUNE in this respect). Hence, a significant difference is visible when half exposure is increased to full exposure in the case of DUNE. However, the combination of both experiments is already performing well with their half exposures combined. While in the middle panel, we observe that the standalone experiments are hinting toward clone solutions --- T2HK in $\delta_{\mathrm{CP}}$ and DUNE in $\sin^{2}\theta_{23}$, the combination is able to precisely measure around the true values. The case in the lower panel is similar. Therefore, with half the exposures in each experiment, the combination is the only solution for better precision around the illustrative true values considered.\\
\section{Allowed regions in ($\sin^{2}\theta_{23}-\Delta m_{31}^{2}$) plane}
\label{appendix1}
\begin{figure}[htb!]
	\centering
	\includegraphics[width=\linewidth]{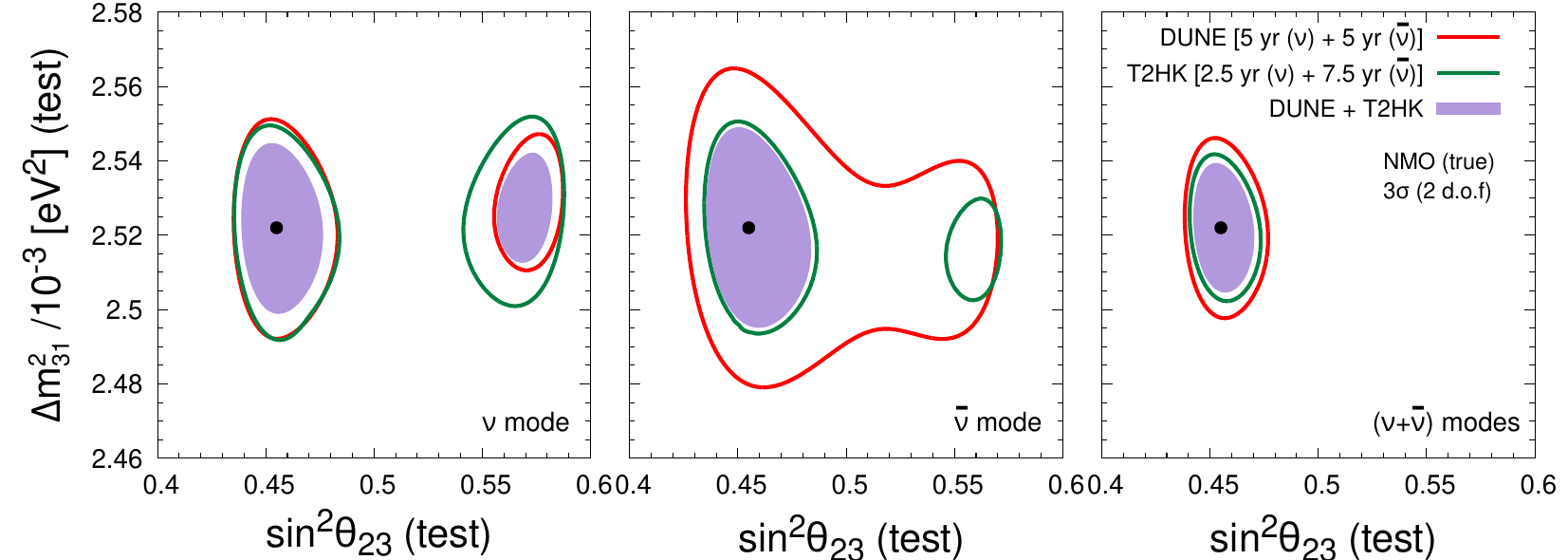}
	\caption{\footnotesize{Allowed regions of the test atmospheric mixing angle, $\sin^2\theta_{23}$  and mass-squared splitting, $\Delta m^2_{31}$. The true values correspond to the benchmark values as mentioned in table~\ref{table:one}. The test-statistics is scanned over $\sin^{2}\theta_{23}$  and $\Delta m^2_{31}$ (refer Eq.~\ref{eq:chi-allowed-ranges-th23-Dm31}). We marginalize over the allowed ranges in $\sin^{2}\theta_{23}$ in the fit. \textit{In only antineutrino mode, DUNE + T2HK is the only solution to exclude the clone solutions in $\sin^2\theta_{23}$.}
}}
        \label{fig:allowed-ranges-nmo}
\end{figure}

We follow
\begin{eqnarray}
    \Delta \chi^{2} &=& \underset{(\vec{\lambda},\, \sin^{2}\theta_{23})}{\mathrm{min}}\left\{ \chi^2\left(\sin^{2}\theta_{23}^{\mathrm{test}}, \Delta m_{31}^{2 ~ \mathrm{test}} \right) - \chi^2 (\sin^{2}\theta_{23}^{\mathrm{true}},\Delta m_{31}^{2 ~ \mathrm{true}} )\right\}\,,
    \label{eq:chi-allowed-ranges-th23-Dm31}
\end{eqnarray}
for generating figure~\ref{fig:allowed-ranges-nmo}, where true values correspond to the benchmark values as mentioned in table~\ref{table:one}, while scanning over the test-statistics for $\sin^{2}\theta_{23}\in [0.4,0.6]$ and allowed ranges of $\Delta m^{2}_{31}$ in table~\ref{table:one}. Further, we also perform marginalization over allowed ranges of $\sin^{2}\theta_{23}$ in the fit. We perform this study, separately for neutrino, antineutrino, and combined modes. 

As discussed previously with respect to DUNE~\cite{DUNE:2020jqi} and many other references in literature~\cite{Agarwalla:2013ju}\,, both neutrino and antineutrino modes are essential for breaking the $\sin^{2}\theta_{23} - \delta_{\mathrm{CP}}$ degeneracy and ruling out the wrong octant solutions in standalone experiments, while running in only antineutrino mode performs differently for the combined DUNE + T2HK. Alone, DUNE, when run in only antineutrino mode, is unable to expel even the MM solution of $\sin^{2}\theta_{23}$, T2HK has a clone solution at higher octant apart from the true octant. However, the complementary features in the combination are sufficient for breaking the $\sin^{2}\theta_{23} - \delta_{\mathrm{CP}}$ degeneracy in only antineutrino mode, which was not possible in standalone experiments. This is because, in the combined scenario, while T2HK has higher $\bar{\nu}$ statistics (refer to section~\ref{sec:2a}) leading to a majority of appearance events free of contamination from matter-induced CP phase, DUNE provides better precision measurement in $\Delta m^{2}_{31}$. The third panel shows the same curves as depicted in figure~\ref{fig:moneyplot} with full exposures. Isolated DUNE and T2HK are already able to achieve good precision around the best-fit breaking the $\sin^{2}\theta_{23} - \delta_{\mathrm{CP}}$ degeneracy, which gets more stringent around the best-fit with the combination.

\section{Sensitivity in ($\sin^{2}\theta_{23}-\Delta m_{31}^{2}$) plane with near-future experiments}
\label{appendix2}

\begin{figure}[htb!]
	\centering
	\includegraphics[width=\linewidth]{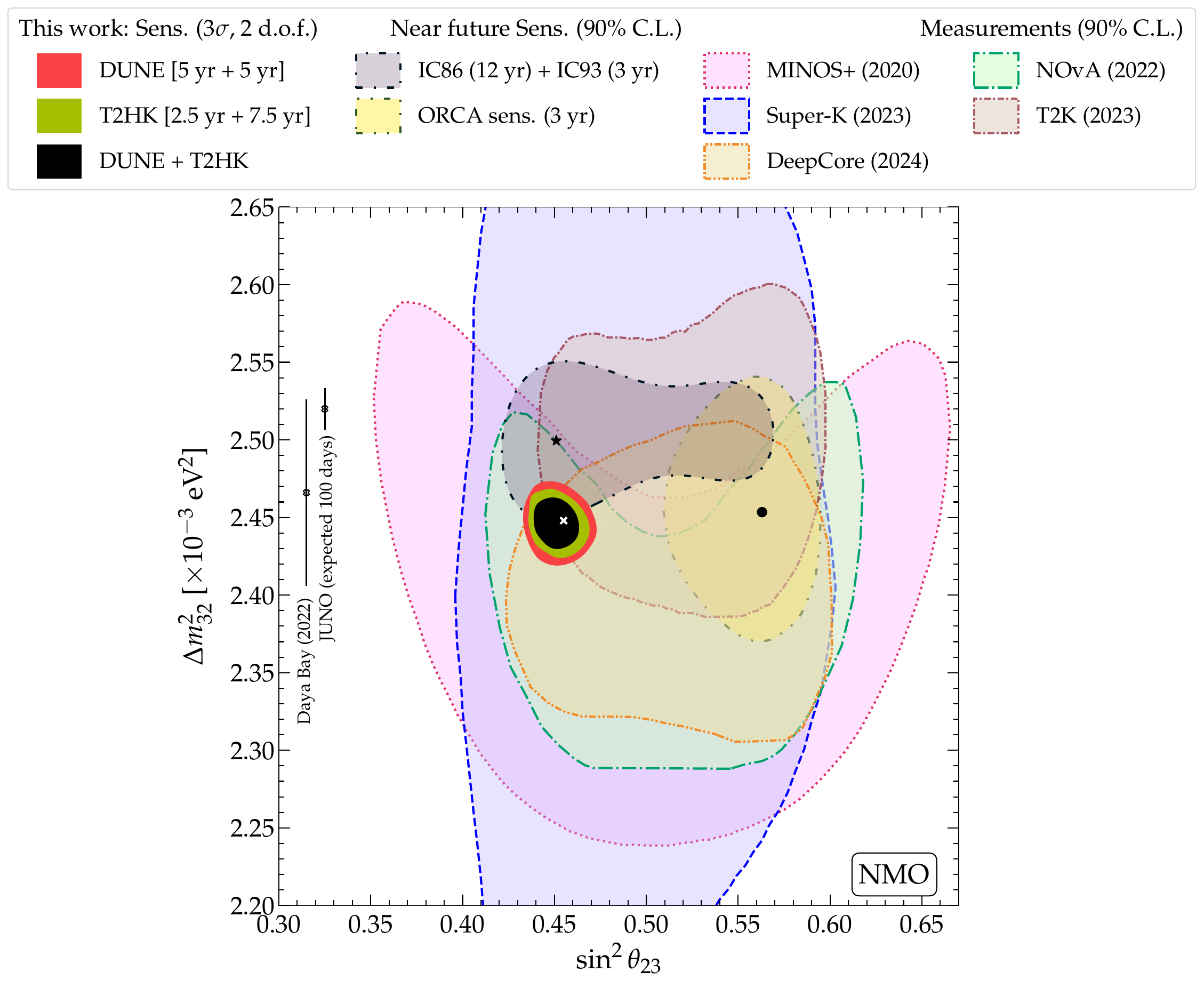}
	\caption{\footnotesize{Allowed ranges at 3$\sigma$ (2 d.o.f.) in the atmospheric mixing parameters, $\sin^{2}\theta_{23}$ and $\Delta m^{2}_{32}$, using DUNE, T2HK, and DUNE + T2HK. DUNE is expected to have an exposure of 480 kt$\cdot$MW$\cdot$yr and T2HK, an exposure of 2431 kt$\cdot$MW$\cdot$yr. Same as figure~\ref{fig:moneyplot}, we show existing allowed ranges from: Super-K~\cite{linyan_wan_2022_6694761}, T2K~\cite{T2K:2023smv}, NO$\nu$A~\cite{NOvA:2021nfi}, MINOS~\cite{MINOS:2020llm}, and IceCube DeepCore~\cite{IceCube:2024xjj}. We also show existing (expected) bounds on $\Delta m^{2}_{32}$ from Daya Bay~\cite{DayaBay:2022orm} JUNO ~\cite{JUNO:2022mxj}). Additionally, we also show expected sensitivity from IceCube Deepcore (12 yr) + Upgrade (3 yr)~\cite{IceCube:2023ins}, and ORCA (3 yr)~\cite{KM3NeT:2021ozk}. Refer to the text for details.}}
	\label{fig:appendix-allowed-ranges-nmo}
\end{figure}

In addition to the details in table~\ref{table:one}, in figure~\ref{fig:appendix-allowed-ranges-nmo}, we also show the expected allowed ranges using the near future development in the two giant atmospheric experiments: IceCube DeepCore and KM3NeT/ORCA. In ref.~\cite{IceCube:2023ins}, the expected sensitivity in atmospheric parameters is studied for 12 years of IceCube with 86 strings along with extra seven strings of IceCube Upgrade from 2026 onwards for three years. Strong improvements can be observed with the expected Upgrade in IceCube DeepCore. In ref.~\cite{KM3NeT:2021ozk}, the expected sensitivity of KM3NeT/ORCA after three years of data taking is studied, which we show in figure~\ref{fig:appendix-allowed-ranges-nmo}. The present scenario hints that the ongoing experiments in the near future will be able to obtain further precision on the $(\Delta m^2_{32} - \sin^{2}\theta_{23})$ plane. For comparison, we also show the assumed best-fit values by the two experiments: IceCube Upgrade ($\sin^{2}\theta_{23} = 0.451, \, \Delta m^2_{32} = 2.49 \times 10^{-3}$ eV$^{2}$) and ORCA ($\sin^{2}\theta_{23} = 0.563, \, \Delta m^2_{32} = 2.45 \times 10^{-3}$ eV$^{2}$). It should be noted that since their benchmark value for $\sin^{2}\theta_{23}$ corresponds to the opposite octant, the allowed region they expect seems complementary. 


\section{Summary and conclusions}
\label{conclusion}
The present generation of neutrino experiments have undoubtedly paved the way for precision studies in neutrino physics. The reactor mixing angle ($\theta_{13}$) measured by Daya Bay has achieved a remarkable precision of 2.8\%~\cite{DayaBay:2022orm}. Solar parameters: $\theta_{12}$ and $\Delta m^{2}_{21}$ have long been measured and stand presently with a relative 1$\sigma$ error of 4.5\% and 2.3\%\,, respectively from the global fit~\cite{Capozzi:2021fjo}. The most precisely measured oscillation parameter is, ironically, the magnitude of atmospheric mass-squared difference ($\Delta m^{2}_{31}$), while determining its sign still remains one of the big questions in neutrino physics left unanswered. The two most uncertain parameters are the atmospheric mixing angle, $\theta_{23}$, and Dirac CP phase, $\delta_{\mathrm{CP}}$. 

In this work, we address the issue of maximal mixing solution of $\sin^{2}\theta_{23}$ and if non-maximal, then ruling out the wrong octant solutions of $\sin^{2}\theta_{23}$. We also study the achievable precision on $(\sin^{2}\theta_{23}-\delta_{\mathrm{CP}})$ and $(\sin^{2}\theta_{23}-\Delta m^{2}_{31})$ planes. To analyze the mentioned issues, we compare and contrast the standalone DUNE and T2HK with their complementarities in the combined DUNE + T2HK setup. While DUNE and T2HK, individually, should be able to improve on the sensitivity studies of deviation from maximal $\sin^{2}\theta_{23}$, exclusion of its wrong octant solutions, and precision measurements, their individual sensitivities are hampered by degeneracies due to uncertainties in $\sin^{2}\theta_{23}$, $\delta_{\mathrm{CP}}$, and $\Delta m^{2}_{31}$.  However, DUNE and T2HK have complementary capabilities: while T2HK is especially well-suited to measure $\sin^{2}\theta_{23}$ and $\delta_{\mathrm{CP}}$, DUNE is especially well-equipped to measure $\Delta m^{2}_{31}$. Thus, combining DUNE and T2HK brings many novelties.
 
We find that the combined DUNE + T2HK increases the sensitivity to establish a deviation from maximality to $\sim 8\sigma$ if in Nature, $\sin^{2}\theta_{23}$ is same as the benchmark value mentioned in table~\ref{table:one}, following present global fit in ref.~\cite{Capozzi:2021fjo}. However, if this present best-fit shifts to its present $1\sigma$ allowed upper bound ($\sim 0.473$), then the combination is the only solution to achieve sensitivity to deviation from maximality greater than 3$\sigma$. The study of sensitivity towards deviation from maximal $\sin^{2}\theta_{23}$ as a function of exposure reveals that following the benchmark values, the discovery potential that combined DUNE + T2HK can reach with just 0.5 times the nominal exposure, is unattainable by either DUNE or T2HK even with their full exposures. In analyzing the sensitivity towards excluding incorrect octant solutions of $\sin^{2}\theta_{23}$, we find that the synergy of combined experiments significantly outperforms individual ones. While no single experiment achieves 5$\sigma$, the combination reaches approximately 8$\sigma$, assuming the true value of $\sin^{2}\theta_{23}$ is the same as the benchmark. Yet again, the discovery potential that standalone experiments attain with full exposure in eliminating the wrong octant, can be achieved by combining with just 0.4 times the full exposure. The estimated precision by alone DUNE and T2HK in both atmospheric parameters: $\sin^{2}\theta_{23}$ and $\Delta m^{2}_{31}$ improves present global fit precision by an approximate factor of 5 and 4, respectively. Furthermore, we find that the range of 5$\sigma$ precision that the combined DUNE + T2HK can achieve with only 0.25 times the exposure in $\Delta m^{2}_{31}$ is a level of precision that individual experiments cannot reach even with full exposures. Moreover, while studying the allowed ranges in $(\sin^{2}\theta_{23}- \delta_{\mathrm{CP}})$ plane, we notice that the weak hint towards clone solutions shown by standalone experiments at 3$\sigma$ can be resolved by the combination with just half the exposure.

Therefore, this study brings about novel perspectives of the upcoming high-precision LBL experiments DUNE and T2HK, stressing how their combination and 
hence, the possible complementarities among them may alleviate the need of a very high exposure from these individual experiments in obtaining the desired
sensitivities towards different oscillation parameters.

\chapter{Impact of flavor-dependent long-range neutrino interactions on the measurements of oscillation parameters using DUNE and T2HK}
\label{C6} 
The interplay between symmetry principles and fundamental interactions has repeatedly guided our journey toward a deeper understanding of nature's workings. From Maxwell's unification of electricity and magnetism \cite{Delphenich:2015rjq, Wilczek:2015rba, Reggia:2024kzh} to the electroweak theory's revelation \cite{Hollik:1995dv, Altarelli:1998xf} of an underlying symmetry between electromagnetic and weak forces, symmetry considerations have proven to be powerful guides in uncovering the fundamental laws of physics. Similarly, the discovery of parity violation in weak interactions \cite{Wu:1957my, Lee:1956qn} and the subsequent development of the Standard Model demonstrate how symmetry breaking \cite{Streater:1977hq} can lead to profound insights into Nature's structure. This historical progression suggests that further explorations of symmetry principles — particularly those that extend beyond the established SM — may continue to yield valuable discoveries.

In the realm of neutrino physics, where three flavor eigenstates undergo quantum mechanical mixing described by the PMNS matrix \cite{Maki:1962mu, Pontecorvo:1967fh}, subtle distortions of oscillation probabilities may reveal hints of new physics. The phenomenon of neutrino oscillations itself emerged as a solution to the solar neutrino problem \cite{Gribov:1968kq, Bellerive:2016byv} and has since provided compelling evidence for physics beyond the SM, establishing that neutrinos possess non-zero masses \cite{Hu:2022ufh}. The precision measurement of the mixing angles $\theta_{12},\, \theta_{13},\, \text{and } \theta_{23}$ as well as the mass-squared differences $\Delta m_{21}^2$ and $\Delta m^2_{31}$ has entered an era of remarkable accuracy \cite{Capozzi:2021fjo, Abada:2025jpk} - with the notable exception of the CP-violating phase $\delta_{\mathrm{CP}}$ and the determination of the octant of $\theta_{23}$ which remain significant challenges.

Beyond the established charged and neutral current interactions, neutrinos could experience additional forces mediated by ultralight bosons associated with broken symmetries in the lepton sector \cite{Fayet:1986vz, Heeck:2011wj, Ferrer:1999ad}. These new interactions could manifest as modifications to the matter potential experienced by neutrinos as they propagate through various media. While the standard matter effect \cite{Wolfenstein:1977ue} arising from coherent forward scattering via W-boson exchange (for electron neutrinos) is well understood, additional flavor-dependent potentials could arise from new gauge bosons coupling distinctively to different lepton flavors \cite{Foot:1994vd, Rodejohann:2005ru, Heeck:2010pg}.

The gauging of lepton flavor differences- specifically $(L_e-L_\mu)$, $(L_e-L_\tau)$, and $(L_\mu-L_\tau)$ \cite{Foot:1994vd} represent a theoretically elegant extension to the SM \cite{Heeck:2014zfa, He:1991qd} that preserves anomaly cancellation \cite{Foot:1994vd, Allanach:2020zna} while introducing distinctive phenomenological signatures. These U(1) gauge symmetries \cite{Rodejohann:2005ru, Heeck:2010pg} stand out among possible extensions because they are anomaly-free \cite{Allanach:2020zna} without requiring additional exotic fermions \cite{Patra:2016ofq}, making them particularly economical additions to the SM gauge structure. When the mediators of these interactions \cite{He:1991qd, Coloma:2020gfv, Farzan:2016wym} possess masses below $10^{-10}$ eV, the resulting force ranges extend across astronomical distances, creating a cosmic web of matter potentials. The characteristic range of a force mediated by a boson of mass $m$ is approximately $\lambda \sim 1/m$, meaning that for mediators with $m \sim 10^{-10}$ eV, the force can act over distances of roughly $10^{5}$ km which is comparable to the Earth-Sun distance. For even lighter mediators with $m \sim 10^{-12}$ eV \cite{Coloma:2020gfv}, the range extends to galactic scales $(\sim 10$ kpc) and beyond \cite{Bustamante:2018mzu, Hogg:1999ad, Giunti:2007ry}. This remarkable reach enables these forces to integrate contributions from matter distributed throughout the solar system \cite{Coloma:2020gfv}, the Milky Way \cite{McMillan:2011wd}, and potentially even neighboring galaxies \cite{Steigman:2007xt, Yuksel:2008cu}. Unlike short-range non-standard interactions \cite{Grossman:1995wx, Proceedings:2019qno} that depend primarily on local matter densities, these ultralight-mediated forces integrate contributions from electrons and neutrons distributed throughout celestial bodies and galactic structures \cite{Coloma:2020gfv}. 

As experimental precision advances into territory \cite{An:2025lws} where subleading effects become discernible, the boundary between parameter measurement and new physics discovery grows increasingly permeable. The three-flavor neutrino oscillation paradigm has successfully explained most experimental results to date, with persistent anomalies at short baselines \cite{Rodrigues:2025tha} potentially pointing to sterile neutrinos or other exotic physics. The next-generation long-baseline experiments — DUNE (Deep Underground Neutrino Experiment in U.S.A)~\cite{DUNE:2020jqi,DUNE:2020ypp} with its broad-band beam and superior energy resolution, and T2HK (Tokai-to-Hyper-Kamiokande in Japan)~\cite{Hyper-KamiokandeProto-:2015xww,Hyper-Kamiokande:2018ofw} with its unprecedented statistical power — stand as complementary probes of neutrino properties. The complementarity in baseline, energy, and detection technology makes their combined analysis particularly robust against systematic uncertainties and parameter degeneracies~\cite{Agarwalla:2024kti}. Both experimental programs aim to resolve persistent questions in the atmospheric sector: whether $\theta_{23}$ deviates from maximal mixing $(\theta_{23} = 45^{\circ})$, which octant houses the true value of $\theta_{23}$ ($\theta_{23} < 45^{\circ}$ or $\theta_{23}>45^{\circ}$), and the magnitude of CP violation in the lepton sector. The determination of deviation of $\theta_{23}$ from maximality~\cite{Gonzalez-Garcia:2004pfd, Antusch:2004yx, Choubey:2005zy, King:2014nza, Agarwalla:2021bzs} has profound theoretical implications, as maximal mixing may point to underlying symmetries in the lepton sector. Similarly, the octant determination could provide crucial input for model builders \cite{Agarwalla:2013ju, Chatterjee:2022pqg} attempting to explain the pattern of neutrino masses and mixings. The measurement of $\delta_{\mathrm{CP}}$, potentially indicating leptonic CP violation~\cite{Agarwalla:2022xdo, Coloma:2012wq}, could shed light on the origin of the matter-antimatter asymmetry in the Universe through the leptogenesis mechanism~\cite{Davidson:2008bu, Buchmuller:2005eh}.

In this work, we examine how the presence of flavor-dependent long-range interactions would reshape the sensitivity landscape of these experimental endeavors. We analyze how these cosmic-scale forces would modify established measurement capabilities. The interference between standard oscillation terms and long-range potential contributions produces a rich phenomenology that may enhance sensitivity to some parameters while degrading the resolution of others, creating a nuanced experimental signature distinct from other new physics scenarios. Our analysis demonstrates that the synergistic combination of DUNE and T2HK data significantly outperforms individual experimental results in constraining oscillation parameters, even when long-range interactions introduce additional complexity to the oscillation probability. This complementarity proves particularly valuable for breaking degeneracies that would otherwise limit experimental reach. By systematically exploring the parameter space of coupling strengths and mediator masses for different gauge symmetries ~\cite{Bustamante:2018mzu, Khatun:2018lzs, Coloma:2020gfv, Singh:2023nek}, we identify regions where these experiments maintain robust sensitivity to standard oscillation parameters and regions where substantial degradation occurs, potentially guiding the development of analysis strategies resilient to new physics effects. 

The architecture of this manuscript is delineated in the following fashion. In section \ref{sec:2}, we present the foundational aspects of long-range interactions (LRI), including the newly proposed gauge symmetries and the associated mediating vector bosons, as well as their implications for standard neutrino oscillations within the $3\nu$ paradigm. In section \ref{sec:effect-on-total-event-rates}, we introduce the rudimentary experimental setups of the impending long-baseline experiments, accentuating their salient complementary attributes pertinent to our analysis. This exposition includes a detailed examination of their total event rates and bi-events under the influence of LRI-induced symmetries. In section \ref{sec:results}, we discuss our results and findings. Lastly, section \ref{sec:summary} distills the essence of our findings, offering a comprehensive summary and concluding reflections.
\label{sec:introduction}
\begin{table}[h]
    \centering
    \resizebox{\textwidth}{!}{
    \begin{tabular}{|c|c|c|c|c|c|c|}
        \hline \hline
        \multirow{2}{*}{\textbf{NMO}} & $\Delta m^2_{21}/10^{-5}$ & \multirow{2}{*}{$\sin^{2}\theta_{12}/10^{-1}$} & \multirow{2}{*}{$\sin^{2}\theta_{13}/10^{-2}$} & \multirow{2}{*}{$\sin^2\theta_{23}/10^{-1}$} & $\Delta m^2_{31}/10^{-3}$  & $\delta_{\text{CP}}$ \\
        & ($\mathrm{eV^{2}}$) & & & & ($\mathrm{eV^{2}}$) & ($^\circ$)\\
        \hline \hline
        Benchmark & \boldmath$7.36$ & \boldmath$3.03$ & \boldmath$2.23$ & \boldmath$4.55$ & \boldmath$2.522$ & \boldmath$223$\\
        \cline{1-7}
        3$\sigma$ range & - & - & - & 4.16 - 5.99 & 2.436 - 2.605 & 139 - 355\\
        \hline\hline
        \multicolumn{7}{|c|}{\hspace{-1.8em}\textbf{LRI Potential} $V_{\alpha\beta}~(\times 10^{-14}~\mathrm{eV})$} \\
        \hline
         & \multicolumn{2}{c|}{$V_{e\mu}$} & \multicolumn{2}{c|}{$V_{e\tau}$} & \multicolumn{2}{c|}{$V_{\mu\tau}$} \\
        \hline
        Benchmark  & \multicolumn{2}{c|}{\boldmath$1.4$} & \multicolumn{2}{c|}{\boldmath$1.0$} & \multicolumn{2}{c|}{\boldmath$0.73$} \\
        \hline
        Range & \multicolumn{2}{c|}{0.1 - 3} & \multicolumn{2}{c|}{0.1 - 3} & \multicolumn{2}{c|}{0.1 - 3} \\
        \hline \hline
    \end{tabular}
    }
    \caption{\footnotesize{The benchmark values and the 3$\sigma$ ranges of the oscillation parameters over which we minimize in our study, assuming normal mass ordering (NMO), following the Ref.~\cite{Capozzi:2021fjo}. Benchmark values of long-range matter potential corresponding to each symmetry, $V_{\alpha\beta}$ to generate data, and the assumed range of minimization over the test statistic, used in the subsection \ref{subsec:DM-fixed-potential}. These values are motivated from Ref.~\cite{Singh:2023nek}. Further details regarding this have been given in sec.~\ref{subsec:simulation}.}}
    \label{tab:oscillation-params}
\end{table}

\section{Introduction to flavor-dependent long-range interaction}
\label{sec:2}
\subsection{Gauging new neutrino interactions under the global $U'(1)$ group via the symmetries of differences between  lepton numbers}
\label{sec:lri_symmetries}
\begin{figure}
\includegraphics[width=\linewidth]{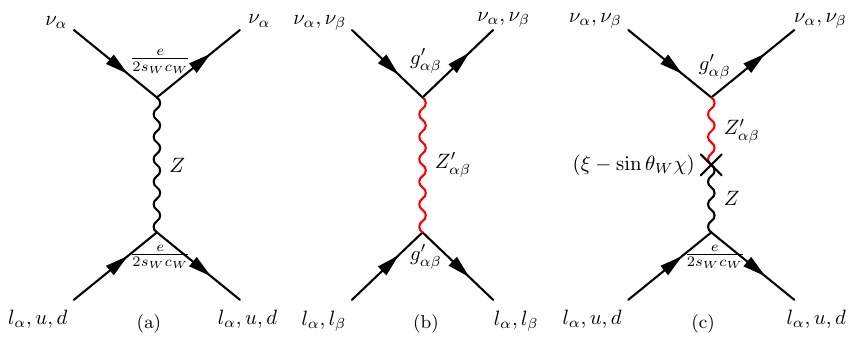}
\caption{\footnotesize{\textbf{\textit{Feynman diagrams for neutrino-matter interactions.}}  Each diagram corresponds to a term in the Lagrangian, \ref{equ:lagrangian_total}: (a) SM contribution mediated by the $Z$ boson, (b) new contribution from the gauge symmetry $U(1)_{L_\alpha-L_\beta}$, mediated by the new boson $Z_{\alpha\beta}^\prime$, and (c) mixing between $Z$ and $Z_{\alpha\beta}^\prime$.  In our analysis, (a) is significant only for neutrinos inside the Earth. For $L_{e} - L_\beta$ symmetries, (b) is the only additional contribution sourced by electrons.  For $L_\mu - L_\tau$, (c) is instead the only additional contribution sourced by neutrons.}}
\label{fig:feynman}
\end{figure}

The edifice of the meticulously constructed Standard Model (SM) is invariant under the Lorentz symmetry ($SO(3,1)$) and the gauge symmetry \textit{SU}$(3)_C\,\otimes\,$\textit{SU}$(2)_L\,\otimes\,$\textit{U}$(1)_Y$. Consequent conserved quantities or quantum numbers maintaining the above-mentioned gauge symmetries are color charge (conserved under $SU(3)$), weak isospin charge (conserved under $SU(2)$), and the hypercharge (conserved under $U(1)$). The last two conserved quantities also point towards the conservation of the electric charge via the Gell-Mann Nishijima Formula ($Q=I_3+\frac{Y}{2}$) ~\cite{Gell-Mann:1956iqa, Nishijima:1955gxk}. Along with these conserved quantities, there are some left-over conserved candidates that are protected by the accidental global $U(1)$ symmetry shown by the SM and those are the individual lepton numbers (electron number ($L_e$), muon number ($L_\mu$), tau number ($L_\tau$) and the total lepton number ($L=L_e+L_\mu+L_\tau$). This also unveils the fact that the difference of the individual lepton numbers is also conserved quantities, $i.e.,\,(L_e-L_\mu)$, $(L_e-L_\tau)$, and $(L_\mu-L_\tau)$ hinting towards the three new symmetries $U(1)_{L_e-L_\mu},\, U(1)_{L_e-L_\tau}, \,\mathrm{and} \,\, U(1)_{L_\mu-L_\tau}$, respectively. The reasons for highlighting these three symmetries are they are economic, simple, and anomaly-free without introducing new fermions or breaking the natural helicity of neutrinos and hence help to gauge the new interaction due to the differences of the individual lepton numbers (although there are other anomaly-free choices of this accidental global $U(1)^\prime$ symmetry \cite{Coloma:2020gfv}). These three newly introduced gauge groups ($i.e.,\, U(1)_{L_e-L_\mu},\, U(1)_{L_e-L_\tau}, \,\mathrm{and}$\\$ \, U(1)_{{L_\mu-L_\tau}}$\,), while being broken locally, give birth to three new neutral gauge vector bosons (or new mediators), $Z^\prime_{e\mu},\, Z^\prime_{e\tau}, \, \text{and} \, Z^\prime_{\mu\tau}$, respectively, and these mediators show new kind of weak interactions between neutrinos and leptons, apart from the standard neutral current (NC) interactions of neutrinos with leptons.
\par The governing Lagrangian density for a given lepton-number symmetry in our study can be depicted as follows.\\
\begin{equation}
 \mathscr{L}_{\rm gov}
=
 \mathscr{L}_{Z}
 +
 \mathscr{L}_{Z^\prime}
 +
 \mathscr{L}_{Z-Z^\prime} \;.
 \label{equ:lagrangian_total} 
\end{equation}
The first term of equation \ref{equ:lagrangian_total} represents the usual weak interaction of neutrinos with leptons as suggested by SM via NC interaction, mediated by the $Z$ boson, $i.e.$,
\begin{equation}
 \mathscr{L}_{Z}
 =
 \frac{e}{\sin \theta_{W} \cos \theta_{W}}Z_\mu 
 \left[-\frac{1}{2}\bar{l}_{\alpha}\gamma^\mu P_{L} l_{\alpha}+\frac{1}{2}\bar{\nu}_{\alpha}\gamma^\mu P_{L} \nu_{\alpha}+\frac{1}{2}\bar{u}\gamma^\mu P_{L} u-\frac{1}{2}\bar{d}\gamma^\mu P_{L} d
 \right] \;,
  \label{equ:lagrangian_sm}
\end{equation}
where $e$ represents the unit elctric charge, $\theta_W$ denotes the Weinberg angle($\sim 29^\circ$) \cite{Weinberg:1967tq,  Fairlie:1979at, Moroi:1993zj}, $\nu_\alpha$ and $l_\alpha$ are a neutrino and charged lepton of flavor $\alpha = e, \mu, \tau$, $P_{L}$ stands for the left-handed projection operator, and $u$, $d$ are up and down quarks, respectively.  Due to the massive  $Z$ boson ($m_Z\sim90$ GeV), the range of the interaction mediated by it is short; in our work, it covers only the distances inside the Earth. Two of the important assumptions at this point for our work are ordinary matter interacting with the neutrinos is electrically neutral ($i.e.$, the presence of equal weightage of electron and proton numbers) and isoscalar ($i.e.$, equal presence of protons and neutrons) except for the Sun \cite{Coloma:2020gfv} and the cosmological matter distributions \cite{McMillan:2011wd, Steigman:2007xt, Yuksel:2008cu}.
\par The second term in \ref{equ:lagrangian_total} tells the interaction between $\nu_\alpha$ and $l_\alpha$ mediated by the flavor-dependent new vector boson $Z_{\alpha\beta}^\prime$ ~\cite{Bandyopadhyay:2021gcn, He:1990pn, He:1991qd, Heeck:2010pg}, $i.e.$, for the $L_\alpha-L_\beta$ symmetry,
\begin{equation}
 \mathscr{L}_{Z^\prime}
 =
 g_{\alpha \beta}^\prime Z^\prime_{\sigma}(\bar{l}_{\alpha}\gamma^{\sigma}l_{\alpha}-\bar{l}_{\beta}\gamma^{\sigma}l_{\beta}+ \bar{\nu}_{\alpha}\gamma^{\sigma}P_{L}\nu_{\alpha}-\bar{\nu}_{\beta}\gamma^{\sigma}P_{L}\nu_{\beta}) \;.
 \label{equ:lagrangian_zprime}
\end{equation}
Here $g_{\alpha\beta}^\prime$ denotes the dimensionless coupling constant of this new interaction. In this case, we showcase the fact that the amount of naturally produced muon or tau in the matter is negligible for interacting with the neutrinos and hence, we only consider the interactions of neutrinos with only abundant electrons present in the matter giving birth the flavor-dependent matter potential. Hence, we neglect the contribution of this term under $L_\mu-L_\tau$ symmetry and consider it only under $L_e-L_\mu$ and $L_e-L_\tau$ symmetries.
\par The penultimate term of equation \ref{equ:lagrangian_total} articulates the hybridization between $Z$ and $Z_{\alpha\beta}^\prime$~\cite{Babu:1997st, Heeck:2010pg, Bhattacharyya:1991rx, Joshipura:2019qxz}, which can be produced naturally or by radiative correction~\cite{Holdom:1985ag, Tomalak:2020zfh}.  In the mass basis, this term is~\cite{Babu:1997st} $\mathscr{L}_{ZZ^\prime} \supset
 (\xi-\sin\theta_W\chi) Z'_{\mu}Z^{\mu}$,
where $\chi$ denotes the kinetic mixing angle between the two gauge bosons and $\xi$ embodies the amount of rotation between weak eigenstates and mass eigenstates. This induces a quartic fermionic interaction between neutrinos and charged leptons, protons, and neutrons via $Z$--$Z_{\alpha\beta}^\prime$ mixing, $i.e.$,
\begin{equation}
 \label{equ:lag_mix}
 \mathcal{L}_{\rm Z-Z^\prime}
 =
 -g_{\alpha\beta}^\prime
 (\xi-\sin\theta_W\chi)\frac{e}{\sin\theta_W \cos\theta_W}
 J'_\rho J_3^\rho \;,
\end{equation}
where $J^\prime_\rho = \bar{\nu}_\alpha \gamma_\rho P_L\nu_\alpha-\bar{\nu}_\beta \gamma_\rho P_L\nu_\beta$ and $J_3^\rho = -\frac{1}{2}\bar{e}\gamma^\rho P_L e+\frac{1}{2}\bar{u}\gamma^\rho P_L u-\frac{1}{2}\bar{d}\gamma^\rho P_L d$ are the conserved currents.
It is pertinent to emphasize that the contribution of electrons is exactly offset by that of protons, resulting in neutrons being the sole effective source for the new interaction arising through mixing between $Z$ and $Z^\prime$. The coefficient $(\xi - \sin\theta_{W}\chi)$ serves as a quantitative measure of the mixing strength between the $Z$ and $Z_{\alpha\beta}^\prime$. While its precise magnitude remains undetermined, there exist upper bounds on its value~\cite{Schlamminger:2007ht, Adelberger:2009zz, Heeck:2010pg}. We consider its value together with $g_{\alpha\beta}^{\prime}$\,, as an effective coupling. It is important to note that the contribution of this term is exclusive under the $L_\mu - L_\tau$ symmetry, and the corresponding matter potential is contributed by interaction with neutrons only.

 \subsection{Genesis of matter potential from the long-range interaction}
\label{sec:lri_matter_potential}
The backbone of this LRI-induced matter potential is nothing but the flavor-dependent Yukawa interaction. Hence, under $L_e-L_\beta$ ($\beta = \mu, \tau$) symmetry, if a neutrino is positioned at a distance $d$ from a cluster of $N_e$ electrons, it feels a matter potential given by
\begin{equation}
 V_{e\beta}
 =
 G_{\alpha\beta}^{\prime 2}
 \frac{N_e}{4\pi d}
 e^{-m'_{e\beta}d} \;.
 \label{equ:potential_e_beta}
\end{equation}

Here, $m_{e\beta}^\prime$ represents the mass of the neutral mediator $Z_{e\beta}^\prime$ boson. Conversely, in the context of $L_\mu-L_\tau$\, symmetry, a neutrino positioned at a distance $d$ from a collection of $N_n$ neutrons is subject to a potential.
\begin{equation}
 V_{\mu\tau}
 =
 G_{\alpha\beta}^{\prime 2}
  \frac{e}{\sin\theta_W\cos\theta_W}
 \frac{N_n}{4 \pi d}
 e^{-m'_{\mu\tau}d} \;,
 \label{equ:potential_mu_tau}
\end{equation}
where $m_{\mu\tau}^\prime$ denotes the mass of the mediator $Z_{\mu\tau}^\prime$ boson. In Eqs.~(\ref{equ:potential_e_beta}) and~(\ref{equ:potential_mu_tau}), the form of the effective flavor-dependent coupling constant can be expressed as,
\begin{equation}
 G_{\alpha \beta}^\prime
 =
 \left\{
  \begin{array}{lll}
   g^{\prime }_{e \mu} & , & ~{\rm for}~\alpha, \beta = e, \mu; \\
   g^{\prime }_{e \tau} & , & ~{\rm for}~\alpha, \beta = e, \tau; \\
   \sqrt{g^{\prime}_{\mu \tau} (\xi-\sin \theta_W \chi)} & , & ~{\rm for}~\alpha, \beta = \mu, \tau. \\
  \end{array}
 \right. \;.
 \label{equ:Gab}
\end{equation}
A significant observation is that this novel interaction potential demonstrates dependence on the spatial separation between the neutrinos and the matter source, signifying the presence of a central force, which is identified as LRI. Consequently, the conservative nature of this force is implied. Typically, the mass scale of the mediator for such interactions spans from $10^{-35}$ to $10^{-10}$~eV, affirming that the interaction range, characterized by $r \propto 1/m^\prime_{\alpha\beta}$, is indeed long-ranged — extending from km to Gpc. In our analysis, we have gradually varied the mass of the mediating bosons within the range of $10^{-15}$~eV to $10^{-13}$~eV to explore its imprints on several aspects of neutrino oscillation phenomenology. This long-range interaction potential originates from electron and neutron distributions across various astrophysical scales, including the Earth matter ($\oplus$), lunar effects ($\leftmoon$), solar contributions ($\astrosun$), the galactic matter profile of the Milky Way (MW), and the cosmological matter distribution (cos) in the local Universe, $i.e.$,
\begin{equation}
 V_{\alpha \beta}
 =
 V_{\alpha \beta}^\oplus + V_{\alpha \beta}^{\leftmoon} + V_{\alpha \beta}^{\astrosun} + V_{\alpha \beta}^{\rm MW} +  V_{\alpha \beta}^{\rm cos} \;.
 \label{equ:pot_total}
\end{equation} 
We assess the resultant value of $m_{\alpha\beta}^\prime$ by incorporating the aggregate contributions imparted by all the aforementioned sources, thereby determining the net matter potential that governs our analysis. In this context, we evaluate the mean potential encountered by neutrinos at their detection site \cite{Bustamante:2018mzu} while disregarding the influence of any fluctuations in the matter potential along their subterranean passages \cite{Coloma:2020gfv}. This approximation holds particularly well for mediators with masses below approximately $10^{-14}$~eV, where the interaction range exceeds the Earth's radius. In this regime, the entire population of electrons and neutrons on Earth contributes uniformly to the potential sensed by a neutrino, irrespective of its location along its course.
\par We envisage the Moon ($N_{e,\leftmoon} = N_{n,\leftmoon} \sim 5 \cdot 10^{49}$) and the Sun ($N_{e,\astrosun} \sim 10^{57}$, $N_{n,\astrosun} = N_{e,\astrosun}/4$) as point sources of electrons and neutrons. In contrast, we treat the Earth ($N_{e, \oplus} \approx N_{n, \oplus} \sim 4 \times 10^{51}$), the Milky Way ($N_{e, {\rm MW}} \approx N_{n, {\rm MW}} \sim 10^{67}$), and the cosmological matter ($N_{e,\mathrm{cos}} \sim 10^{79}$, $N_{n,\mathrm{cos}} \sim 10^{78}$) as extended objects with continuous mass distributions. For a detailed computation of \ref{equ:pot_total}, we refer to \cite{Bustamante:2018mzu}, but we adopt one contrasting aspect introduced by \cite{Agarwalla:2023sng}.  In \cite{Bustamante:2018mzu}, only the potential is shown to be generated by electrons for $L_e-L_\mu$ and $L_e-L_\tau$ symmetries, we additionally compute the potential engendered by neutrons for $L_\mu-L_\tau$ symmetry also.

\subsection{Analyzing the neutrino oscillation probability in the presence of long-range interactions}
We proceed with our discourse in this section on the texture of the well-established $3\nu$ paradigm. For the $L_\alpha-L_\beta$ symmetry in Nature, the governing Hamiltonian of the neutrinos while traveling through a medium, in the weak basis, can be written as
\begin{equation}
 \label{equ:hamiltonian_tot}
 \mathbf{H}_{\rm gov} = \mathbf{H}_{\rm free} + \mathbf{V}_{\rm matt} + \mathbf{V}_{\alpha\beta} \;.
\end{equation} 
The first two candidates on the right-hand side of this equation showcase the neutrino oscillations in vacuum and through the standard interactions with matter and in vacuum, whereas the third piece elucidates oscillations due to the new long-range interactions.  
\par 
The standard form of $\mathbf{H}_{\textrm{free}}$, representing the conventional Hamiltonian for neutrino oscillations in a vacuum, can be expressed as follows,
\begin{equation}
 \mathbf{H}_{\rm free} = \frac{1}{2 E} \mathbf{U}~{\rm diag}(0, \Delta m^2_{21}, \Delta m^2_{31})~\mathbf{U}^{\dagger} \;,
 \label{equ:total_hamiltonian}
\end{equation}
where $E$ denotes the true neutrino energy, $\Delta m^2_{ij} \equiv m^2_i-m^2_j$ are the mass-squared differences between two neutrino mass eigenstates, and $\mathbf{U}$ represents the Pontecorvo-Maki-Nakagawa-Sakata (PMNS) matrix~\cite{Maki:1962mu, Pontecorvo:1967fh, ParticleDataGroup:2022pth}. In our whole analysis, we consider the normal mass ordering (NMO) in Nature ($i.e.$, $\Delta m^2_{31}>0$), and the choices of the benchmark values of the neutrino oscillation parameters are shown in the table~\ref{tab:oscillation-params}.
\par The second term of equation \ref{equ:total_hamiltonian} tells us the standard charged-current (CC) interaction of neutrinos with the ambient electrons of earth matter via $W^\pm$ boson of the SM and its form can be written in flavor basis as,
\begin{equation}
 \mathbf{V}_{\rm matt}
 =
 {\rm diag}(V_{\rm CC}, 0, 0) \;,
\end{equation}
where $V_{\rm CC} = \sqrt{2} G_F n_e \simeq 7.6\,\times Y_e\,\times[ \rho / (10^{-14}~{\rm g}~{\rm cm}^{-3})]$~eV is matter potential arising from CC interactions, $G_F$ represents the Fermi coupling constant, governing the weak interaction strength. The quantity $n_e$ denotes the density of electron numbers, $Y_e \equiv n_e / (n_p + n_n)$ characterizes the fractional contribution of ambient electrons, encapsulating its relative prevalence compared to protons and neutrons, $n_p$ and $n_n$, and $\rho$ corresponds to the mass density, averaged along the line of sight, serving as a crucial factor in neutrino-matter interactions. This matter potential changes its sign while considering the interactions with antineutrinos. In our investigation, we fix the value of $\rho$ as 2.848~g/$\mathrm{cm}^3$ for DUNE and 2.8~g/$\mathrm{cm}^3$ for T2HK experiment referring to the Preliminary Reference Earth Model (PREM profile)~\cite{Dziewonski:1981xy}. 
\par The penultimate term in equation \ref{equ:total_hamiltonian} characterizes the matter potential arising from the novel interaction governed by the $L_\alpha-L_\beta$ symmetry and is formulated as follows,
\begin{equation}
 \label{equ:pot_lri_matrix}
 \mathbf{V}_{\alpha\beta}
 =
 \left\{
  \begin{array}{ll}
   {\rm diag}(V_{e\mu}, -V_{e\mu}, 0), & {\rm for}~ \alpha, \beta = e, \mu; \\
   {\rm diag}(V_{e\tau}, 0, -V_{e\tau}), & {\rm for}~ \alpha, \beta = e, \tau; \\
   {\rm diag}(0, V_{\mu\tau}, -V_{\mu\tau}), & {\rm for}~ \alpha, \beta = \mu, \tau. \\   
  \end{array}
 \right. \;.
\end{equation}
There are two control parameters of this new kind of matter potential
$V_{\alpha\beta}$ , as referred in equation~\ref{equ:pot_total} --- they are coupling constant $G_{\alpha\beta}^\prime$ focusing the strength of the interaction and the mediator mass $m_{\alpha\beta}^\prime$ specifying the range of this interaction. Like the ordinary matter potential ($i.e.$, $\mathbf{V}_{\rm matt}$), the new matter potential $V_{\alpha\beta}$ also shows the property of flipping its polarity for the case of the interactions of matter with antineutrinos.
\par Upon incorporating the modified matter potential into the formalism of \ref{equ:total_hamiltonian},  the resultant theoretical formulation governing the oscillation probability (in $3\nu$ paradigm) of a neutrino of flavor $\alpha$ to a flavor $\beta$ after covering a distance $L$ km with a true energy $E$ GeV can be expressed as, 
\begin{equation}
 P_{\nu_{\alpha}\rightarrow \nu_{\beta}}
 =
 \Bigg| 
 \sum_{i=1}^{3} 
 U^\prime_{\alpha i}
 \exp \left( \dfrac{\Delta \tilde{m}^2_{i1}L}{2E} \right)
 U^{\prime \ast}_{\beta i}
 \Bigg|^{2} \;,
 \label{equ:osc_prob}
\end{equation}
where $\Delta \tilde{m}^2_{ij} \equiv \tilde{m}_i^2 - \tilde{m}_j^2$, with $\tilde{m}_i^2/2E$ represent the eigenvalues of the effective Hamiltonian $\mathbf{H}_{\rm gov}$ \ref{equ:hamiltonian_tot}. $\mathbf{U}^\prime$ is the unitary matrix that diagonalizes the Hamiltonian. To ensure a parametrization of $\mathbf{U}^\prime$ that retains the structural form of the PMNS matrix, the mixing parameters must be redefined in the presence of the effective matter potential $V_{\alpha \beta}$ \ref{equ:pot_lri_matrix} from Eq. \ref{equ:pot_lri_matrix}, leading to the modified parameters $\theta_{12}^m$, $\theta_{23}^m$, $\theta_{13}^m$, and $\delta^m_{\mathrm{CP}}$. The exact expressions for these matter-modified oscillation parameters are adopted from the reference~\cite{Agarwalla:2021zfr}.
\par We consider well-motivated selections of the new matter potential values, ensuring a meaningful comparison with the standard neutrino oscillations in vacuum, thereby allowing it to have a significant impact on our oscillation-based analysis. Since the sensitivity of any oscillation study is predominantly driven by the first oscillation maximum — where the neutrino flux is at its peak (located at 2.5 GeV for DUNE and 0.6 GeV for T2HK) — the characteristic value of the frequency term of the neutrino oscillation is  $\frac{\Delta m^2_{31}}{2E}\sim\mathcal{O}(10^{-13}\, \mathrm{eV})$, which is comparable to the magnitude of the effective matter potential $V_{\alpha\beta}$. Consequently, we set the upper bound of our benchmark choices for the LRI potential approximately around $10^{-13}$ eV. For a detailed listing of the benchmark values of the LRI potential corresponding to the three considered symmetries, refer to Table 3 in \cite{Singh:2023nek}.

\label{sec:lri_osc_prob}

\section{Effect of long-range interactions on total event rates}
\label{sec:effect-on-total-event-rates}

\subsection{Experimental details of DUNE and T2HK}
\label{subsec:exp-details}
Long-baseline neutrino experiments are at the forefront of exploring some of the most intriguing questions in neutrino physics~\cite{Feldman:2012jdx, Agarwalla:2014fva, Diwan:2016gmz, Giganti:2017fhf}. By using powerful, precisely tuned beams of neutrinos over distances of hundreds to thousands of kilometers, these experiments can study how neutrinos change as they travel through matter. This offers a unique opportunity to investigate phenomena like CP violation, the order of neutrino masses, and the influence of matter on oscillations. Furthermore, they also provide a vast window into potential physics beyond the Standard Model~\cite{Mena:2005ek, Feldman:2012jdx, Agarwalla:2014fva, Diwan:2016gmz, Farzan:2017xzy, Giganti:2017fhf}. Current experiments like T2K~\cite{T2K:2023smv} and NO$\nu$A~\cite{NOvA:2021nfi} have already provided valuable data for understanding these processes. In the coming decade, new experiments such as DUNE~\cite{DUNE:2015lol, DUNE:2020lwj, DUNE:2020ypp, DUNE:2020jqi, DUNE:2021cuw, DUNE:2021mtg}, T2HK~\cite{Hyper-Kamiokande:2016srs, Hyper-Kamiokande:2018ofw}, and the European Spallation Source neutrino Super Beam (ESS$\nu$SB)~\cite{Blennow:2019bvl, ESSnuSB:2021azq}, promise to take this research even further~\cite{Mena:2005ek, Feldman:2012jdx, Agarwalla:2014fva, Diwan:2016gmz, Farzan:2017xzy, Giganti:2017fhf}, offering the potential for groundbreaking discoveries. 

At Fermilab, the Long Baseline Neutrino Facility (LBNF) generates the neutrino beam for DUNE. This is achieved by the Main Injector, which propels an on-axis 1.2 MW proton beam with an energy of 120 GeV onto a graphite target. This collision produces charged mesons that decay in flight, creating a wide-band neutrino flux. The flux ranges from a few hundred MeV to several tens of GeV, peaking at 2.5 GeV. We generate statistics in the 0.5 - 18 GeV range. The planned run time is 5 years for both neutrino and antineutrino modes, resulting in approximately $1.1 \times 10^{21}$ protons on target (P.O.T.) per year and a total exposure of 480 kton MW year. DUNE incorporates two key detectors: a near detector and a far detector. The near detector, positioned 600 m downstream from the neutrino production point at Fermilab, is designed to monitor and characterize the neutrino beam, while also possessing notable physics capabilities. The far detector, located 1285 kilometers away and 1.5 kilometers underground at the Sanford Underground Research Facility in South Dakota, is central to our analysis due to its exceptional sensitivity to neutrino oscillations. This detector, a liquid argon time projection chamber with a net volume of 40 kt, is considered for single-phase detection only in our study. Neutrino detection occurs via both neutral-current (NC) and charged-current (CC) neutrino interactions. While the deployment of the detector will be phased~\cite{DUNE:2020jqi,DUNE:2021mtg}, our simulations account for the total final detector volume. Our results are based on the DUNE simulation configuration from Ref.~\cite{DUNE:2021cuw}.

T2HK will use the 2.5$^\circ$ off-axis JPARC neutrino beam, similar to its predecessor, T2K (Tokai-to-Kamioka). JPARC generates the beam by firing a 1.3 MW proton beam at 30 GeV onto a graphite target~\cite{McDonald:2001mc}. This process creates a narrow-band neutrino flux, ranging from a few MeV to a few GeV, and peaking at 600 MeV, with most neutrinos within the 0.1 - 3 GeV range. According to Ref.~\cite{Hyper-Kamiokande:2016srs}, the planned run time is 2.5 years in neutrino mode and 7.5 years in antineutrino mode, resulting in approximately $2.7 \times 10^{22}$ P.O.T. and a total exposure of 2431 kton MW year. T2HK comprises both near and far detectors. The near detectors are located approximately 280 m downstream from the neutrino production point at the Japan Proton Accelerator Research Complex (JPARC) and are tasked with monitoring and characterizing the neutrino beam. The far detector, located 295 km away and 1.7 km underground in the Tochibora mines of Japan is our primary focus due to its high sensitivity. This detector consists of a tank filled with purified water with a net volume of 187 kton. Neutrino detection occurs via quasielastic charged-current scattering and charged-current deep inelastic scattering. Our results are based on the T2HK simulation configuration from Ref.~\cite{Hyper-Kamiokande:2016srs}.

\begin{figure}[htb!]
	\centering
	\includegraphics[width=\linewidth]{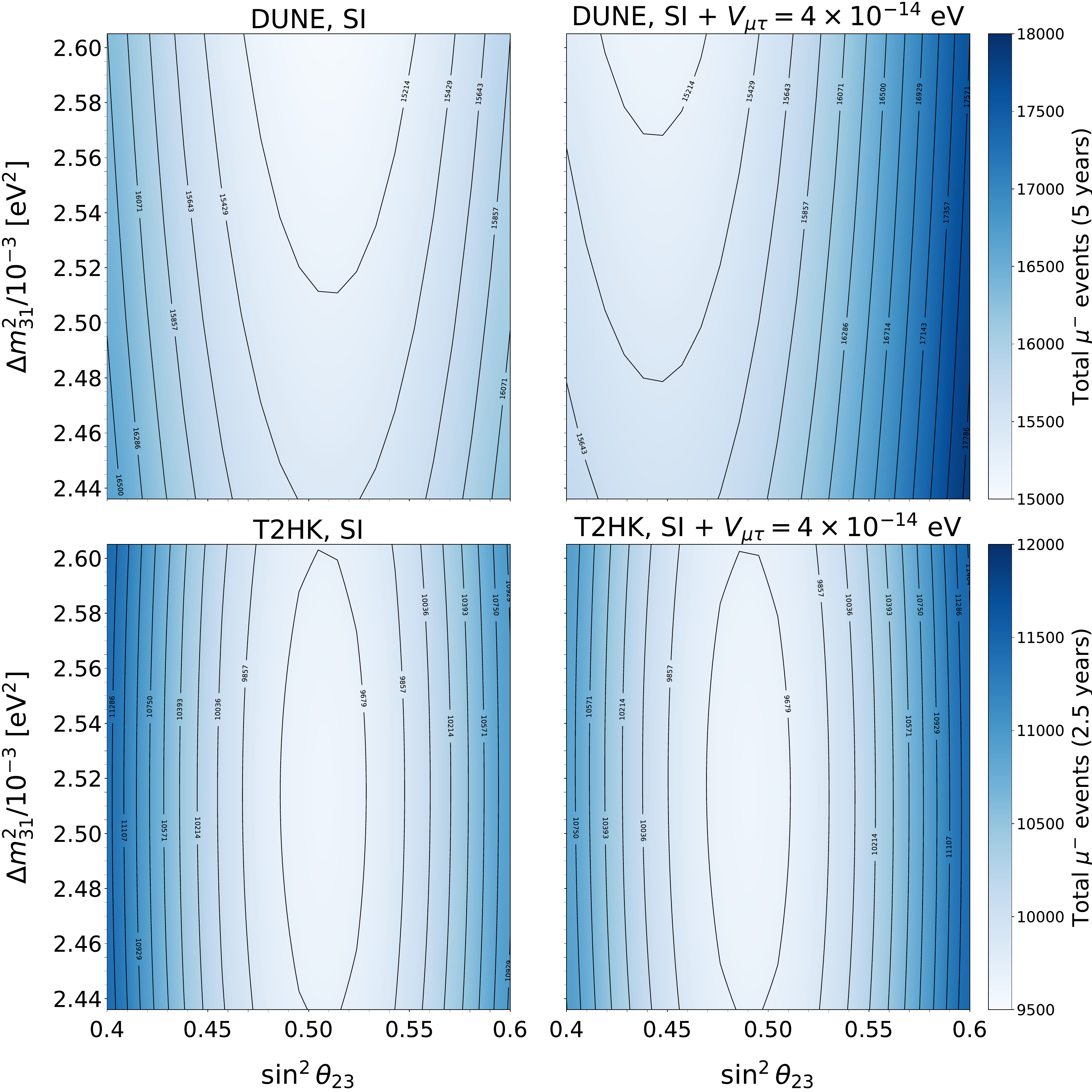}

	\caption {\footnotesize{Total $\nu_\mu\rightarrow\nu_\mu$ disappearance event rates (signal + background) as a function of $\sin^2\theta_{23}$ and $\Delta m^2_{31}$ using DUNE and T2HK considering standard interaction (left panel) and in presence of $L_{\mu}-L_{\tau}$ (right panel) assuming $V_{\mu\tau} = 4 \times 10^{-14}$~eV illustratively in the top and bottom panels, respectively. The benchmark values correspond to Table~\ref{tab:oscillation-params}, assuming NMO.}}
   
 \label{fig:Oscillogram_disapp_nu}
 \end{figure}

 \begin{figure}[htb!]
	\centering
	\includegraphics[width=\linewidth]{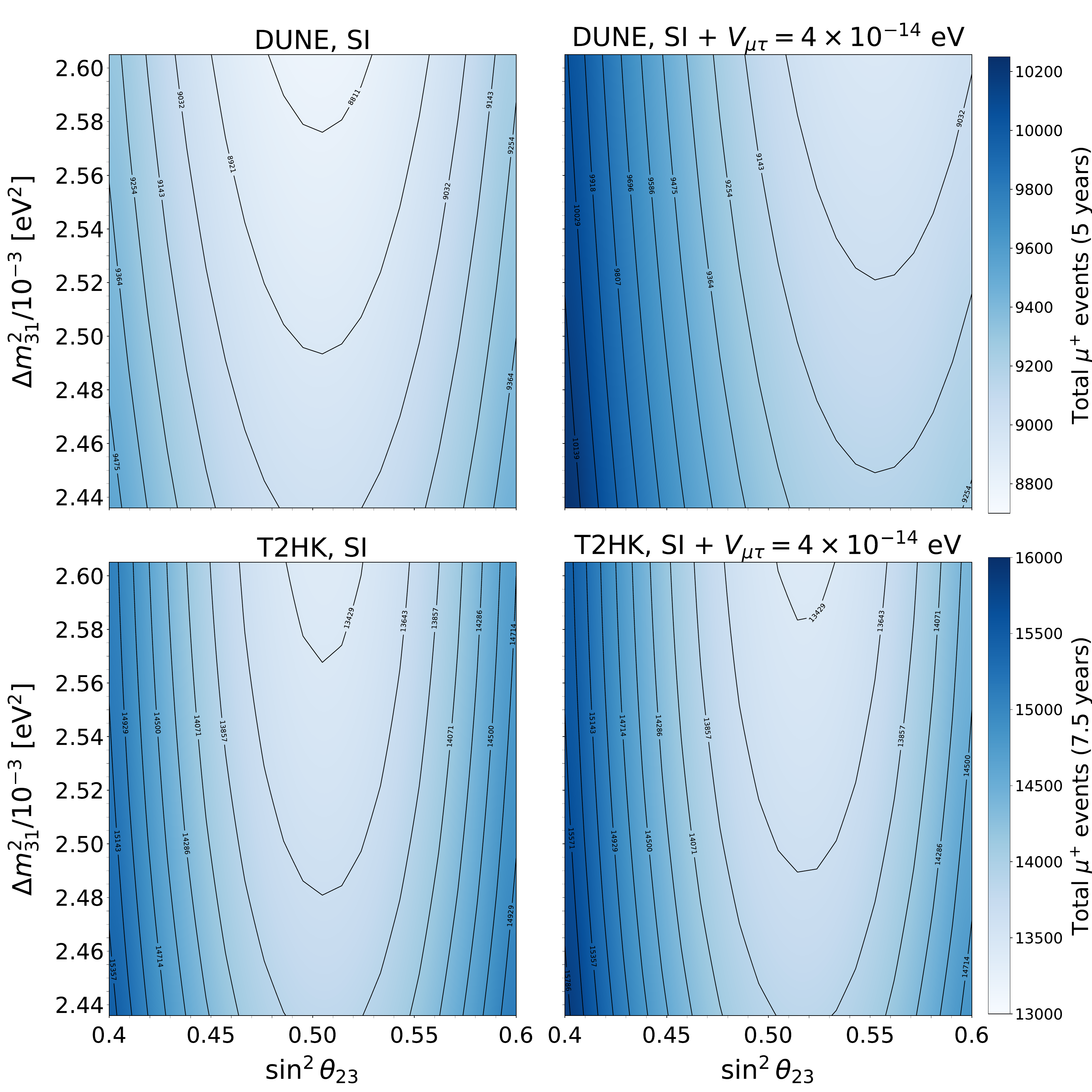}

	\caption{\footnotesize{Total $\bar{\nu}_\mu\rightarrow\bar{\nu}_\mu$ disappearance events (signal + background) of DUNE (T2HK) in upper (lower) panel in $\sin^2\theta_{23}-\Delta m^2_{31}$ plane considering standard interaction (left panel) and in presence of $L_{\mu}-L_{\tau}$ (right panel) assuming $V_{\mu\tau} = 4 \times 10^{-14}$ eV illustratively in the top and bottom panels, respectively. The benchmark values correspond to Table~\ref{tab:oscillation-params}, assuming NMO.}}
 \label{fig:Oscillogram_disapp_anti}
 \end{figure}

\subsection{Effect on disappearance statistics}
\label{subsec:disapp-stats}

In this section, we study the effect of the presence of LRI under $L_{\mu}-L_{\tau}$ symmetry in disappearance statistics. For completion, we show the effect under $L_{e}-L_{\mu}$ and $ L_{e} - L_{\tau}$ symmetries in Appendix... For studying the effect in disappearance statistics, we generate total (signal+background) event rates as a function of atmospheric mixing angle, $\sin^{2}\theta_{23}$ and mass-squared splitting, $\Delta m^{2}_{31}$. Illustratively, we depict here the effects assuming the strength of $V_{\mu\tau} = 4 \times 10^{-14}$ eV. The choice of this value is explained in Sec.~\ref{subsec:simulation}. 

We expect the rates to decrease around the maximal mixing solutions ($\sin^2\theta_{23} = 0.5$) as they are $\propto (1 - \sin^2 \theta_{23})$~\cite{Agarwalla:2021bzs, Agarwalla:2024kti}. This trend is consistent in the presence of LRI. This U-shaped nature under SI does not have its minimum exactly at $\sin^2\theta_{23} = 0.5$ due to $\theta_{13}$ corrections; however, the presence of $L_{\mu}-L_{\tau}$ symmetry shifts this minimum to $\sin^2\theta_{23} \sim \text{LO}$. We checked that depending upon the strength of LRI potential, this minima shifts. This can be explained by examining the evolution of the $\theta_{13}$ parameter in the presence of matter (referred to as $\theta^{\rm m}_{13}$) under SI and  $V_{\mu\tau}$ [we can refer to fig if we plan to give or we can refer to our previous works in the literature]. In the presence of LRI, $\theta^{\rm m}_{13}$ achieves a higher value much more rapidly than under SI. Furthermore, we observe that in the higher octant (HO) ($\sin^2\theta_{23} > 0.5$), neutrino event rates are higher with LRI compared to SI. For a given value of $\Delta m^{2}_{31}$, the increase in the event rates on either side of the maximal mixing (MM) solutions of $\sin^2\theta_{23}$ occurs more rapidly in the presence of LRI than in the presence of SI. In the case of antineutrinos, for any combination of $\sin^2\theta_{23}$ and $ \Delta m^{2}_{31}$, there are significantly higher event rates in the presence of LRI than SI. This can be explained by examining the evolution of the $\theta_{23}$ parameter in the presence of matter (denoted as $\theta^{\rm m}_{23}$), which attains a higher value at lower energies in the presence of LRI than SI. 

\subsection{Effect on appearance statistics}
\label{subsec:app-stats}

\begin{figure}[htb!]
	\centering 
 \includegraphics[width=\linewidth]{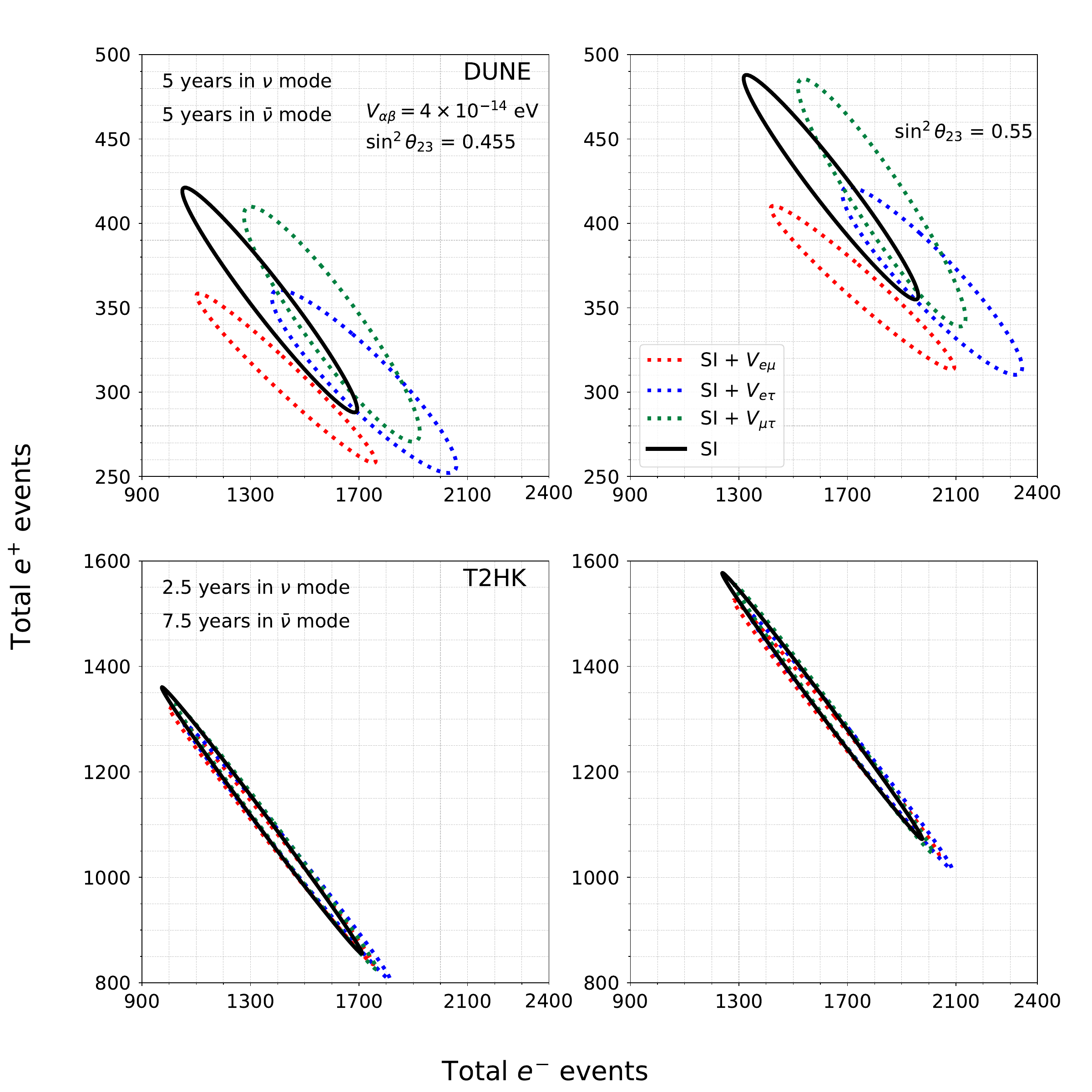}
	\caption{\footnotesize{The top (bottom) panel depicts the bi-events in the $\nu_\mu\rightarrow\nu_e$ and $\bar{\nu}_\mu\rightarrow\bar{\nu}_e$ appearance channel in DUNE (T2HK). Red, blue, and green curves are for $L_e-L_\mu$, $L_e-L_\tau$, and $L_\mu-L_\tau$ symmetries, respectively, assuming $V_{\alpha\beta} = 4 \times 10^{-14}$ eV, whereas the black curve depicts the bi-events in the standard interaction case.  Left and right panels represent two different values of $\sin^{2}\theta_{23}$ = 0.455 and 0.55, respectively. We vary $\delta_{\mathrm{CP}}$ from $-180^\circ$ to $+180^\circ$ for the appearance bi-events. The benchmark oscillation parameters are from Table \ref{tab:oscillation-params}. Please note different y-ranges in the top and bottom panels.}} 
	\label{fig:App_Event_spectra}
\end{figure}

Bi-probability in appearance channel is a crucial pictorial representation of three major effects: intrinsic or genuine CP-violating phase $\delta_{\mathrm{CP}}$\,, induced due to the presence of $\sin \delta_{\mathrm{CP}}$ term; a CP-conserving effect due to the presence of $\cos\delta_{\mathrm{CP}}$ and the matter effect. CP-trajectory forms an ellipse and this shape has been previously 
justified~\cite{Arafune:1996bt, Minakata:1997td, Minakata:1998bf,
 Tanimoto:1996ky, Tanimoto:1996by, Bilenky:1997dd, Yasuda:1999uv, Koike:1999hf}. In Fig.~\ref{fig:App_Event_spectra}\, we extend this discussion in terms of appearance event rates under $L_{e}-L_{\mu}$, $ L_{e} - L_{\tau}$, and $L_{\mu}-L_{\tau}$ symmetries and the SI. For elaboration, we depict these bievents for two choices of $\sin^2\theta_{23}$, one in the lower octant (LO) (left panel), which is also the benchmark value in Table~\ref{tab:oscillation-params}, and other in HO corresponding to $\sin^2\theta_{23} = 0.55$ (right panel). The solid and dotted curves represent the bi-events in the presence of only SI and SI+LRI, respectively. These are generated for some illustrative choice, $V_{e\mu}$ =  $V_{e\tau}$ = $V_{\mu\tau}$ = $4.0\times10^{-14}$ eV. The choice of the strength of these potentials is elaborated in Sec.~\ref{subsec:simulation}.

As discussed in disappearance statistics, the standard appearance rates also exhibit greater deviations due to LRI under DUNE compared to T2HK. In the presence of all the three symmetries, $L_{e}-L_{\mu}$, $ L_{e} - L_{\tau}$, and $L_{\mu}-L_{\tau}$ under DUNE, the maximum achievable neutrino event statistics surpass those of SI. This can be explained from the fact that in the presence of LRI, the probability of $\nu_{\mu} \rightarrow \nu_{e}$ attains oscillation maximum at lower energies than the SI~\cite{Singh:2023nek, Agarwalla:2024ylc}. The appearance statistics is proportional to $\sin^2\theta_{23}\cdot\sin^22\theta_{13}$. Upon examining the evolution of the parameters $\theta_{23}^{\mathrm{m}}$ and $\theta_{13}^{\mathrm{m}}$  we observe that they attain higher values at lower energies compared to SI. However, the highest attainable antineutrino event statistics occur under SI, rather than in the presence of any of the considered LRI symmetries. This nature of ellipses is consistent with the choice of the octant of $\sin^2 \theta_{23}$. As expected, the overall appearance statistics are higher in HO compared to LO.

\section{Projected sensitivities in the presence of long-range interactions}
\label{sec:results}

In the subsequent subsections, we give a detailed description of the numerical analysis we employ to estimate our results. We forecast the effect of long-range interaction under $(L_{e}-L_{\mu})$, $(L_{e} - L_{\tau})$, and $(L_{\mu}-L_{\tau})$ symmetries on the projected sensitivity in establishing: CP coverage, deviation from maximal mixing, and exclusion of wrong octant of $\sin^2\theta_{23}$. We show our results in two ways: (i) assuming a fixed 3$\sigma$ discoverable value of $V_{\alpha\beta}$ corresponding to each symmetry and then studying its effect in the above-mentioned physics potentials. (ii) depicting the fraction of sensitivity in achieving the corresponding physics potential at a designated confidence level as a function of $V_{\alpha\beta}$. 

\subsection{Simulation Details}
\label{subsec:simulation}

In this subsection, we give a detailed analysis of the statistical treatment that we adopt to quantify our achievable constraints from the upcoming long-baseline experiments DUNE and T2HK. To perform all our simulations, we make use of publicly available GLoBES software~\cite{Huber:2004ka,Huber:2007ji}. Further, extended snu library~\cite{Kopp:2006wp,Kopp:2007ne} is used to include the effect of long-range interactions. We generate our data using the benchmark values of oscillation parameters from Table~\ref{tab:oscillation-params} following Ref.~\cite{Capozzi:2021fjo}, assuming NMO. In the fit, we minimize over the consecutive mentioned 3$\sigma$ ranges in $\sin^{2}\theta_{23}\,, \delta_{\mathrm{CP}\,,}$ and $ \Delta m^{2}_{31}$ (refer to Table~\ref{tab:oscillation-params}). Furthermore, we also vary $V_{\alpha\beta}$ in the range $ [1\times10^{-15}, 3 \times 10^{-14}]$ eV in different phenomenological aspects addressed in the subsection \ref{subsec:DM-fixed-potential}. This range is chosen to encompass the region where measurable impacts on neutrino oscillation phenomena could arise in experiments like DUNE and T2HK. The selected interval provides a robust foundation for placing constraints or making statistically significant measurements, ensuring the analysis remains sensitive to any potential deviations introduced by the long-range potential as explored previously in~\cite{Singh:2023nek}. For an elaborate checking we extend this range to $V_{\alpha\beta} \in [1\times10^{-15}: 1\times 10^{-13}]$ eV in the different phenomenological aspects discussed in the subsection \ref{subsec:frac-DM} for the test statistic, where the true value of LRI potential strength is not fixed while generating data, but instead varied within the range $V_{\alpha\beta} \in [5 \times 10^{-14}: 5\times 10^{-14}]$ eV. We keep $\theta_{13}$ fixed at the benchmark value as Daya Bay has already achieved an unprecedented precision of 2.8\%~\cite{DayaBay:2022orm}. We do not vary solar parameters as we do not expect them to affect our results. We do not minimize over the wrong mass ordering. We only show the effect assuming NMO in both data and fit. We expect the upcoming long-baseline experiment DUNE to fix the mass ordering issue in their earlier years of data-taking~\cite{DUNE:2020jqi}. 
Following the detailed discussions in references~\cite{Baker:1983tu,
Cowan:2010js,Blennow:2013oma}, we use the Poissonian $\chi^{2}$ function as:
\begin{equation}
\chi^{2} = \underset{\xi_{s}, \, \xi_{b}}{\mathrm{min}} \left[ 2\sum^{n}_{i=1}(\Tilde{y_{i}} - x_{i} - x_{i}ln\frac{\Tilde{y}_{i}}{x_{i}}) + \xi^{2}_{s} + \xi^{2}_{b} \right]\,,
\label{eq:poisonian-eq}
\end{equation}
where $n$ gives the total reconstructed energy bins, and $\xi_{s}$ and $\xi_{b}$ are the pull parameters on signal and background, respectively. True-statistic enters the above Eq.~\ref{eq:poisonian-eq} through $x_{i}$, where $x_{i} = N^{\mathrm{ex}}_{i} + N^{b}_{i}$. Here, $N^{\mathrm{ex}}_{i}$ denotes the predicted true event rates in the $i$-th bin, and $N^{b}_{i}$ provides the corresponding background event rates. The value of $\Tilde{y}_{i}$ comes from the predicted set of test event rates (denoted by $N^{th}_{i}\{\omega\}$) for the $i-$th energy bin. Both $x_{i}$ 
and $\Tilde{y}_{i}$ are a function of $\omega$, which is the set of 
oscillation parameters (elaborated later). 
The test-statistic $\Tilde{y}_{i}$ is computed as,
\begin{equation}
\Tilde{y}_{i}(\{\omega\}, \{\xi_{s}, \xi_{b}\}) = 
N^{th}_{i} (\{\omega\})\left[1+\pi^{s}\xi_{s}\right] + 
N^{b}_{i} (\{\omega\})\left[1+\pi^{b}\xi_{b}\right]\,,
\end{equation}
where $N^{b}_{i}(\{\omega\})$ depicts the total number of background events, dependent upon systematic uncertainty $\pi^{b}$. In principle, an experiment can have a number of detector systematic uncertainties contributing to background, for a given channel probed as signal. However, for simplicity we stick to one uncertainty in the background, corresponding to each signal: for appearance events, we assume 5\% systematic uncertainties in the background, while for disappearance events, we consider 10\%. $\pi^{s}$ quantifies the systematic errors on signal events. We consider 2\% and 5\% systematic uncertainties on appearance signal events and 5\% and 3.5\% on disappearance signal events, in both neutrino and antineutrino modes in in DUNE and T2HK, respectively. However, these uncertainties in neutrino and antineutrino modes are fully uncorrelated.

For obtaining the full $\chi^{2}$, we sum over the $\chi^{2}$ contributions from individual oscillation channels in both neutrino and antineutrino modes as described below:
\begin{equation}
\chi^{2}_{\mathrm{total}} = 
\chi^{2}_{\nu_{\mu} \rightarrow{ \nu_{e}}} + \chi^{2}_{\bar{\nu}_{\mu} 
\rightarrow{\bar{\nu}_{e}}} + \chi^{2}_{\nu_{\mu} \rightarrow{ \nu_{\mu}}} + 
\chi^{2}_{\bar{\nu}_{\mu} \rightarrow{\bar{\nu}_{\mu}}}\,.
\end{equation}
In the above expression, oscillation channels and the systematic 
uncertainties on both signal and backgrounds are fully uncorrelated, 
whereas in a given channel, all the energy bins are fully correlated. 
Further, while minimizing the $\chi^{2}_{\mathrm{total}}$\,, 
first it is minimized with respect to the pull variables ($\xi_{s}$ and $\xi_{b}$). This is then further minimized over a subset of all oscillation parameters in their defined allowed ranges in the theory (refer to Table~\ref{tab:oscillation-params} for allowed ranges). 

\subsection{Influence of fixed LRI in data}

\subsubsection{Impact on deviation from maximal $\sin^{2}\theta_{23}$ }
\label{subsec:DM-fixed-potential}
\begin{figure}[htb!]
	\centering 
	\includegraphics[width=\linewidth]{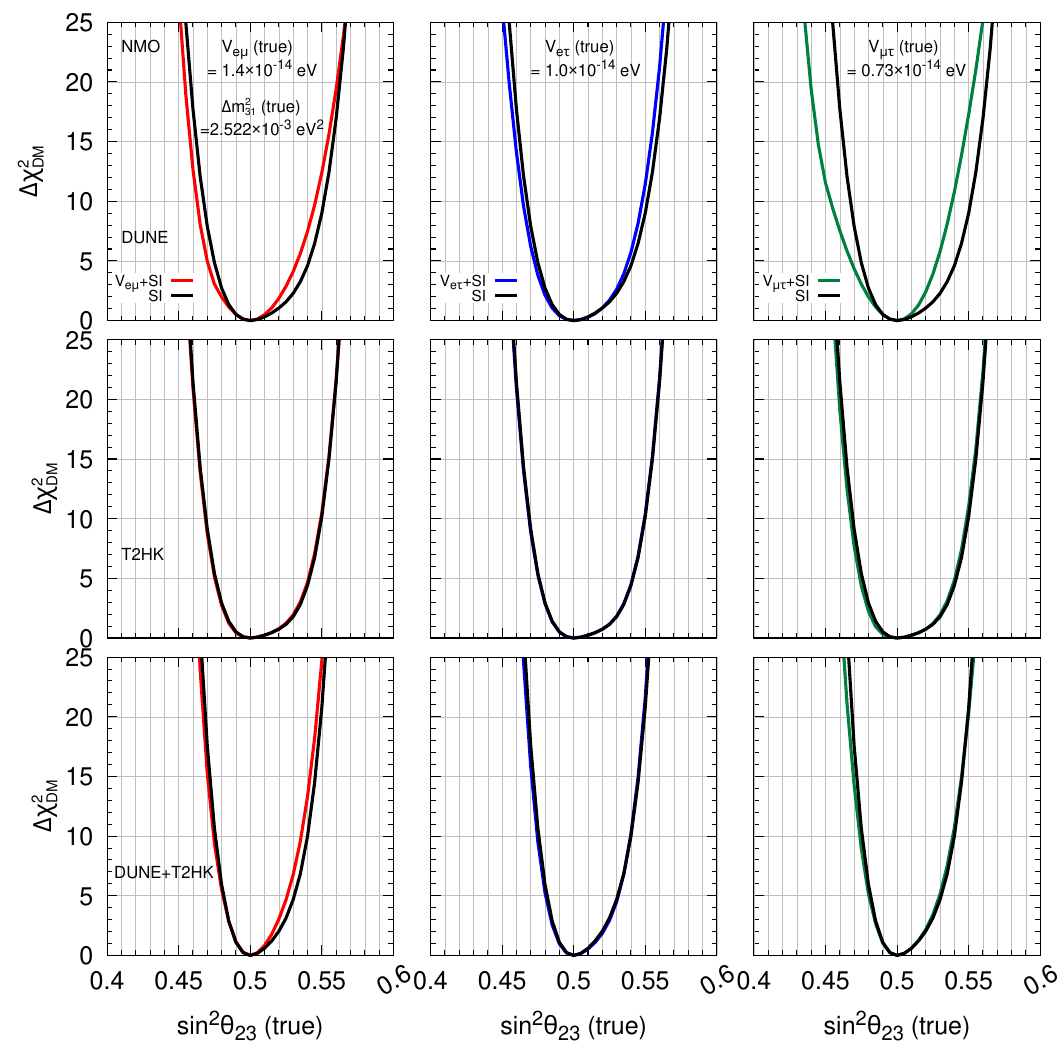}
	\caption{\footnotesize{Effect of $L_e-L_\mu$ (left panel), $L_e-L_\tau$ (middle panel), and $L_\mu-L_\tau$ (right panel) symmetries on the sensitivity to deviations from maximal mixing using DUNE (top), T2HK (middle), and their combination DUNE + T2HK (bottom) under NMO.  The black and colored curves represent the absence and presence of LRI in SI with a fixed potential strength in data, using benchmark oscillation parameters and minimization ranges for $\delta_{\mathrm{CP}}$, $\Delta m^2_{31}$, and $V_{\alpha\beta}$ from Table~\ref{tab:oscillation-params}. \textit{The sensitivity decreases in the presence of LRI compared to SI when data is generated with LO but increases with HO, with DUNE exhibiting the most significant effect due to its large matter influence. As the study is predominantly driven by disappearance statistics, $L_{\mu}-L_{\tau}$ symmetry impacts the sensitivity the most.}}
     } 
	\label{fig:DM_LRI}
\end{figure}

Fig. \ref{fig:DM_LRI} illustrates the sensitivity of DUNE and T2HK in identifying deviations of $\sin^2\theta_{23}$ from its maximal mixing value, both individually and in combination, under the influence of LRI. We compute the sensitivity using

\begin{equation}
    \resizebox{\textwidth}{!}{$
    \Delta \chi^2_{\mathrm{DM}} = \underset{(\delta_{\mathrm{CP}},\,\Delta m^{2}_{31},\, V_{\alpha\beta})}{\mathrm{min}} \bigg[
    \chi^2(\sin^2\theta^{\mathrm{test}}_{23} = 0.5, (\rm{SI} + V_{\alpha\beta}))-\chi^2(\sin^2\theta^{true}_{23}\in[0.4:0.6],(\rm{SI} + V_{\alpha\beta}^{\rm {fixed}}))\bigg]$}
    \label{eq:chi2-sensitivity-DM}.
\end{equation}

As extensively discussed in existing studies, the sensitivity to deviations from MM primarily relies on disappearance-driven analyses~\cite{Agarwalla:2024kti}. The characteristic U-shaped nature of the sensitivity curves reaffirms this, as it reflects the variation of disappearance statistics proportional to $1- \sin^2 2\theta_{23}$.  A narrower and more sharply defined U-shape indicates enhanced precision in determining deviations from MM, emphasizing the experiment's ability to tightly constrain $\sin^2\theta_{23} = 0.5$. The substantial disappearance statistics and the lower disappearance systematic uncertainties of 3.5\% in T2HK enable superior sensitivity in identifying deviations from MM compared to DUNE, both in the presence and absence of LRI. Notably, the $L_{\mu}-L_{\tau}$ symmetry significantly influences the $\nu_{\mu} \rightarrow \nu_{\mu}$ and $\bar{\nu}_{\mu} \rightarrow \bar{\nu}_{\mu}$ disappearance channels, resulting in notable deviations in sensitivity when compared with the SI case under DUNE. This effect is more pronounced in DUNE due to its relatively higher disappearance systematic uncertainties of 5\%\,, as illustrated in Fig.~\ref{fig:DM_LRI}. To provide a quantitative measure of our findings, we introduce a variable, $\zeta_{\rm DM}$, which captures the deviation in sensitivity observed in the presence of LRI compared to the SI scenario in Table~\ref{table:DM_LRI_2}. Illustratively, we evaluate  for $\sin^2\theta_{23} = 0.48$ (a choice in LO) and $\sin^2\theta_{23} = 0.52$ (a choice in HO). 
\begin{equation}
\zeta_{\rm DM}\, =\, \dfrac{(\Delta\chi^2_{\mathrm{SI}}-\Delta\chi^2_{\mathrm{LRI}})}{\Delta\chi^2_{\mathrm{SI}}}\times \, 100\%\, .
    \label{eq:deflection-DM}
\end{equation}

\begin{table}[htb!]
    \centering
    {\footnotesize  
    \renewcommand{\arraystretch}{2.0} 
    \begin{tabular}{|c|c|c|c|}
        \hline
        \multirow{2}{*}{Sym.} & \multicolumn{3}{|c|}{Departure from SI, $\zeta_{\rm DM}$(\%) at $\sin^2\theta_{23}=0.48\,(0.52)$} \\
        \cline{2-4}
        & DUNE & T2HK & DUNE + T2HK \\
        \hline
        $L_e-L_\mu$ & 27.2 ($-\,77.2$) & 6.8 ($-\,6.5$) & $\,7.0$ ($-50.4$) \\
        \hline
        $L_e-L_\tau$ & 21.8 (-0.4) & 1.8 (-0.1) & 11.6 (9.1) \\
        \hline
        $L_\mu-L_\tau$ & 27.1 ($-\,129.2$) & 23.6 (-0.6) & 13.4 (-7.0) \\
        \hline
    \end{tabular}
    \caption{\footnotesize{Deflection in sensitivity for establishing deviation from maximal $\sin^2\theta_{23}$, assuming true $\sin^2\theta_{23} = 0.48\,(0.52)$ under different LRI symmetries relative to the SI case. \textit{Positive values indicate a deterioration in sensitivity, while negative values signify an improvement}.}}
    \label{table:DM_LRI_2}
    }
\end{table}

From Table \ref{table:DM_LRI_2}, we see that the positive values of $\zeta_{\rm DM}$ indicate a decrease in sensitivity when probing a non-maximal $\theta_{23}$ in the presence of LRI, compared to the SI case. This holds true for all LRI symmetries, with the true value of $\sin^2\theta_{23}$ set at 0.48 (LO). The largest deflection from SI occurs in DUNE for the $L_\mu - L_\tau$ symmetry, with a deflection of $\sim 129\%$. In contrast, the smallest deflection is seen in T2HK for the $L_e - L_\mu$ symmetry (1.7\%). Interestingly, we find that when the data is generated with $\sin^2\theta_{23}$ in the HO, the presence of both $L_\mu - L_\tau$ and $L_{e} - L_{\mu}$ symmetries in DUNE improves the sensitivity significantly compared to the SI case. This improvement is meagre in the case of $L_e - L_\tau$ symmetry. The improvement is more pronounced for $L_\mu - L_\tau$ symmetry than for $L_{e} - L_{\mu}$ symmetry. This enhancement occurs because the neutrino disappearance statistics under these symmetries are higher than in the SI case (refer Fig.~\ref{fig:Oscillogram_disapp_nu}), with the highest statistics observed under the $L_\mu - L_\tau$ symmetry, followed by $L_e - L_\mu$ and the least under $L_e - L_\tau$ symmetry. We observe that among DUNE and T2HK, T2HK generally provides better sensitivity to non-maximal $\theta_{23}$ than DUNE. This is because T2HK benefits from huge disappearance statistics and reduced disappearance systematic uncertainties, both of which are crucial for this study. Moreover, when combining results from DUNE and T2HK, the sensitivity under $L_\mu - L_\tau$ symmetry improves as compared to their individual sensitivities; however, it remains lower than in the SI case. This is attributed to the behavior of how $\theta_{23}$ evolves in the presence of matter ($\theta_{23}^{\rm m}$) with $E$ under the three LRI symmetries considered, as shown in Fig. \ref{fig:running_th23}. 


\subsubsection{Effect on exclusion of wrong octant of $\sin^{2}\theta_{23}$ }
\label{subsec:wrong-octant-fixed-potential}
\begin{figure}[htb!]
	\centering 
	\includegraphics[width=\linewidth]{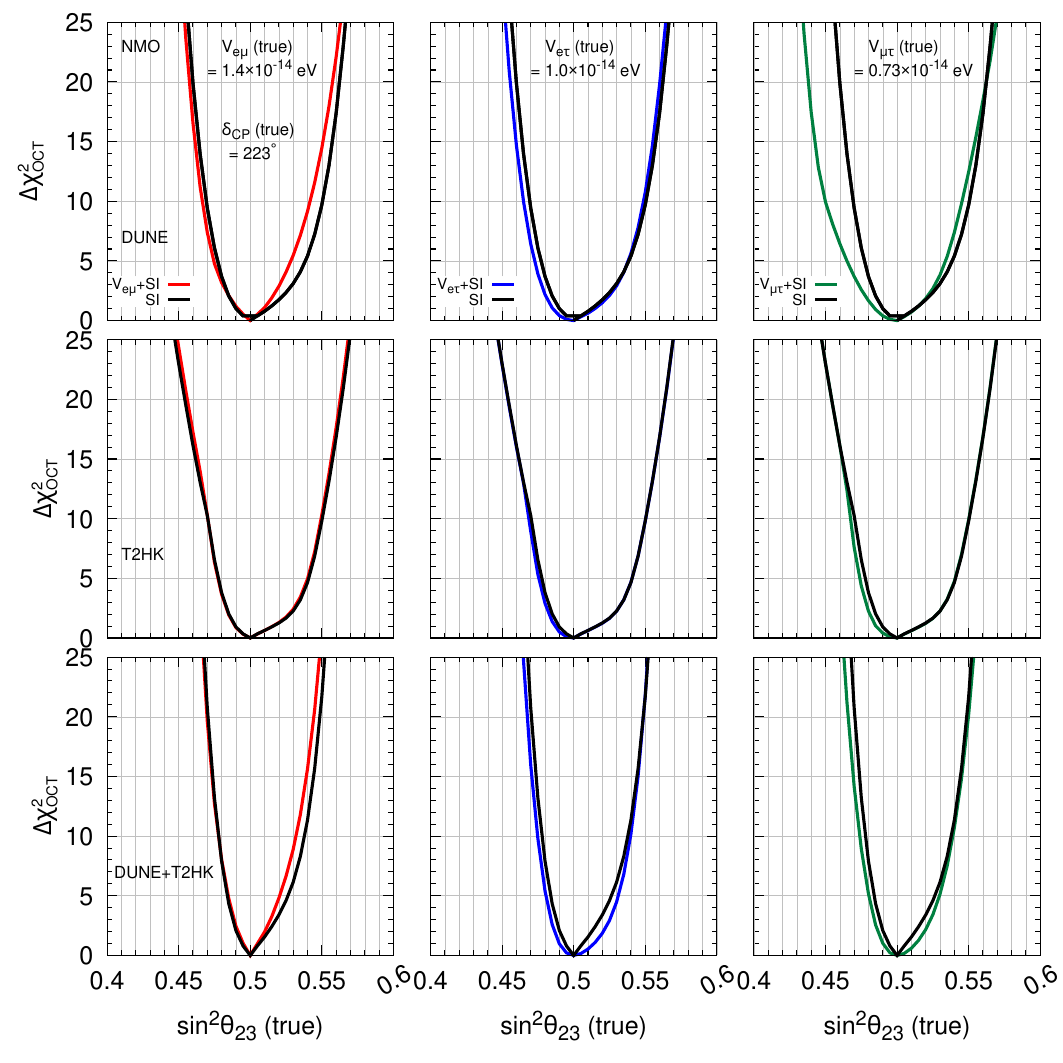}
	\caption{\footnotesize{Effect of the presence of $L_e-L_\mu$ (left panel), $L_e-L_\tau$ (middle panel), and $L_\mu-L_\tau$ (right panel) symmetries in excluding the wrong octant solutions using DUNE (top), T2HK (in between), and their combination DUNE + T2HK (bottom), assuming NMO. The black and colored curves represent the absence and presence of LRI in SI with a fixed potential strength (Refer Table~\ref{tab:oscillation-params}). The benchmark choices of oscillation parameters used to generate the data, as well as the ranges of $\delta_{\mathrm{CP}}$ and $\Delta m^2_{31}$ over which the test statistic is minimized, are taken from Table~\ref{tab:oscillation-params}. Additionally, we also minimize test statistic over $V_{\alpha\beta}$ (Refer Table~\ref{tab:oscillation-params}). The rest of the oscillation parameters remain fixed at their benchmark. \textit{In DUNE, the presence of $L_{\mu}-L_{\tau}$ symmetry, the sensitivity is significantly affected, decreasing for true LO and enhancing for true HO compared to SI, followed by the effects of $L_{e}-L_{\mu}$ symmetry.}
    }} 
	\label{fig:LRI_oct_excl}
\end{figure}
Fig.~\ref{fig:LRI_oct_excl} illustrates the sensitivity of DUNE and T2HK in excluding the wrong octant of $\sin^2\theta_{23}$ in isolation and combination, under the influence of LRI. This sensitivity is computed using
\begin{equation}
    \Delta \chi^2_{\mathrm{oct}} = \underset{(\delta_{\mathrm{CP}},\,\Delta m^{2}_{31},\, V_{\alpha\beta})}{\mathrm{min}} \bigg[
    \chi^2(\sin^2\theta^{\mathrm{test}}_{23}, (\rm{SI} + V_{\alpha\beta}))-\chi^2(\sin^2\theta^{true}_{23},(\rm{SI} + V_{\alpha\beta}^{\rm {fixed}}))\bigg]
    \label{eq:chi2-sensitivity-octant}\,,
\end{equation}
where $\sin^2\theta_{23}^{\rm test}$ and $\sin^2\theta_{23}^{\rm true}$ corresponds to the opposite octant. The ranges of minimization over the uncertain parameters are taken from Table~\ref{tab:oscillation-params}. Under SI, the sensitivity to excluding the wrong octant solutions is similar for DUNE and T2HK individually, but it improves when combined, highlighting the complementarity between these experiments, as studied in Ref.~\cite{Agarwalla:2024kti}. 

Under the influence of new LRI potential, we observe very little effect under T2HK, only in LO, where the sensitivity decreases in the presence of LRI. However, under DUNE, we observe significant deviation from the SI sensitivities in the presence of each LRI symmetry considered. The presence of $L_{\mu}-L_{\tau}$ symmetry significantly reduces the sensitivity to exclude the wrong octant solutions if the correct octant in Nature is LO, with a similar but lesser effect observed for $L_{e}-L_{\tau}$ symmetry. However, in the presence of $L_{e}-L_{\mu}$ symmetry, the sensitivity does not get affected. This can be attributed to the overall reduction in total neutrino rates under $L_{\mu}-L_{\tau}$ and $L_{e}-L_{\tau}$ symmetries compared to SI, whereas in the case of $L_{e}-L_{\mu}$ symmetry, the neutrino rates remain comparable to SI when LO is assumed in Nature (Refer fig.~\ref{fig:Oscillogram_disapp_nu}). This can be understood by observing the nature of the evolution of $\theta_{23}^{\rm m}$ with matter in the presence of these three symmetries in Fig.~\ref{fig:running_th23}. If $\sin^2\theta_{23}$ is in HO in Nature, the sensitivity increases in the presence of $L_{e}-L_{\mu}$ symmetry, followed by $L_{\mu}-L_{\tau}$ symmetry, while $L_{e}-L_{\tau}$ symmetry has little effect. This enhancement arises because, under $L_{e}-L_{\mu}$ symmetry, the total antineutrino rates increase significantly compared to SI. This behavior can be understood by examining the evolution of $\theta_{23}^{\rm m}$ in matter for antineutrinos under this LRI symmetry in fig~\ref{fig:running_th23}, which differs notably from SI in a manner opposite to that observed for $L_{\mu}-L_{\tau}$ and $L_{e}-L_{\tau}$ symmetries. In general, combining DUNE and T2HK does not significantly alter the sensitivity in excluding the wrong octant solutions of $\sin^2\theta_{23}$ under any symmetry. This highlights the complementarity between DUNE and T2HK, ensuring a similar sensitivity to excluding the wrong octant solutions in the presence of LRI symmetries - an outcome that was not observed with DUNE alone. 

To provide a quantitative measure of our findings, we introduce a variable, $\zeta_{\rm oct}$, which captures the deviation in sensitivity observed in the presence of LRI compared to the SI scenario in Table~\ref{table:Oct_LRI_2}. Illustratively, we evaluate  for $\sin^2\theta_{23} = 0.48$ (a choice in LO) and $\sin^2\theta_{23} = 0.52$ (a choice in HO).

\begin{table}[htb!]
    \centering
    {\footnotesize  
    \renewcommand{\arraystretch}{2.0} 
    \begin{tabular}{|c|c|c|c|}
        \hline
        \multirow{2}{*}{Sym.} & \multicolumn{3}{|c|}{Departure from SI, $\zeta_{\rm oct}$(\%) at $\sin^2\theta_{23}=0.48\,(0.52)$} \\
        \cline{2-4}
        & DUNE & T2HK & DUNE + T2HK \\
        \hline
        $L_e-L_\mu$ & 13.8 ($-\,64.9$) & 3.7 ($-\,3.2$) & $-\,1.9$ ($-41.4$) \\
        \hline
        $L_e-L_\tau$ & 41.0 (14.3) & 23.7 (1.5) & 31.7 (45.3) \\
        \hline
        $L_\mu-L_\tau$ & 54.9 ($-\,6.1$) & 40.8 (2.5) & 34.6 (36.0) \\
        \hline
    \end{tabular}
    }
    \caption{\footnotesize{Departure in sensitivity for excluding the wrong octant solutions of $\sin^2\theta_{23}$, assuming true $\sin^2\theta_{23} = 0.48\,(0.52)$ under different LRI symmetries relative to the SI case. \textit{Positive values indicate a deterioration in sensitivity, while negative values signify an improvement}.}}
    \label{table:Oct_LRI_2}
\end{table}

The numerical values align with the above explanation, where a positive $\zeta_{\rm oct}$ indicates a deterioration in sensitivity, while a negative value signifies an improvement. As discussed, DUNE exhibits the strongest impact of LRI among the two experiments. Under $L_{\mu}-L_{\tau}$ , DUNE experiences the greatest deterioration in sensitivity for excluding the wrong octant solutions, deviating by approximately 55\% from the SI case when in Nature $\sin^2\theta_{23} = 0.48$. Conversely, if $\sin^2\theta_{23} = 0.52$, the same symmetry enhances the sensitivity by  6\% compared to the SI case.

\subsubsection{Influence on CP violation }
\label{subsec: CPV-fixed-potential}
\begin{figure}[htb!]
	\centering
	\includegraphics[width=\linewidth]{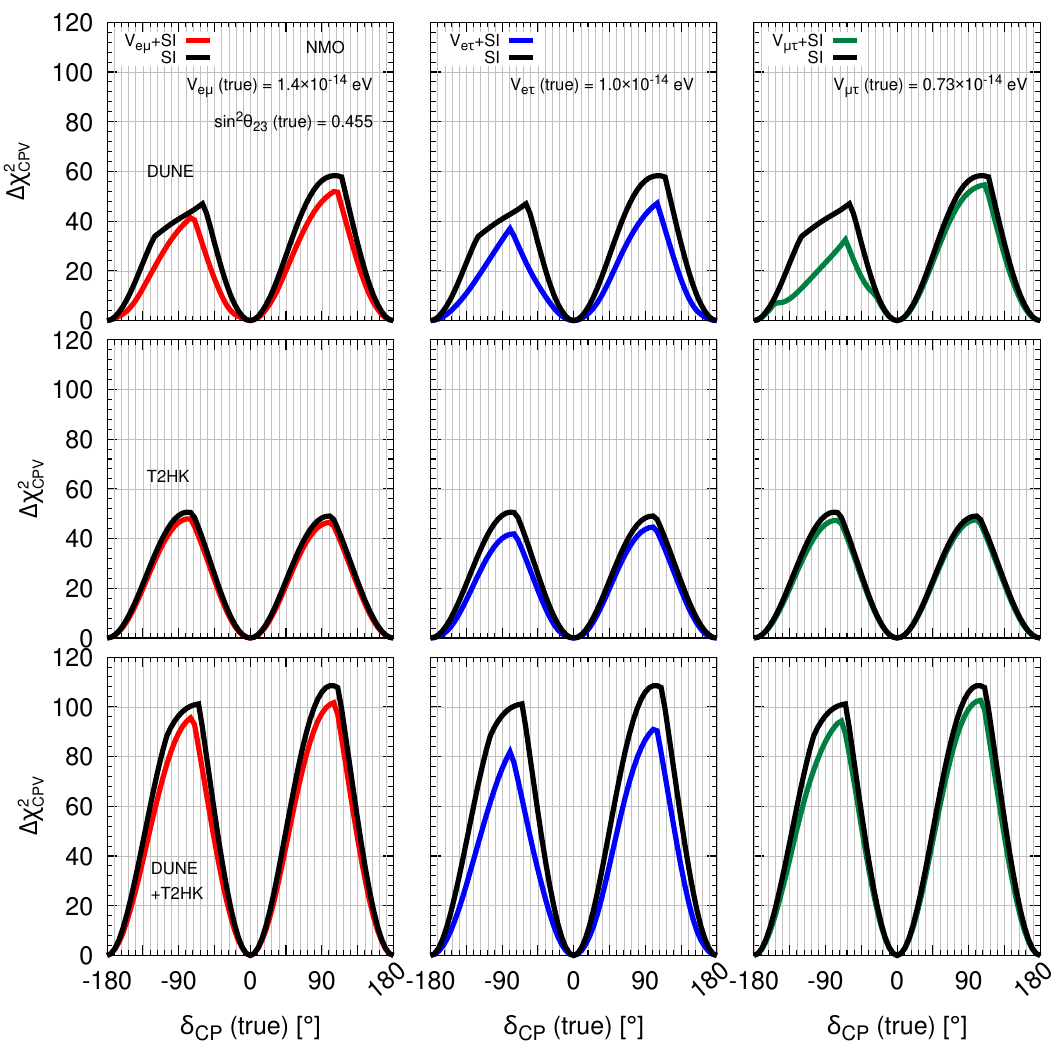}
	\caption{\footnotesize{Effect of the presence of $L_e-L_\mu$ (left panel), $L_e-L_\tau$ (middle panel), and $L_\mu-L_\tau$ (right panel) symmetries on CP violation sensitivity using DUNE (top), T2HK (middle), and their combination DUNE + T2HK (bottom) under NMO. The black and colored curves represent the absence and presence of LRI in SI with potential strengths taken from Table~\ref{tab:oscillation-params} using benchmark oscillation parameters and minimizing the test statistic over $\sin^2\theta_{23}$, $\Delta m^2_{31}$, and $V_{\alpha\beta}$ within the specified ranges. \textit{The presence of each LRI symmetry decreases the CP violation sensitivity, particularly near maximal CP-violating values, with $L_{e}-L_{\tau}$ symmetry causing the most significant reduction.}
    }}
 \end{figure}
 \label{fig:CPV}
Fig.~\ref{fig:CPV} illustrates the sensitivity of DUNE and T2HK in determining CP violation both individually and in combination, under the influence of LRI. This sensitivity is computed using
\begin{equation}
 \Delta \chi^2 = \underset{(\delta_{\rm CP}^{\rm test} = 0^{\circ} \text{ and } 180^{\circ},\,\sin^2\theta_{23},\,\Delta m^{2}_{31}, \, V_{\alpha\beta})}{\mathrm{min}} \bigg[
    \chi^2(\delta^{\mathrm{test}}_{\mathrm{CP}}, (\rm{SI} + V_{\alpha\beta}))-\chi^2(\delta^{\mathrm{true}}_{\mathrm{CP}} = 223^\circ, (\rm{SI} + V_{\alpha\beta}^{\rm fixed})\bigg],
    \label{eq:chi2-sensitivity-cpv}
\end{equation}
where the ranges over which minimization is performed is taken from Table~\ref{tab:oscillation-params}. T2HK, in isolation, shows better sensitivity to determining CP violation than DUNE under SI, particularly in the favorable region of NMO and the lower half-plane of $\delta_{\mathrm{CP}}$. This is due to its narrow-band energy beam and reduced matter effects from a shorter baseline, compared to DUNE’s wide-band beam and stronger matter effects from a longer baseline~\cite{Singh:2024nvt}. However, DUNE outperforms T2HK in the upper half-plane of $\delta_{\mathrm{CP}}$, thanks to its superior appearance channel systematic uncertainties. Combining both experiments leverages their complementarity, significantly enhancing sensitivity under SI~\cite{Singh:2024nvt}. 

The presence of LRI symmetry reduces the sensitivity to CP violation for both DUNE and T2HK, whether in isolation or combination, with the most significant impact occurring around maximal CP-violating values. As observed in previous sensitivity results, this effect is more pronounced in DUNE than in T2HK due to DUNE's greater sensitivity to LRI effects. Among the symmetries, $L_{\mu}-L_{\tau}$ symmetry causes the strongest deterioration around $\delta_{\mathrm{CP}} = -90^{\circ}$, followed by $L_{e}-L_{\tau}$ symmetry and then $L_{e}-L_{\mu}$ symmetry. This deterioration arises from degeneracies when the test statistic is minimized over both $\sin^2\theta_{23}$ and $V_{\alpha\beta}$. T2HK, with its large disappearance statistics and reduced matter effects, helps resolve these degeneracies, leading to less deterioration than in DUNE. The complementarity between DUNE and T2HK preserves CP violation sensitivity under $L_{e}-L_{\mu}$ and $L_{\mu}-L_{\tau}$ symmetries, keeping it close to the sensitivity seen under SI. However, the presence of $L_{e}-L_{\tau}$ symmetry still reduces sensitivity around maximal CP-violating values. This is because the presence of $L_{e}-L_{\tau}$ symmetry primarily affects $\nu_{\mu} \rightarrow \nu_{e}$ and $\bar{\nu}_{\mu} \rightarrow \bar{\nu}_{e}$ appearance channels. Since CP violation sensitivity is largely driven by appearance statistics, it remains impaired even with the combined complementarity of DUNE and T2HK.

\subsubsection{Consequence in computing allowed ranges in $(\sin^{2}\theta_{23}- \delta_{\mathrm{CP}})$ plane}
\label{subsec: allowed-ranges-th23-dcp}
\begin{figure}[htb!]
	\centering
	\includegraphics[width=\linewidth]{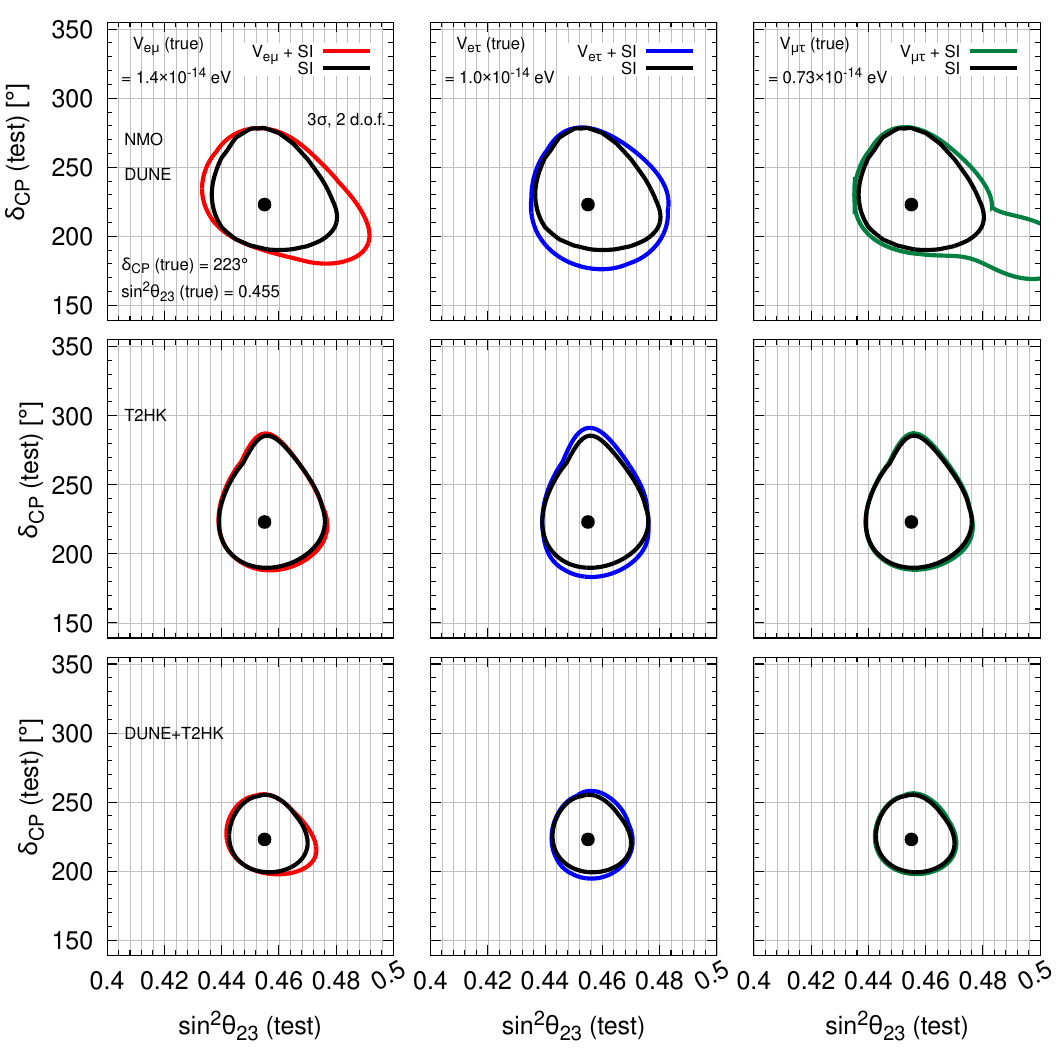}
	\caption{\footnotesize{Effect of $L_e-L_\mu$ (left), $L_e-L_\tau$ (middle), and $L_\mu-L_\tau$ (right) symmetries on allowed ranges in the $\sin^2\theta_{23}-\delta_{\mathrm{CP}}$ plane at $3\sigma$ C.L. (2 d.o.f.) assuming NMO, true $\delta_{\mathrm{CP}}=223^\circ$, and true $\sin^2\theta_{23}=0.455$.The black and colored curves represent the absence and presence of LRI in SI with potential strengths from Table~\ref{tab:oscillation-params} using DUNE (top), T2HK (middle), and DUNE + T2HK (bottom). The test statistic is scanned over $3\sigma$ ranges of $\delta_{\mathrm{CP}}$, $\sin^2\theta_{23}$, and $\Delta m^2_{31}$ and also $V_{\alpha\beta}$ from Table~\ref{tab:oscillation-params}. \textit{The presence of LRI relaxes constraints in the $\sin^2\theta_{23}-\delta_{\mathrm{CP}}$ plane: $L_\mu - L_\tau$ and $L_e - L_\mu$ symmetries (dominating disappearance channels) weaken $\sin^2\theta_{23}$ constraints in DUNE, while $L_e - L_\tau$ symmetry (dominating appearance channels) loosens $\delta_{\mathrm{CP}}$ constraints in T2HK. DUNE and T2HK together maintain constraints under LRI nearly as stringent as in the SI case.}}}
	\label{fig:th23-dcp-allowed-contours}
\end{figure}

 Fig. \ref{fig:th23-dcp-allowed-contours} illustrates the impact of long-range interactions on depicting the allowed ranges in $(\sin^2\theta_{23}-\delta_{\mathrm{CP}})$ plane. We compute it using

\begin{equation}
   \Delta \chi^2 = \underset{(\Delta m^{2}_{31}, \, V_{\alpha\beta})}{\mathrm{min}} \bigg[
    \chi^2(\delta^{\mathrm{test}}_{\mathrm{CP}}, \sin^2\theta^{\mathrm{test}}_{23})-\chi^2(\delta^{\mathrm{true}}_{\mathrm{CP}},  \sin^2\theta^{\mathrm{true}}_{23}) = (223^\circ, 0.455)\bigg],
    \label{eq:chi2-sensitivity-contours_left}
\end{equation}
where the $\delta^{\mathrm{test}}_{\mathrm{CP}}$ and $\sin^2\theta^{\mathrm{test}}_{23}$ corresponds to their corresponding 3$\sigma$ ranges. Data is generated by considering a fixed true value of $V_{e\mu} = 1.4\times10^{-14}$ eV, $V_{e\tau} = 1.0\times10^{-14}$ eV, and $V_{\mu\tau} = 0.73\times10^{-14}$ eV in the top, middle, and bottom panels, respectively. These values are motivated from Ref.~\cite{Singh:2023nek}, following the constraints established by the combined DUNE + T2HK on $V_{e\mu},\, V_{e\tau}$, and $V_{\mu\tau}$ at 2$\sigma$ confidence level.

The inclusion of LRI does not significantly improve the constraints on the $(\sin^2\theta_{23}-\delta_{\mathrm{CP}})$ plane. Among the three LRI symmetries considered, the tightest constraints are observed under the $L_{e} - L_{\tau}$ symmetry. This is because it has the most pronounced effect on neutrino appearance statistics, as highlighted in the bievents in Sec.~\ref{subsec:app-stats}. The presence of $L_{e} - L_{\mu}$ symmetry follows while $L_{\mu}-L_{\tau}$ symmetry yields the loosest constraints. In the case of $L_{\mu} - L_{\tau}$, DUNE exhibits degeneracies in $\sin^2\theta_{23}$, suggesting solutions in both the lower octant (LO) and maximal mixing (MM). Constraints improve when $\sin^2\theta_{23} = 0.581$ (HO) is used to generate data due to increased neutrino statistics in the HO compared to the LO. However, even in this scenario, a degenerate solution at MM persists. In contrast, T2HK achieves tighter constraints along $\sin^2\theta_{23}$ owing to its substantial disappearance statistics and superior disappearance systematic uncertainties, although its constraints on $\delta_{\mathrm{CP}}$ remain less stringent. The complementarity between DUNE and T2HK is evident in their respective strengths. The wide-band beam, diverse $L/E$ ratios, and superior appearance systematic uncertainties in DUNE enable stronger constraints on $\delta_{\mathrm{CP}}$. Meanwhile, the large disappearance statistics and enhanced disappearance systematics in T2HK deliver tighter bounds on $\sin^2\theta_{23}$. Together, these strengths produce a well-constrained region in the $(\sin^2\theta_{23}-\delta_{\mathrm{CP}})$ parameter space, demonstrating the synergy of combining data from both experiments.

\subsection{Impact of scanning LRI in data}


\subsubsection{Effect on fractional coverage of $\sin^2\theta_{23}$ differentiable from MM }
\label{subsec:frac-DM}
\begin{figure}[htb!]
	\centering 
	\includegraphics[width=\linewidth]{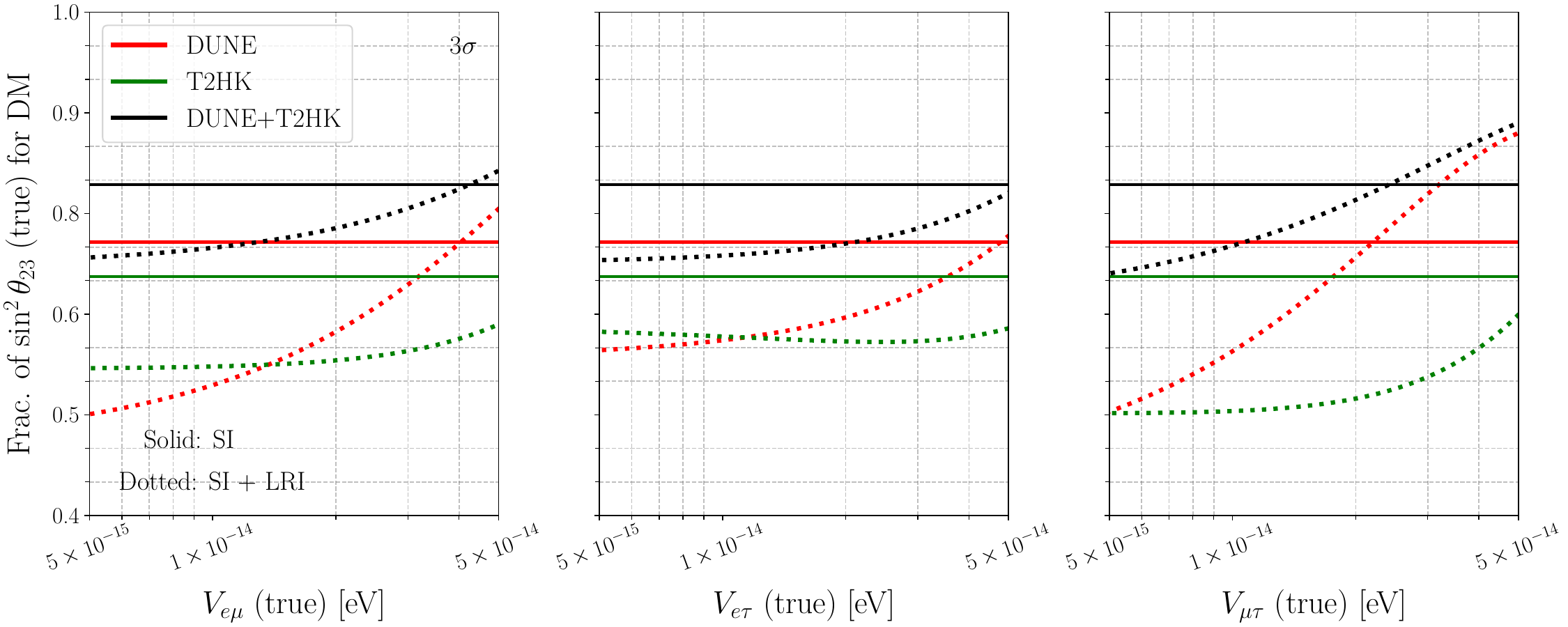}
	\caption{\footnotesize{Fraction of $\sin^2\theta_{23}$ in Nature that can be differentiated from MM at $\geq 3\sigma$ using DUNE, T2HK, and DUNE + T2HK in the presence of  $(L_e-L_\mu)$, $(L_e-L_\tau)$, or $(L_\mu-L_\tau)$ symmetries (dotted curves) in left, middle, and right panels, respectively. We generate these curves following Eq.~\ref{eq:chi2-sensitivity-DM-2}, using the benchmark oscillation parameters and allowed ranges from Table~\ref{tab:oscillation-params}. We assume the benchmark choice of $\Delta m^2_{31}=2.522\times10^{-3}$ eV$^2$ and consider NMO in Nature. For reference, we show fractional contributions of $\theta_{23}$ for our study under standard neutrino interaction (solid line) following the same benchmark values. The test statistic is minimized over 3$\sigma$ ranges in $\delta_{\mathrm{CP}}$, $\Delta m^2_{31}$ (Refer Table~\ref{tab:oscillation-params}), and $V_{\alpha\beta} \in [1 \times 10^{-15}, 1 \times 10^{-13}]$ eV. \textit{This sensitivity, being predominantly driven by disappearance statistics, is most affected by the presence of $L_{\mu}-L_{\tau}$ symmetry. 
    } }}
	\label{fig:frac_th23_DM}
\end{figure}
We revisit the sensitivity studies to establish the deviation from maximality using DUNE, T2HK, and their combination by computing the fractional percentage of $\sin^2\theta_{23}$ that can be differentiated from MM solutions. We use Eq.~\ref{eq:chi2-sensitivity-DM-2}, and generate data by scanning over $V_{\alpha\beta}$ between [$5.0\times10^{-15}$ eV : $5.0\times10^{-14}$ eV] and $\sin^2\theta_{23} \in [0.4 : 0.6]$, fixing all other oscillation parameters to their benchmark values. In the fit, we minimize over 3$\sigma$ uncertainty in $\delta_{\mathrm{CP}}$, $\Delta m^{2}_{31}$) (refer Table~\ref{tab:oscillation-params}), and $V_{\alpha\beta} \in [1.0\times10^{-15} : 1.0\times10^{-13}]$ eV.  For quantification, we consider only those true values of $\sin^2\theta_{23}$ that reach a 3$\sigma$ level of statistical significance. 

\begin{equation}
    \resizebox{\textwidth}{!}{%
    $
    \Delta \chi^2 = 
    \underset{(\delta_{\mathrm{CP}},\,\Delta m^{2}_{31},\, V_{\alpha\beta})}{\mathrm{min}} 
    \left[
    \chi^2\bigl(\sin^2\theta^{\mathrm{test}}_{23} = 0.5, (\mathrm{SI} + V_{\alpha\beta})\bigr) 
    - 
    \chi^2\bigl(\sin^2\theta^{\mathrm{true}}_{23} \in [0.4:0.6], (\mathrm{SI} + V_{\alpha\beta})\bigr)
    \right]
    $
    }
    \label{eq:chi2-sensitivity-DM-2}
\end{equation}


Figure~\ref{fig:frac_th23_DM} illustrates the fractional contributions of $\sin^2\theta_{23}$ in Nature that are responsible for establishing non-maximality at the $3\sigma$ confidence level, as envisioned by DUNE and T2HK, both individually and in combination in both presence and absence of LRI. We expect that when the effect of LRI is minimal ( $ <10^{-14}$ eV), the test statistic will closely match the data, resulting in similar sensitivity with and without the presence of LRI. However, we observe that the presence of LRI suppresses the sensitivity due to the degeneracy arising from the minimization over the uncertainties in oscillation parameters and $V_{\alpha\beta}$. When LRI becomes sub-dominant (beyond $10^{-14}$ eV), the sensitivity to establishing deviations from the MM scenario increases compared to the SI case. This increment is most pronounced in the presence of $L_{\mu}- L_{\tau}$ symmetry, followed by $L_{e}- L_{\mu}$ symmetry, consistent with the trend observed in Fig.~\ref{fig:DM_LRI}. The increase in sensitivity is notably more significant in DUNE than in T2HK. This behavior can be explained by analyzing the evolution of $\theta_{23}^{\rm m}$ in Fig.~\ref{fig:running_th23}, where the atmospheric mixing angle gets enhanced from $45^\circ$ at 1st oscillation maxima of DUNE in the presence of LRI compared to the case where effect of LRI is absent. This rapid attainment of the maximal mixing angle in the presence of LRI, as opposed to SI, is critical. Since the disappearance channel probability is directly proportional to $ (1-\sin^2 2\theta_{23}^{\rm m})$, achieving the higher value of $\theta_{23}^{\rm m }$ leads to $\sin^2 2\theta_{23}^{\rm m} \rightarrow 0$, thereby increasing the leading-order disappearance probability and consequently enhancing the sensitivity. As the presence of  $L_{\mu}- L_{\tau}$ symmetry predominantly influences the disappearance channels, it enhances the sensitivity most significantly. Leveraging the complementarity between DUNE and T2HK, their combined analysis exhibits the most pronounced effect.

\subsubsection{Effect on excluding HO solutions of $\sin^2\theta_{23}$ }
\label{subsec:frac-octant}

\begin{figure}[htb!]
	\centering 
	\includegraphics[width=\linewidth]{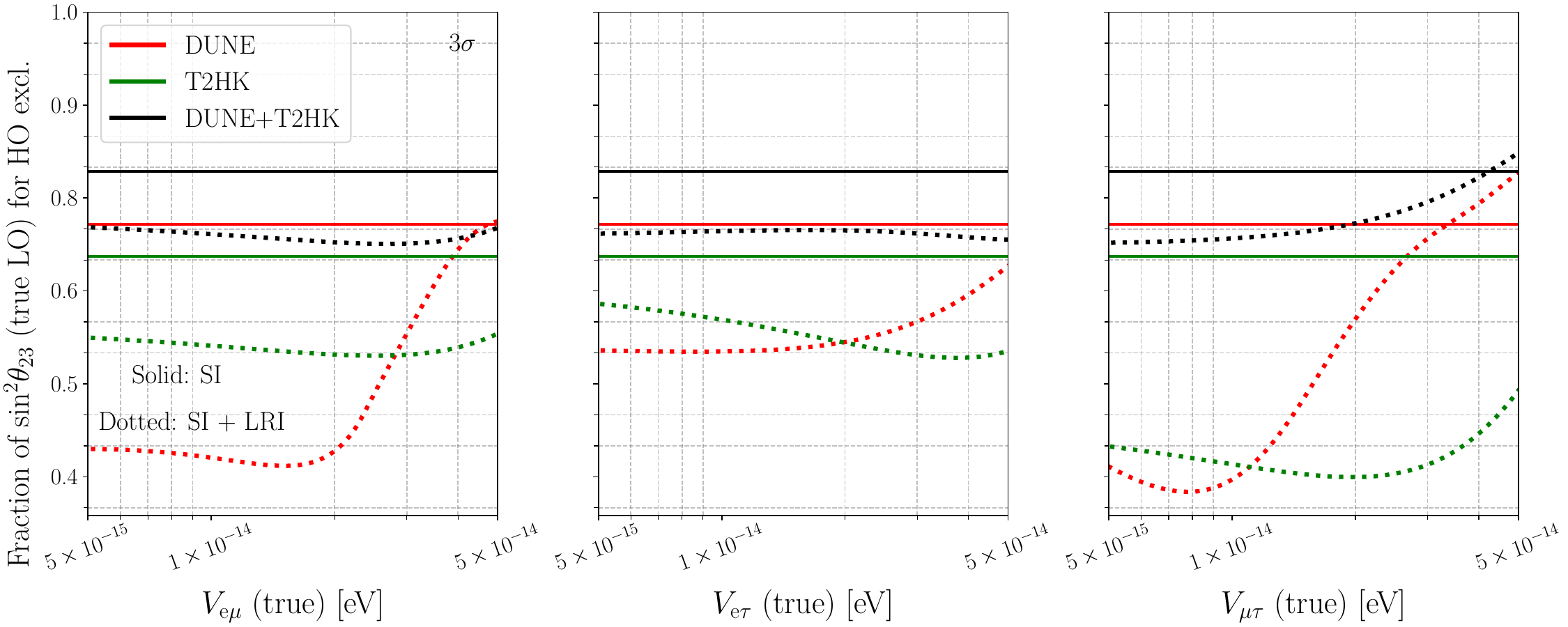}
	\caption{\footnotesize{Fraction of $\sin^2\theta_{23}$ in Nature that can exclude wrong octant solutions of $\sin^2\theta_{23}$ at $\geq 3\sigma$ using DUNE, T2HK, and DUNE + T2HK in the presence of  $(L_e-L_\mu)$, $(L_e-L_\tau)$, or $(L_\mu-L_\tau)$ symmetries (dotted curves) in left, middle, and right panels, respectively. We generate these curves following Eq.~\ref{eq:chi2-sensitivity-oct-2}, using the benchmark oscillation parameters and allowed ranges from Table~\ref{tab:oscillation-params}. We assume the benchmark choice of $\delta_{\mathrm{CP}}$ as $223^\circ$ and consider the LO ($i.e., 0.4\leq\sin^2\theta_{23}\leq0.5$) as a true octant of $\theta_{23}$ and excluding the HO ($i.e., 0.5\leq\sin^2\theta_{23}\leq0.6$), treating NMO as the underlying mass hierarchy in nature. For reference, we show this fraction under standard neutrino interaction (solid line) following the same benchmark values. The test statistic is minimized over 3$\sigma$ ranges in $\delta_{\mathrm{CP}}$, $\Delta m^2_{31}$ (Refer Table~\ref{tab:oscillation-params}), and $V_{\alpha\beta} \in [1 \times 10^{-15}, 1 \times 10^{-13}]$ eV. \textit{This sensitivity, being predominantly driven by appearance statistics, is most affected by the presence of $L_{\mu}-L_{\tau}$ symmetry. 
    } }}
	\label{fig:frac_th23_oct}
\end{figure}

We revisit the sensitivity studies to exclude the wrong octant solutions of $\sin^2\theta_{23}$ using DUNE, T2HK, and their combination by computing the fractional percentage of $\sin^2\theta_{23}$ in LO that can be differentiated from HO solutions. For simplicity, we focus solely on generating data with LO and evaluating the sensitivity in excluding HO solutions in the test statistic. This approach is justified since the best-fit scenario falls within LO when data is generated under NMO. We use Eq.~\ref{eq:chi2-sensitivity-oct-2}, and generate data by scanning over $V_{\alpha\beta}$ between [$5\times10^{-15}$ eV : $5\times10^{-14}$ eV] and $\sin^2\theta_{23} \in [0.4 : 0.5]$, fixing all other oscillation parameters to their benchmark values. In the fit, we scan over $\sin^2\theta_{23} \in [0.5 : 0.6]$ and minimize over 3$\sigma$ uncertainty in $\delta_{\mathrm{CP}}$ and $\Delta m^{2}_{31}$) (refer Table~\ref{tab:oscillation-params}), and $V_{\alpha\beta} \in [1\times10^{-15} : 1\times10^{-13}]$ eV.  For quantification, we consider only those true values of $\sin^2\theta_{23}$ that reach a 3$\sigma$ level of statistical significance. 

\begin{equation}
    \Delta \chi^2_{\mathrm{oct}} = \underset{(\delta_{\mathrm{CP}},\,\Delta m^{2}_{31},\, V_{\alpha\beta})}{\mathrm{min}} \bigg[
    \chi^2(\sin^2\theta^{\mathrm{test}}_{23} \in \text{HO}, (\rm{SI} + V_{\alpha\beta}))-\chi^2(\sin^2\theta^{true}_{23} \in \text{LO},(\rm{SI} + V_{\alpha\beta}))\bigg]
    \label{eq:chi2-sensitivity-oct-2}\,,
\end{equation}

Figure~\ref{fig:frac_th23_oct} presents the fractional contributions of $\sin^2\theta_{23}$ in the lower octant (LO) that enable the exclusion of incorrect higher octant (HO) solutions at the $3\sigma$ confidence level, as predicted by DUNE and T2HK—both individually and in combination, with and without the presence of long-range interactions (LRI). The sensitivity under standard interactions (SI), assuming $\sin^2\theta_{23} \in$ LO and other parameter values from Table~\ref{tab:oscillation-params} while generating data, is comparable for DUNE and T2HK when scanning over $\sin^2\theta_{23} \in$ HO and minimizing over $\delta_{\mathrm{CP}}$ and $\Delta m^{2}_{31}$ in the test statistic.

When the effect of LRI is minimal (below $10^{-14}$ eV), the test statistic closely aligns with the data, leading to similar sensitivity with and without LRI. However, the presence of LRI reduces sensitivity due to degeneracies arising from minimization over uncertainties in the oscillation parameters and $V_{\alpha\beta}$. As LRI becomes subdominant (beyond $10^{-14}$ eV), sensitivity to excluding HO solutions begins to increase. This effect is most pronounced under the $L_{\mu} - L_{\tau}$ symmetry, surpassing the SI case, followed by $L_{e} - L_{\mu}$ symmetry, in agreement with the trend observed in Fig.~\ref{fig:LRI_oct_excl}. The enhancement in sensitivity is notably stronger in DUNE than in T2HK.

This behavior can be understood by examining the evolution of $\theta_{23}^{\rm m}$ and $\theta_{13}^{\rm m}$ in Fig.~\ref{fig:running_th23} and Fig.~\ref{fig:running_th13}, where the matter-modified mixing angles ($i.e., \theta^m_{23}\,\,\text{and}\,\,\theta^m_{13}$) increase at the 1st oscillation maxima of DUNE in the presence of LRI compared to the absence of the LRI. Since the disappearance probability is proportional to $(1-\sin^2 2\theta_{23}^{\rm m})$, $\theta_{23}^{\rm m}$ increase as $\sin^2 2\theta_{23}^{\rm m}\rightarrow0$ and consequently increases the disappearance probability through the leading-order term, thereby improving sensitivity. However, the appearance probability is proportional to $(\sin^2\theta_{23}^{\rm m}\cdot\sin^22 \theta_{13}^{\rm m})$, meaning that achieving higher values for both $\theta_{23}^{\rm m}$ and $\theta_{13}^{\rm m}$ reduce the leading term, mitigating the sensitivity increase due to disappearance statistics. Since the $L_{e} - L_{\tau}$ symmetry primarily impacts the appearance channels, its contribution to the sensitivity increase in the presence of subdominant LRI is relatively small. In contrast, the $L_{\mu} - L_{\tau}$ and $L_{e} - L_{\mu}$ symmetries primarily affect the disappearance channel, leading to a significant enhancement in the exclusion sensitivity of the wrong octant.

By leveraging the complementarity between DUNE and T2HK, their combined analysis exhibits the strongest effect, particularly under the $L_{\mu} - L_{\tau}$ symmetry.

\subsubsection{Impact on CP coverage }
\label{subsec:frac-CPV}
\begin{figure}[htb!]
	\centering
\includegraphics[width=\linewidth]{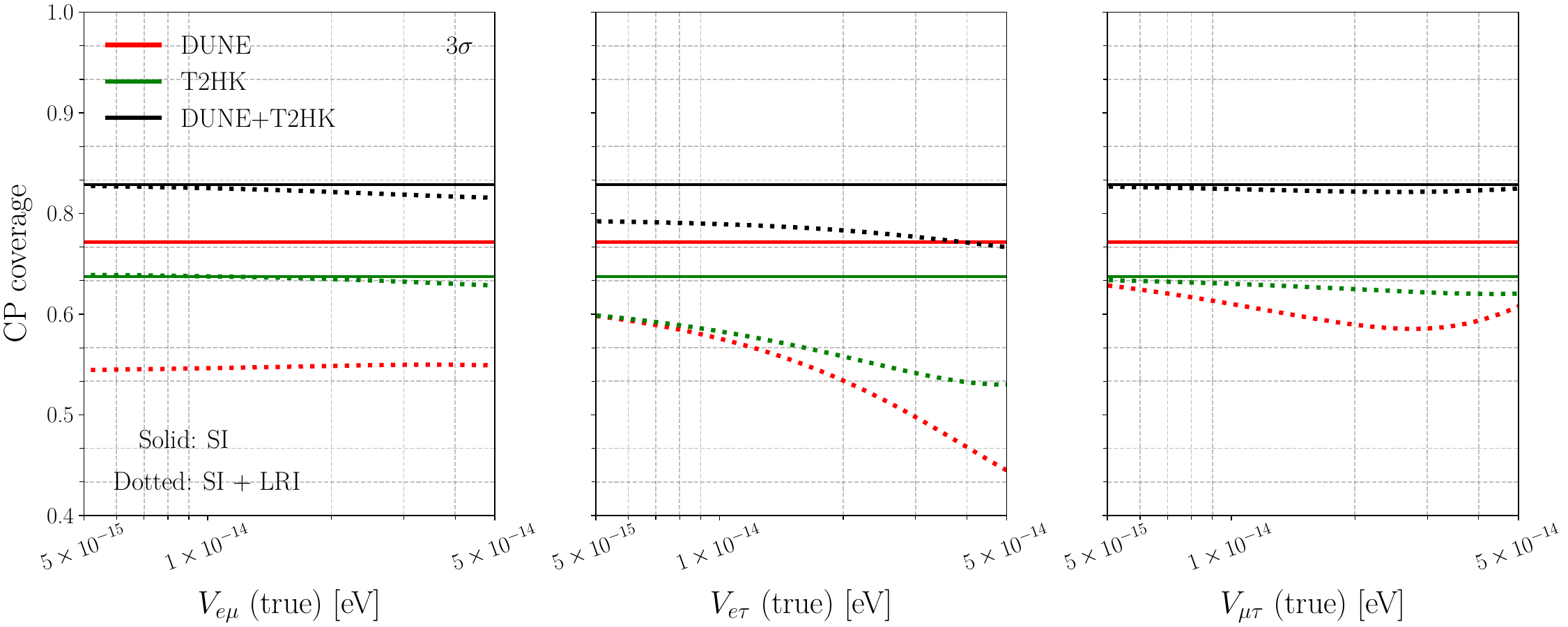}
	\caption{\footnotesize{Fraction of $\delta_\mathrm{CP}$ in Nature that can establish CPV at $\geq 3\sigma$ using DUNE, T2HK, and DUNE + T2HK in presence of  $(L_e-L_\mu)$, $(L_e-L_\tau)$, or $(L_\mu-L_\tau)$ symmetries in dotted curves in left, middle, and right panels, respectively. We generate these curves following Eq.~\ref{eq:chi2-sensitivity-to-cp-coverage}, using the benchmark oscillation parameters and allowed ranges from Table~\ref{tab:oscillation-params} assuming NMO in Nature. We assume the benchmark choice of $\sin^2\theta_{23}$ as $0.455$ ($i.e.,$ in LO) and marginalize over $3\sigma$ range of $\Delta m^2_{31}$ and $\sin^2\theta_{23}$ as mentioned in \ref{tab:oscillation-params}. For reference, we show CP coverage under standard neutrino interaction (solid line) following the same benchmark values. \textit{If subdominant long-range interaction exists in Nature, then under each symmetry, expected CP coverage in leptonic CP violation at $\geq 3\sigma$ will decrease due to the degeneracies arising by minimizing over $V_{\alpha\beta} \text{ and } \sin^2\theta_{23}$ in test statistic.}}}
 \label{fig:CPC_LRI}
 \end{figure}
 
In this subsection, we revisit the sensitivity studies in establishing CPV using DUNE, T2HK, and their combination. For computing it, we use

\begin{equation}
    \resizebox{\textwidth}{!}{$
    \Delta \chi^2 = \underset{(\delta^{\mathrm{test}}_{\mathrm{CP}}, \,\theta_{23},\,\Delta m^{2}_{31}, V_{\alpha\beta})}{\mathrm{min}} \bigg[
    \chi^2(\delta^{\mathrm{test}}_{\mathrm{CP}}=0^{\circ}\, \text{and}\, 180^{\circ}, (\mathrm{SI + LRI}))-\chi^2(\delta^{\mathrm{true}}_{\mathrm{CP}}\in[0^{\circ}, 360^{\circ}], (\mathrm{SI + LRI}))\bigg]
    $}.
    \label{eq:chi2-sensitivity-to-cp-coverage}
\end{equation}

 For quantifying our results, we follow the convention of CP coverage~\cite{Nath:2015kjg, Machado:2015vwa, Agarwalla:2022xdo}, which is defined as the fraction of true $\delta_{\mathrm{CP}}$ that can establish CPV at 3$\sigma$ (refer to Fig.~\ref{fig:CPC_LRI}), studied under NMO. For this, we follow Eq.~\ref{eq:chi2-sensitivity-to-cp-coverage}, and generate data by scanning over $V_{\alpha\beta}$ between [$5\times10^{-15}$ eV : $5\times10^{-14}$ eV] 
 and $\delta_{\mathrm{CP}}$ in the range [$-180^\circ$ : $+180^\circ$], fixing all other oscillation parameters to their benchmark values. In the fit, we minimize over the $\delta_{\mathrm{CP}} = 0^{\circ}$ and $180^{\circ}$, considering only the minimum. We also minimize over 3$\sigma$ uncertainty in $\theta_{23},\,\Delta m^{2}_{31}$) \ref{tab:oscillation-params}, and $V_{\alpha\beta} \in [1 \times 10^{-15}: 1 \times 10^{-13}]$ eV. 
 
 In Fig.~\ref{fig:CPC_LRI}, we observe that depending upon the strength of LRI potential, the sensitivity in establishing CP coverage is affected in the presence of each symmetry. For $V_{\alpha\beta}$ close to $10^{-15}$ eV and smaller, the test statistic matches with the data, and hence, the sensitivity in both the presence and absence of LRI remains similar, suppressing a little because of the minimization over the uncertainties in oscillation parameters and $V_{\alpha\beta}$. Following this, in the region where $V_{\alpha\beta}$ gets sub-dominant, there is visible suppression in the CP coverage fraction. This is mostly because of the degeneracy arising when minimized over the uncertainty in $\sin^2\theta_{23}$ and $V_{\alpha\beta}$. 
 In the presence of $L_e-L_\tau$ symmetry, the sensitivities corresponding to DUNE and T2HK in isolation reduces because of the above-mentioned degeneracy. Yet when these experiments are combined, the DUNE + T2HK configuration effectively breaks these degeneracies in the sub-dominant parameter space, resulting in enhanced overall sensitivity. The $L_\mu-L_\tau$ symmetry exhibits analogous behavior, though its impact is less pronounced. This difference primarily stems from the fact that CP coverage sensitivity is predominantly determined by the
 $\nu_{\mu}\rightarrow\nu_{e}$ and $\bar{\nu}_{\mu}\rightarrow\bar{\nu}_{e}$ appearance channels. Since $L_e-L_\tau$ symmetry has its strongest influence on the appearance channel, its impact on CP coverage sensitivity is particularly pronounced. 

\section{Summary and conclusion}
\label{sec:summary}

The next-generation long-baseline experiments, equipped with their excellent precision, reserve adequate room for exploring new kinds of neutrino interactions in Nature. The long-range interactions are too subtle to be detected due to the massive mediating new gauge boson \( Z^\prime \) responsible for this interaction. In our work, we portray the illuminating prospects of the future long-baseline experiments DUNE and T2HK to detect this interaction through phenomenological studies ($i.e.$, non-maximal \( \theta_{23} \), resolution of the octant ambiguity of \( \theta_{23} \), and CP violation sensitivity), either in isolation or in unison.

Here, we observe that the sensitivity of DUNE to establish the deviation from maximal \( \theta_{23} \) at its first oscillation maximum is enhanced in the presence of LRI arising from the \( L_\mu - L_\tau \) symmetry. Since this study is statistics-dominated, DUNE outperforms T2HK in this domain, as per the benchmark values given in Table~\ref{tab:oscillation-params}. However, based on this set of benchmark choices, the performance of both DUNE and T2HK deteriorates compared to the SI case. On the other hand, if we shift the best-fit value of \( \theta_{23} \) from the lower octant (LO) to the higher octant (HO), the performance of DUNE significantly improves (especially for the \( L_\mu - L_\tau \) symmetry), leaving a striking footprint of LRI in Nature. The fraction of true \( \sin^2 \theta_{23} \) that can be segregated from the MM solution also increases in DUNE for the \( L_\mu - L_\tau \) symmetry at \( 3\sigma \) C.L., while considering the LRI potential in the sub-dominant range (\( > 10^{-14} \) eV). In the dominant LRI range (\( < 10^{-14} \) eV), however, the sensitivity of our study remains largely unaffected by the presence of LRI. The synergistic setup of DUNE and T2HK does not show very promising results in this context at the dominated LRI range (\( > 10^{-14} \) eV), primarily due to the greater systematic uncertainty (5\%) of T2HK in the disappearance channel. Next, we show that the effect of LRI due to \( L_\mu - L_\tau \) is most clearly manifested in DUNE when testing the exclusion sensitivity of the higher octant solution of \( \theta_{23} \) for a benchmark LO choice. The sensitivity in the presence of LRI for \( L_\mu - L_\tau \) symmetry is significantly lower than in the SI case, while for the \( L_e - L_\mu \) and \( L_e - L_\tau \) symmetries, the sensitivity remains similar to SI. This behavior can be attributed to the fact that at the first oscillation maximum of DUNE, the matter-modified \( \theta_{23} \) and \( \theta_{13} \) are more affected in the presence of LRI compared to SI, which in turn significantly impacts the \( \nu_\mu \rightarrow \nu_e \) appearance probability and the corresponding total event rates. In this scenario, assuming the best-fit \( \theta_{23} \) in the HO, we find that the sensitivity to exclude LO increases across all experimental configurations and LRI symmetries. Here as well, standalone DUNE shows almost equal potential to detect LRI for the \( L_e - L_\mu \) symmetry to the combined DUNE+T2HK setup, due to its good control over systematics in the \( \nu_\mu \rightarrow \nu_e \) appearance channel, which is crucial for excluding the opposite octant of \( \theta_{23} \). However, DUNE’s matching performance with the combined DUNE+T2HK setup in excluding HO becomes significant (at \( 3\sigma \) C.L.) only in the sub-dominant regime of \( V_{\mu\tau} \) (\( \geq 1.0 \times 10^{-14} \) eV). The incorporation of LRI symmetry attenuates the sensitivity to CP violation in both DUNE and T2HK—whether considered independently or synergistically—with the most profound suppression manifesting near maximally CP-violating phases. This suppression is more accentuated in DUNE compared to T2HK, owing to DUNE’s heightened responsiveness to LRI effects. Among the symmetries considered, the \( L_{\mu} - L_{\tau} \) symmetry induces the most pronounced degradation near \( \delta_{\mathrm{CP}} = -90^{\circ} \), followed sequentially by the \( L_{e} - L_{\tau} \) and \( L_{e} - L_{\mu} \) symmetries. The observed deterioration stems from parameter degeneracies emerging during the minimization over both \( \sin^2\theta_{23} \) and \( V_{\alpha\beta} \). Due to its substantial disappearance statistics and minimal matter effects, T2HK mitigates these degeneracies more effectively than DUNE, thereby exhibiting a comparatively milder decline in CP sensitivity. Lastly, we show, among the three LRI symmetries, the $L_e-L_\tau$ symmetry imposes the strongest constraint on $(\sin^2\theta_{23}-\delta_{\mathrm{CP}})$ plane, due to its most significant impact on neutrino appearance statistics. The complementarity of DUNE  and T2HK plays a prime role in constraining this parameter space by breaking the degeneracies that are negligible in the individual setups.

Hence, this study shows a promising avenue to detect the existence of the long-range neutrino interaction in the mirrors of the upcoming highly precise long-baseline experiments DUNE and T2HK. The possible complementarities between them also help to decipher the oscillation phenomenology in the sub-dominant regime of LRI in Nature.

\chapter{Summary and Conclusions}
\label{C7}
\par This thesis addresses some important issues in three-flavor neutrino oscillation in the context of upcoming high-precision long-baseline experiments. Neutrinos are omnipresent and they are the second most abundant particles in the Universe after photons. There are a variety of neutrino sources (both natural and artificial) producing neutrinos over a wide range of energies in the range of $10^{-4}$ to $10^{18}$ eV or so and traveling distances from a few meters to hundreds of Gpc. Neutrinos were postulated first by Wolfgang Pauli in 1930 to explain how beta-decay could conserve energy, momentum, and angular momentum. In 1934, Enrico Fermi named the new particle as ``neutrino", which means the ``little neutral one" in the Italian language. In 1956, Clyde L. Cowan and Frederick Reines first observed neutrinos in a nuclear reactor. There are three (3) light active flavor neutrinos: $\nu_e$, $\nu_\mu$, and $\nu_\tau$ - experimentally confirmed by the precision data on the \textit{Z}-decay width at the $e^+e^-$ collider at LEP \cite{ALEPH:2005ab}. They only take part in weak interactions through massive \textit{W}$^\pm$ and \textit{Z}$^0$ gauge bosons. They are electrically neutral, left-handed, and form \textit{SU}$(2)$ doublets in the Standard Model with the corresponding left-handed charged leptons. In the basic SM of particle physics, neutrinos are massless due to the absence of right-handed neutrinos. \\
\par Over the past two decades, data from several world-class neutrino experiments firmly established that neutrinos change flavor after propagating a finite distance in space and time - a phenomenon known as ``neutrino oscillation". Neutrino oscillation demands non-zero neutrino mass and mixing. Hence, non-zero neutrino mass is the first experimental proof for physics beyond the Standard Model for which the Nobel Prize was awarded to Takaaki Kajita and Arthur McDonald in 2015. Neutrino oscillation, in the $3\nu$ paradigm, is mainly governed by six oscillation parameters, viz., solar mixing angle ($\theta_{12}$), atmospheric mixing angle ($\theta_{23}$), reactor mixing angle ($\theta_{13}$), Dirac CP-violating phase ($\delta_{\mathrm{CP}}$), solar mass-squared splitting ($\Delta m^2_{21} \equiv m^2_{2}-m^2_{1}$), and atmospheric mass-squared splitting ($\Delta m^2_{31} \equiv m^2_{3}-m^2_{1}$). Still, there are a few fundamental issues that need to be resolved in neutrino oscillation experiments, namely, i) the neutrino mass ordering (whether $\Delta m^2_{31}>0$, known as normal mass ordering - NMO, or $\Delta m^2_{31}<0$, termed as inverted mass ordering - IMO), ii) the octant of $\theta_{23}$ (whether $\theta_{23}<45^\circ$ known as lower octant - LO, or $\theta_{23}>45^\circ$ termed as higher octant - HO), and iii) whether the CP is violated in the neutrino sector like quark sector and if so, then what is the precise value of the CP phase? The next-generation high-precision long-baseline (LBL) experiments are capable to address these issues at a high confidence level. \\\vspace{0.1cm}
In this thesis, we study i) the physics reach of the upcoming LBL experiment, Deep Underground Neutrino Experiment (DUNE) in the US, to establish the possible deviation from maximal mixing of $\theta_{23}$, ii) to achieve an improved precision on the 2-3 oscillation parameters using the synergy between the DUNE and Tokai-to-Hyper-Kamiokande (T2HK) experiments, and iii) the possible impact of the flavor-dependent long-range neutrino interactions on the measurements of neutrino oscillation parameters in DUNE and T2HK.\\
\section{Establishing the deviation from maximal mixing of $\theta_{23}$ in DUNE:}\hspace{0.2cm}The robustness of the three-flavor neutrino oscillation paradigm got established on a strong footing after the discovery of non-zero $\theta_{13}$ in the Daya Bay experiment and opened the door for searching CP violation in currently running and upcoming LBL experiments. The recent global fit studies ~\cite{Capozzi:2021fjo,deSalas:2020pgw,Esteban:2020cvm,NuFIT} of the three-flavor neutrino oscillation parameters indicate that $\sin^2\theta_{23}\neq0.5$ or $\sin^22\theta_{23}\neq1$. This gives rise to the issue of non-maximal $\theta_{23}$, which is quite important from the neutrino mass-mixing models point of view. In this work, we study the prospect of the next-generation long-baseline (LBL) experiment DUNE to establish the non-maximal $\theta_{23}$ in light of the current neutrino oscillation data.\\
DUNE \cite{DUNE:2020jqi} is an upcoming LBL experiment with a baseline of 1285 km where neutrinos travel from Fermilab to Homestake Mine in South Dakota. It will use an on-axis, wide-band neutrino beam with high intensity covering both first and second oscillation maxima. Some of its notable features are the excellent energy resolution of the Liquid Argon TPC (LArTPC) having 40 kt fiducial volume, large matter effect due to a longer baseline, and less systematic uncertainties in the appearance channel. The neutrino flux peaks around the first oscillation maxima which occurs at 2.5 GeV for 1285 km.\\ 
The study of non-maximal $\theta_{23}$ is mostly driven by the disappearance channel, and hence the uncertainty in $\delta_{\mathrm{CP}}$ does not affect our results much. As the disappearance channel is statistics-driven, it also helps to improve the precision on $\theta_{23}$. Our benchmark choices of the oscillation parameters are from Capozzi $et$ al. \cite{Capozzi:2021fjo}, where we have only considered the NMO as the true mass ordering in Nature. In our benchmark choices, $\sin^2\theta_{23} = 0.455$ (in LO), and $\delta_{\mathrm{CP}} = 223^\circ$ (belongs to lower-half-plane). We fix the value of $\theta_{12},\, \theta_{13}$, and $\Delta m^2_{21}$ as they are already measured with high precision using solar and reactor experiments. In our analysis, we have allowed the values of $\theta_{23}, \Delta m^2_{31}$, and $\delta_{\mathrm{CP}}$ to vary within their $3\sigma$ allowed range as mentioned in \cite{Capozzi:2021fjo}.\\
As the disappearance probability $\propto$ $\sin^22\theta_{23}$, degeneracies persist at the total event level. However, due to the superior energy resolution of DUNE, this degeneracy can be lifted using spectral information alongwith the total event rates. Some energy bins around the 1st oscillation maxima ($i.e.$, at 2.5 GeV) show the maximum sensitivity towards the non-maximal $\theta_{23}$ whereas, the performance of lower and higher energy bins deteriorates for the benchmark value of $\Delta m^2_{31}$. We show, with [3.5 yr $\nu$ + 3.5 yr $\bar{\nu}$] run, DUNE can establish non-maximal $\theta_{23}$ at $\sim$ $5\sigma$ C.L. with total rate + shape analyses whereas, it gets dropped to $1.5\sigma$ C.L. with an analysis based on only total event rates.\\
\par We see that with 7 years of run time equally shared by neutrino and antineutrino, DUNE can establish non-maximal $\theta_{23}$ with $\sim$ 4.2$\sigma$ C.L. at the benchmark choice of true $\sin^2\theta_{23}$. It is obvious that the more the true (benchmark) value of $\sin^2\theta_{23}$ shifts towards the maximal mixing value, the more it will be difficult to establish non-maximality of $\theta_{23}$. It is noteworthy to mention that if the true value of $\sin^2\theta_{23}$ shifts towards the upper range of the $1\sigma$ uncertainty of $\sin^2\theta_{23}$ ($i.e.$, maximum limit within $1\sigma$ tolerance in the vicinity of MM), DUNE can still establish non-maximality of $\sin^2\theta_{23}$ with $\sim 2.1\sigma$ C.L. However, this study is improved for the enhanced run time [$i.e.$, 5 yr $\nu$ + 5 yr $\bar{\nu}$] as it is mainly governed by the statistics. We observe that a $3\sigma\, (5\sigma)$ determination of non-maximal $\theta_{23}$ is possible in DUNE with an exposure of 336 kt$\cdot$MW$\cdot$years if true $\sin^2\theta_{23}\lesssim 0.465\, (0.450)$ or $\sin^2\theta_{23}\gtrsim0.554\, (0.572)$ for any value of true $\delta_{\mathrm{CP}}$ in its present $3\sigma$ range and true NMO.\\
\section{Improved precision on the 2-3 oscillation parameters using the synergy between DUNE and T2HK:} \hspace{0.2cm} A high-precision measurement of $\Delta m^2_{31}$ and $\theta_{23}$ is inevitable to estimate the Earth's matter effect in long-baseline experiments which in turn plays an important role in addressing the issue of neutrino mass ordering and to measure the value of CP phase in $3\nu$ framework. Following the footprints of present-generation measurements and near-future sensitivities from different oscillation experiments, our work examines how next-generation experiments, DUNE and T2HK, bring improvement in the 2-3 sector which will also help to decipher the long-standing flavor problem glorifying the oscillation-based neutrino research in the precision era. We highlight the relevance of the complementarity between DUNE and T2HK in determining the sensitivity towards deviation from maximal $\theta_{23}$, excluding the wrong octant of $\theta_{23}$, and establishing the precision on atmospheric parameters.\\
\par The next-generation long baseline experiment in Japan, T2HK, \cite{Hyper-KamiokandeProto-:2015xww, Hyper-Kamiokande:2018ofw} will use an off-axis, narrow-band neutrino beam that will travel through a distance of 295 km from J-PARC to Hyper-Kamiokande. This experiment has several features that are complementary to DUNE. Due to its short baseline, T2HK will experience less matter effect than DUNE and can have better prospects in measuring intrinsic CP violation due to $\delta_{\mathrm{CP}}$. T2HK's larger detector volume than DUNE can improve any disappearance-driven phenomena with the power of more enriched statistics. To compensate for the low antineutrino cross section, T2HK has been planned with 3 times higher run time in antineutrino mode than neutrino [2.5 yr $\nu$ + 7.5 yr $\bar{\nu}$]. T2HK's larger systematic uncertainty in appearance channel ($5\%$) than DUNE ($2\%$) is responsible for reduced sensitivity in any analysis driven by appearance channel, whereas, its improved disappearance systematics ($3.5\%$) than DUNE ($5\%$) along with larger detector volume helps to improve any disappearance-driven sensitivity.\\
We show that T2HK, with its benchmark exposure of 2431 kt$\cdot$MW$\cdot$yr, can establish non-maximal $\theta_{23}$ at $5.6\sigma$ C.L. which deteriorates a bit for DUNE to $5\sigma$ with its benchmark exposure of 480 kt$\cdot$MW$\cdot$yr at the benchmark choice of $\sin^2\theta_{23}$ ($i.e.$, 0.455) \cite{Capozzi:2021fjo} assuming NMO. T2HK's enriched statistics and less systematic uncertainty in the disappearance channel enhance its performance over DUNE. The performance is much improved ($\sim 7.6\sigma$) in the combined setup of DUNE + T2HK as compared to their standalone setups. The notable finding is that, in Nature, if true $\sin^2\theta_{23}$ attains the upper boundary of the current $1\sigma$ uncertainty (0.473) which is close to MM, only DUNE + T2HK can achieve $3\sigma$ sensitivity of non-maximal $\theta_{23}$ with the present best-fit values of oscillation parameters. Based on the current benchmark choices of oscillation parameters \cite{Capozzi:2021fjo}, the range of true values of $\sin^{2}\theta_{23}$ that can be differentiated from MM choices, by DUNE + T2HK with just half of their nominal exposures, cannot be achieved by either of the individual experiments even with their respective full projected exposures.\\
We study one of the most pressing unsolved issues in $3\nu$ paradigm, $i.e.$, the octant ambiguity of $\theta_{23}$. As it is driven by appearance channel, DUNE performs better ($5.2\sigma$) in excluding the wrong octant solution of $\theta_{23}$ with its nominal exposure than T2HK ($4.4\sigma$) at benchmark value of $\sin^2\theta_{23} = 0.455$ for true $\delta_{\mathrm{CP}} = 223^\circ$ and true NMO. However, the joint setup of DUNE + T2HK outperforms ($7.4\sigma$) in our study than their solo performances. Here also, our finding show that the joint performance of DUNE + T2HK can give $\sim 4\sigma$ sensitivity to our study at the upper edge of $1\sigma$ uncertainty in $\sin^2\theta_{23}$ ($i.e.$, in the vicinity of MM). At lower confidence, T2HK wins due to larger statistics, whereas, at higher confidence, DUNE wins due to lesser systematics in the appearance channel. One more noteworthy finding in this context is that, with just 0.25 times the benchmark exposure of the individual experiments, the combined setup can exclude $60\%$ of $\theta_{23}$ from the wrong octant solutions at $3\sigma$ C.L.\\
\par We address one more pertinent issue in the precision domain of neutrino oscillation, viz., precision measurements of $\sin^2\theta_{23}$ and $\Delta m^2_{31}$ around their benchmark choices ($i.e.$, 0.455 and $2.522\times10^{-3}$ eV$^2$) for NMO. Due to the statistical abundance, the standalone T2HK shows more precise measurements than DUNE and, of course, their combined setup enhances the performance in these measurements than their individual setups at their nominal exposures. The highlighted result from this section is that, the combined setup improves the present achievable precision of $\sin^2\theta_{23}$ and $\Delta m^2_{31}$, by a factor of 7 and 5, respectively. While standalone DUNE and T2HK with low exposures ($\sim 0.25$ times nominal exposure) cannot rule out clone solutions in $\sin^2\theta_{23}$ at $3\sigma$, the combined DUNE + T2HK provide degeneracy-free measurements. In the case of $\Delta m^2_{31}$, using approximately $20\%$ of the individual exposures of DUNE and T2HK together can achieve an impressive relative $1\sigma$ precision of $0.25\%$.\\
The combination of DUNE and T2HK can exclude the HO only in antineutrino
mode at $3\sigma$ C.L. breaking $(\sin^2\theta_{23} - \delta_{\mathrm{CP}})$ degeneracy due to higher $\bar{\nu}$ statistics in T2HK. So, the majority of the appearance events are free from fake (matter-induced) CP-phase. However, for the combined neutrino and antineutrino modes, we can give more precise measurements of $\sin^2\theta_{23}$ and $\Delta m^2_{31}$ by squeezing the parameter space irrespective of all experimental setups under discussion. DUNE's excellent energy-resolution helps to precisely measure $\Delta m^2_{31}$ whereas, T2HK's statistical enrichment assists to precisely measure $\sin^2\theta_{23}$. We also show the reduction in the parameter space $(\sin^2\theta_{23} - \delta_{\mathrm{CP}})$ using DUNE and T2HK, either in isolation or in combination. We find that, at half-exposure, the synergy between DUNE and T2HK is the only hope to exclude the MM solution and disfavor the CP-conserving region.\\
\section{Impact of long-range neutrino interactions on the measurements of oscillation parameters at DUNE and T2HK :} \hspace{0.2cm} Neutrino oscillation, having the bright prospect of measuring oscillation parameters with high precision in the next-generation experiments, may unravel novel signatures of BSM physics. The symmetries of the Standard Model are preserved under the gauge group \textit{SU}$(3)_C\,\otimes\,$\textit{SU}$(2)_L\,\otimes\,$\textit{U}$(1)_Y$. Now, a new, anomaly-free interaction can be added through a simple extension of the gauge group, viz., \textit{U}$(1)^\prime$, which brings some new symmetries in the Standard Model. These symmetries are chosen as the difference between the lepton numbers, $i.e.$, $(L_e-L_\mu)$, $(L_e-L_\tau)$, and $(L_\mu-L_\tau)$, manifesting the new interactions in Nature. Now, if the mass of the corresponding gauge boson $Z^\prime$ is very light ($\leq\,10^{-10}$ eV), the new interaction is long-ranged and characterized as a long-range interaction (LRI). The next-generation LBL experiments DUNE and Hyper-K may feel the presence of these new interactions \cite{Singh:2023nek}. In this work, we showcase the prominent effect of LRI along with the standard neutrino interaction (SI) in DUNE, T2HK, and their combined setups on measuring non-maximal $\theta_{23}$ and leptonic CP violation. The value of $(\Delta m^2_{31}/2E)$ for DUNE at 1st oscillation maxima is $\sim$ $10^{-13}$ eV which is compatible with the order of magnitude of LRI potential and hence DUNE and T2HK are good portals in probing the existence of LRI potential through oscillation. We assume NMO as a true mass ordering in Nature and the true LRI potential as $1.4\times10^{-14}$ eV, $1.0\times10^{-14}$ eV, and $0.73\times10^{-14}$ eV for $(L_e-L_\mu)$, $(L_e-L_\tau)$, and $(L_\mu-L_\tau)$ symmetries, respectively \cite{Singh:2023nek} throughout our analyses. In general, $L_e-L_\mu$ and $L_e-L_\tau$ are sensitive to $\delta_{\mathrm{CP}}$ and $\theta_{23}$ whereas, 
$L_\mu-L_\tau$ symmetry is  more sensitive to $\Delta m^2_{31}$ and $\theta_{23}$.\\
When the strength of long-range interactions increases ($V_{\rm{LRI}} > 10^{-14}$ eV), the distinguishable fraction of $\sin^{2}\theta_{23}$ values in Nature, deviating from maximal mixing, increases significantly. This enhancement is more pronounced under the $L_{\mu} - L_{\tau}$ symmetry, where the combined capabilities of DUNE and T2HK can achieve a coverage of approximately 90\% if $V_{\mu\tau}\,(\mathrm{true}\sim5\times10^{-14})$ eV. This improvement is attributed to the substantial disappearance event statistics in both DUNE and T2HK. For an illustrative true LRI value of $\sim 10^{-14}$ eV, we find that if the true octant of $\sin^{2}\theta_{23}$ in Nature corresponds to the lower octant (LO), the presence of LRI interactions under all the three symmetries reduces the sensitivity to deviations from maximal mixing compared to the SI scenario. Conversely, the sensitivity improves when $\sin^{2}\theta_{23}$ resides in the higher octant (HO) and is close to maximal mixing. We observe similar findings when studying the sensitivity towards the exclusion of the wrong octant. $L_{\mu} - L_{\tau}$ symmetry affects the most. We also investigate the impact of subdominant LRI on establishing leptonic CP violation (CPV). Our analysis reveals that if subdominant LRI exists in Nature, the fraction of $\delta_{\rm{CP}}$ values capable of establishing CPV at $\geq 3\sigma$ under each symmetry decreases significantly. This reduction arises from the critical role of $\sin^{2}2\theta_{13}$ evolution as a function of neutrino energy in the presence of matter. For an illustrative value of $V_{\rm{LRI}} \sim 10^{-14}$ eV, we find that the presence of LRI has a relatively greater impact on CPV sensitivity in DUNE compared to T2HK. This can be attributed to the higher prevalence of fake CP asymmetry in DUNE due to its longer baseline. Additionally, the $L_\mu - L_\tau$ symmetry introduces the most significant degradation in CPV sensitivity. This deterioration is primarily driven by the degeneracy between $V_{\rm{LRI}}$ and the uncertainty in $\sin^{2}\theta_{23}$. We also analyze the impact of LRI on the allowed parameter ranges in $(\sin^2\theta_{23}-\delta_{\mathrm{CP}})$ plane and observe varying effects across different symmetries. Under the $L_{e}-L_{\tau}$ symmetry, the allowed ranges remain largely consistent with those under SI, as sensitivity in this case is predominantly driven by the appearance statistics. For the $L_{e}-L_{\mu}$ symmetry, a noticeable deterioration in the precision of $\sin^{2}\theta_{23}$ is observed in DUNE when the data is generated with $\sin^{2}\theta_{23}$ in the LO. However, the complementarity between DUNE and T2HK resolves these degeneracies, resulting in stringent allowed regions under LRI that are comparable to those in the SI scenario. To conclude, our findings emphasize that while subdominant LRIs may influence the sensitivity and precision measurements of oscillation parameters, the complementary capabilities of DUNE and T2HK mitigate these challenges to a large extent. This synergy underscores the importance of a combined experimental approach in disentangling the impact of LRIs from neutrino oscillation under SI.

\appendix

\renewcommand\thechapter{A}

\renewcommand\thesection{\thechapter.\arabic{section}}
\renewcommand\theequation{\thechapter.\arabic{equation}}
\renewcommand\thefigure{\thechapter.\arabic{figure}}
\renewcommand\thetable{\thechapter.\arabic{table}}
\counterwithin{equation}{chapter}

\chapter*{Appendix}
\addcontentsline{toc}{chapter}{Appendix}

\markboth{Appendix}{Appendix}

\setcounter{section}{0}
\setcounter{chapter}{1} 


\chapter{Symmetry-driven deviations of atmospheric and reactor mixing angles in LRI perspective}
\label{app:running_osc_params}

\section{Impact of LRI on $\theta_{23}$ as a function of  neutrino energy}

\begin{figure}[htb!]
 \includegraphics[width=\linewidth]{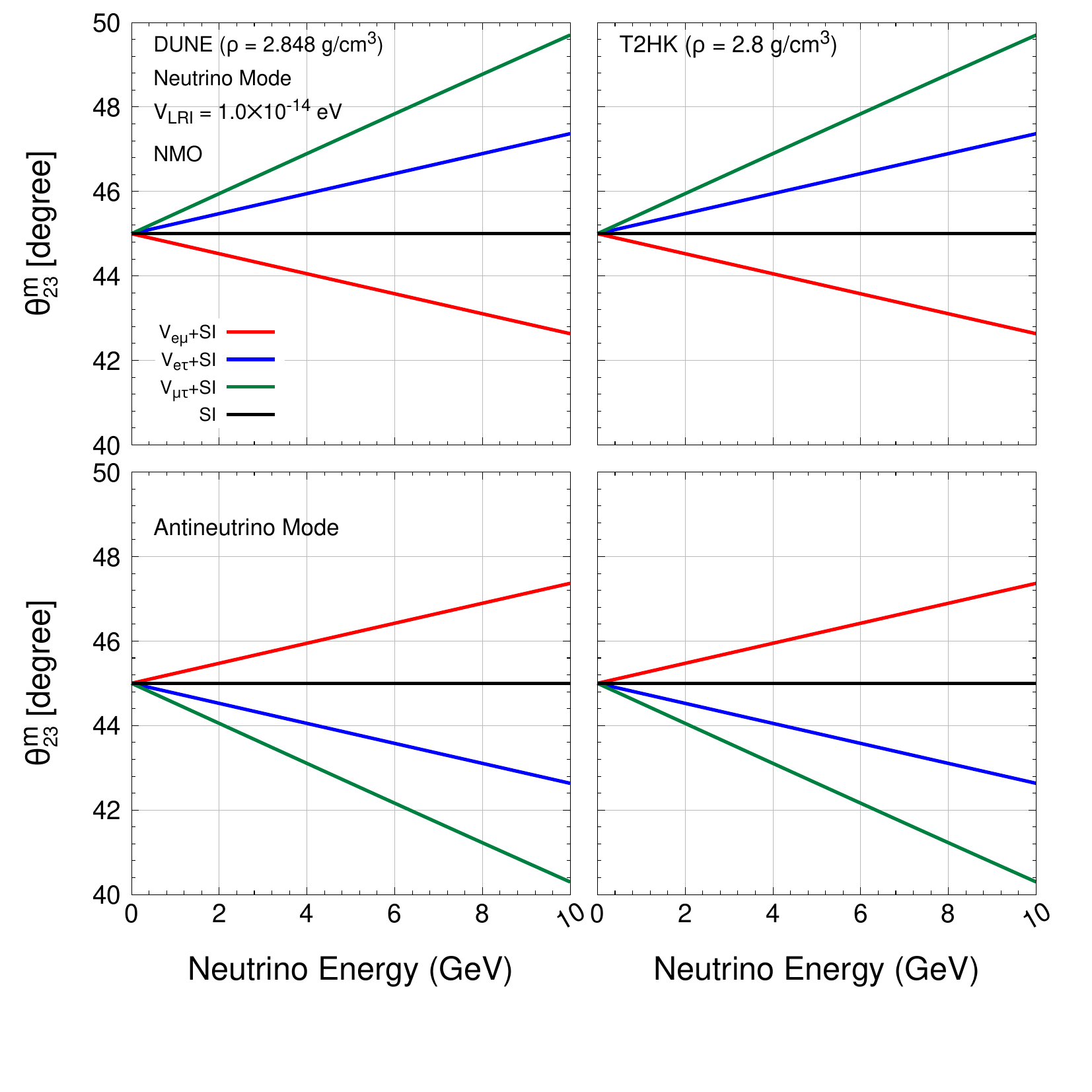}
 \caption{\footnotesize{\textbf{\textit{Impact on the matter-modified atmospheric mixing angle ($\theta_{23}$) arising from the lepton-number symmetries; $i.e.$,  $L_e - L_\mu$ (red line), $L_e - L_\tau$ (blue line), and $L_\mu - L_\tau$ (green line),  as a function of neutrino energy (in GeV).}} The results presented in this figure correspond to baselines, line-averaged matter potentials, and energy regimes reflective of the DUNE and T2HK experimental setups (see Section~\ref{subsec:exp-details}). An indicative value of the LRI potential is chosen as $V_{\mathrm{LRI}} = 1.0 \times 10^{-14}$ eV across all symmetry cases. In contrast, the impact on $\theta_{23}$ only due to standard Earth matter effects is also shown (black line). The upper (lower) panel depicts the case of neutrinos (antineutrinos), whereas the left (right) panel represents the scenario in the DUNE (T2HK) experiment. The reference values for the vacuum oscillation parameters are listed in Table \ref{tab:oscillation-params}, while the energy dependence of these parameters are elaborated here.}}
 \label{fig:running_th23}
 \end{figure}
Figure \ref{fig:running_th23} illustrates the influence of $L_e - L_\mu$, $L_e - L_\tau$, and $L_\mu - L_\tau$ symmetries, induced by long-range interactions (LRI) between neutrinos in the detectors (DUNE and T2HK) and electrons, protons, and neutrons distributed across various astrophysical and terrestrial sources (such as the Earth, the Sun, the Moon, the Milky Way, and beyond), on the effective atmospheric mixing angle $\theta_{23}$ as a function of neutrino energy assuming NMO in Nature. Here, for comparison purposes, we have given equal weightage to the magnitude of these three LRI potentials in Nature ($i.e.$, $1.0\times10^{-14}$ eV). We observe that $\theta^m_{23}$ (effective atmospheric mixing angle under standard Earth-matter potential; $i.e.$, the black line) is almost independent of the energy of neutrinos (or antineutrinos). The maximum deviation from the standard value of $\theta_{23}$ is observed in the presence of $L_\mu - L_\tau$ symmetry, in both DUNE and T2HK (even in both neutrino or antineutrino mode). The expression of the effective $\theta_{23}$ can be written as,
\begin{equation}
    \tan 2\theta^m_{23} = \frac{(c^2_{13}-\alpha c^2_{12})\times\Delta m^2_{31}\times10^{-18}}{2E(V_{33}-V_{22})}
    \label{equn:running_th23}
\end{equation}
In equation \ref{equn:running_th23}, $\alpha=\dfrac{\Delta m^2_{21}}{\Delta m^2_{31}}<<1$ and $c_{ij}\equiv \cos\theta_{ij}$. For $L_e\ - L_\tau$ symmetry, $V_{22}= 0$ and $V_{33}=V_{e\tau}$, so ($V_{33}-V_{22}= +V_{e\tau}$). This is same but with opposite polarity for $L_e - L_\mu$ symmetry. Hence, the value of effective $\theta^m_{23}$ gets equally deviated towards opposite to each other from standard interaction under the influences of $L_e - L_\mu$ and $L_e - L_\tau$ symmetries, respectively. But it gets deviated maximally from the standard case for $L_\mu - L_\tau$ symmetry as, $V_{22}= V_{\mu\tau}$ and $V_{33}=-V_{\mu\tau}$, so ($V_{33}-V_{22}=-2V_{\mu\tau}$). If we focus on the 1st oscillation maxima of DUNE (at around 2.5 GeV) and T2HK (at around 0.6 GeV), we see that the deviation of the value of $\theta_{23}$ from the standard Earth-matter case is more pronounced in DUNE than T2HK. For the antineutrinos, the entries in the LRI potential matrix (diagonal) will change their polarities from the neutrino case, hence the red and blue line will interchange their positions.

\section{Impact of LRI on $\theta_{13}$ as a function of neutrino energy}
\begin{figure}[t!]
 \includegraphics[width=\linewidth]{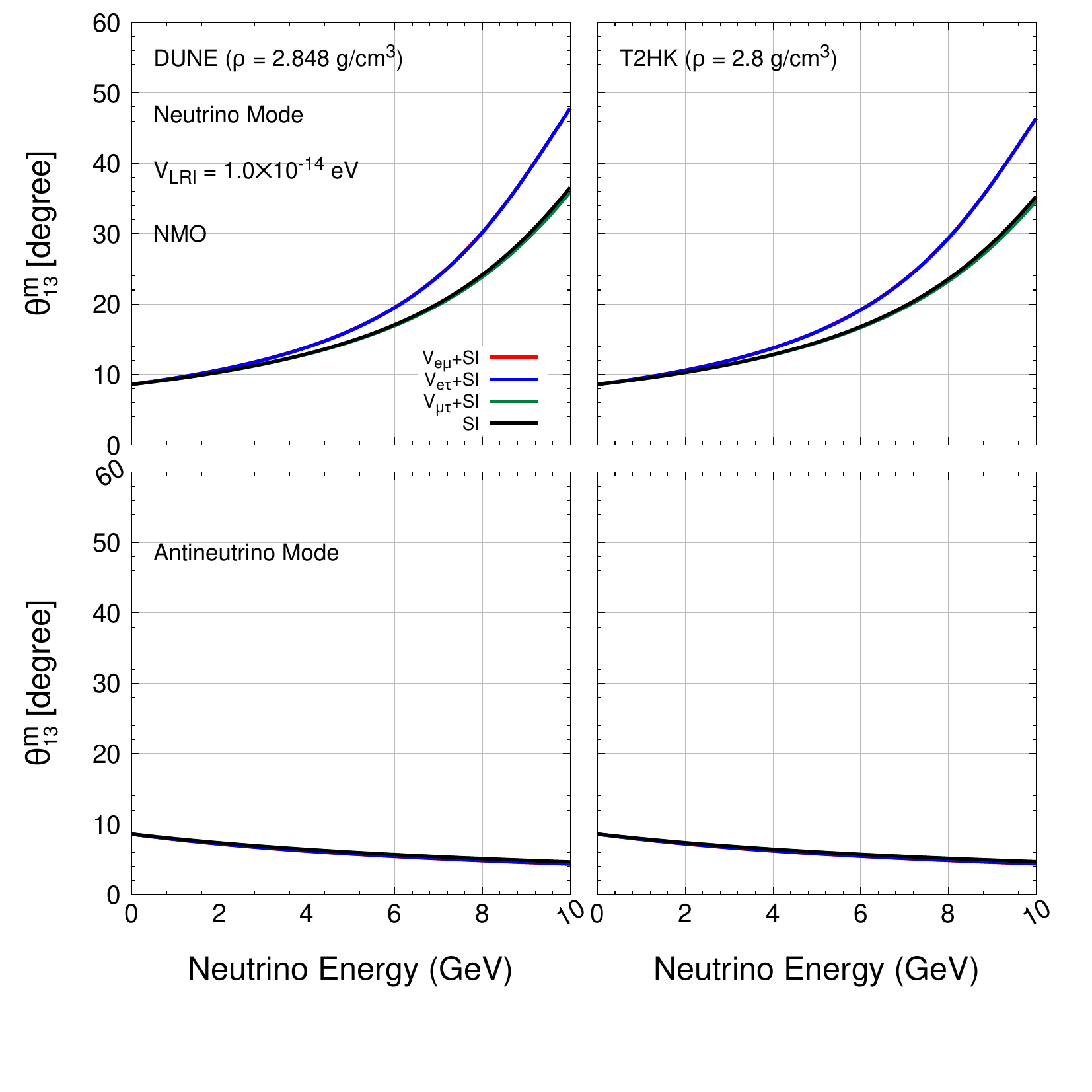}
 \caption{\footnotesize{\textbf{\textit{Impact on the matter-modified atmospheric mixing angle ($\theta_{13}$) arising from the lepton-number symmetries; $i.e.$,  $L_e - L_\mu$ (red line), $L_e - L_\tau$ (blue line), and $L_\mu - L_\tau$ (green line),  as a function of neutrino energy (in GeV).}} The results presented in this figure correspond to baselines, line-averaged matter potentials, and energy regimes reflective of the DUNE and T2HK experimental setups (see Section~\ref{subsec:exp-details}). An indicative value of the LRI potential is chosen as $V_{\mathrm{LRI}} = 1.0 \times 10^{-14}$ eV across all symmetry cases. In contrast, the impact on $\theta_{13}$ only due to standard Earth matter effects is also shown (black line). The upper (lower) panel depicts the case of neutrinos (antineutrinos), whereas the left (right) panel represents the scenario in the DUNE (T2HK) experiment. The reference values for the vacuum oscillation parameters are listed in Table \ref{tab:oscillation-params}, while the energy dependence of these parameters is elaborated here.}}
 \label{fig:running_th13}
\end{figure}
 Figure \ref{fig:running_th13} illustrates the influence of $L_e - L_\mu$, $L_e - L_\tau$, and $L_\mu - L_\tau$ symmetries, sourced by the long-range interaction between neutrinos detected at DUNE and T2HK and the ambient matter content — electrons, protons, and neutrons — distributed across various terrestrial and astrophysical sources, on the effective reactor mixing angle $\theta_{13}$ as a function of neutrino energy assuming NMO. Here, for comparison purposes, we have given equal weightage to the magnitude of these three LRI potentials in Nature ($i.e.$, $1.0\times10^{-14}$ eV). We observe that, $\theta^m_{13}$ (effective reactor mixing angle under standard Earth-matter potential; $i.e.$, the black line) is almost independent of the energy of neutrinos (or antineutrinos). The maximum departure from the standard value of $\theta_{13}$ is observed in the presence of $L_\mu - L_\tau$ symmetry, in both DUNE and T2HK (even in both neutrino or antineutrino mode). The expression of the effective $\theta_{13}$ can be written as,
\begin{align}
\tan 2\theta^m_{13} &= \frac{1}{
    \sqrt{2} \left( \lambda_3 - A_{\mathrm{CC}} - A_{11} - s^2_{13} - \alpha s^2_{12} c^2_{13} \right)
} \times \Bigg[ \notag \\
& \quad \left\{
    \sin(2\theta_{13}) \cos\delta_{\mathrm{CP}} (1 - \alpha s^2_{12})(c^m_{23} + s^m_{23})
    - \alpha c_{13} \sin(2\theta_{12})(c^m_{23} - s^m_{23})
\right\}^2 \notag \\
& \quad + \left\{
    \sin(2\theta_{13})(1 - \alpha s^2_{12}) \sin\delta_{\mathrm{CP}} (c^m_{23} + s^m_{23})
\right\}^2 \Bigg]^{1/2}.
\label{equn:running_th13}
\end{align}

In equation \ref{equn:running_th13}, $\lambda_3= 0.5\times\left[c^2_{13}+\alpha c^2_{12}+A_{22}+A_{33}+\sqrt{(c^2_{13}-\alpha c^2_{12})^2+(A_{33}-A_{22})^2}\right]$. Here, $A_{\mathrm{CC}}=2EV_{\mathrm{CC}}/\Delta m^2_{31}$, $A_{ij}= 2EV_{\mathrm{ij}}\times10^{18}/\Delta m^2_{31}$, $c_{ij}\equiv\cos\theta_{ij}$, and $s_{ij}\equiv\sin\theta_{ij}$. By the same prescription used in the previous subsection, we see that the influence of effective $\theta_{13}$ gets influenced more in $L_e - L_\tau$ symmetry than the other two cases. DUNE shows more deviation than T2HK from the effect only due to the standard Earth-matter effect due to its greater averaged Earth-matter density. It is noted that for neutrinos, the value of $\theta^m_{13}$ gets drastically enhanced for the higher values of neutrino energy as the equation \ref{equn:running_th13} indicates higher sensitivity of $\theta^m_{13}$ to the neutrino energy.

\addcontentsline{toc}{chapter}{REFERENCES}
\bibliographystyle{hcdas}
\bibliography{thesis}
\end{document}